\documentclass[prd,aps,notitlepage,twocolumn,floatfix,preprintnumbers,nofootinbib,superscriptaddress]{revtex4-2} 
\usepackage{amsfonts,amssymb,amsmath,mathrsfs,graphicx,xcolor,bm,ulem,mathtools,orcidlink}
\definecolor{ultramarine}{rgb}{0.07, 0.04, 0.56}
\definecolor{cadmiumgreen}{rgb}{0.0, 0.42, 0.24}
\definecolor{indigo(dye)}{rgb}{0.0, 0.25, 0.42}
\usepackage{hyperref}
\hypersetup{
colorlinks=true,
citecolor=ultramarine,
linkcolor=cadmiumgreen,
urlcolor=indigo(dye),
}

\newcommand{\f}[2]{\frac{#1}{#2}}  
\newcommand{\mk}[1]{\left( #1 \right)}

\newcommand{\be}{\begin{equation}}  
\newcommand{\ee}{\end{equation}}

\newcommand{\mO}{\mathcal{O}}
\renewcommand{\Re}{{\rm Re}\,}
\renewcommand{\Im}{{\rm Im}\,}

\newcommand{\fit}{{\rm fit}}
\newcommand{\iter}{\textrm{iter}}
\newcommand{\sxs}{\textrm{SXS}}
\newcommand{\fid}{{\rm fid}}
\newcommand{\peak}{{\rm peak}}
\newcommand{\ext}{{\rm ext}}

\newcommand{\M}{\mathcal{M}}

\newcommand{\la}{\langle}
\newcommand{\ra}{\rangle}

\begin{document}

\title{
Iterative extraction of overtones from black hole ringdown
}

\author{Kazuto Takahashi}
\email{takahashi.k.ds@m.titech.ac.jp}
\thanks{Corresponding author}
\affiliation{Department of Electrical and Electronic Engineering, Tokyo Institute of Technology, 2-12-1 Ookayama, Meguro-ku, Tokyo, 152-8552, Japan}

\author{Hayato Motohashi}
\email{motohashi@cc.kogakuin.ac.jp}
\affiliation{Division of Liberal Arts, Kogakuin University, 2665-1 Nakano-machi, Hachioji, Tokyo, 192-0015, Japan}

\begin{abstract}
Extraction of multiple quasinormal modes from ringdown gravitational waves emitted from a binary black hole coalescence is a touchstone to test whether a remnant black hole is described by the Kerr spacetime in general relativity.
However, it is not straightforward to check the consistency between the ringdown signal and the quasinormal mode frequencies predicted by the linear perturbation theory. 
While the longest-lived mode can be extracted in a stable manner, the higher overtones damp more quickly and hence the fitting of overtones tends to end up with the overfit.
To improve the extraction of overtones, we propose an iterative procedure consisting of fitting and subtraction of the longest-lived mode of the ringdown waveform in the time domain.
Through the analyses of the mock waveform and numerical relativity waveform, we clarify that the iterative procedure allows us to extract the overtones in a more stable manner.
\end{abstract}

\maketitle

\section{Introduction}
\label{sec:intro}

At the final stage of a binary black hole coalescence, where a remnant black hole is formed, emitted gravitational waves have a peak strain amplitude followed by damped oscillations.
Ringdown is such regime of the damped oscillations of gravitational waves emitted during the remnant black hole going down to a stationary state.
Black hole linear perturbation theory predicts that the ringdown gravitational waves are described by a superposition of exponentially damped sinusoids characterized by complex frequencies known as the quasinormal modes (QNMs)~\cite{Nollert:1999ji,Kokkotas:1999bd,Berti:2009kk,Konoplya:2011qq}.
The QNMs are discrete complex frequencies $\omega_{\ell m n}$, which are, in addition to the two angular indices $(\ell, m)$, labeled by overtone index $n$ according to the damping time $-1/\Im(\omega_{\ell mn})$.
The longest-lived mode $n=0$ is called the fundamental mode, and modes with $n\geq 0$ are $n$-th overtones and have shorter damping time.
Since the QNM frequencies of a Kerr black hole are fully determined by mass and spin~\cite{Detweiler:1977gy}, the measurement of the frequency and damping time of the fundamental mode would provide an estimation of the mass and spin of the remnant black hole~\cite{Echeverria:1989hg,Finn:1992wt}.
Further, an independent measurement of multiple modes would allow us to test whether the remnant black hole is consistent with the Kerr spacetime, known as the black hole spectroscopy~\cite{Dreyer:2003bv,Berti:2005ys}.

It is known that the QNM spectrum is unstable with respect to a small change of the effective potential~\cite{Nollert:1996rf,Barausse:2014tra,Daghigh:2020jyk,Jaramillo:2020tuu,Qian:2020cnz,Cheung:2021bol,Jaramillo:2021tmt}.
Such small perturbation would be caused by nonlinear effects in dynamical formation of a deformed black hole and/or surrounding material.
However, the ringdown waveform in the time domain is affected only at late time, which would be actually subject to noises in the observational signals, and hence it is well described by the unperturbed QNMs~\cite{Nollert:1996rf,Nollert:1998ys,Barausse:2014tra,Berti:2022xfj,Kyutoku:2022gbr}.
In this sense, the black hole spectroscopy program is robust and still works.  
The apparent discrepancy is due to the fact that the frequency domain analysis and time domain analysis are not necessarily equivalent when the full data is not available. 
Thus, given that we cannot observe the signal for a very long time and with sufficiently high precision, it is important to analyze both the frequency domain and the time domain in a complementary manner. 

Since the merger of binary black holes is highly nonlinear process, it is in principle nontrivial if the ringdown gravitational waves can be described by a superposition of QNM damped sinusoids obtained by the linear perturbation theory.
Nevertheless, from the firstly observed ringdown gravitational waves GW150914~\cite{Abbott:2016blz}, the fundamental mode of $(\ell,m)=(2,2)$ mode is robustly extracted~\cite{LIGOScientific:2016lio}.
However, the extraction of the overtones is not straightforward. 
The possibility of detection of the overtones from the GW150914 is still under debate~\cite{Isi:2019aib,LIGOScientific:2020tif,CalderonBustillo:2020rmh,Cotesta:2022pci,Isi:2022mhy,Finch:2022ynt,Ma:2023cwe,Ma:2023vvr,Crisostomi:2023tle,Wang:2023xsy}. 
The ringdown analyses of GW190521 has also been focused recently~\cite{LIGOScientific:2020iuh,LIGOScientific:2020ufj,LIGOScientific:2020tif,Capano:2021etf,Capano:2022zqm,Siegel:2023lxl}.
Since the observational ringdown data is inevitably subject to noises, it would be worthwhile to revisit a simpler and ideal setup such as mock waveform and/or numerical relativity waveform.

Actually, there has also been an active discussion on the possibility of the extraction of the overtones from numerical relativity waveform.
It was shown in \cite{Giesler:2019uxc} that the fit starting from the peak of the strain amplitude improves as one includes more overtones in the fitting function.
However, the significance of the overtones in numerical relativity waveform was questioned in \cite{Baibhav:2023clw,Nee:2023osy} and the possibility of the overfit was pointed out. 
It was clarified that the fit of the overtones is sensitive to the start time of the data interval used for the fit.
Further, it was shown that one can improve the mismatch even with overtones of frequencies irrelevant to the Kerr QNMs corresponding to the remnant spin.
The difficulty of extraction of overtones is also pointed out in \cite{Mourier:2020mwa} with an analysis of simulation of head-on collision of two nonspinning black holes.
An interesting approach was considered in \cite{Ma:2022wpv}, where the fundamental mode and lower overtones are filtered out to reveal subdominant effects in the ringdown such as nonlinear effects and spherical-spheroidal mode mixing~\cite{Berti:2014fga}. 
Such an approach would be useful to clarify the stability of the extraction of the overtones, but the frequency domain filter not only removes the mode of interest but also modifies the contribution from other modes.
It would be worthwhile to consider a time domain approach to remove the dominant mode to help the extraction of subdominant modes.

The aim of the present paper is to scrutinize the extraction of the overtones from the ringdown gravitational waves.
In contrast to the conventional fitting method, where one extracts the coefficients of all the modes of interest from a single fitting, we consider an iterative procedure of fitting and subtraction to extract the longest-lived mode at each step.
Starting from analyses of mock waveform, we shall see that we can fit the longest-lived mode in the most stable manner compared to the subdominant modes, and when we subtract a damped sinusoid of the longest-lived mode from the time domain waveform, this nature can be taken over to the next-longest-lived mode.
Hence, this procedure improves the stability of the fit of the subdominant modes.
We clarify that, by making use of iterative fitting and subtractions of the longest-lived mode, we can improve the stability of the overtones compared to the conventional fit.

The rest of the paper is organized as follows. 
In \S\ref{sec:fit}, we review the fitting algorithm used throughout the present paper. 
We investigate the iterative fitting of mock waveforms in \S\ref{sec:mock} and \S\ref{sec:dawnoise}. 
First, in \S\ref{sec:mock}, we deal with a mock waveform composed of a superposition of pure damped sinusoids.
Second, in \S\ref{sec:dawnoise}, we add a mock waveform a small constant mimicking noise and/or tail in the ringdown signal.
For these toy examples, we shall see how we can improve the fit of overtones by iteratively extracting the longest-lived mode.
In \S\ref{sec:nr}, we apply the iterative fitting method to numerical relativity waveform in the Simulating eXtreme Space-times (SXS) catalog~\cite{Mroue:2013xna,Boyle:2019kee} and show an improvement of the extraction of overtones.
\S\ref{sec:conc} is devoted to conclusion and discussion.

Throughout the present paper, we work in the natural unit where $c=G=1$, and employ the total binary mass $M$ before merger to normalize physical quantities such as time and frequency.
We mainly focus on $\ell=m=2$ mode, so we often omit the subscript $\ell$ and $m$ for $\omega_{\ell m n}$ etc.\ when no confusion occurs.
Numerical calculations are performed with machine precision but when we present specific numbers, values are rounded to four significant digits unless otherwise noted.

\section{Fitting algorithm}
\label{sec:fit}

In this section, we briefly review the fitting algorithm known as the eigenvalue method~\cite{Cook:2020otn}, which we exploit throughout the present paper.
While the optimization algorithms such as {\tt curve\_fit} in {\tt scipy} or {\tt NonLinearModelFit} in {\sc Mathematica} are widely used in the literature, it is known that for these algorithms the results depend on the initial guess for the fitting parameters.
Such ambiguities are not ideal when we discuss the possibility of overfit.
The eigenvalue method does not require the initial guess and hence provides a robust result.

Let $\Psi(t)$ be a given ringdown waveform of gravitational wave strain which we would like to fit by a fitting function.
Regarding $\Psi(t)$, we shall consider mock waveform in \S\ref{sec:mock} and \S\ref{sec:dawnoise}, and numerical relativity waveform in \S\ref{sec:nr}.
Regarding the fitting function, we use the following function:
\begin{equation}
    \label{eq:fit}
    \psi_{[K,N]}^\fit(t) = \sum_{n=K}^{N} C_n \psi_n(t),
\end{equation}
which is a superposition of the damped sinusoids $\psi_n(t) = e^{-i\omega_n t}$ 
with complex frequencies $\omega_n$. 
In the present paper, we perform the frequency-fixed fitting, i.e., we determine complex coefficient for each given QNM mode by the fitting algorithm.
We thus set $\omega_n$ the QNM frequencies corresponding to the given waveform $\Psi(t)$.
Specifically, we use the fiducial value frequencies for the analysis of the mock waveform, and the Kerr QNM frequencies associated with the remnant black hole for the analysis of the numerical relativity waveform. 
The fitting parameters are complex coefficients $C_n$.
For the following, we shall also use the notation $C_n=A_n e^{i\phi_n}$ with amplitude $A_n$ and phase $\phi_n$ being real parameters.

To quantify the goodness of fit, let us define the overlap $\rho$ between the ringdown waveform $\Psi$ and the fitting function $\psi^{\fit}$ by
\begin{equation}
\label{eq:rho2}
    \rho^2 = \frac{ | \la \psi^{\fit} | \Psi \ra |^2}
    {\la \Psi | \Psi \ra \la \psi^{\fit} | \psi^{\fit} \ra} ,
\end{equation}
where
\begin{equation}
    \la \psi | \phi \ra
    =\int_{t_i}^{t_e} \psi^* \phi ~dt .
\end{equation}
Here, $\psi^*$ is the complex conjugate of $\psi$, $t_i$ is the start time of the fit, and $t_e$ is the end time of the fit.
By introducing the notation $B_n = \la \psi_n | \Psi \ra$ and $D_{nm} = \la \psi_n | \psi_m \ra = (e^{i(\omega_n^*-\omega_m)t_e} - e^{i(\omega_n^*-\omega_m)t_i})/[i(\omega_n^*-\omega_m)]$, we can rewrite Eq.~\eqref{eq:rho2} as
\begin{equation}
    \rho^2 = 
    \frac{ | \sum_n C_n^* B_n |^2 }
    { \la \Psi | \Psi \ra 
    \sum_{n,m} C_n^* D_{nm} C_m } .
\end{equation}

We determine the coefficient $C_n$ by maximizing the overlap.
Requiring a derivative of $\rho^2$ with respect to $C_n$ to vanish, we obtain
\begin{equation}
    C_n = \sum_m (\mathbb{D}^{-1})_{nm} B_m .
\end{equation}
Here, $\mathbb{D}^{-1}$ is the inverse of a matrix $\mathbb{D}$, whose argument is $D_{nm}$. 
The maximum $\rho_{\max}^2$ of the overlap is then given by
\begin{equation} 
\label{eq:rho_max}
    \rho_{\max}^2 = \frac{\sum_{n,m} B^*_n (\mathbb{D}^{-1})_{nm} B_m}{ \la \Psi | \Psi \ra } .
\end{equation}
We can quantify the goodness of fit by using the mismatch defined by 
\be \M = 1-\rho_{\max}. \ee 

Since the ringdown waveform is given as a discrete data of $(t_a,\Psi(t_a))$, we need to evaluate the integrals $B_n$ and $\la \Psi | \Psi \ra$ in Eq.~\eqref{eq:rho_max} numerically.
In doing so, we exploit the Romberg integration $T_{2,2}$~\cite{Stoer2002}. 
Although the Simpson integration $T_{1,1}$ is commonly used for the numerical integration, we find that it does not provide a sufficient accuracy for our purpose.
Namely, the mismatch calculated by the Simpson's rule sometimes becomes negative. 
Therefore, following the Romberg's method with the Richardson extrapolation, we improve the accuracy of numerical integration by employing $T_{2,2}$, which we use for the numerical integration throughout the present paper.

\section{Pure damped sinusoids}
\label{sec:mock}

Before considering the extraction of QNMs from numerical relativity ringdown waveform, we study mock ringdown waveform.
For the following, we shall study the fitting of two kinds of mock waveforms.
In this section, we consider a mock waveform composed of a superposition of fiducial damped sinusoids.
In \S\ref{sec:dawnoise}, we shall add the mock waveform a constant to mimic the noise and/or tail of the ringdown signal. 

Let us consider a mock waveform of a superposition of fiducial damped sinusoids: 
\begin{equation}
\label{eq:mock}
\Psi_{[0,7]}(t) = \sum_{n=0}^{7} a_n e^{-i(\omega_n t-\theta_n)} .
\end{equation}
As a representative mock waveform, we consider a waveform mimicking the GW150914-like numerical relativity waveform SXS:BBH:0305 in the Simulating eXtreme Space-times (SXS) catalog~\cite{Mroue:2013xna,Boyle:2019kee}. 
We set $\omega_n$ the QNM frequencies for $(\ell,m)=(2,2)$ mode of Kerr black hole with the spin parameter $0.6921$, which corresponds to the remnant dimensionless spin of SXS:BBH:0305. 
The QNM frequencies are listed in Table~\ref{tab:0305_QNMs} in Appendix~\ref{sec:qnms}.
The $n=0$ mode is known as the fundamental mode, and $n\geq 1$ is the $n$-th overtone.
The damping time $-1/\Im(\omega_n)$ decreases as the overtone index $n$ increases.
We set the fiducial value for $a_n$ and $\theta_n$ as listed in Table~\ref{tab:mock_fiducial},  
which we determine by the best-fit value obtained by the conventional fit (see Table~\ref{tab:0305_bestfit_An_phin}).
Namely, we fit the gravitational strain $\Psi^\sxs = (h_+ - i h_\times)r/M$ of $(\ell,m)=(2,2)$ mode of SXS:BBH:0305 by the fitting algorithm described in \S\ref{sec:fit} with a fitting function $\psi_{[0,7]}^{\fit}$.
Here, $h_+$ and $h_\times$ are two orthogonal polarization modes of gravitational waves, and we normalize the strain by multiplying $r/M$, where $r$ is the distance between the binary and observer and $M$ is the total binary mass before merger. 
We generate a discrete data of $(t_a,\Psi(t_a))$ using the same sequence of sampling time $t_a$ of the simulation SXS:BBH:0305 with the resolution Lev6 and the extrapolation order $N_{\ext}=2$.
Typical time data spacing is about $\Delta t_a/M=\mO(10^{-1})$ for the ringdown phase. 

\begin{table}[h]
    \centering
    \caption{Fiducial values used for the mock waveforms $\Psi_{[0,7]}$ in \eqref{eq:mock} and $\Psi^{(c)}_{[0,7]}$ in \eqref{eq:wave_mock_const}.}
    \begin{tabular}{ccc} \hline\hline
       $n$ ~&~ $a_n$ ~&~ $\theta_n/\pi$ \\ \hline
        $0$ ~&~ $0.9683$ ~&~ $-0.4697$ \\
        $1$ ~&~ $4.180$ ~&~ $0.2134$ \\
        $2$ ~&~ $11.21$ ~&~ $-0.9134$ \\
        $3$ ~&~ $22.59$ ~&~ $0.08648$ \\
        $4$ ~&~ $32.36$ ~&~ $-0.8415$ \\
        $5$ ~&~ $28.93$ ~&~ $0.2520$ \\
        $6$ ~&~ $14.14$ ~&~ $-0.6589$ \\
        $7$ ~&~ $2.921$ ~&~ $0.4015$ \\
         \hline\hline
    \end{tabular}
    \label{tab:mock_fiducial}
\end{table}

\begin{figure}[t]
    \centering
    \includegraphics[width=\columnwidth]{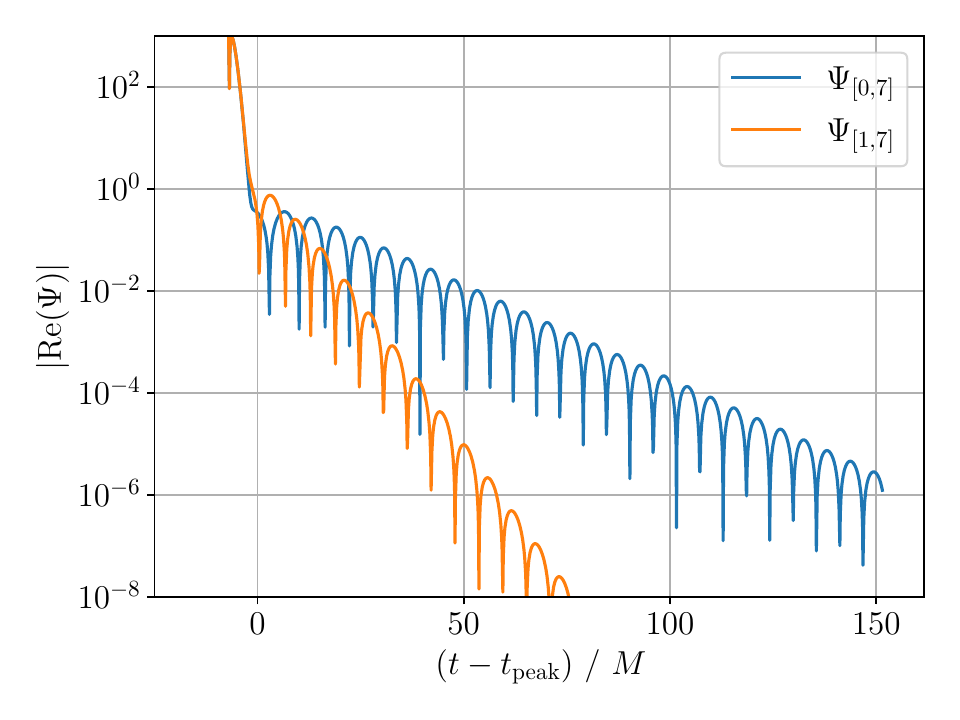}
    \caption{Mock waveform $\Psi_{[0,7]}$ (blue) given in Eq.~\eqref{eq:mock} and $\Psi_{[1,7]}$ (orange) after the subtraction of the fundamental mode.}
    \label{fig:0305_wave_mock_sub_const}
\end{figure}

The mock waveform $\Psi_{[0,7]}$ given in \eqref{eq:mock} is dominated by the fundamental mode at late time, whereas the contribution of the overtones mainly affect the early time waveform.
Figure~\ref{fig:0305_wave_mock_sub_const} depicts $|\Re(\Psi_{[0,7]}(t-t_\peak))|$ (blue) and $|\Re(\Psi_{[1,7]}(t-t_\peak))|$ (orange) as a function of time $t-t_{\peak}$, 
where $t_{\rm peak}$ is the peak time at which the strain amplitude $|\Psi^\sxs|=\sqrt{h_+^2+h_\times^2}$ of the numerical relativity waveform has a maximum.
To compare with the fit of the simulated ringdown waveform which we shall analyze in \S\ref{sec:nr}, we write the time $t-t_{\peak}$ measured from the peak time.
The late-time waveform of $|\Re(\Psi_{[0,7]})|$ is dominated by a single damped sinusoid, which is the fundamental mode $a_0 e^{-i[\omega_0 (t-t_\peak)-\theta_0]}$.
On the other hand, the late-time waveform of $|\Re(\Psi_{[1,7]})|$ is dominated by the first overtone $a_1 e^{-i[\omega_1 (t-t_\peak)-\theta_1]}$, whose damping time is shorter than the fundamental mode.
Since $\Psi_{[0,7]}$ is purely a superposition of damped sinusoids, for $t<t_{\peak}$ the waveform diverges and does not have physical meaning.

\subsection{Conventional fit}
\label{sec:con_fit}

First, we apply the conventional fitting method to the mock waveform.
Namely, we fit the mock waveform $\Psi_{[0,7]}(t-t_\peak)$ by using the fitting algorithm described in \S\ref{sec:fit} and the fitting function $\psi_{[0,N]}^\fit(t-t_\peak)$,
and obtain the best-fit values $C_n$ for all modes $n=0,\cdots,N$ from a single fitting.
In doing so, we need to specify the region of the data used for the fit.
Let us denote the time interval for the region for the fit by $[t_i,t_e]$.
It is important to clarify how much the fitting is sensitive to the choice of the start time of the fit $t_i$ and the end time of the fit $t_e$.
As we can see from Fig.~\ref{fig:0305_wave_mock_sub_const}, the damped sinusoids continue to the end of the plotted region.
We confirm that the result is insensitive to the choice of the end time $t_e$ of the fit so long as we set $t_e$ sufficiently late, say $(t_e-t_{\peak})/M\gtrsim 100$. 
As a representative value, we set $(t_e-t_{\peak})/M=150$.

\begin{figure}[t]
    \centering
    \includegraphics[width=\columnwidth]{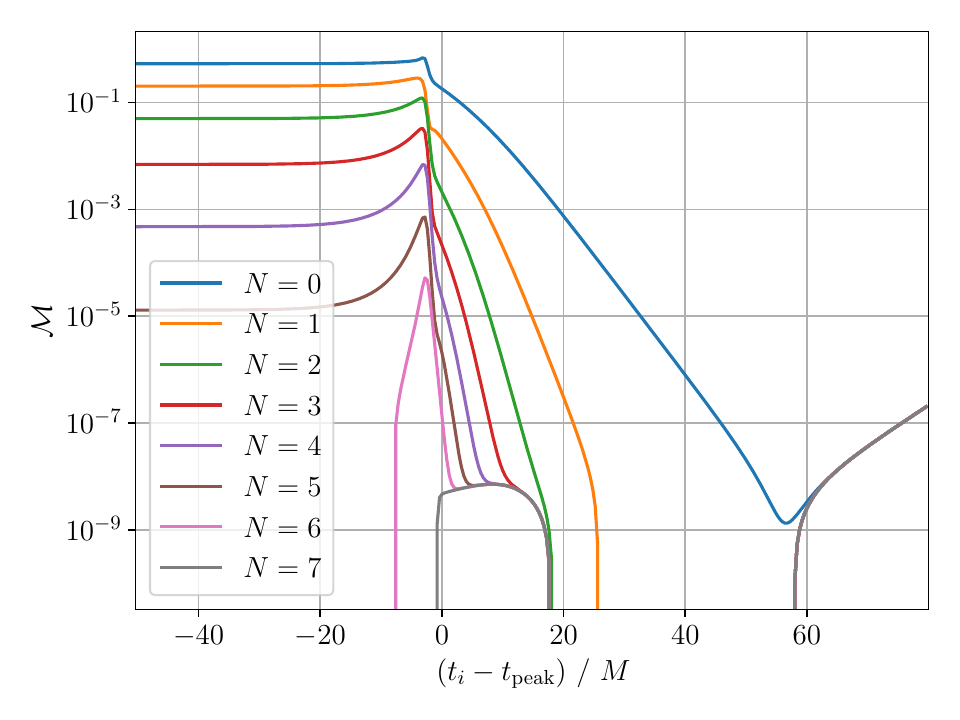}
    \caption{Mismatch $\M$ between the mock waveform $\Psi_{[0,7]}$ and the fitting function $\psi^\fit_{[0,N]}$.}
    \label{fig:0305_M_mock_sub_const}
\end{figure}

On the other hand, as we shall see below, the goodness of the fit depends on the choice of the start time $t_i$ of the fit.
Figure~\ref{fig:0305_M_mock_sub_const} shows the mismatch $\M$ between the mock waveform~\eqref{eq:mock} and the fitting function~\eqref{eq:fit} as a function of the start time of the fit $t_i$.
For the fitting function~\eqref{eq:fit}, we consider the sum from the fundamental mode $K=0$ to the $N$-th overtone.
For $t_i-t_{\peak}>0$, the slope of $\M$ are different for each $N$.
The larger $N$, the earlier start time of the fit $t_i$ at which the mismatch $\M$ reduces.
This trend is consistent with \cite{Giesler:2019uxc}.
This is because the overtones contribute to the waveform at earlier times.
In particular, the lower overtones improve the fit significantly even when they are subdominant.
For the case with $N=7$, since the fitting function is given by a superposition of the the same damped sinusoids used in the mock waveform, we expect the mismatch becomes tiny.
Indeed, in Fig.~\ref{fig:0305_M_mock_sub_const}, the mismatch for the case with $N=7$ remains $\mO(10^{-8})$ with fluctuations, and this behavior is expected to be due to errors in the numerical integration.

\begin{figure*}[t]
    \centering
    \includegraphics[width=\textwidth]{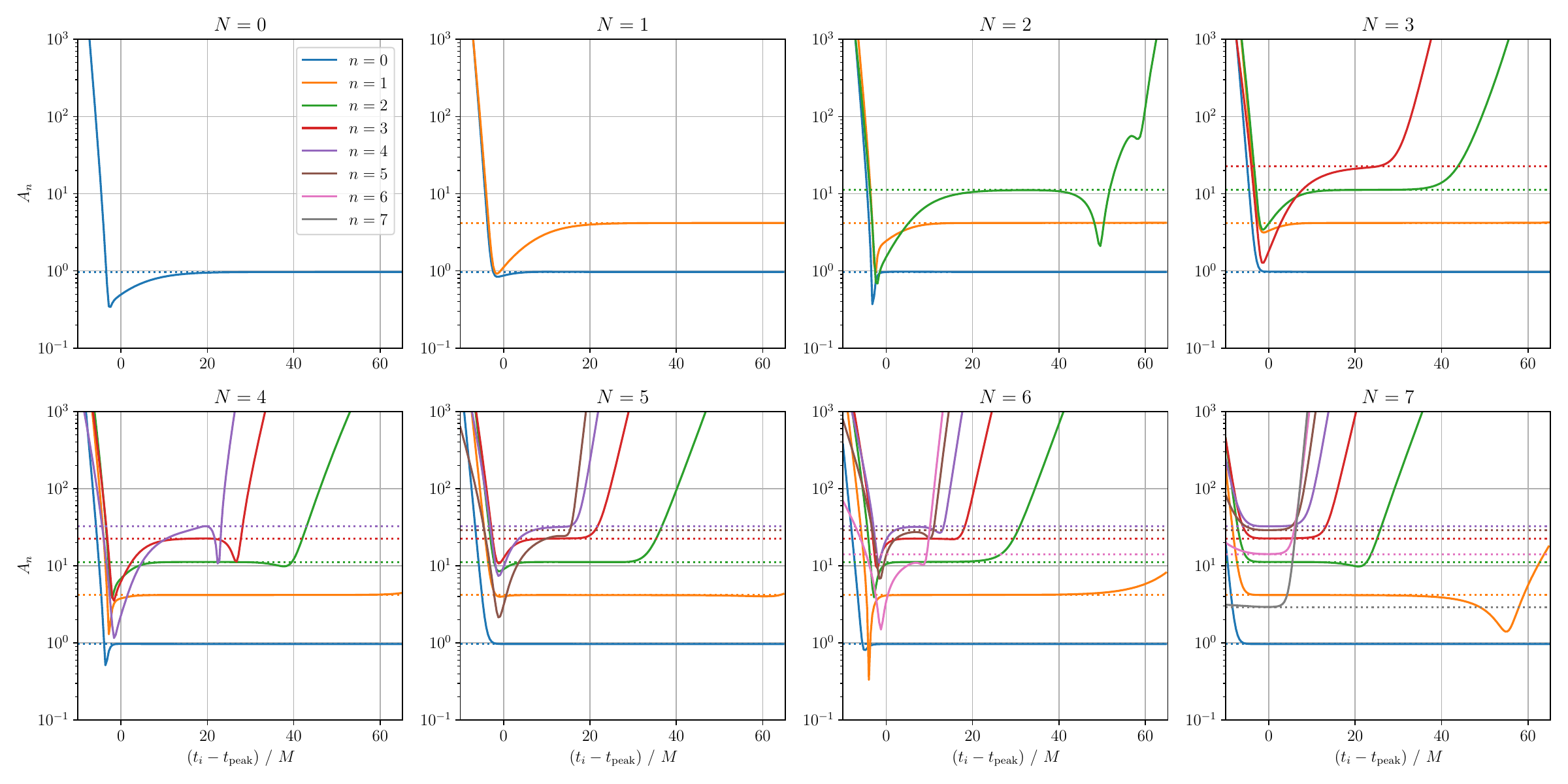}
    \caption{Amplitude $A_n$ for the fit of the mock waveform $\Psi_{[0,7]}$ by the fitting function $\psi^\fit_{[0,N]}$.}
    \label{fig:0305_A_mock_sub_const}
\end{figure*}

\begin{figure*}[t]
    \centering
    \includegraphics[width=\textwidth]{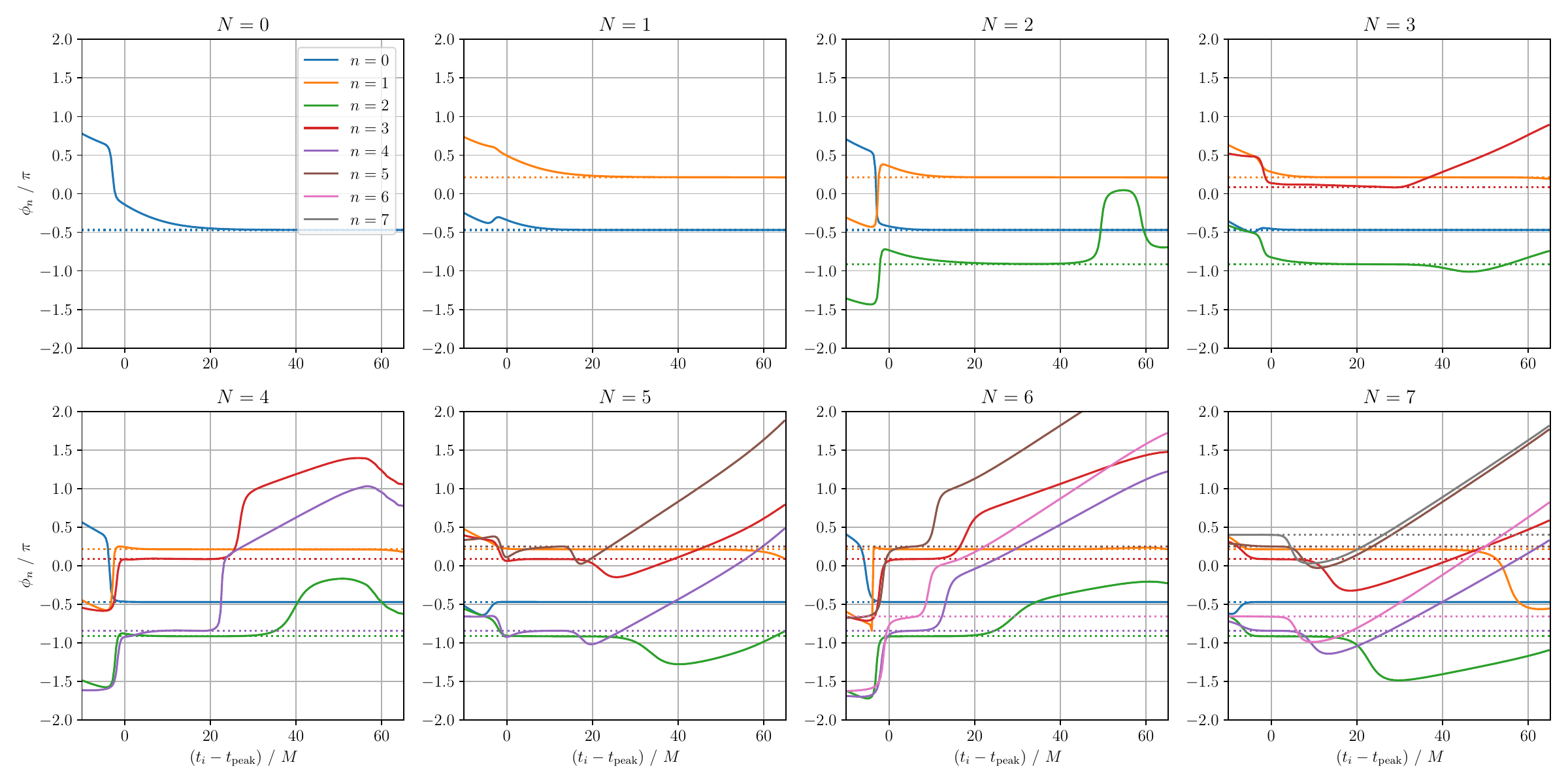}
    \caption{Phase $\phi_n$ for the fit of the mock waveform $\Psi_{[0,7]}$ by the fitting function $\psi^\fit_{[0,N]}$.}
    \label{fig:0305_phi_mock_sub_const}
\end{figure*}

Figures~\ref{fig:0305_A_mock_sub_const} and \ref{fig:0305_phi_mock_sub_const} respectively show the amplitudes $A_n$ and phases $\phi_n$ for the coefficients in the fitting function~\eqref{eq:fit} for each $N$, where the dotted lines indicate the fiducial values.
While the mismatch $\M$ monotonically improves as we increase the number of overtones included in the fitting function, 
as demonstrated in \cite{Mourier:2020mwa,Baibhav:2023clw}, the $t_i$-dependence of the fit sheds light on different aspects of the fitting.
It helps us to understand overfit, stability, and correctness of the fit of overtones.

We can see from Fig.~\ref{fig:0305_A_mock_sub_const} that 
for the early time, the overtones in the fitting function~\eqref{eq:fit} can fit the overtones in the mock waveform~\eqref{eq:mock} correctly, matching the amplitude $A_n$ to the fiducial value.
However, for the late time, the overtones becomes much smaller than the envelope of the waveform, and the fitting becomes more difficult.  
As we move the start time of the fit $t_i$ later, the amplitudes for the higher overtones obtained from the fit diverge, and this behavior can be interpreted as the overfit.
Since the divergence of $A_n$ compensates the damping of the overtone $\psi_n$, the contribution of the overtone in the fitting function remains roughly constant even for the fit starting from the late time.
Such behavior is not consistent with the expected decay of the contribution of the overtones in the mock waveform.
Therefore, if we set $t_i$ late time, the higher overtones in the fitting function cannot extract the overtones in the mock waveform correctly, but overfit the mock waveform.

We can see the same trend in the phase $\phi_n$ in Fig.~\ref{fig:0305_phi_mock_sub_const}.
The region of the plateau consistent with the fiducial values in Fig.~\ref{fig:0305_phi_mock_sub_const} matches to the one in Fig.~\ref{fig:0305_A_mock_sub_const}. 
There are some places where the value of $\phi_n$ suddenly jumps by $\pi$, such as $n=1$ in $N=6$. 
This is due to the fact that the coefficients $C_n=A_ne^{i\phi_n}$ actually pass near the origin in the complex plane.
Indeed, we can observe the corresponding sharp dips in Fig.~\ref{fig:0305_A_mock_sub_const}.

Further, Figs.~\ref{fig:0305_A_mock_sub_const} and \ref{fig:0305_phi_mock_sub_const} allow us to check the stability and correctness of the fit.
If there is a flat region, or plateau, where the variations of $A_n$ and $\phi_n$ are small with respect to $t_i$, we can say that the fit is stable.
In addition, if the values of $A_n$ and $\phi_n$ at the plateau are consistent with the fiducial values $a_n$ and $\theta_n$, we can say that the fit extracts the correct amplitude of the mode.
We can see that as we increase the number of overtones included in the fitting function, the fitting of the fundamental mode and lower overtones becomes more stable.
Specifically, the plateau of $n=0$ and $n=1$ is extended as we increase $N$, and $(A_0,\phi_0)$ and $(A_1,\phi_1)$ are consistent with the fiducial values.
However, if one includes too many overtones in the fitting function, the fitting of the lower overtones are actually destabilized, as we shall discuss below.

The $N=7$ case corresponds to the fit of the mock waveform~\eqref{eq:mock} by the fitting function~\eqref{eq:fit} with the same number of the overtones.
Naively, one may expect that the fitting works well for this case since the mismatch remains indeed small as we saw in Fig.~\ref{fig:0305_M_mock_sub_const}.
However, in Figs.~\ref{fig:0305_A_mock_sub_const} and \ref{fig:0305_phi_mock_sub_const}, while the fitting reproduces the fiducial value of $(A_n,\phi_n)$ around $t_i=t_{\peak}$, the plateau is not so long for higher overtones. 
This is consistent with the results obtained in \cite{Baibhav:2023clw}.
As we stressed in \S\ref{sec:fit}, in our case the fitting algorithm does not require the initial guess, and there are no ambiguities. 
Nevertheless, higher overtones cannot be extracted at late time. 
This may be due to the fact that the higher modes damp faster and become negligible at the later stage.
Therefore, even if we fit the mock waveform of $n=0$ to $7$ modes by the fitting function composed of the same modes, the fit does not reproduce the correct value for all region of $t_i$, especially for higher overtones.

It is also worthwhile to note that the fit of lower overtones are actually destabilized as we increase the number of overtones included in the fitting function.
For instance, in Fig.~\ref{fig:0305_A_mock_sub_const}, the time when $n=2$ curve starts to diverge is $(t_i-t_\peak)/M \simeq 40$ for $N=3,4$, but it becomes earlier for $N\geq 5$.
The fit of the first overtone $n=1$ is also destabilized in $N=6$ and $7$.
Although a seemingly counterintuitive result, this observation indicates that increasing the number of modes used in the fitting function does not always result in a better fitting.

On the other hand, at least in the linear perturbation theory, the ringdown waveform includes an infinite number of QNM damped sinusoids. 
Thus, we cannot include the same number of modes in the fitting function of a superposition of a finite number of modes.
So long as we deal with the fitting function with the finite number modes, we always fit the ringdown waveform with fewer number modes.
To consider a similar situation in the analysis of the mock waveform~\eqref{eq:mock} up to the seventh overtone, it would be reasonable to focus on the case where the fitting function~\eqref{eq:fit} includes the fewer number modes.
Specifically, we focus on the results with $N\leq 5$.

\begin{figure*}[t]
    \centering
    \includegraphics[width=\textwidth]{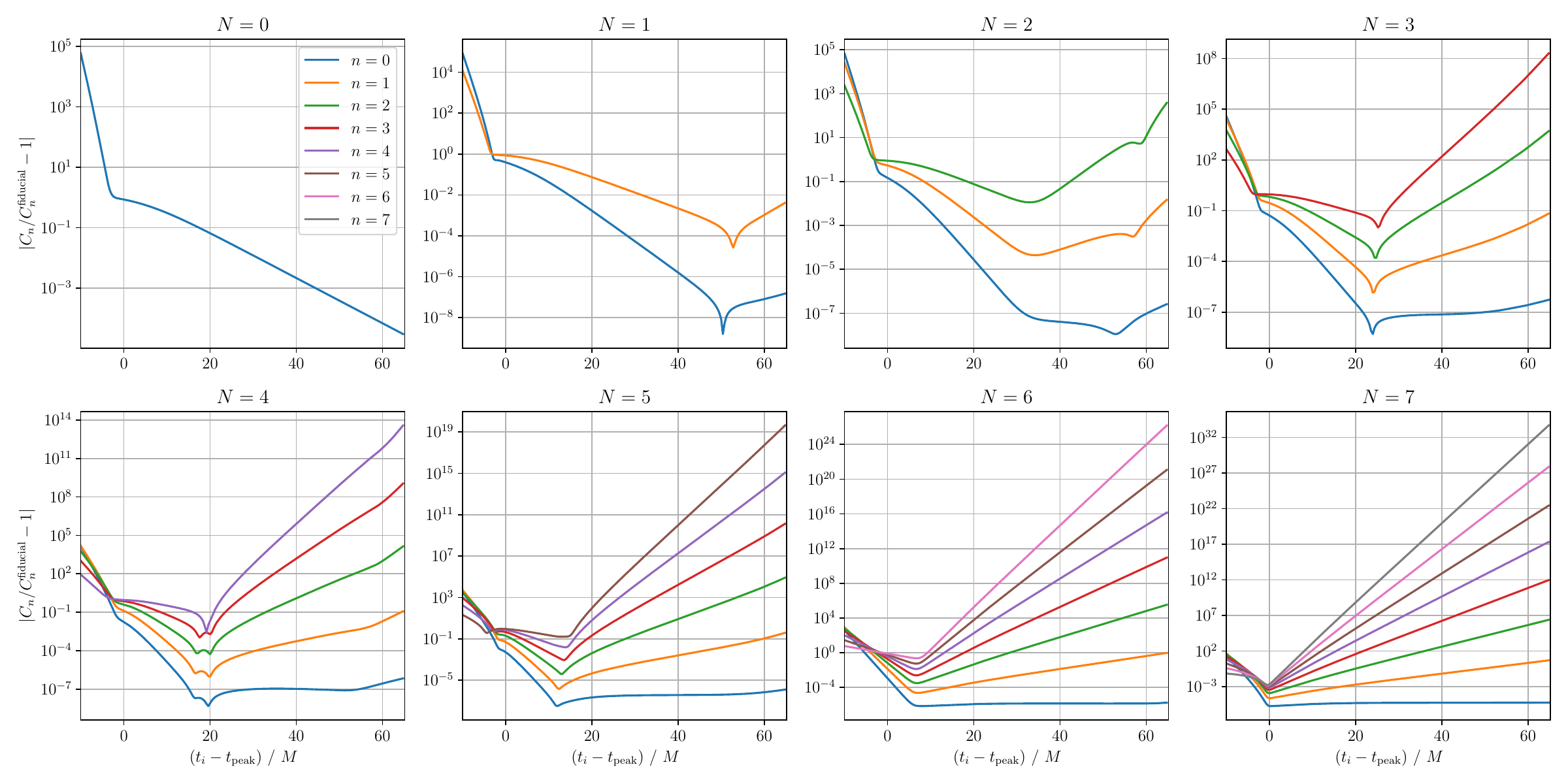}
    \caption{Relative error of $C_n$ for the fit of the mock waveform $\Psi_{[0,7]}$ by the fitting function $\psi^\fit_{[0,N]}$.}
    \label{fig:0305_Cre_mock_sub_const}
\end{figure*}

\begin{figure*}[t]
    \centering
    \includegraphics[width=\textwidth]{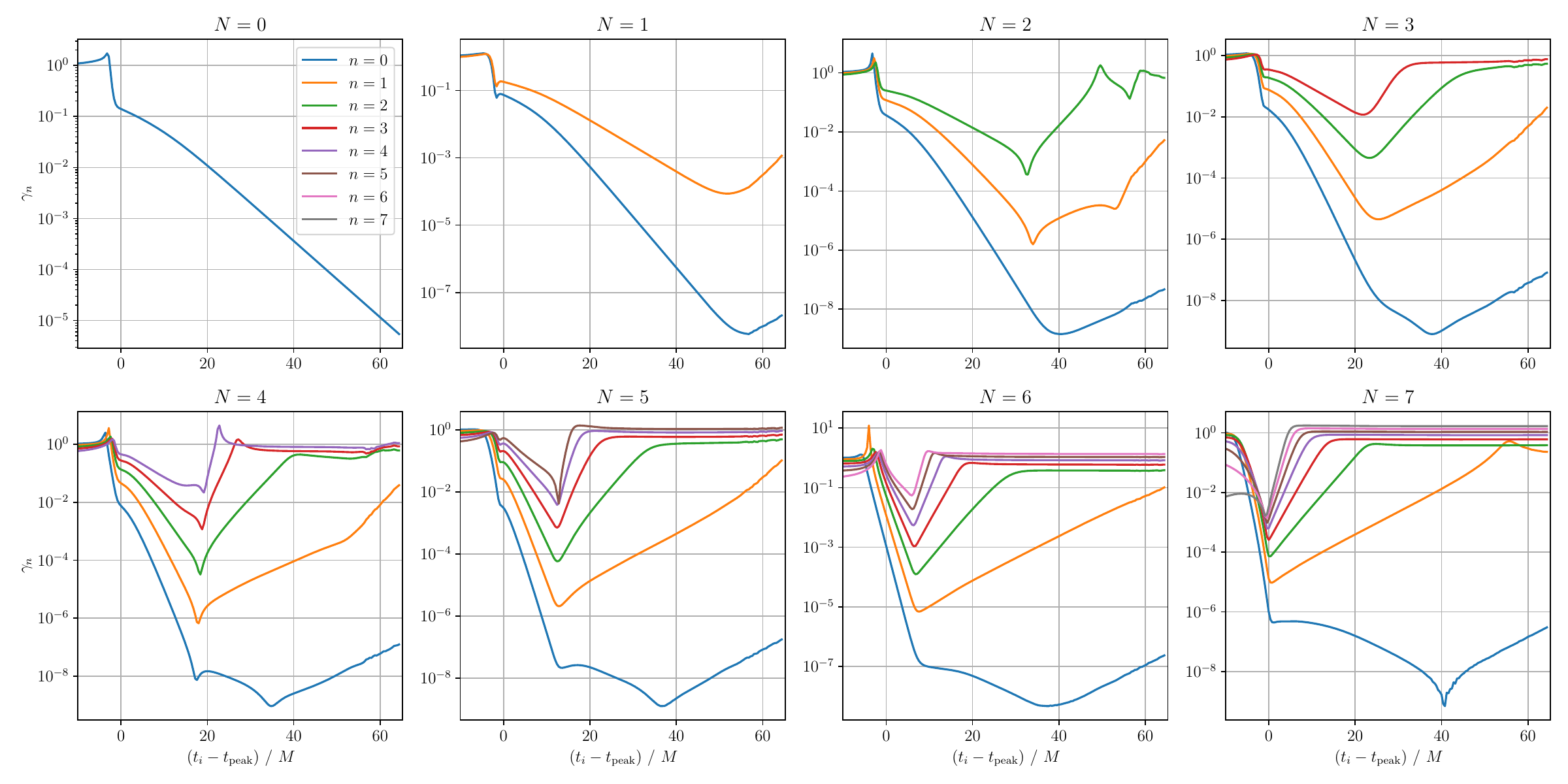}
    \caption{Rate of change $\gamma_n$ for the fit of the mock waveform $\Psi_{[0,7]}$ by the fitting function $\psi^\fit_{[0,N]}$.}
    \label{fig:0305_gamma_mock_sub_const}
\end{figure*}

To extract coefficients consistent with the fiducial values, we need to choose a suitable start time of the fit.
It is common to adopt the start time of the fit as the time when the decrease of the mismatch saturates. 
However, as we saw above, the mismatch itself is not sufficient to check if the overtones are overfitted. 
We would like to develop an alternative criterion to choose a start time of the fit taking into account the stability of the fit.

To visualize the accuracy of the fit, let us consider the relative error between $C_n$ obtained by the fit and the fiducial value $C^{\fid}_{n}=a_ne^{i\theta_n}$, i.e.,
\begin{equation}
    \label{eq:Cre}
    \left | \frac{C_{n}}{C^{\fid}_{n}} - 1 \right |~.
\end{equation}
In Fig.~\ref{fig:0305_Cre_mock_sub_const}, we present the relative error as a function of the start time of the fit $t_i$.
As $N$ increases, the relative error has a minimum at earlier time.
For the mock waveform analysis, we can identify the most favorable start time of the fit as a start time of the fit $t_i$ with which the relative error from the fiducial value is minimum.
However, we are ultimately interested in the fitting of the ringdown signal where the fiducial value is unknown.
We thus would like to consider an indicator that helps us to extract such start time of the fit that gives the best-fit value without using the knowledge of the fiducial value.

As such an indicator, it is useful to quantify the flatness of the plateau of the coefficient $C_n=A_n e^{i\phi_n}$.
Let us introduce the rate $\gamma_n$ of change of the coefficient $C_n$ with respect to the start time of the fit $t_i$ as 
\begin{equation}
\label{eq:gamma}
    \gamma_n = \left| \frac{1}{C_n}\frac{dC_n}{dt_i} \right|~.
\end{equation}
A time $t_i$ when $\gamma_n$ takes a minimum value indicates that the rate of change of $C_n$ is smallest.
We expect that we can use such a $t_i$ to extract the best-fit values close to the fiducial values.
In practice, since the data is discrete, the derivative in \eqref{eq:gamma} is understood as a finite difference $\Delta C_n/\Delta t_i$.

Figure~\ref{fig:0305_gamma_mock_sub_const} shows $\gamma_n$ as a function of the start time of the fit $t_i$.
If $\gamma_n$ persists small for some range of $t_i$, we can say that the $A_n$ and $\phi_n$ have a plateau there.
As a concrete example, let us define the plateau region as a region that satisfies the criterion $\gamma_n < 10^{-2}$.
For instance, for the case of $N=6$, based on the criterion, we can say from Fig.~\ref{fig:0305_gamma_mock_sub_const} that there exists plateau up to fourth overtone, whereas there is no plateau in the fifth and sixth overtone.
Indeed, we confirm that there is a correspondence between the plateau in Figs.~\ref{fig:0305_A_mock_sub_const} and \ref{fig:0305_phi_mock_sub_const} and the region where $\gamma_n$ is small in Fig.~\ref{fig:0305_gamma_mock_sub_const}.
As we increase $N$, we can see that $\gamma_n$ becomes smaller at earlier start time of the fit $t_i$, which is consistent with the behavior of the plateau in $A_n$.

From Figs.~\ref{fig:0305_A_mock_sub_const}--\ref{fig:0305_gamma_mock_sub_const}, 
we can infer that it is reasonable to adopt the best-fit value of $C_n$ by the fitting starting from $t_i$ where $\gamma_1$ is minimum. 
We see that there is a correspondence between $t_i$ where $\gamma_1$ is minimum in Fig.~\ref{fig:0305_gamma_mock_sub_const} and $t_i$ when the relative error of $C_n\, (n=0,\cdots N)$ have the minimum in Fig.~\ref{fig:0305_Cre_mock_sub_const}. 
We also confirm that there exist the plateau consistent with the fiducial value in Figs.~\ref{fig:0305_A_mock_sub_const} and \ref{fig:0305_phi_mock_sub_const} around $t_i$ where $\gamma_1$ is minimum in Fig.~\ref{fig:0305_gamma_mock_sub_const}.
The $t_i$ when $\gamma_{n\geq 2}$ takes the minimum is almost the same as the $t_i$ when $\gamma_1$ takes the minimum, and there are little changes in $C_n$ around those times.

Note that the start time of the fit $t_i$ when $\gamma_n$ is minimum is more or less the same, except the longest-lived mode, which is $n=0$ in the present case.
From Fig.~\ref{fig:0305_gamma_mock_sub_const}, we see that $\gamma_0$ takes the minimum at later start time of the fit $t_i$ compared to $\gamma_{n\geq 1}$.
In such late times, we expect that only the fundamental mode contributes to the fit. 
Indeed, for instance, as we can see in the panel $N=3$ of Fig.~\ref{fig:0305_gamma_mock_sub_const}, $\gamma_0$ is minimum at $(t_i-t_\peak)/M\simeq 40$, but there is no corresponding plateau in $A_n$ in Fig.~\ref{fig:0305_A_mock_sub_const} and the relative error of $C_n$ is not minimum in Fig.~\ref{fig:0305_Cre_mock_sub_const}.
Thus, we adopt $\gamma_1$ as a more appropriate indicator than $\gamma_0$ to determine the best-fit value of $C_n$, including $C_0$.

Comparing the panels for $N=0,\cdots, 7$ in Fig.~\ref{fig:0305_gamma_mock_sub_const}, we note that the minimum of $\gamma_n$ for each $n$ decreases as we increase $N$, but more or less saturates when we superpose three or four modes. 
Therefore, we expect that, in order to extract $C_0$ close to the fiducial values, it would be reasonable to use the fitting function superposing four modes, and to read off the best-fit values by the fitting starting from $t_i$ where $\gamma_1$ is minimum. 

Also, note that $\gamma_n$ for $n\geq 1$ asymptotically approaches to almost constant for the late start time of the fit.
This originates from the divergence of $A_n$ due to the overfit, which simply compensates the exponential growth of $\psi_n=e^{-i\omega_n t}$.
The compensation implies $C_n\sim e^{i\omega_n t_i}$, for which $\gamma_n = |d\ln C_n/dt_i|$ remains an almost constant.
Indeed, the asymptotic value of $\gamma_n$ for $n\geq 1$ roughly coincides with the damping rate $|\Im(\omega_n)|$.
Therefore, we can use $\gamma_n$ as an indicator for both of the stable fit and overfit.

In summary, we conclude that we can improve the conventional fit and extract the best-fit value of $C_n$ close to the fiducial values by the fitting starting from $t_i$ when $\gamma_n$ of the next longest-lived mode takes the minimum value.
In particular, the fit of the longest-lived mode is most stable and gives the best-fit value close to the fiducial value.
To extract the longest-lived mode and lower overtones in a stable manner from the mock waveform $\Psi_{[0,7]}$, we should superpose four damped sinusoids as the fitting function.
Superposition of five or more modes do not improve the relative error significantly.
Rather, increasing the number of modes used in the fitting function sometimes destabilizes the fitting of the lower overtones.
We expect that this setup captures the qualitative behavior of a more realistic situation, where one tries to fit the actual ringdown signal composed of a superposition of an infinite number of QNMs, whose fiducial values are unknown, by a superposition of the finite number of damped oscillations.

\subsection{Iterative fit and subtractions}
\label{sec:dosub}

\begin{figure}[t]
    \centering
    \includegraphics[width=\columnwidth]{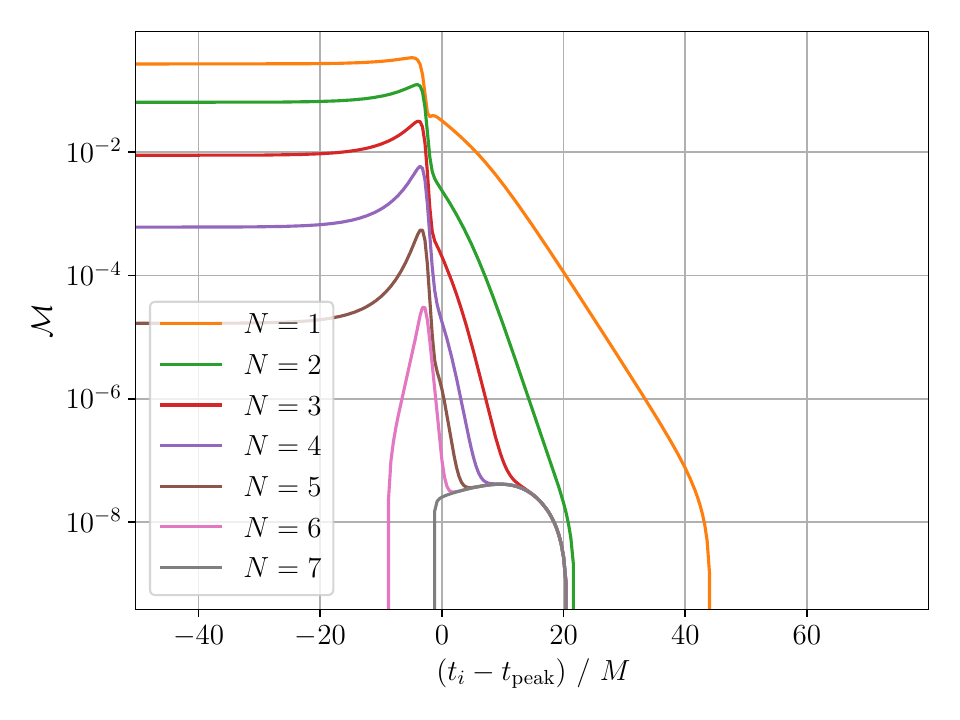}
    \caption{Mismatch $\M$ between the mock waveform $\Psi_{[1,7]}$ and the fitting function $\psi^\fit_{[1,N]}$.}
    \label{fig:0305_M_mock_sub_const0}
\end{figure}

So far, we consider the conventional way of fitting, i.e., the fit of the mock waveform $\Psi_{[0,7]}$ given in Eq.~\eqref{eq:mock} by the fitting function~\eqref{eq:fit}.
In particular, we can extract the longest-lived mode close to the fiducial value in a stable manner.
However, we found that the plateau for the higher overtone with respect to the start time of the fit is short and hence the fitting of overtones is still not robust.
This trend is consistent with the previous work~\cite{Baibhav:2023clw}.
Usually, the fundamental mode $n=0$ is the longest-lived mode, and hence the extraction of overtones that have shorter damping time is challenging.

Suppose that the first overtone $n=1$ was the longest-lived mode of the mock waveform.
Then, it is natural to expect that we would be able to extract the first overtone stably.
Therefore, if we subtract the best-fit fundamental mode from the original waveform, the first overtone becomes the longest-lived mode and we expect that we can extract the first overtone more stably.
This is the main idea of the iterative extraction method which we propose in the present paper.

In this section, from the above point of view, we consider the fitting of a mock waveform, where the fundamental mode is subtracted, and examine how the extraction of overtones can be improved.
In practice, we cannot extract the longest-lived mode without numerical errors.
Before performing a more realistic subtraction of the best-fit longest-lived mode in the subsequent sections, in this section, we assume an ideal subtraction and consider a mock waveform $\Psi_{[K,7]}$ composed of a superposition of the modes from $n=K$ to $7$ with $K\geq 1$.

\begin{figure*}[t]
    \centering
    \includegraphics[width=\textwidth]{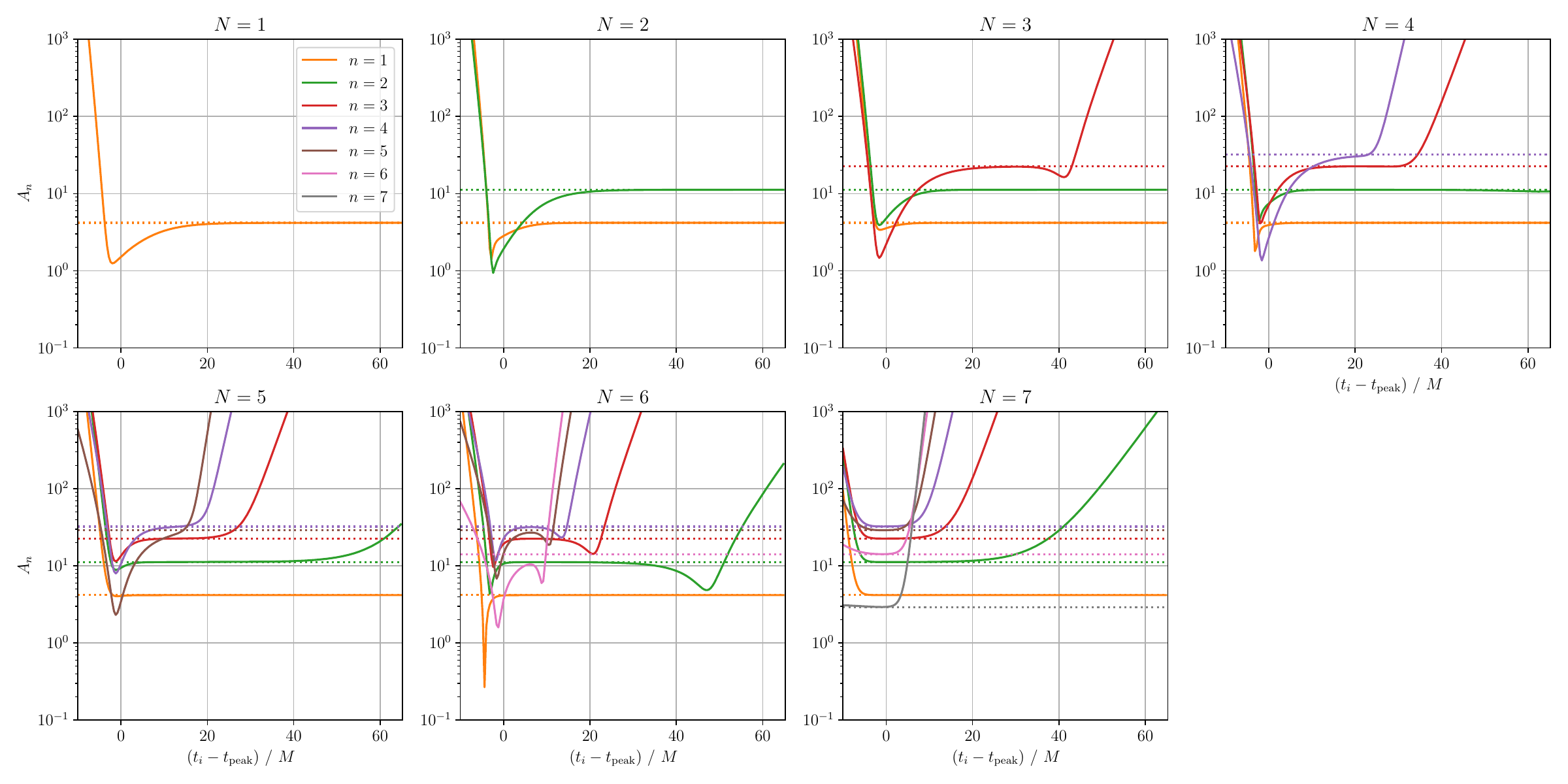}
    \caption{Amplitude $A_n$ for the fit of the mock waveform $\Psi_{[1,7]}$ by the fitting function $\psi^\fit_{[1,N]}$.}
    \label{fig:0305_A_mock_sub_const0}
\end{figure*}

\begin{figure*}[t]
    \centering
    \includegraphics[width=\textwidth]{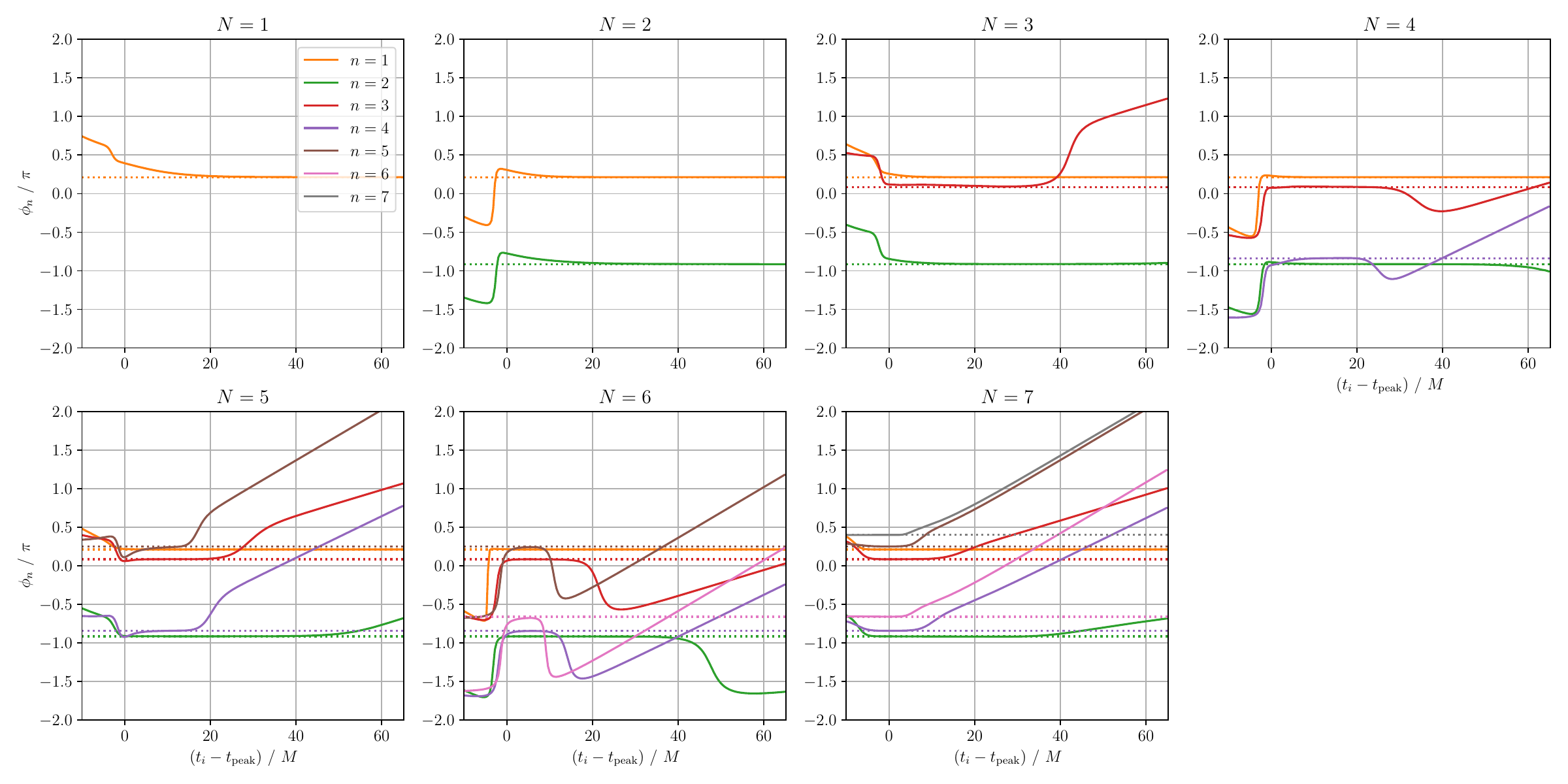}
    \caption{Phase $\phi_n$ for the fit of the mock waveform $\Psi_{[1,7]}$ by the fitting function $\psi^\fit_{[1,N]}$.}
    \label{fig:0305_phi_mock_sub_const0}
\end{figure*}

First, let us consider a situation that we subtract the fundamental mode from the waveform $\Psi_{[0,7]}$.
Assuming an ideal subtraction without errors, we consider the mock waveform $\Psi_{[1,7]}$ without the fundamental mode, which is shown by the orange curve in Fig.~\ref{fig:0305_wave_mock_sub_const}.

We then perform the fit of the mock waveform $\Psi_{[1,7]}$ by the fitting function $\psi^{\fit}_{[1,N]}$.
Figures~\ref{fig:0305_M_mock_sub_const0} shows the mismatch $\M$ between $\Psi_{[1,7]}$ and $\psi^{\fit}_{[1,N]}$ as a function of the start time of the fit $t_i$.
Compared to the case of the mock waveform $\Psi_{[0,7]}$ in Fig.~\ref{fig:0305_M_mock_sub_const}, we can see some common features.
Specifically, the mismatch does not reduce until we move the start time of the fit $t_i$ sufficiently late if we fit the mock waveform by a single damped sinusoid, and it reduces for the earlier $t_i$ if we increase the number of overtones included in the fitting function.
As expected, the first overtone in Fig.~\ref{fig:0305_M_mock_sub_const0} takes over typical features of the longest-lived mode in Fig.~\ref{fig:0305_M_mock_sub_const}.

Figures~\ref{fig:0305_A_mock_sub_const0} and \ref{fig:0305_phi_mock_sub_const0} respectively show the amplitude $A_n$ and phase $\phi_n$ for the fit of the mock waveform $\Psi_{[1,7]}$ by the fitting function~$\psi^{\fit}_{[1,N]}$ as a function of the start time of the fit $t_i$.
Compared to the case of the mock waveform $\Psi_{[0,7]}$ in Figs.~\ref{fig:0305_A_mock_sub_const} and \ref{fig:0305_phi_mock_sub_const}, we can see the same qualitative behavior.
In particular, comparing the $N=3$ panels in Figs.~\ref{fig:0305_A_mock_sub_const} and \ref{fig:0305_phi_mock_sub_const} and the $N=4$ panels in Figs.~\ref{fig:0305_A_mock_sub_const0} and \ref{fig:0305_phi_mock_sub_const0},
$A_2$ and $\phi_2$ start to diverge at $(t_i-t_{\peak})/M\simeq 40$ in Figs.~\ref{fig:0305_A_mock_sub_const} and \ref{fig:0305_phi_mock_sub_const}, but in Figs.~\ref{fig:0305_A_mock_sub_const0} and \ref{fig:0305_phi_mock_sub_const0}, $A_2$ and $\phi_2$ do not diverge up to $(t_i-t_{\peak})/M\simeq 50$ and the plateau is extended.
The same improvement applies to the plateau of $A_3$ and $\phi_3$.
We see that the subtraction process stabilizes the fit of overtone coefficients.

\begin{figure*}[t]
    \centering
    \includegraphics[width=\textwidth]{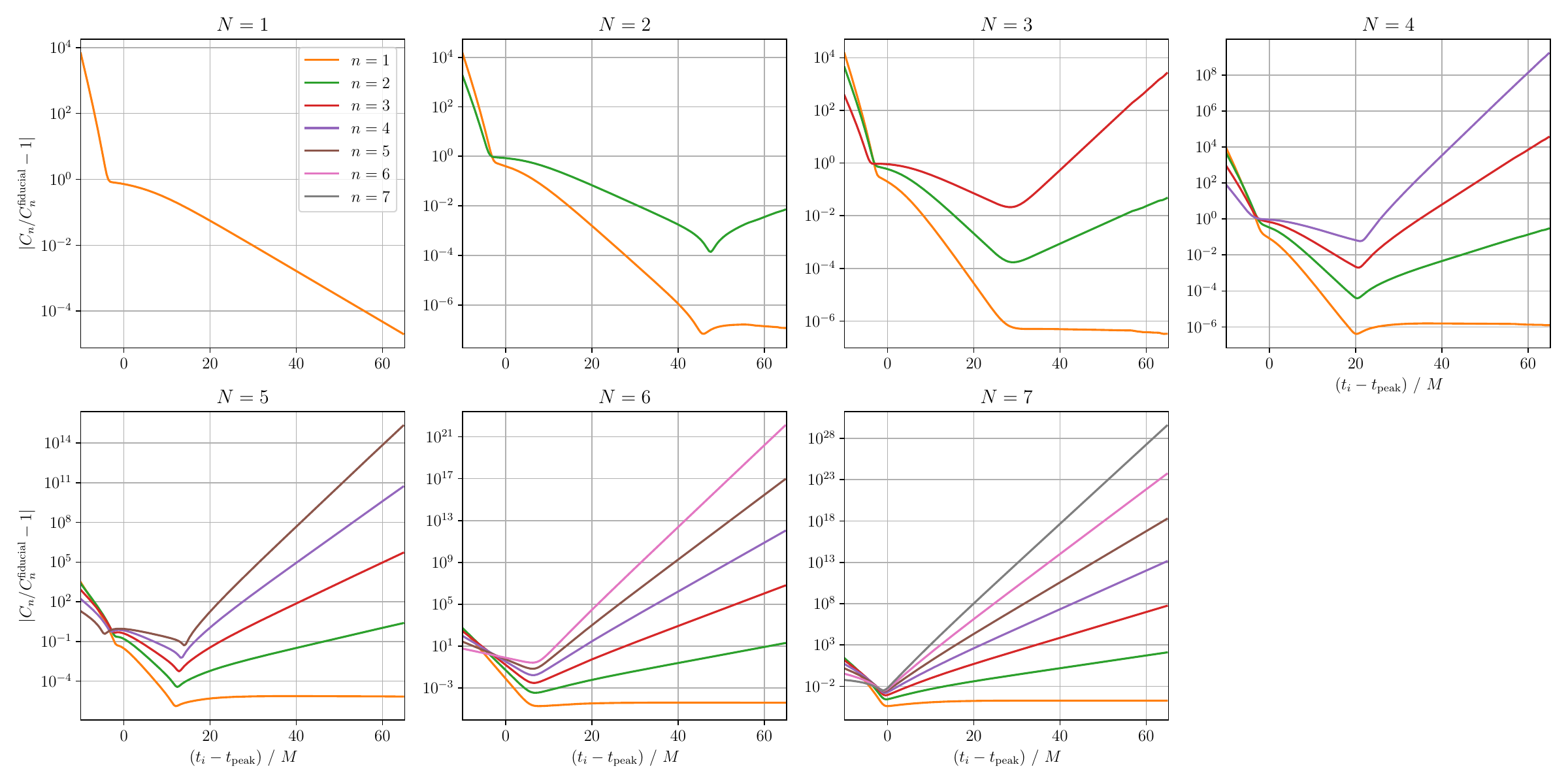}
    \caption{Relative error of $C_n$ for the fit of the mock waveform $\Psi_{[1,7]}$ by the fitting function $\psi^\fit_{[1,N]}$.}
    \label{fig:0305_Cre_mock_sub_const0}
\end{figure*}

\begin{figure*}[t]
    \centering
    \includegraphics[width=\textwidth]{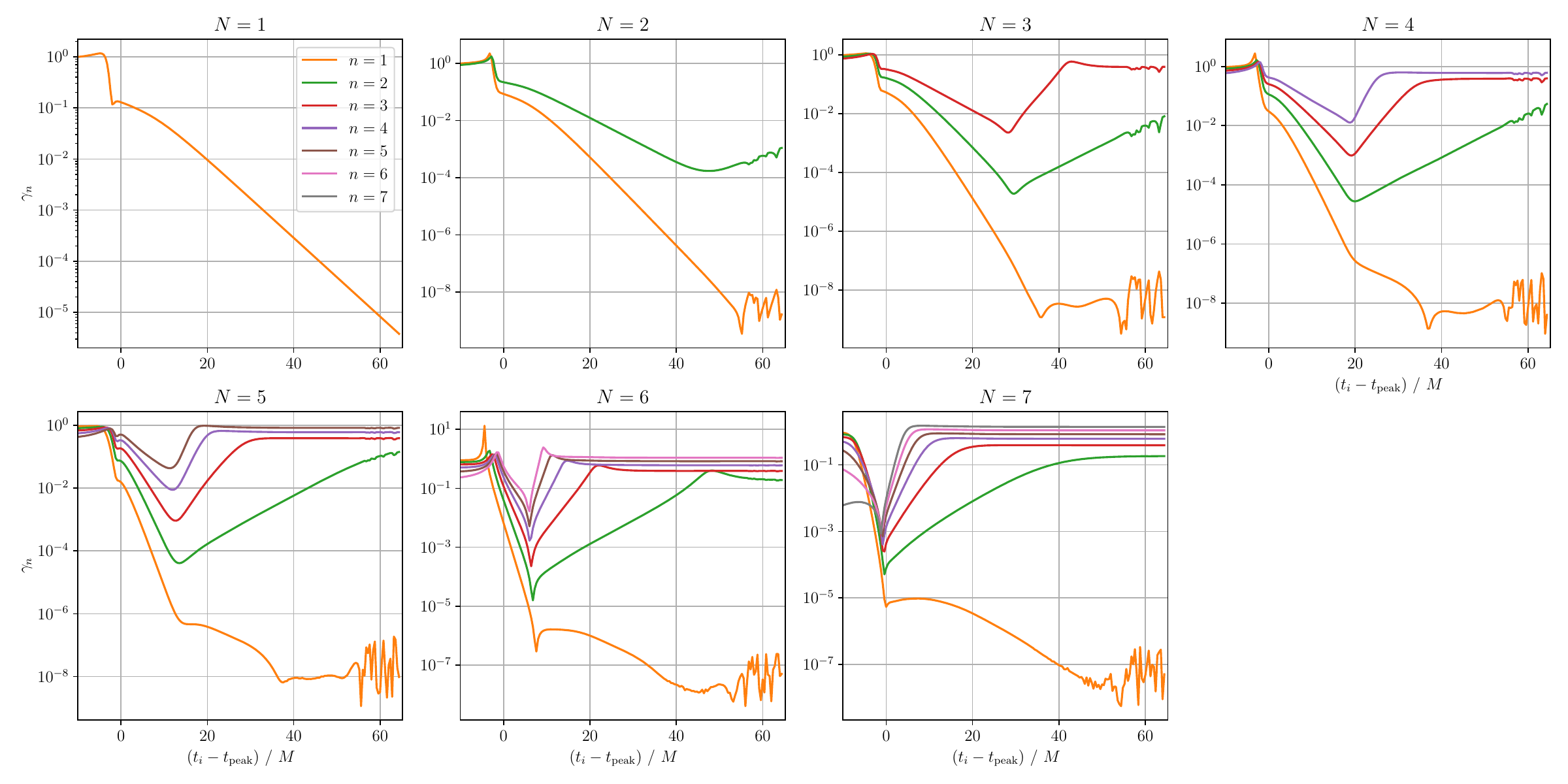}
    \caption{Rate of change $\gamma_n$ for the fit of the mock waveform $\Psi_{[1,7]}$ by the fitting function $\psi^\fit_{[1,N]}$.}
    \label{fig:0305_gamma_mock_sub_const0}
\end{figure*}

Let us check the relative error of $C_n$ and the rate of change $\gamma_n$, which are shown in 
Figs.~\ref{fig:0305_Cre_mock_sub_const0} and \ref{fig:0305_gamma_mock_sub_const0} as a function of $t_i$, respectively.
Again, we can see qualitatively same behavior as in Figs.~\ref{fig:0305_Cre_mock_sub_const} and \ref{fig:0305_gamma_mock_sub_const} for the case of the mock waveform $\Psi_{[0,7]}$.
In particular, $\gamma_1$ in Fig.~\ref{fig:0305_gamma_mock_sub_const0} now has a global minimum, which is located at a different $t_i$ of the minimum for other modes $\gamma_{n\geq 2}$.
This precisely matches the behavior of $\gamma_0$ in Fig.~\ref{fig:0305_gamma_mock_sub_const}.
Further, we see that our indicator for the best-fit values works well: We use $t_i$ when $\gamma_n$ for the next-longest-lived mode, which is $\gamma_2$ for the present case, is minimum.
It matches the start time of the fit $t_i$ when the relative error for $C_n$ is minimum.

Therefore, as we expected, if the longest-lived mode is completely subtracted, the original next-longest-lived mode takes over the role of the longest-lived mode.
Since we can extract the longest-lived mode most stably, this process improves the stability of the extraction of the original next-longest-lived mode.
Indeed, the plateau of $A_n$ and $\phi_n$ is extended, and the minimum of the relative error of $C_n$ is improved.
Comparing the fit by the four lowest modes, i.e., the fit of $\Psi_{[0,7]}$ by $\psi^\fit_{[0,3]}$ and the fit of $\Psi_{[1,7]}$ by $\psi^\fit_{[1,4]}$, the relative error of $C_1$ at the best-fit value improves from 
$6.8 \times 10^{-6}$ to $4.1 \times 10^{-7}$.

\begin{figure*}[t]
    \centering
    \includegraphics[width=0.32\textwidth]{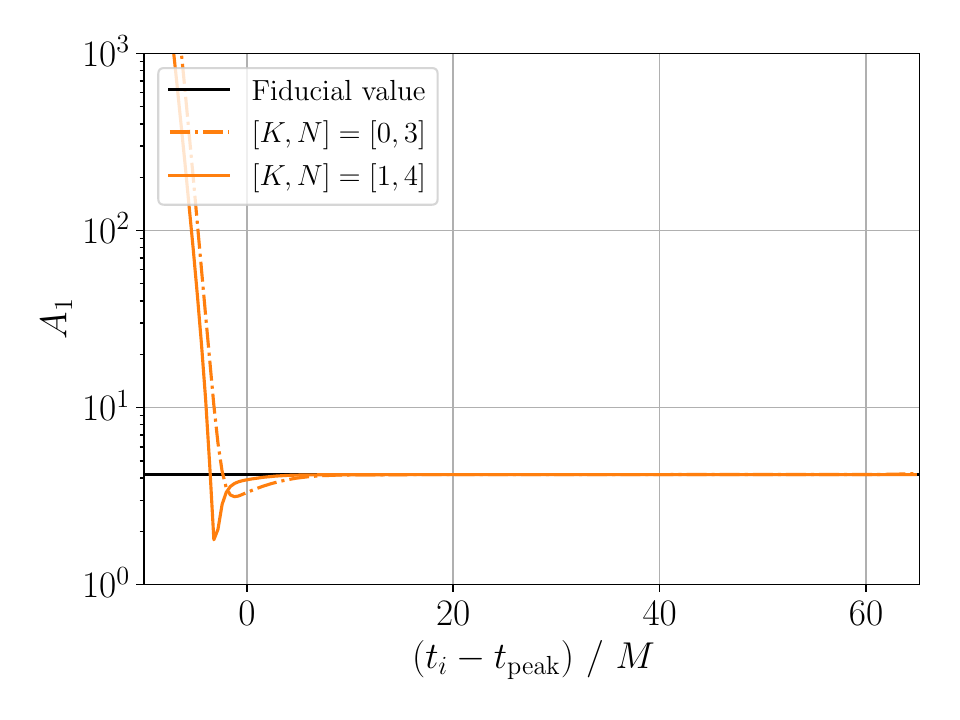}
    \includegraphics[width=0.32\textwidth]{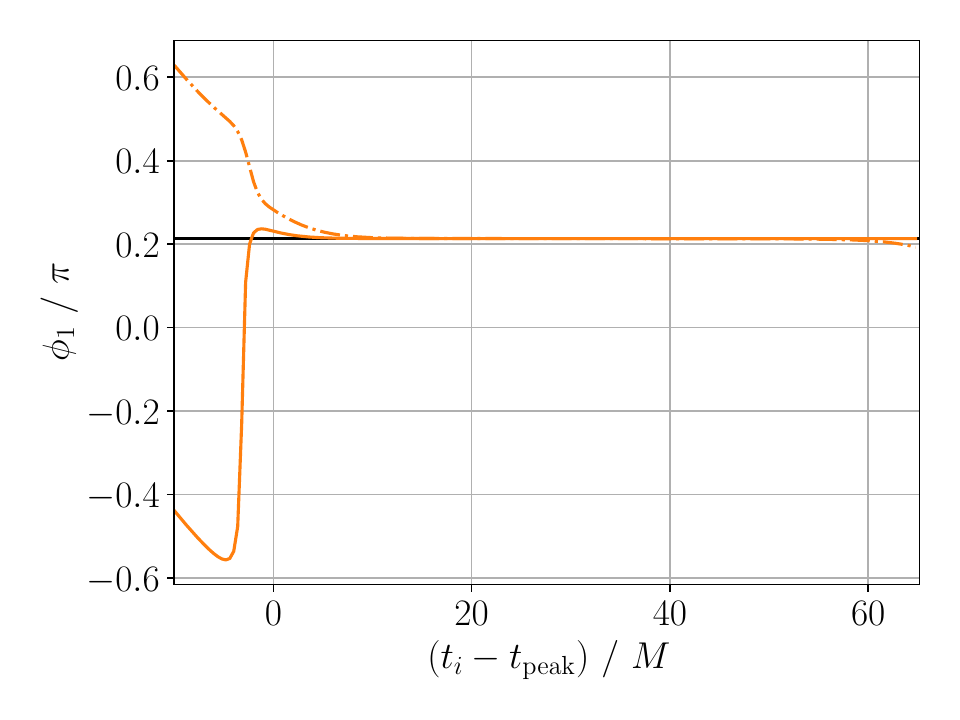}
    \includegraphics[width=0.32\textwidth]{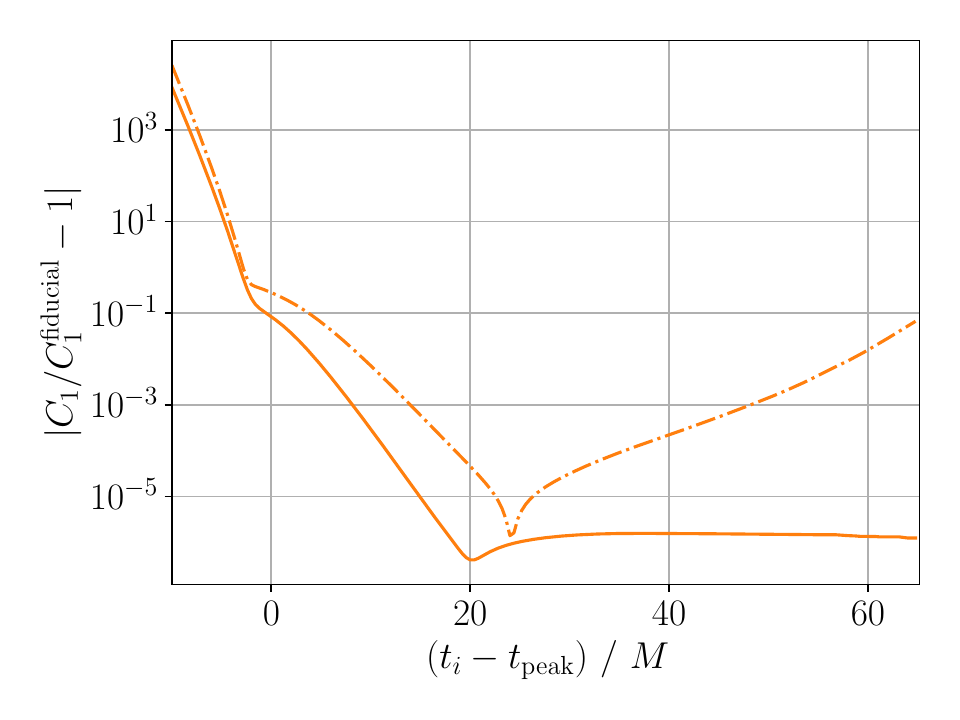}
    
    \includegraphics[width=0.32\textwidth]{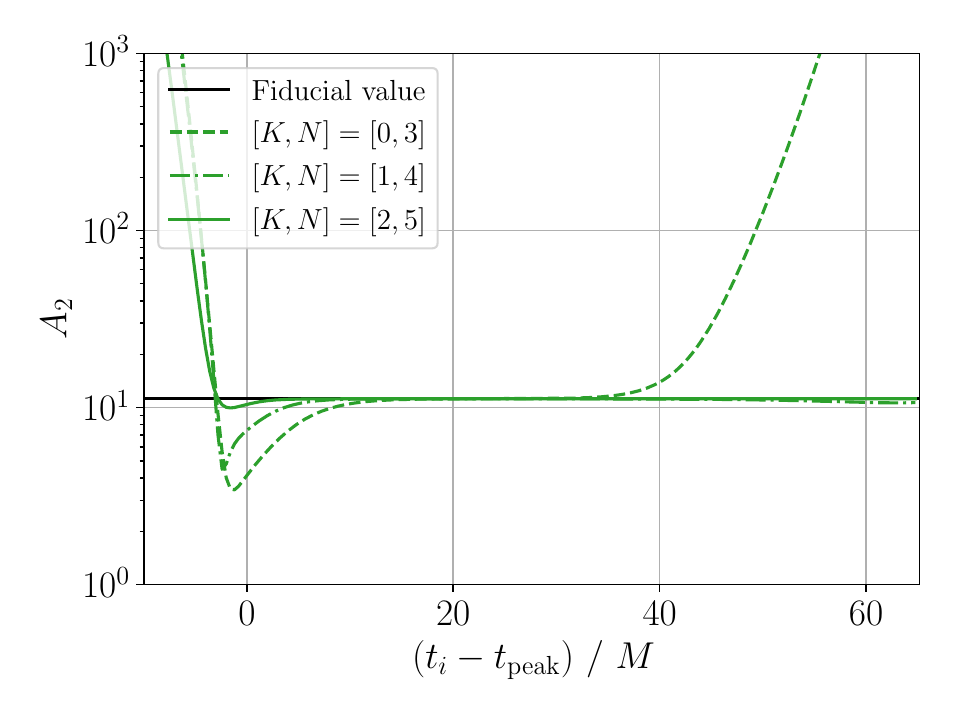}
    \includegraphics[width=0.32\textwidth]{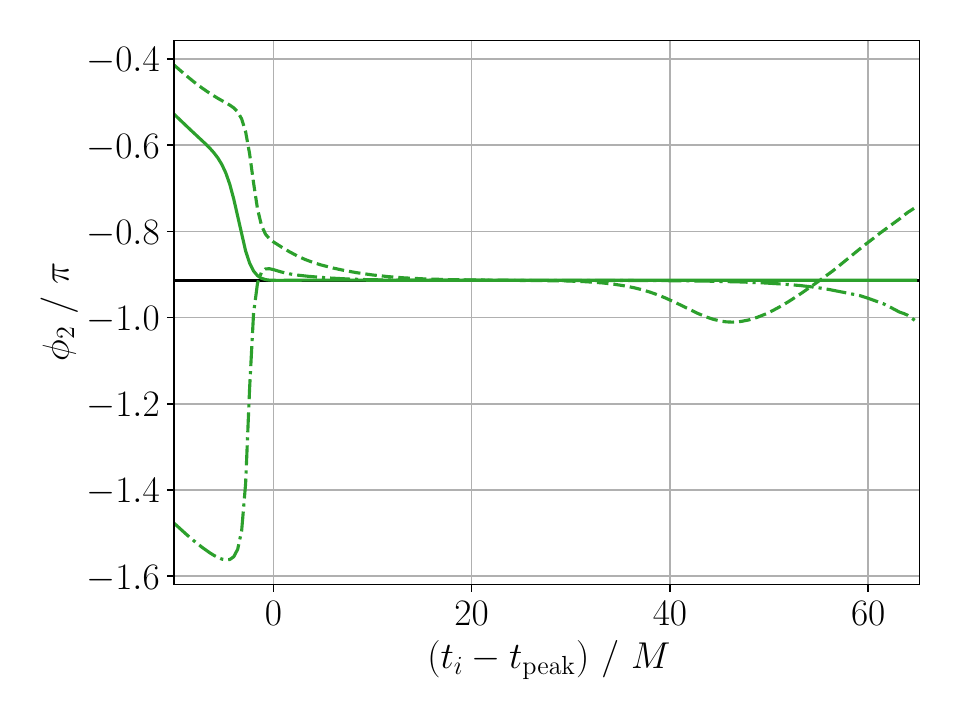}
    \includegraphics[width=0.32\textwidth]{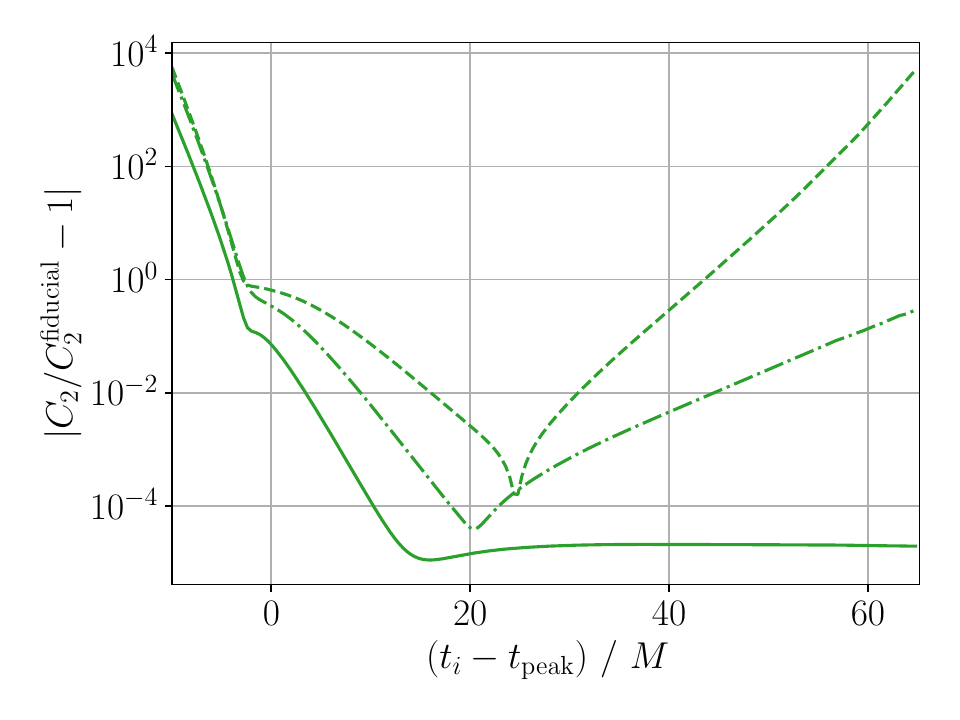}
    
    \includegraphics[width=0.32\textwidth]{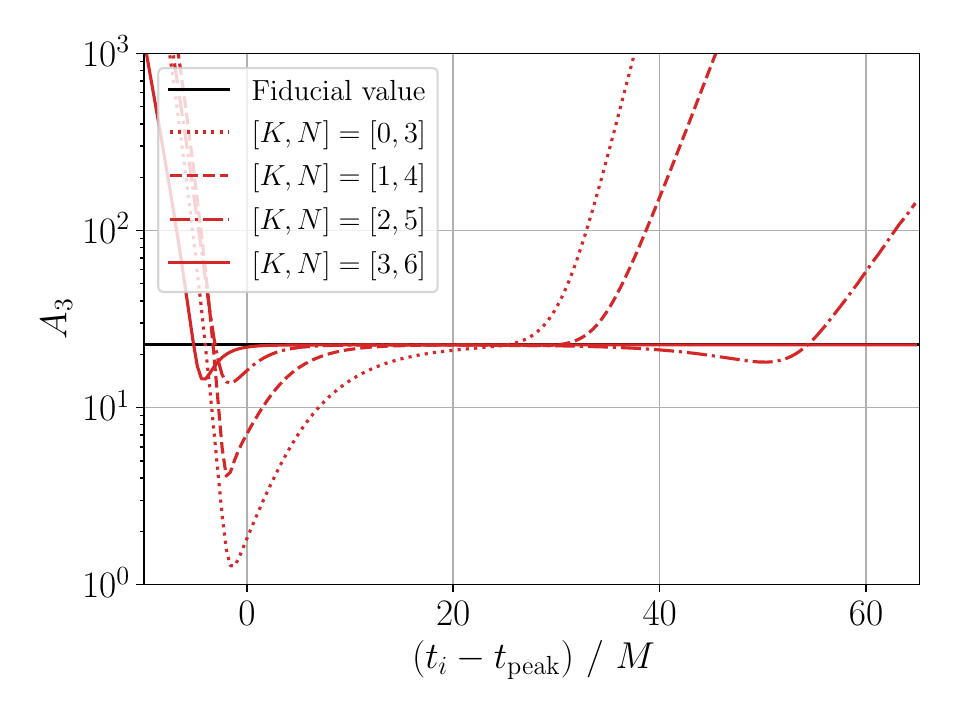}
    \includegraphics[width=0.32\textwidth]{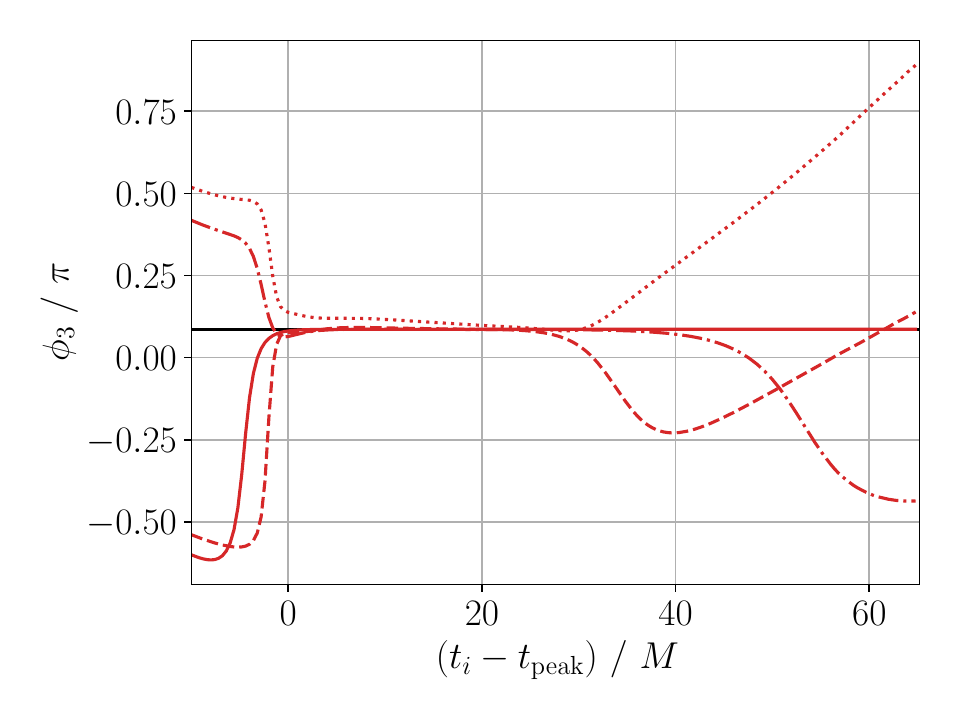}
    \includegraphics[width=0.32\textwidth]{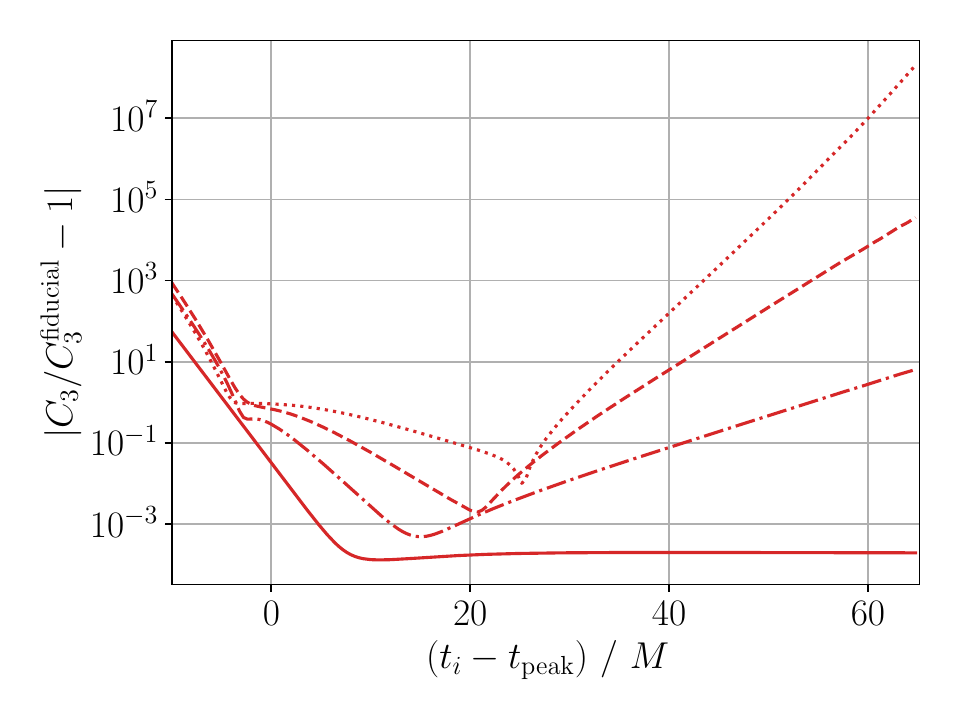}
    
    \includegraphics[width=0.32\textwidth]{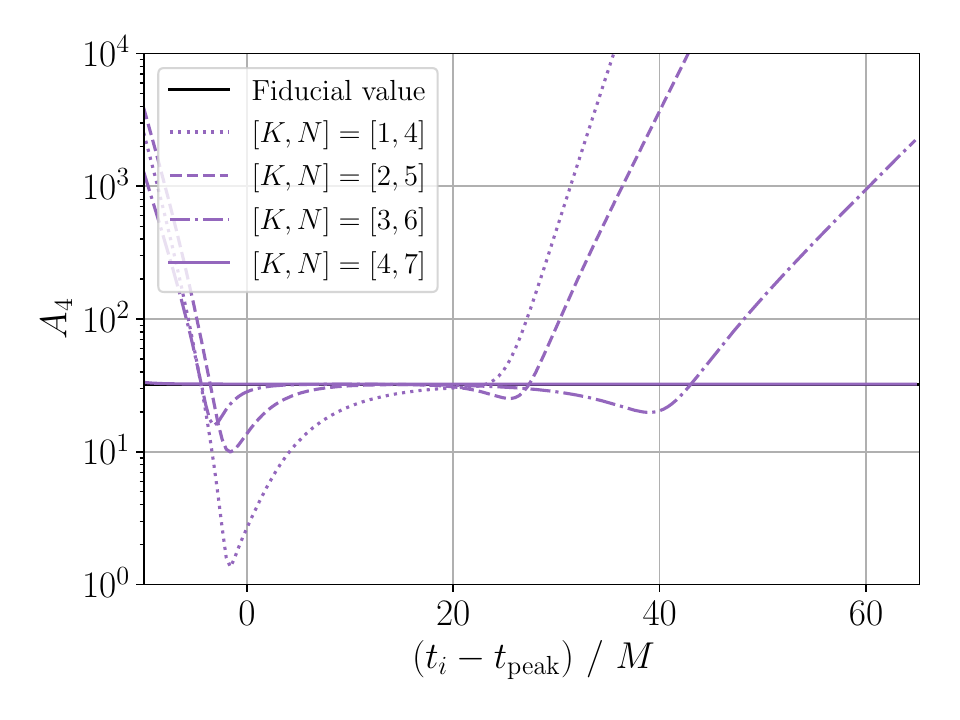}
    \includegraphics[width=0.32\textwidth]{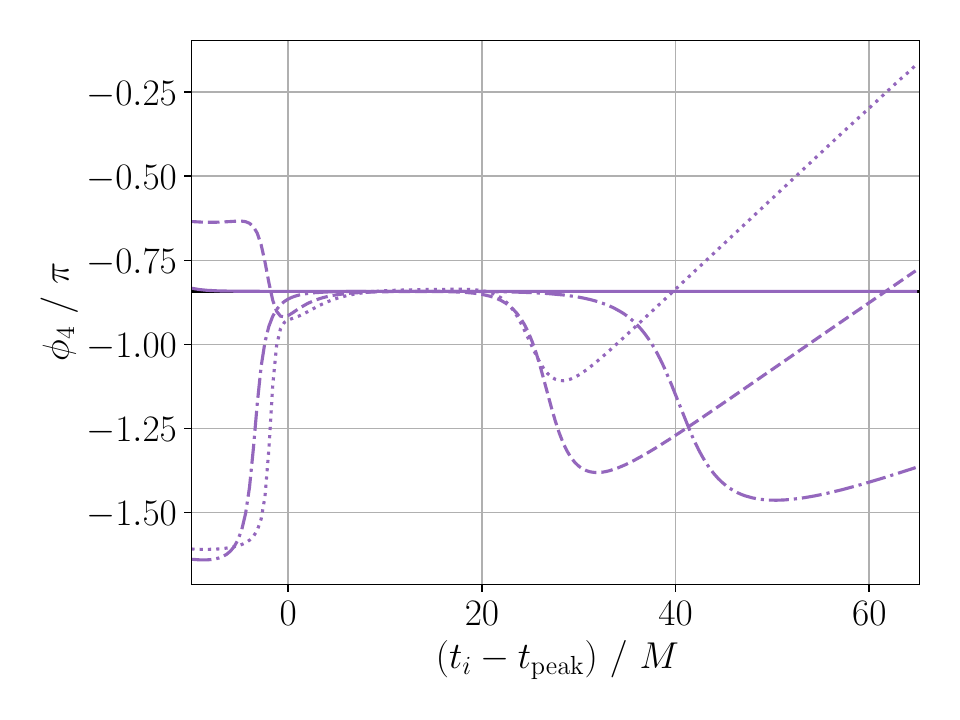}
    \includegraphics[width=0.32\textwidth]{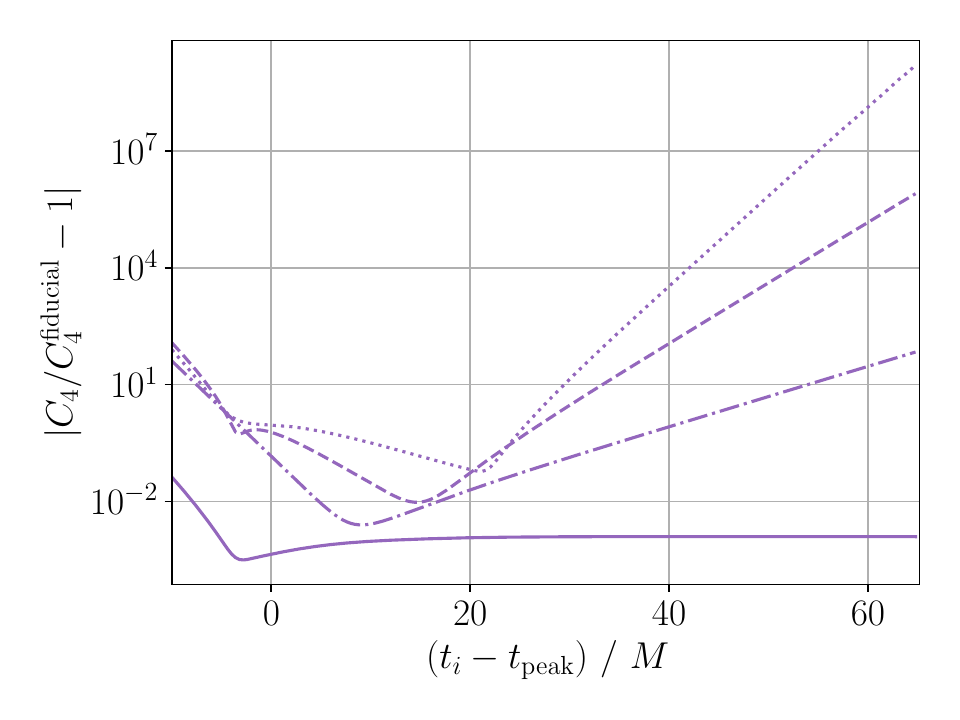}
    \caption{The improvement of the amplitude $A_n$ (left column), phase $\phi_n$ (middle column), and relative error of $C_n$ (right column) from $n=1$ mode (first row) to $n=4$ mode (fourth row) for the iterative fit of the mock waveform $\Psi_{[K,7]}$ by the fitting function $\psi^{\fit}_{[K,N]}$ with $N=K+3$.}
    \label{fig:0305_n4_improve_mock_sub_const}
\end{figure*}

We expect that, by iterating the subtraction of the longest-lived mode, we can improve the fit of higher overtones step by step.
Namely, the first subtraction of $n=0$ mode makes $n=1$ mode the longest-lived mode, and the second subtraction  of $n=1$ mode makes $n=2$ mode the longest-lived mode, and so on.
We then adopt the best-fit value when the mode of interest becomes the longest-lived mode.
To examine the efficiency of the fit based on the above procedure, assuming the ideal subtraction, we investigate the fit of $\Psi_{[K,7]}$ by $\psi^\fit_{[K,N]}$ subsequently for $K=0,\cdots, 4$.
We present the results in Fig.~\ref{fig:0305_n4_improve_mock_sub_const} to see how the fit of $n=1,\cdots, 4$ modes is improved by the subtraction procedure.
The first row of Fig.~\ref{fig:0305_n4_improve_mock_sub_const} shows the amplitude $A_1$, the phase $\phi_1$, and relative error of $C_1$ as a function of the start time of the fit $t_i$ for the iterative fit of the mock waveform $\Psi_{[K,7]}$ by the fitting function $\psi^{\fit}_{[K,K+3]}$. 
Specifically, in the first row of Fig.~\ref{fig:0305_n4_improve_mock_sub_const}, the dot-dashed curve corresponds to the fit of $\Psi_{[0,7]}$ by $\psi^{\fit}_{[0,3]}$, and the solid curve corresponds to the fit of $\Psi_{[1,7]}$ by $\psi^{\fit}_{[1,4]}$.
We can see that as we subtract the longest-lived mode, the fitting improves; the plateau in $A_1$ and $\phi_1$ becomes longer and the relative error from the fiducial value becomes smaller.
This improvement due to the iterative fit and subtraction procedure is more manifest for $n=2,3$, and $4$, which are presented in the second, third, and fourth row of Fig.~\ref{fig:0305_n4_improve_mock_sub_const}, respectively.
For instance, let us focus on the case of $n=3$ in the third row of Fig.~\ref{fig:0305_n4_improve_mock_sub_const}. 
It may be difficult to extract the correct value from the fit of $\Psi_{[0,7]}$ by $\psi^{\fit}_{[0,3]}$ shown by the dotted curve as the plateau is marginal and the fitting is not so stable. 
However, iterating the (ideal) subtractions, we can obtain a sufficiently long plateau for the fit of $\Psi_{[3,7]}$ by $\psi^{\fit}_{[3,6]}$ shown by the solid curve, which is also consistent with the fiducial value $C_3^\fid$.

\begin{figure}[t]
    \centering
    \includegraphics[width=\columnwidth]{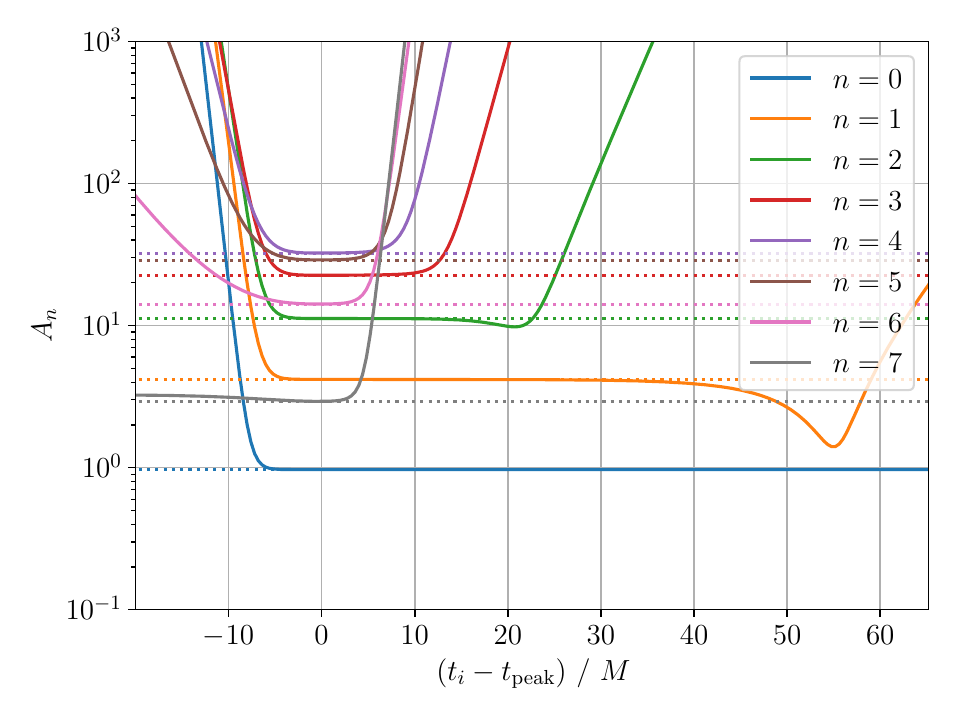}
    \includegraphics[width=\columnwidth]{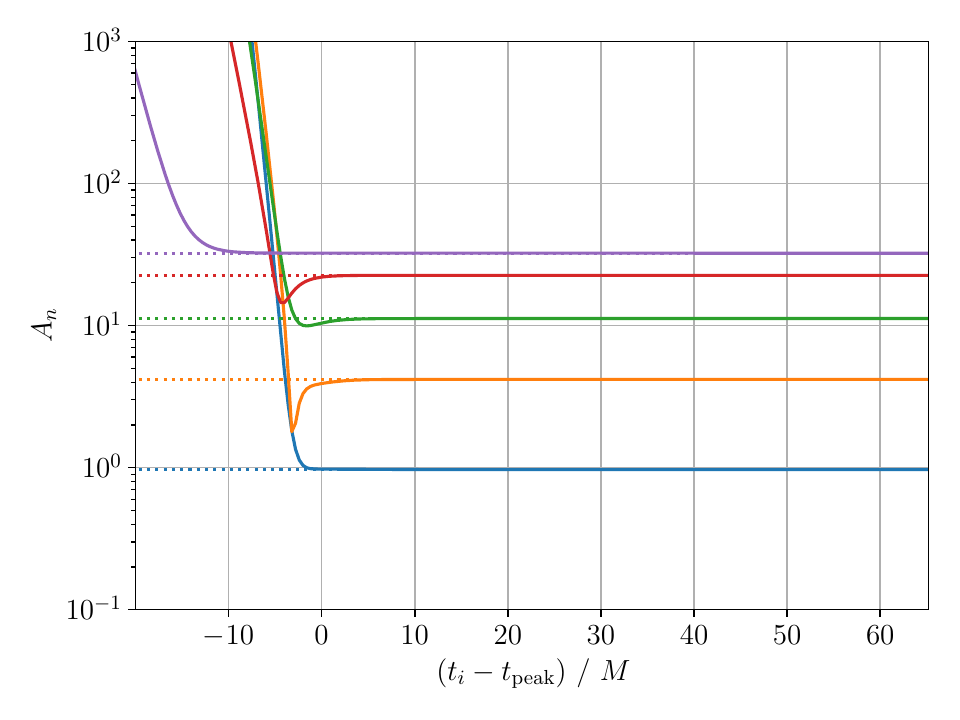}
    \caption{Comparison of the amplitude $A_n$ obtained by fitting the mock waveform $\Psi_{[0,7]}$ by the conventional method (top) with the fitting function $\psi^\fit_{[0,7]}$ and the iterative method (bottom) with the fitting function $\psi^\fit_{[K,N]}$ with $K=0,\cdots, 4$ and $N=K+3$. The dotted lines represent the fiducial values.}
    \label{fig:0305_An_improve_mock_sub_const}
\end{figure}

\begin{figure}[t]
    \centering
    \includegraphics[width=\columnwidth]{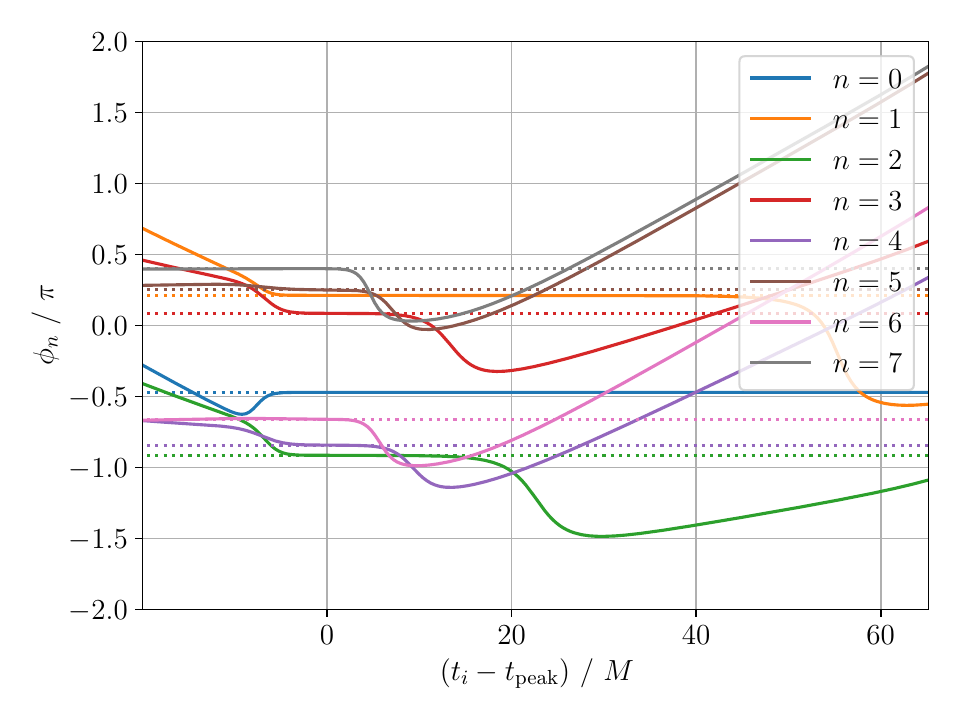}
    \includegraphics[width=\columnwidth]{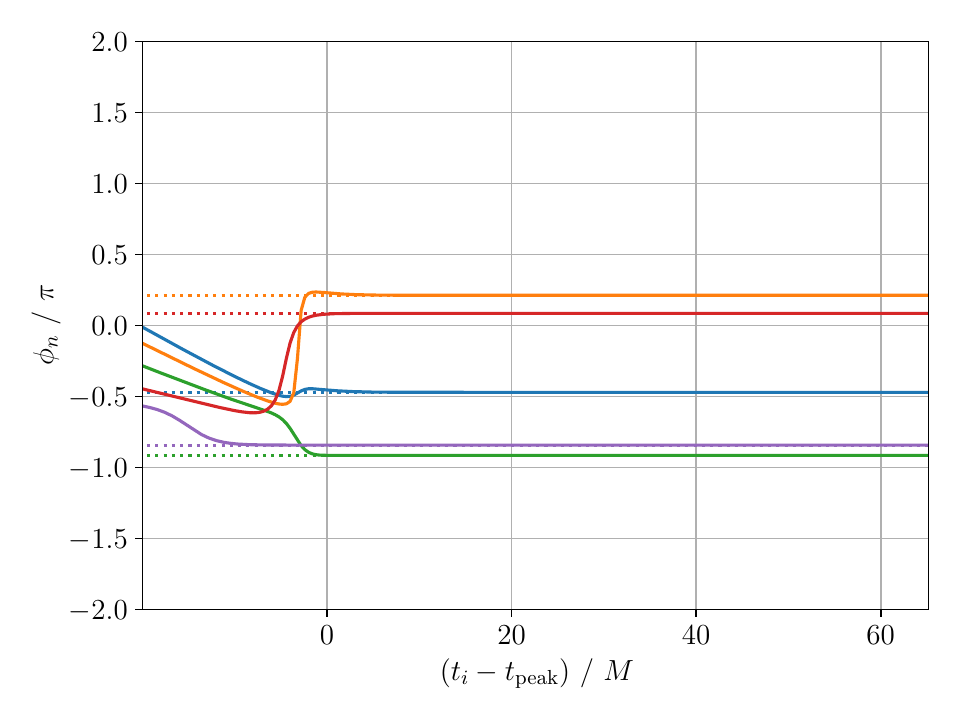}
    \caption{Comparison of the phase $\phi_n$ obtained by fitting the mock waveform $\Psi_{[0,7]}$ by the conventional method (top) with the fitting function $\psi^\fit_{[0,7]}$ and the iterative method (bottom) with the fitting function $\psi^\fit_{[K,N]}$ with $K=0,\cdots, 4$ and $N=K+3$. The dotted lines represent the fiducial values.}
    \label{fig:0305_phin_improve_mock_sub_const}
\end{figure}

Let us compare the amplitude $A_n$ and phase $\phi_n$ obtained by the conventional fit and iterative fit of the mock waveform $\Psi_{[0,7]}$.
Here we provide two ways of comparison.
First, in Figs.~\ref{fig:0305_An_improve_mock_sub_const} and \ref{fig:0305_phin_improve_mock_sub_const}, we compare the results of the conventional fit with the fitting function $\psi^\fit_{[0,7]}$ and the results of the iterative fit with the fitting function $\psi^\fit_{[K,N]}$ with $K=0,\cdots, 4$ and $N=K+3$.
This is because, as we stressed above, the longest-lived mode is more stably extracted when we fit it by a superposition of several modes.
As a representative number of the superposition, we adopt the four modes.
Consequently, while both methods use $n=0,\cdots, 7$ overtones for the fitting function, we obtain $C_n$ with $n=0,\cdots, 7$ for the conventional method and $C_n$ with $n=0,\cdots, 4$ for the iterative method.
The top panels show the result of the conventional fitting, which is the same as the panels for $N=7$ in Figs.~\ref{fig:0305_A_mock_sub_const} and \ref{fig:0305_phi_mock_sub_const}.
We see that while the fundamental mode can be extracted in a stable manner, the overtones have shorter plateau and are sensitive to the choice of the start time of the fit.
The bottom panels show the result of the fitting based on the iterative fitting method.
Namely, we present the result of the subsequent fit of the mock waveform $\Psi_{[n,7]}$ by the fitting function $\psi^{\fit}_{[n,n+3]}$ for $n=0,\cdots, 4$, where the $n$-th mode is extracted when it is the longest-lived mode.
We see that for both of $A_n$ and $\phi_n$, the plateau consistent with the fiducial value is extended for the overtones.
Thus, the iterative fitting method improves the stability of the extraction of the higher overtones.

\begin{figure}[t]
    \centering
    \includegraphics[width=\columnwidth]{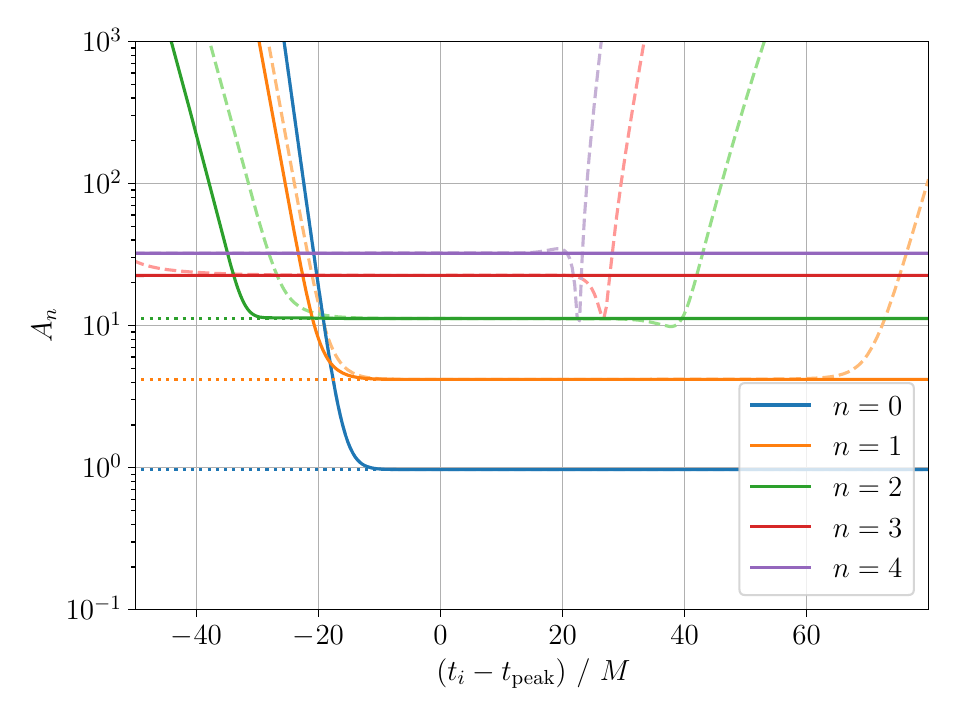}
    \includegraphics[width=\columnwidth]{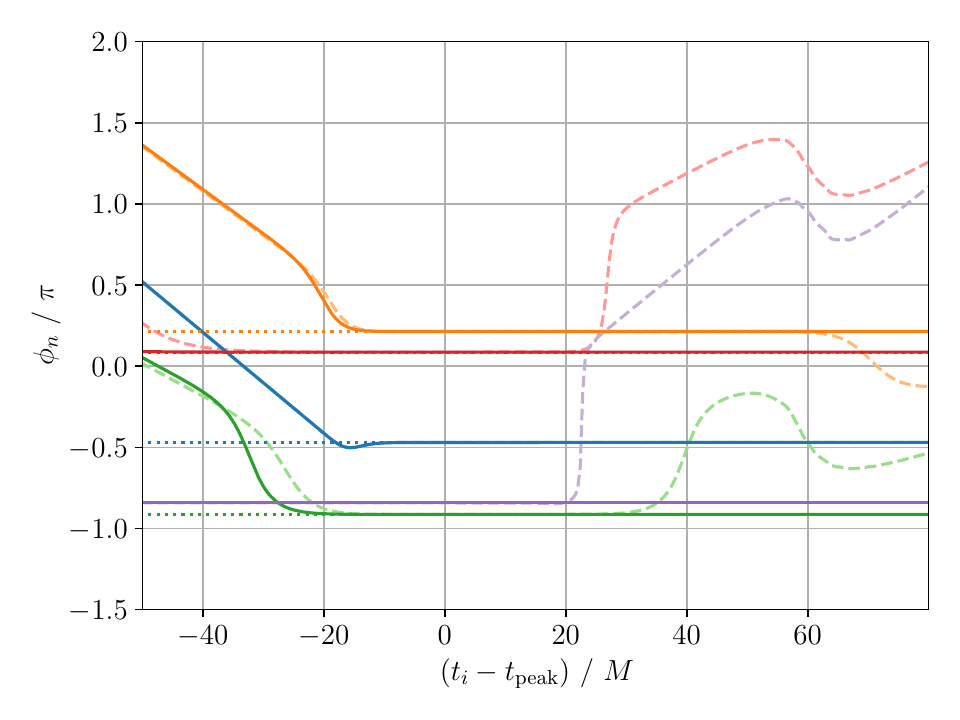}
    \caption{Comparison of the amplitude $A_n$ (top) and phase $\phi_n$ (bottom) obtained by fitting the mock waveform $\Psi_{[0,7]}$ by the conventional method (dashed) with the fitting function $\psi^\fit_{[0,4]}$ and the iterative method (solid) with the fitting function $\psi^\fit_{[K,4]}$ with $K=0,\cdots, 4$. The dotted lines represent the fiducial values.}
    \label{fig:0305_An_improve_mock_sub_const_K0-4_N4}
\end{figure}

As an alternative way of comparison of the conventional and iterative fitting methods, in Fig.~\ref{fig:0305_An_improve_mock_sub_const_K0-4_N4}, we compare the results of the conventional fit with the fitting function $\psi^\fit_{[0,4]}$ and the results of the iterative fit with the fitting function $\psi^\fit_{[K,4]}$ with $K=0,\cdots, 4$.
For both methods, we use $n=0,\cdots, 4$ overtones for the fitting function, and obtain the same number of $C_n$.
We see that the iterative fitting method provides a longer plateau in $A_n$ and $\phi_n$ for overtones, and hence improves the stability of their extraction.

In summary, through the analysis of a simple mock waveform of a superposition of damped sinusoids, we see that the longest-lived mode can be extracted in a stable manner, while a stable extraction of the overtones is in general difficult.  
We can improve the fit of the overtones by subtracting the longest-lived mode. 
After the subtraction, the next-longest-lived mode in the original waveform plays the role of the longest-lived mode, and its fit is stabilized.
By iteratively subtracting the longest-lived modes, we can realize a stable extraction of the overtones.
The subtraction of the longest-lived mode does not affect the contribution from other modes, and we can directly extract the fiducial value from the mock waveform.
However, we should be bear in mind that we treated an ideal mock waveform without noise nor power-law tail and assumed an ideal subtraction without errors.
In \S\ref{sec:dawnoise}, we consider another mock waveform taking into account these points.

\section{Damped sinusoids with constant}
\label{sec:dawnoise}

In \S\ref{sec:mock}, we analyze the mock waveform of a superposition of pure damped sinusoids assuming the ideal subtraction.
By subtracting the longest-lived mode, the next-longest-lived mode can be extracted in a more stable manner.
By iteratively subtracting the longest-lived mode from the waveform, we see that we can extract higher overtones such as $n=3$ mode.

In this section, we consider a bit more realistic analysis with increasing complexity.
We consider a mock waveform with a constant added and perform the actual subtraction of the longest-lived mode by using the best-fit values.
We shall also discuss how the stability of the fit will change for the data with different sampling rates at the end of this section. 
We shall see that the iterative fitting method is still efficient for waveform data with low sampling rates.

We add a constant to the mock waveform to mimic a more realistic situation.
For the numerical relativity waveform, which is the main focus on the present paper, it is known that the waveform data mostly approaches a small constant at the late time due to numerical errors.
The observed ringdown waveform is also contaminated by noises, which may not be removed completely.
Even if noises were completely removed, we would not be able to observe damped sinusoids very long time, since the ringdown waveform would be dominated by the power-law tail at the late time.
Thus, it would be inevitable that these effects hide the damped sinusoids at the late time below a certain order of threshold.

\begin{figure}[t]
    \centering
    \includegraphics[width=\columnwidth]{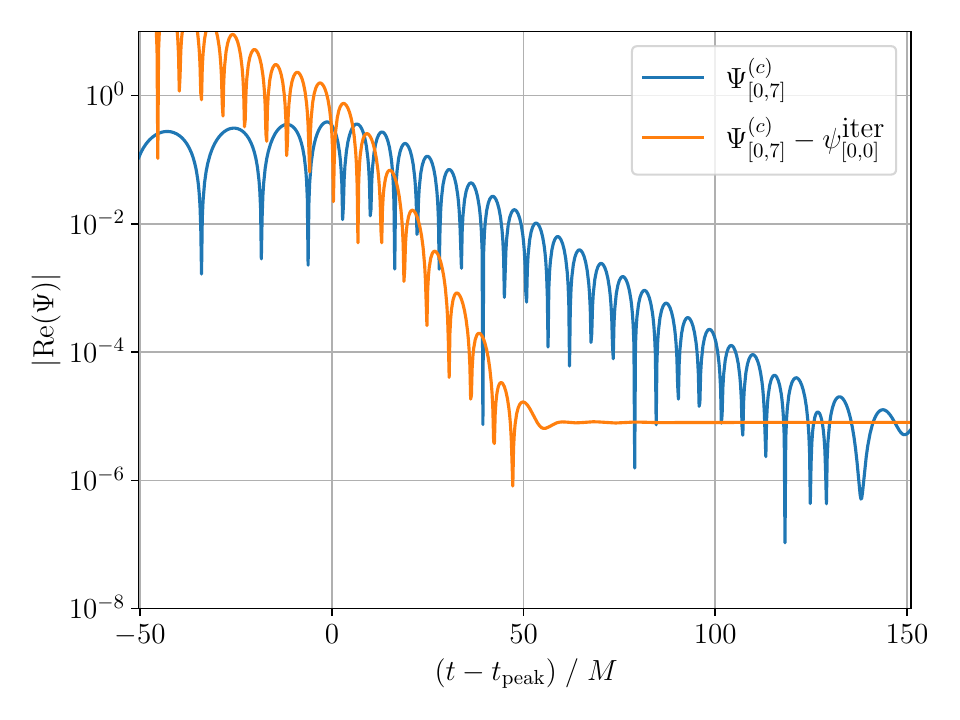}
    \caption{Mock waveform $\Psi_{[0,7]}^{(c)}$ (blue) with a constant given in Eq.~\eqref{eq:wave_mock_const} and the waveform $\Psi_{[0,7]}^{(c)}-C^{\iter}_{0}\psi_{0}$ (orange) after the first subtraction, which is $m=1$ case of Eq.~\eqref{eq:wave_mock_subn}.}
    \label{fig:0305_wave_mock_naive}
\end{figure}

Taking into account the above facts, we consider the following mock waveform for $t\geq t_{\peak}$: 
\begin{equation}
\label{eq:wave_mock_const}
    \Psi_{\rm{[0,7]}}^{(c)}(t)
    = \sum_{n=0}^{7} a_n e^{-i(\omega_n t-\theta_n)} + c,
\end{equation}
where we set $a_n$ and $\theta_n$ the same fiducial value as in \eqref{eq:mock}, which is listed in Table~\ref{tab:mock_fiducial}.
Here, we introduced a small complex constant $c=-(0.8+1.8i)\times10^{-5}$ as a ``noise'', whose value is the same as the constant appearing in the late time waveform of the numerical simulation SXS:BBH:0305.
For the region $t<t_{\peak}$, to make the mock waveform closer to those of the numerical simulation, we use the waveform of SXS:BBH:0305.

\begin{figure}[t]
    \centering
    \includegraphics[width=0.49\columnwidth]{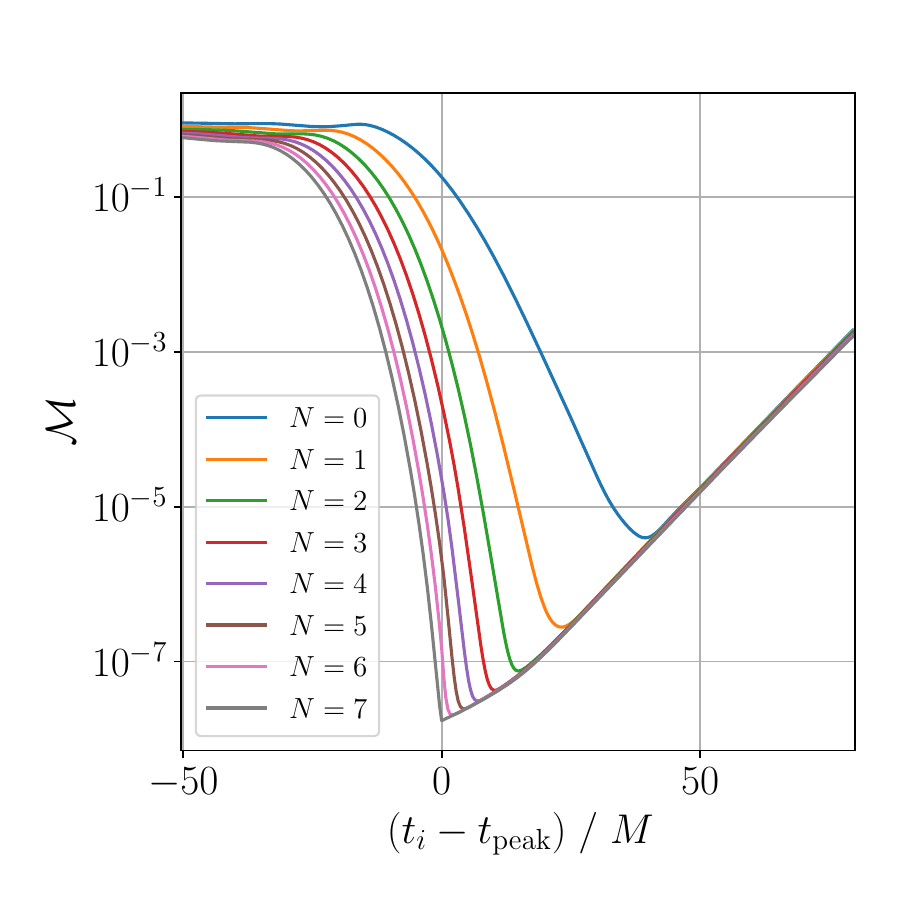}
    \includegraphics[width=0.49\columnwidth]{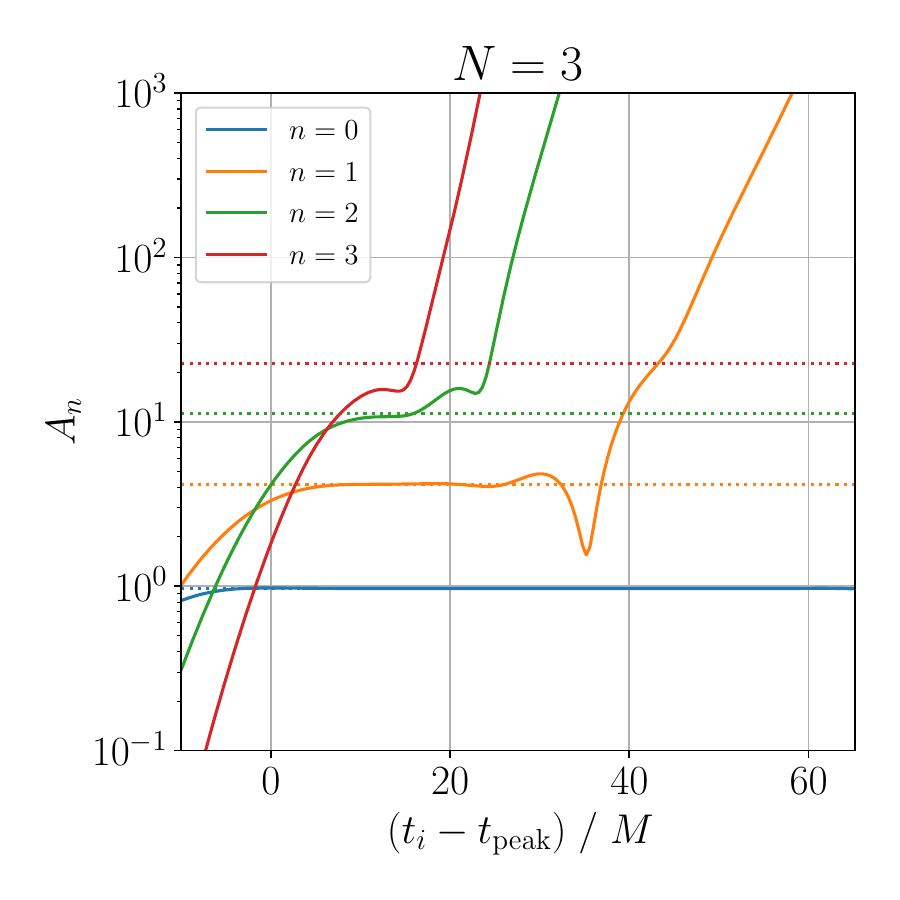}
    \includegraphics[width=0.49\columnwidth]{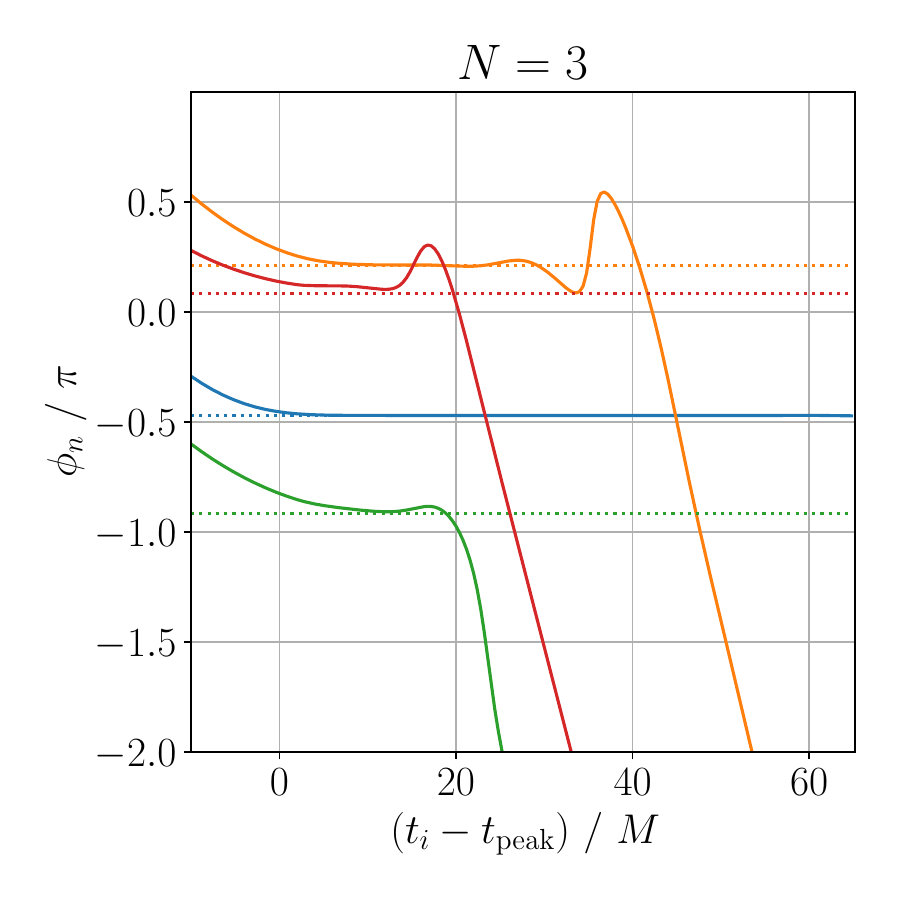}
    \includegraphics[width=0.49\columnwidth]{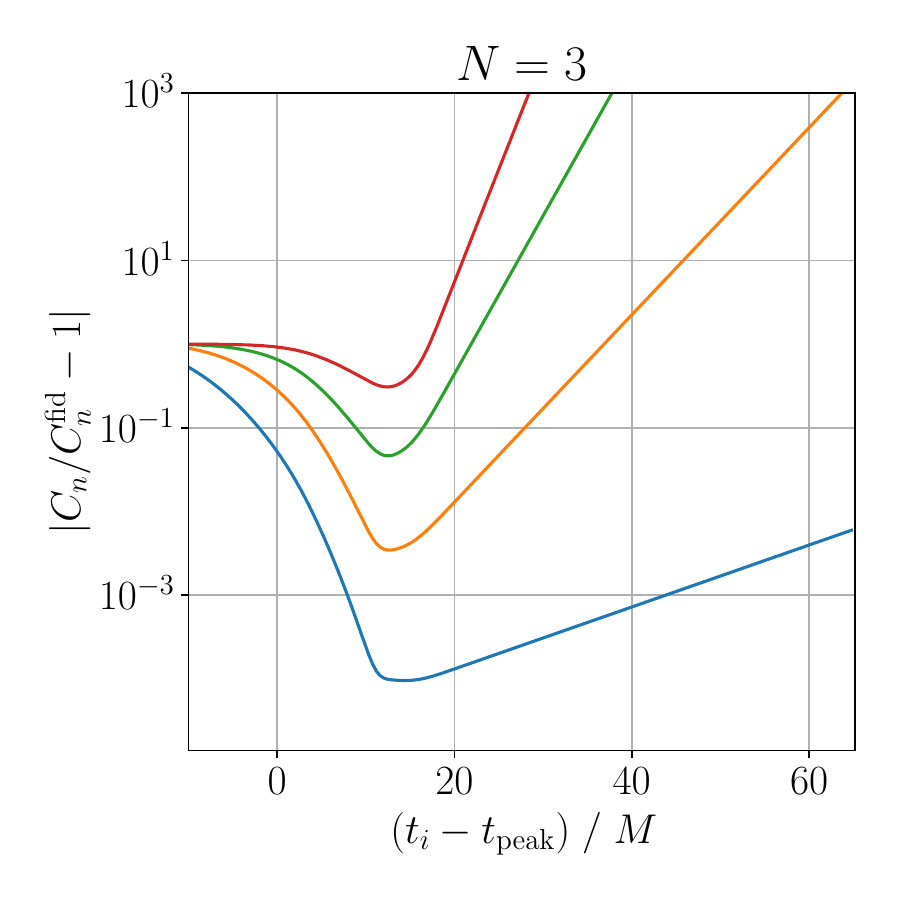}
    \includegraphics[width=0.49\columnwidth]{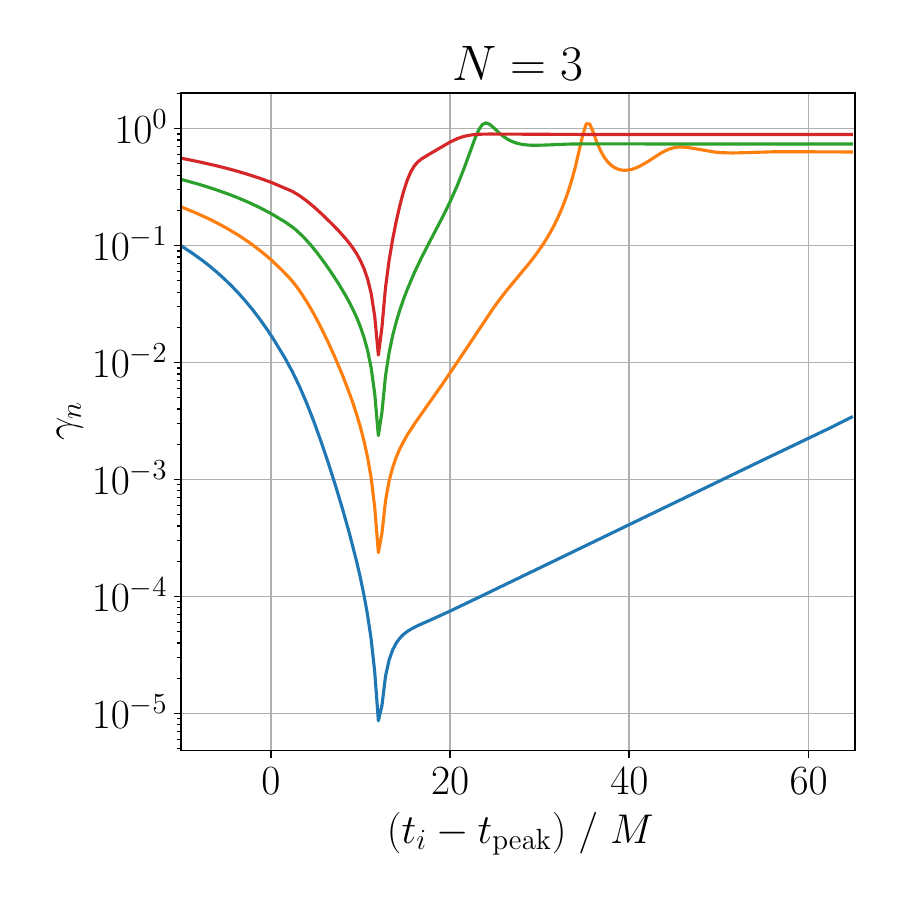}
    \caption{Mismatch $\M$ (left top), amplitude $A_n$ (right top), phase $\phi_n$ (left middle), relative error of $C_n$ (right middle), and rate of change $\gamma_n$ (bottom) for the fit of the mock waveform $\Psi_{[0,7]}^{(c)}$ by the fitting function $\psi_{[0,N]}^{\fit}$ with $N=3$.}
    \label{fig:0305_M_A_Cre_gamma_mock_naive}
\end{figure}

\begin{figure}[t]
    \centering
    \includegraphics[width=0.49\columnwidth]{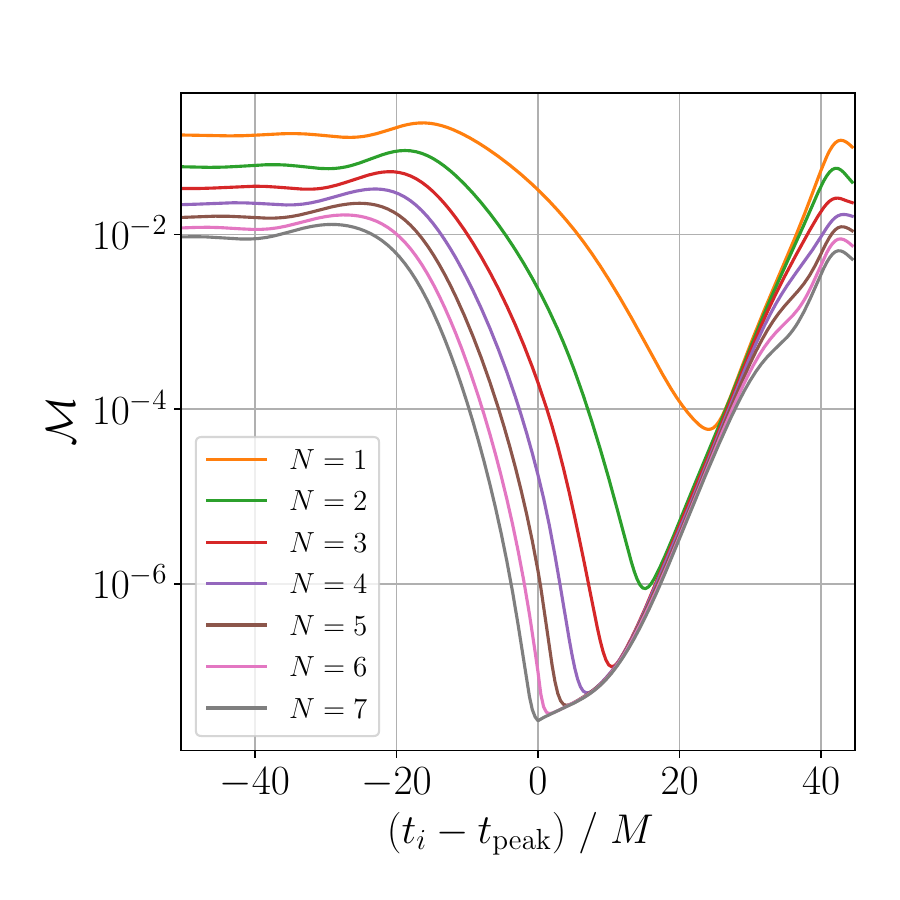}
    \includegraphics[width=0.49\columnwidth]{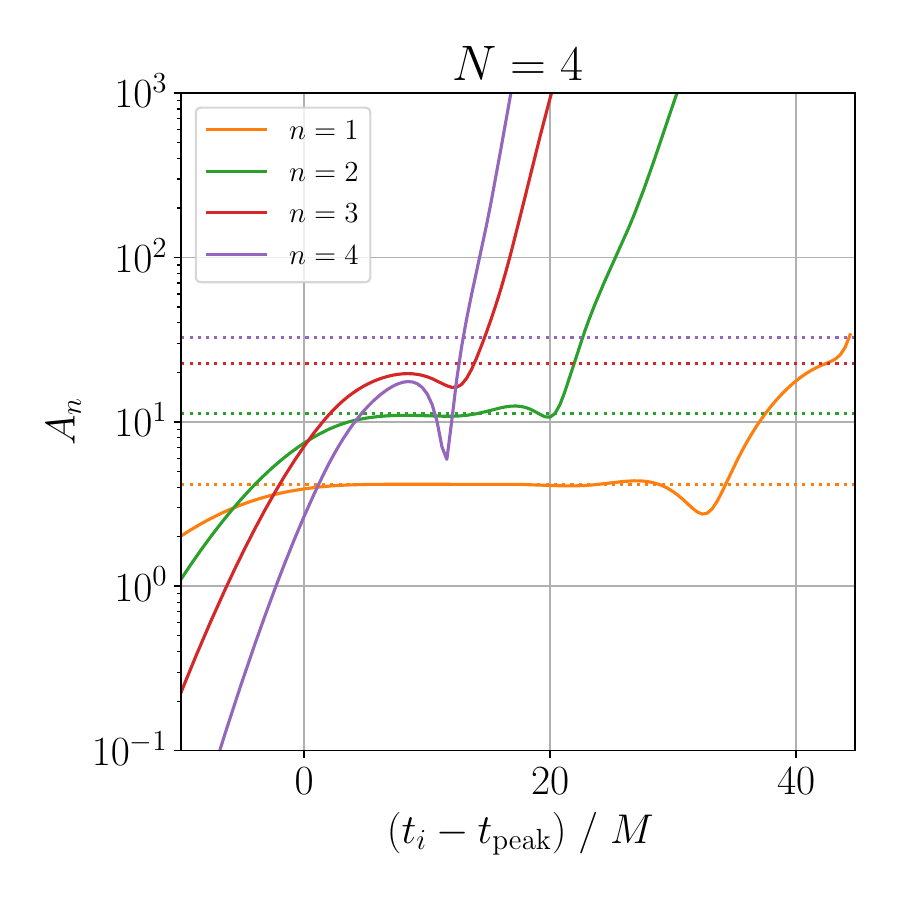}
    \includegraphics[width=0.49\columnwidth]{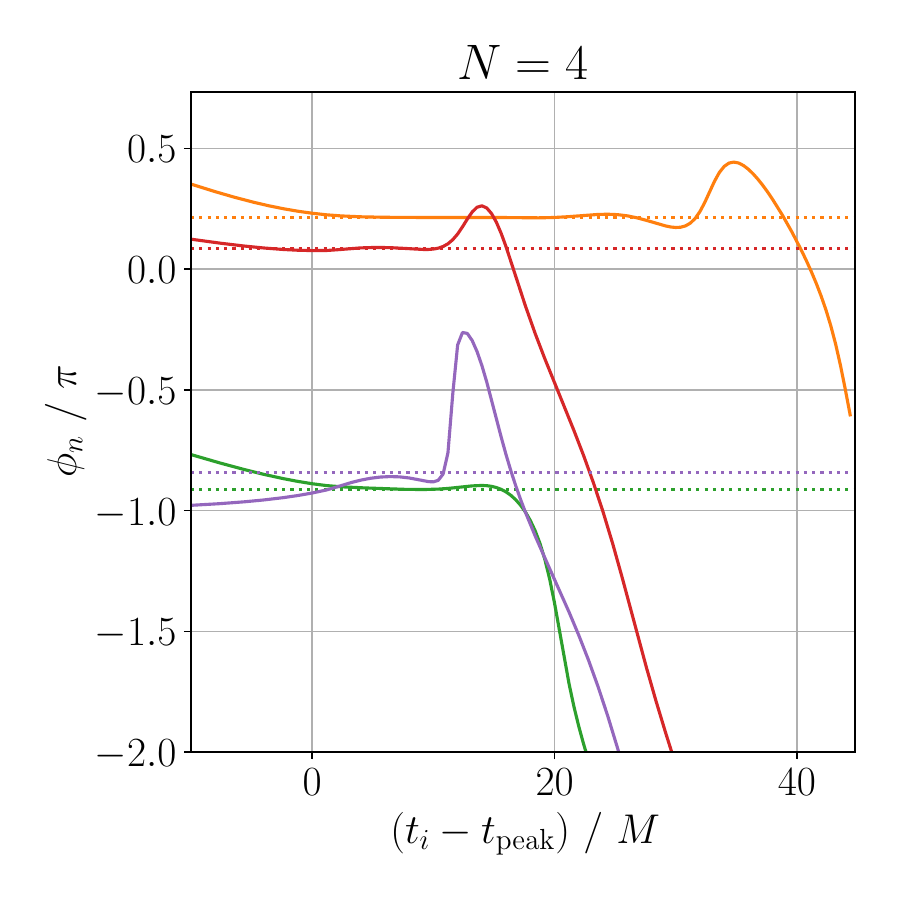}
    \includegraphics[width=0.49\columnwidth]{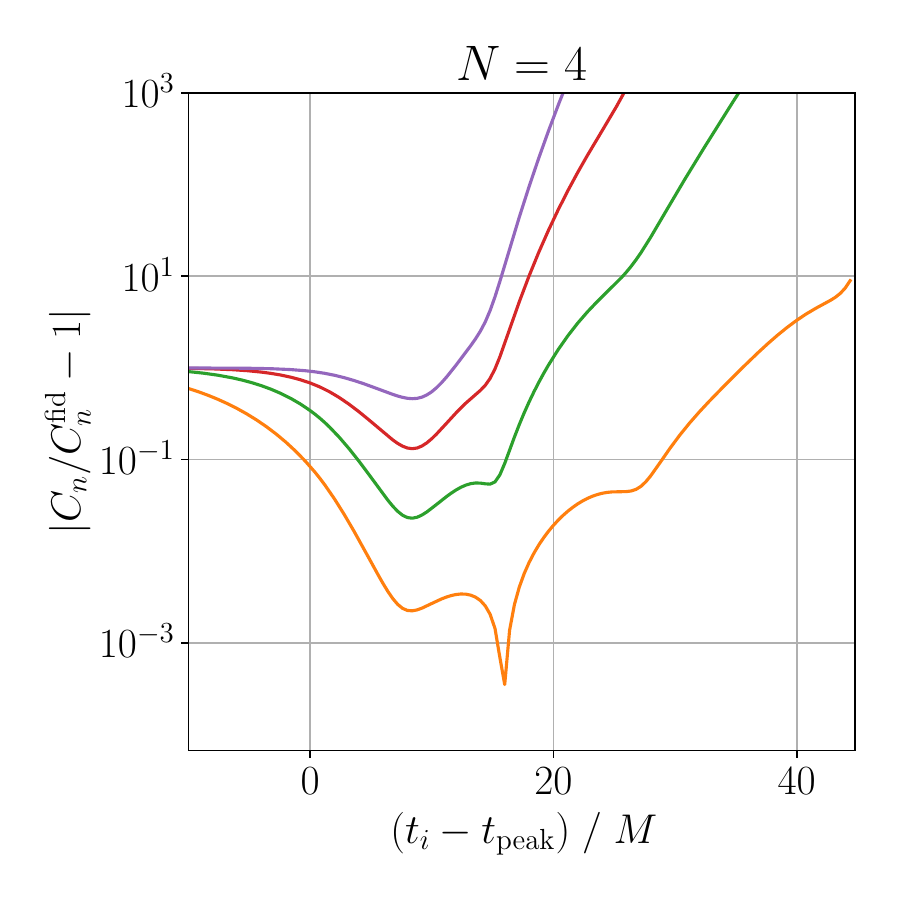}
    \includegraphics[width=0.49\columnwidth]{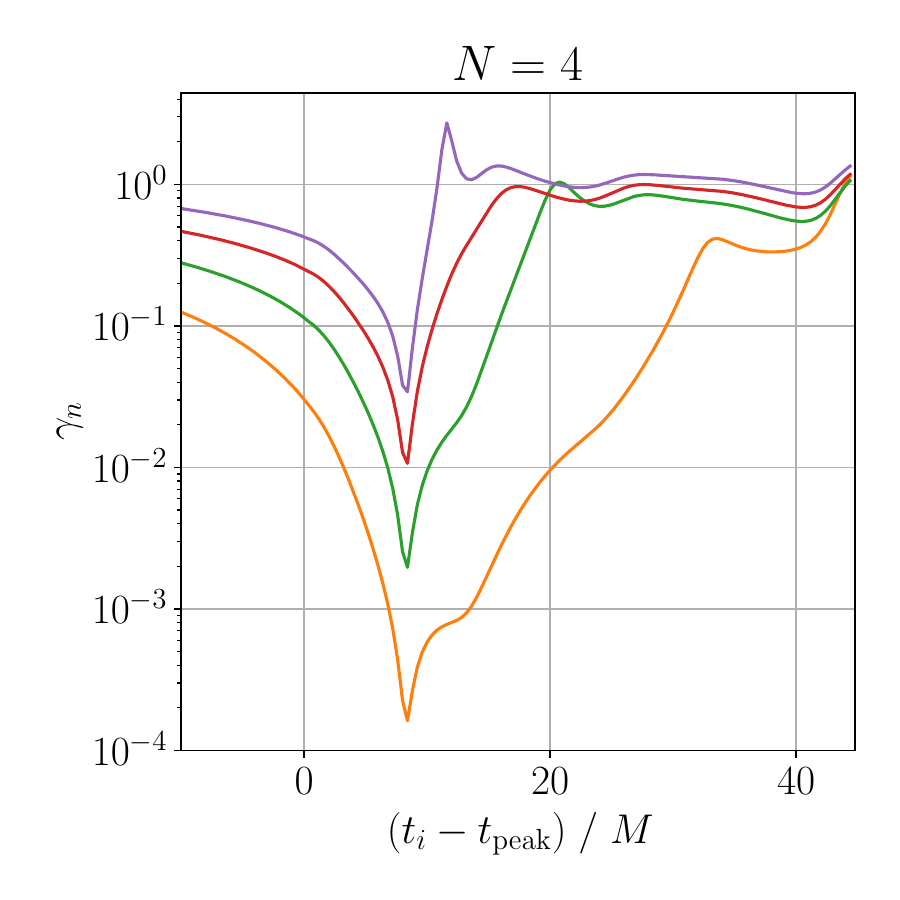}
    \caption{Mismatch $\M$ (left top), amplitude $A_n$ (right top), phase $\phi_n$ (left middle), relative error of $C_n$ (right middle), and rate of change $\gamma_n$ (bottom) for the fit of the mock waveform $\Psi_{[0,7]}^{(c)}-C^{\iter}_{0}\psi_{0}$ by the fitting function $\psi_{[1,N]}^{\fit}$ with $N=4$.}
    \label{fig:0305_M_A_Cre_gamma_mock_sub_0}
\end{figure}

While the actual noise and tail are not constant, the qualitative behavior that the damped sinusoids become subdominant at the late time is well captured by this mock waveform.
Actually, the conventional fit of a mock waveform with/without tail and numerical relativity waveform is extensively studied in \cite{Baibhav:2023clw}, and we shall see that our results of the conventional fit of $\Psi_{\rm{[0,7]}}^{(c)}$ are qualitatively consistent with the previous results of the fit of the mock waveform with tail. 

We present the mock waveform $\Psi_{\rm{[0,7]}}^{(c)}(t-t_\peak)$ by a blue curve in Fig.~\ref{fig:0305_wave_mock_naive}.
The damped sinusoids begin to be hidden for $(t-t_{\peak})/M\gtrsim 125$ as they reach to the order of the added constant $\simeq 10^{-5}$.
Compared to the case of the mock waveform $\Psi_{\rm{[0,7]}}$ without noise in Fig.~\ref{fig:0305_wave_mock_sub_const}, due to the constant, we have less data available for the fitting, and hence expect that the efficiency of the fit becomes worse.

We perform the fitting of the mock waveform $\Psi_{\rm{[0,7]}}^{(c)}(t-t_\peak)$ by the fitting function $\psi_{[0,N]}^{\fit}(t-t_\peak)$, and present the results in Fig.~\ref{fig:0305_M_A_Cre_gamma_mock_naive}.
For the amplitude, relative error, and the rate of change in Fig.~\ref{fig:0305_M_A_Cre_gamma_mock_naive}, 
we display only $N=3$ as the representative case with the four modes in the fitting function as we discussed in \S\ref{sec:mock}.
The all cases $N=0,\cdots,7$ are supplemented in Figs.~\ref{fig:0305_A_mock_naive} in Appendix~\ref{sec:plots}.

The first panel of Fig.~\ref{fig:0305_M_A_Cre_gamma_mock_naive} shows the mismatch.
Unlike the fit of the mock waveform without constant shown in Fig.~\ref{fig:0305_M_mock_sub_const}, we see that the mismatch in Fig.~\ref{fig:0305_M_A_Cre_gamma_mock_naive} is bounded from below due to the existence of the noise in the mock waveform.
The lower bound increases for the fitting starting from later time.
This behavior is typical for the fit of the waveform with noise or tail, originating from the larger ratio of the noise or tail to the damping waveform at later time. 
In the case of the mock waveform without noise in Fig.~\ref{fig:0305_M_mock_sub_const}, we obtain smaller value of mismatch at earlier $t_i$ reaching the order of the numerical errors.
Such behavior of the mismatch does not occur for a more realistic case with noises, and the fitting becomes more difficult in general.

\begin{figure*}[t]
    \centering
    \includegraphics[width=0.32\textwidth]{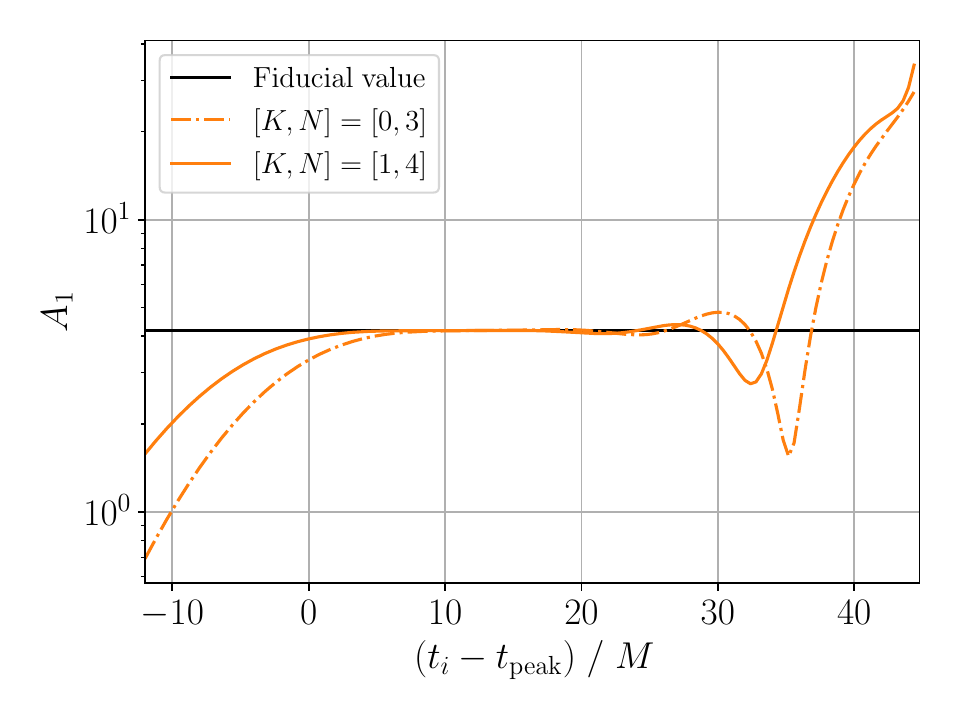}
    \includegraphics[width=0.32\textwidth]{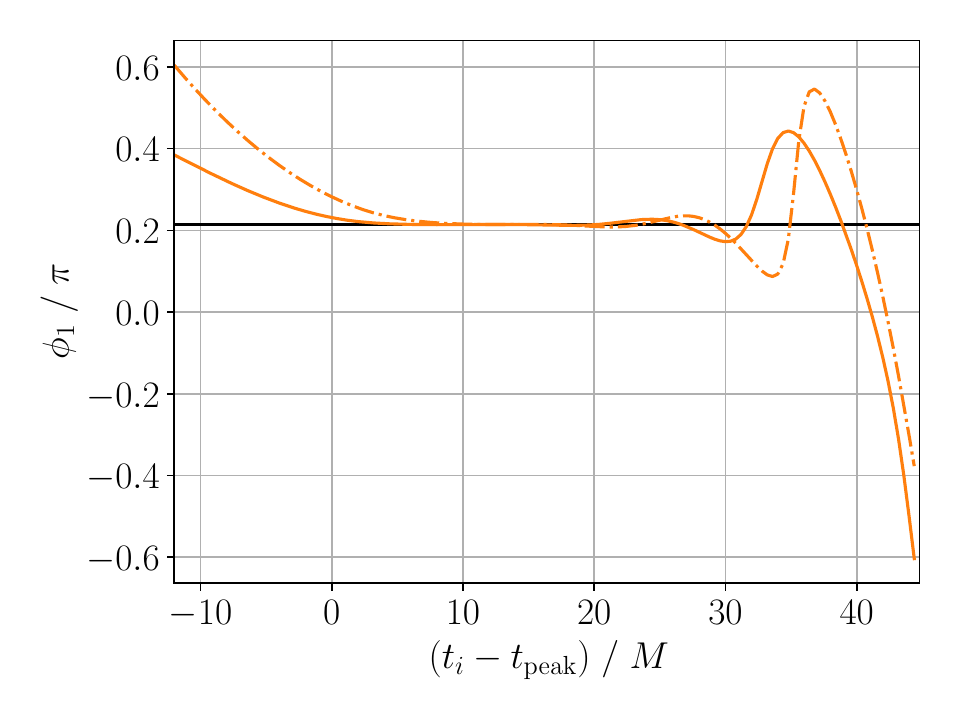}
    \includegraphics[width=0.32\textwidth]{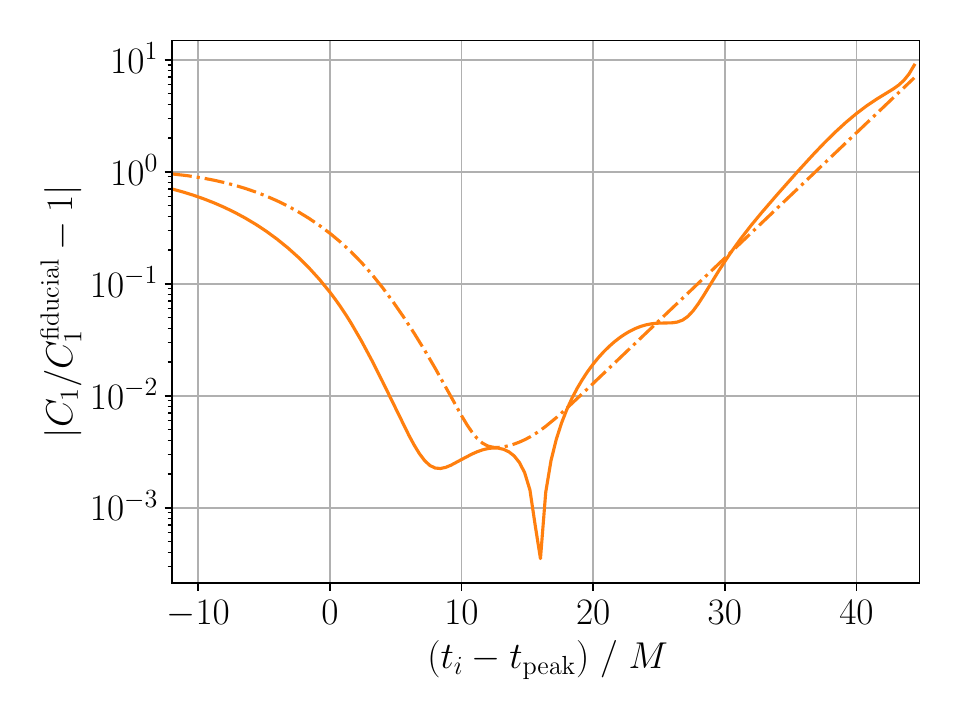}
    
    \includegraphics[width=0.32\textwidth]{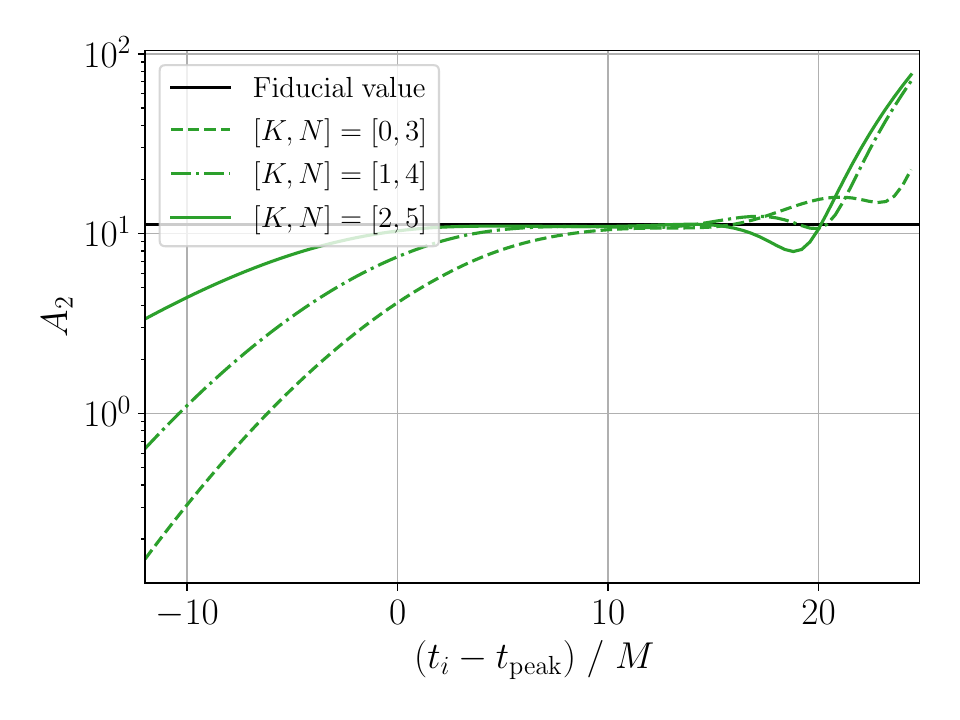}
    \includegraphics[width=0.32\textwidth]{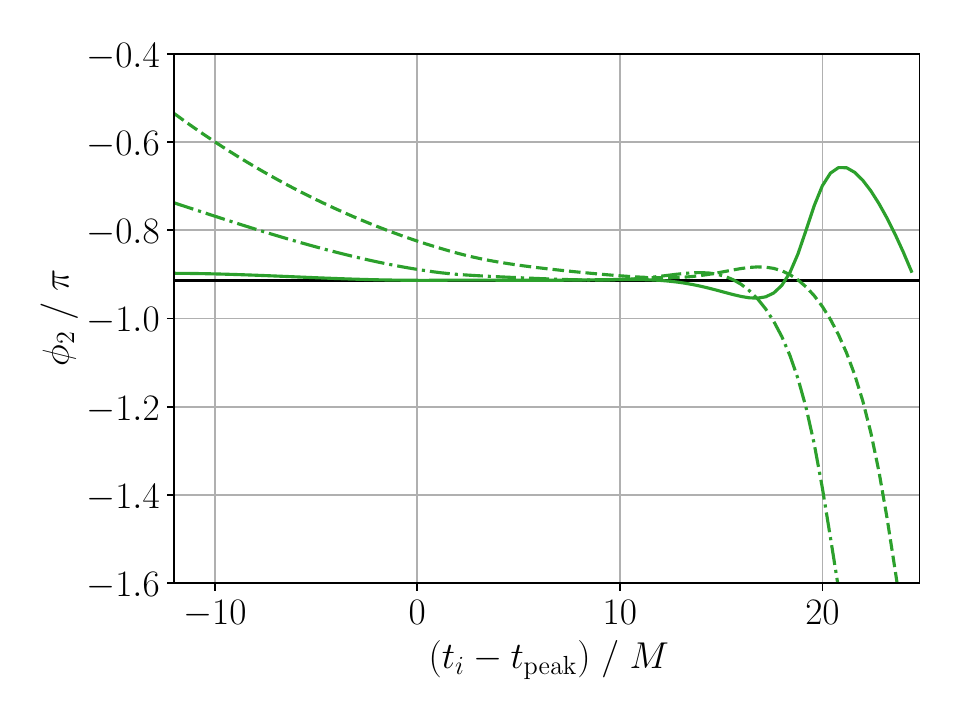}
    \includegraphics[width=0.32\textwidth]{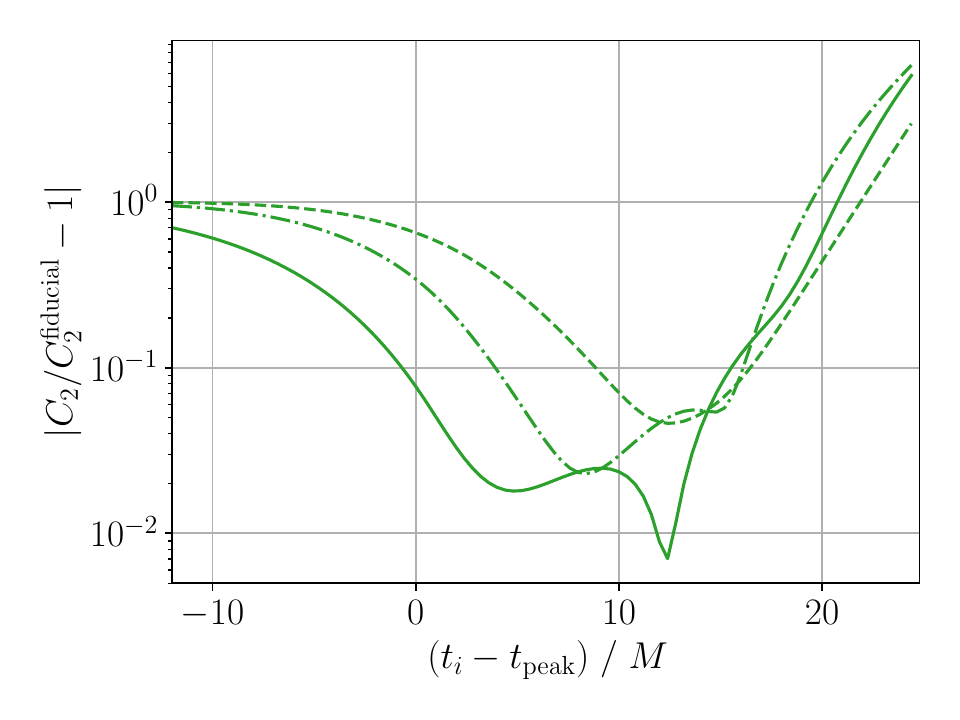}

    \includegraphics[width=0.32\textwidth]{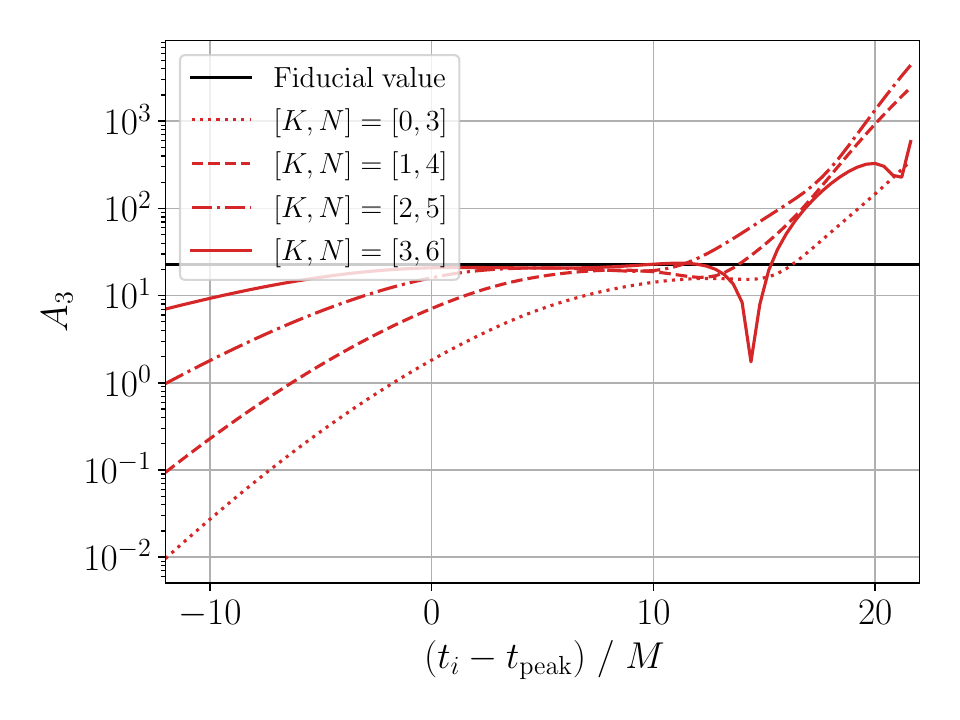}
    \includegraphics[width=0.32\textwidth]{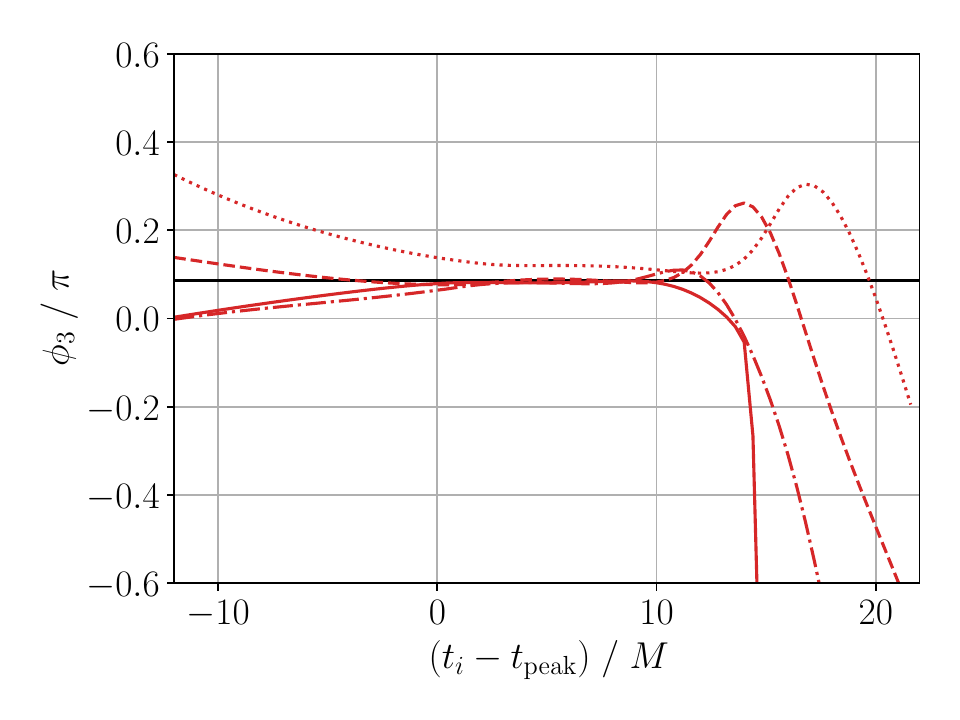}
    \includegraphics[width=0.32\textwidth]{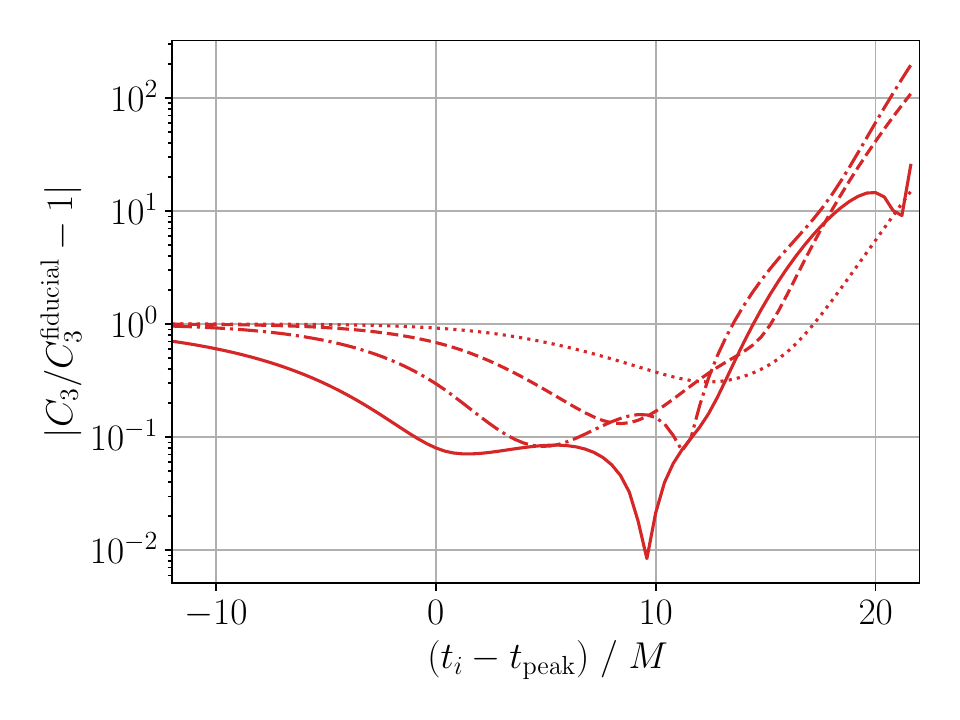}
    
    \includegraphics[width=0.32\textwidth]{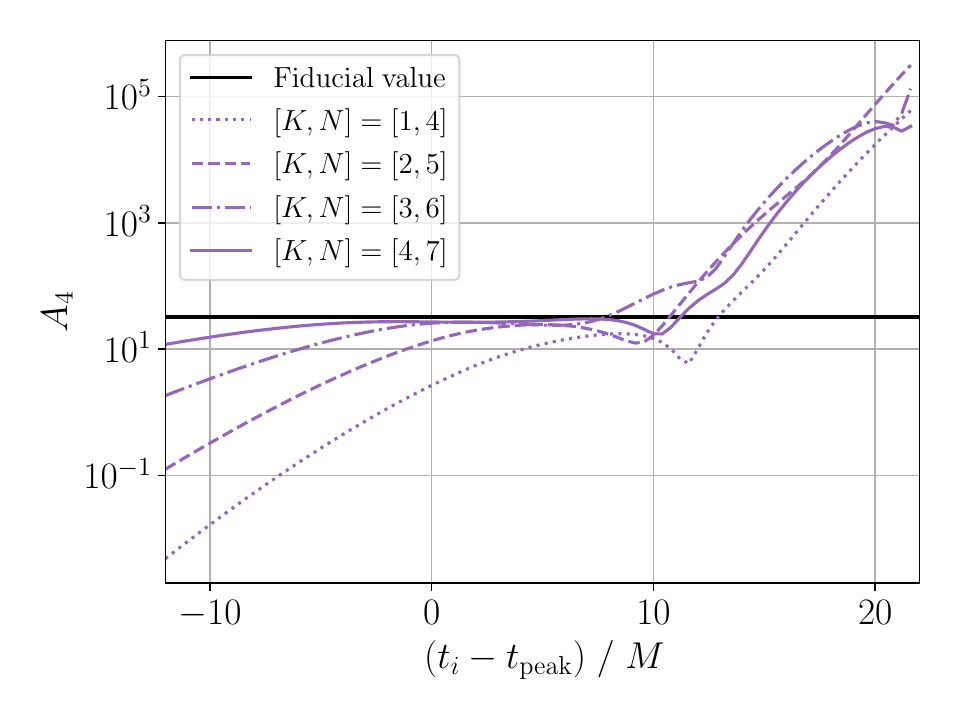}
    \includegraphics[width=0.32\textwidth]{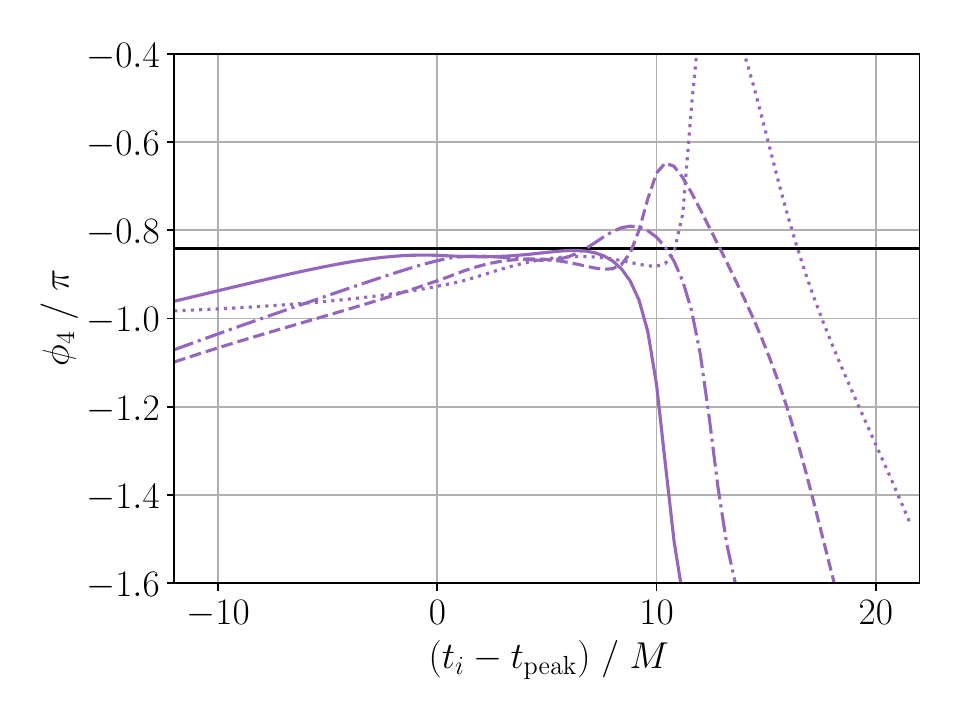}
    \includegraphics[width=0.32\textwidth]{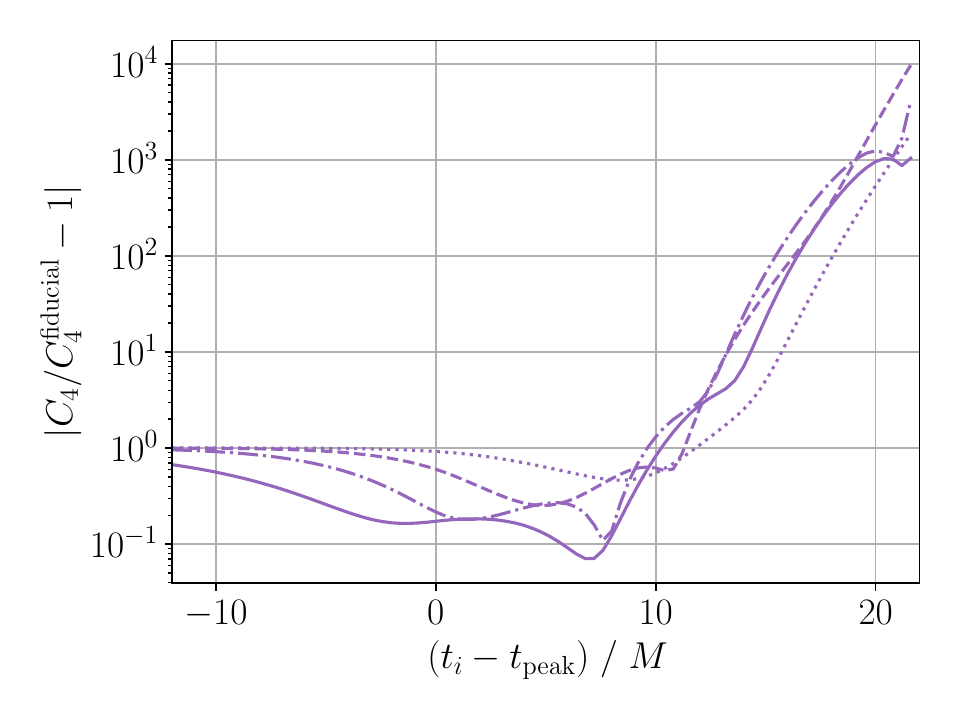}
    \caption{The improvement of the amplitude $A_n$ (left column), phase $\phi_n$ (middle column), and relative error of $C_n$ (right column) from $n=1$ mode (first row) to $n=4$ mode (fourth row) for the iterative fit of the mock waveform given in \eqref{eq:wave_mock_subn} by the fitting function $\psi^{\fit}_{[K,N]}$ with $N=K+3$.}
    \label{fig:0305_n4_improve_mock_naive}
\end{figure*}

The second and third panels of Fig.~\ref{fig:0305_M_A_Cre_gamma_mock_naive} show the amplitude $A_n$ and phase $\phi_n$ as a function of the start time of the fit $t_i$ for the fit of the mock waveform $\Psi_{[0,7]}^{(c)}$ by the fitting function $\psi_{[0,N]}^{\fit}$.
Compared to the fit of $\Psi_{[0,7]}$ without noise in Figs.~\ref{fig:0305_A_mock_sub_const} and \ref{fig:0305_phi_mock_sub_const}, the plateau is shortened due to the existence of noise.
For the $N=3$ case, the length of the plateau for $A_2$ and $\phi_2$ differs by about a factor of 2 with and without noise.
The plateau is also shortened for other $N$ cases shown in Fig.~\ref{fig:0305_A_mock_naive}. 
For instance, for the $N=1$ case, $A_1$ and $\phi_1$ remains almost constant until $(t_i-t_{\peak})/M\simeq 50$ in Figs.~\ref{fig:0305_A_mock_sub_const} and \ref{fig:0305_phi_mock_sub_const} but $(t_i-t_{\peak})/M\simeq 40$ in Fig.~\ref{fig:0305_A_mock_naive}.
The divergence tends to begin earlier with higher overtones.
As pointed out in \cite{Baibhav:2023clw}, the introduction of noise shortens the plateau and reduces the stability of the fit.

We present the relative error of $C_n$ in the fourth panel of Fig.~\ref{fig:0305_M_A_Cre_gamma_mock_naive} for $N=3$ and Fig.~\ref{fig:0305_A_mock_naive} for the all cases $N=0,\cdots,7$.
For the most modes, the behavior of the relative error of $C_n$ is similar to that of Fig.~\ref{fig:0305_Cre_mock_sub_const}.
For instance, $t_i$ at which the relative error reaches a minimum is roughly the same regardless of $N$, and the larger $N$ is, the earlier $t_i$ at which the relative error reaches a minimum becomes in the early ringdown period.
However, the differences show up in the fundamental mode for the fitting starting from late time.
For the $N=2$ case in Fig.~\ref{fig:0305_Cre_mock_sub_const} for the mock waveform without noise, the relative error in the fundamental mode has the minimum at $t_i$ that were significantly different from the minima in the other modes.
In contrast, in Fig.~\ref{fig:0305_A_mock_naive} for the mock waveform with noise, the relative error has minima at similar $t_i$ in all modes.
This originates from the fact that the waveform in the late time is contaminated by the noise. 
The tendency of the lower bound for the fitting starting from late time $t_i$ is also consistent with the behavior of the mismatch in Fig.~\ref{fig:0305_M_A_Cre_gamma_mock_naive}.

The bottom panels of Fig.~\ref{fig:0305_M_A_Cre_gamma_mock_naive} 
and Fig.~\ref{fig:0305_A_mock_naive} show the rate of change $\gamma_n$, where we see several differences from Fig.~\ref{fig:0305_gamma_mock_sub_const} for the case without noise.
The minima of $\gamma_n$ appear more sharply when the noise is present than when the noise is absent.
Also, a lower bound shows up at the late time, which is similar to the mismatch.
Further, there is no exceptional behavior of the minimum of $\gamma_0$, and all $\gamma_n$ take the minima in the same start time of the fit.

Even with the noise in the mock waveform, we confirm that the rate of change $\gamma_n$ for the next-longest-lived mode works well as an indicator of the optimal start time of the fit with the minimum relative error.
In Fig.~\ref{fig:0305_M_A_Cre_gamma_mock_naive},
we see that there exists a clear correspondence between the minimum of the relative error of $C_n$ 
and the minimum of the rate of change $\gamma_n$ for the next-longest-lived mode.
The correspondence also holds for other $N$ cases shown in Fig.~\ref{fig:0305_A_mock_naive}.
Compared to the amplitude $A_n$ and phase $\phi_n$ in Fig.~\ref{fig:0305_A_mock_naive}, we see that in each $N$, the minimum of $\gamma_n$ is located just before the highest overtone $A_n$ and $\phi_n$ begin to diverge.
It would be reasonable to consider that the minimum of $\gamma_n$ is located before the contamination of the fit by noise.
Thus, it is still efficient to use the minimum of $\gamma_n$ as an indicator for the survey of the best-fit value even when the noise is present.
Incidentally, there exist relatively sharp peaks just after $\gamma_n$ reaches a minimum, which are caused by the passage of $C_n$ near the origin on the complex plane, as confirmed by the dips in $A_n$ and the sudden changes in $\phi_n$.

Next, we perform the subtraction of the longest-lived mode.
While in \S\ref{sec:dosub} we assumed the ideal subtraction and used $\Psi_{[1,7]}$, in this section we subtract the longest-lived mode from the mock waveform by using the best-fit value of $C_n$.
Namely, the waveform after the $m$-th subtraction reads
\begin{equation}
\label{eq:wave_mock_subn}
    \Psi_{[0,7]}^{(c)}-\sum_{n=0}^{m-1} C^{\iter}_{n}\psi_{n} .
\end{equation}
Here, $C^{\iter}_{n}$ is the best-fit value of $C_n$ obtained by the fitting of the waveform after the $n$-th subtraction, where $n$-th overtone becomes the effective longest-lived mode.

The waveform after the first subtraction is shown by orange curve
in Fig.~\ref{fig:0305_wave_mock_naive}, where we can see that the damped sinusoid corresponding to the first overtone.
Furthermore, we can clearly see that the constant mimicking the noise and tail becomes dominant at the late time.
Compared to the mock waveform $\Psi_{[1,7]}$ in \S\ref{sec:mock} assuming the absence of the noise/tail and the ideal subtraction, it is clear that the noise shortens the region where we can observe the damped sinusoids.
We cannot fit the first overtone for the region where it is subdominant compared to the noise, and hence we have less data available for the fit.
Specifically, for the original mock waveform $\Psi_{[0,7]}^{(c)}$ before the subtraction we can observe the damped sinusoids up to $(t-t_{\peak})/M \simeq 140$, whereas for the waveform $\Psi_{[0,7]}^{(c)}-C^{\iter}_{0}\psi_{0}$ we can only see the damped sinusoids up to $(t-t_{\peak})/M \simeq 50$. 
This is because the first overtone appeared after the subtraction has shorter damping time than the fundamental mode and hence becomes comparable to the noise more quickly.
Thus, for the waveform with the noise, iterating the subtraction of the longest-lived mode, we have less data available for the fit of overtones. 
Consequently, for the iterative fitting method, we gradually decrease the end time of the fit $t_e$ at each step.
Specifically, we set $(t_e-t_{\peak})/M = 50$ for the fit of the waveform $\Psi_{[0,7]}^{(c)}-C^{\iter}_{0}\psi_{0}$. 

To the mock waveform $\Psi_{[0,7]}^{(c)}-C^{\iter}_{0}\psi_{0}$ obtained after the first subtraction, we apply the fitting algorithm with the fitting function $\psi_{[1,N]}^{\fit}$, and present the results in Fig.~\ref{fig:0305_M_A_Cre_gamma_mock_sub_0}.
Again, for the amplitude, phase, relative error, and the rate of change, we display $N=4$ case in Fig.~\ref{fig:0305_M_A_Cre_gamma_mock_sub_0} and the all cases $N=1,\cdots,7$ in Fig.~\ref{fig:0305_A_mock_sub_0} in Appendix~\ref{sec:plots}.

The first panel of Fig.~\ref{fig:0305_M_A_Cre_gamma_mock_sub_0} shows the mismatch.
Compared to the fit of the original waveform before the subtraction of the fundamental mode in Fig.~\ref{fig:0305_M_A_Cre_gamma_mock_naive}, there are no significant difference for the range $0<(t_i-t_{\peak})/M\lesssim 20$.
However, the tilt of the lower bound of the mismatch becomes larger around $(t_i-t_{\peak})/M\simeq 20$. 
This is caused by the shortening of the interval available for the fit due to the noise appeared after the subtraction of the fundamental mode.
The position of the start time of the fit $t_i$ where the mismatch takes a minimum value is almost unchanged.

We present the amplitude $A_n$ and phase $\phi_n$ in the second and third panels of Fig.~\ref{fig:0305_M_A_Cre_gamma_mock_sub_0} for $N=4$ and Fig.~\ref{fig:0305_A_mock_sub_0} for the all cases $N=1,\cdots,7$.
Compared to the fit of the original mock waveform before the subtraction of the fundamental mode, we see that the plateau of the first overtone is slightly extended.
As for the second overtone, the plateau is even more extended.
However, the improvement is not as large as that in the case of the waveform without noise, i.e., from Figs.~\ref{fig:0305_A_mock_sub_const} and \ref{fig:0305_phi_mock_sub_const} to Figs.~\ref{fig:0305_A_mock_sub_const0} and \ref{fig:0305_phi_mock_sub_const0}.
This is because at the late time we are actually trying to fit the constant noise, which is dominant over the fiducial damped sinusoids, by the fitting function of a superposition of damped sinusoids.
Such inappropriate fitting inevitably leads to exponential divergence of the coefficients determined by the damping rate of each mode.
While we observed this phenomenon for the fit of the higher overtones of the mock waveform without noise, in the case of the mock waveform with noise, it also occurs for the fit of the longest-lived mode, which prevents a significant extension of the plateau.

\begin{figure}[t]
    \centering
    \includegraphics[width=\columnwidth]{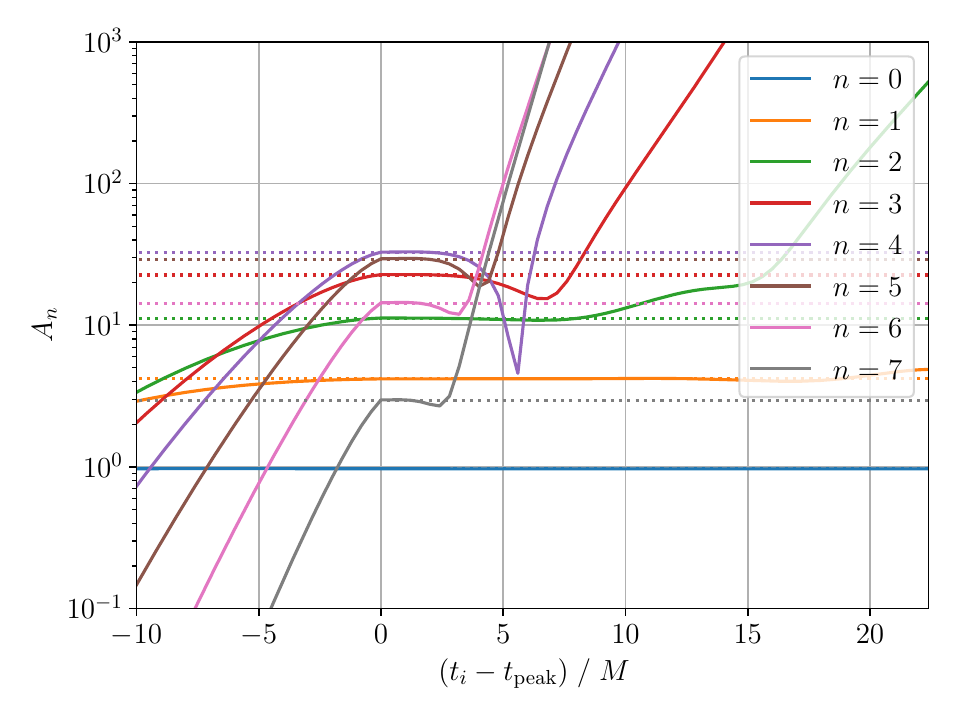}
    \includegraphics[width=\columnwidth]{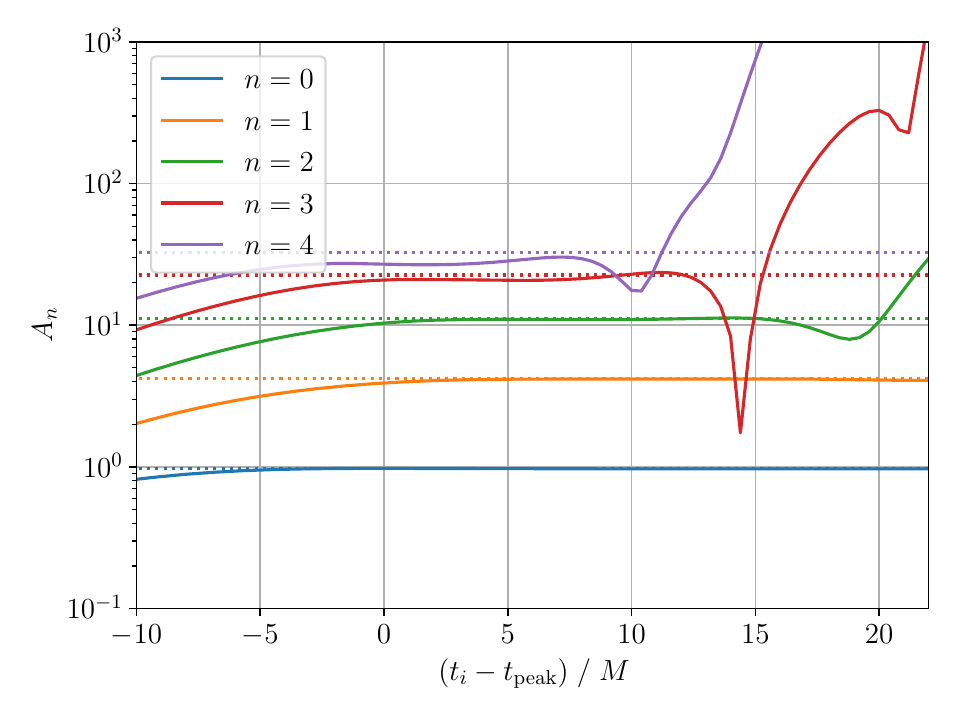}
    \caption{Comparison of the amplitude $A_n$ obtained by fitting the mock waveform $\Psi^{(c)}_{[0,7]}$ by the conventional method (top) with the fitting function $\psi^\fit_{[0,7]}$ and the iterative method (bottom) with the fitting function $\psi^\fit_{[K,N]}$ with $K=0,\cdots, 4$ and $N=K+3$. The dotted lines represent the fiducial values.}
    \label{fig:0305_An_improve_mock_naive_N4}
\end{figure}

\begin{figure}[t]
    \centering
    \includegraphics[width=\columnwidth]{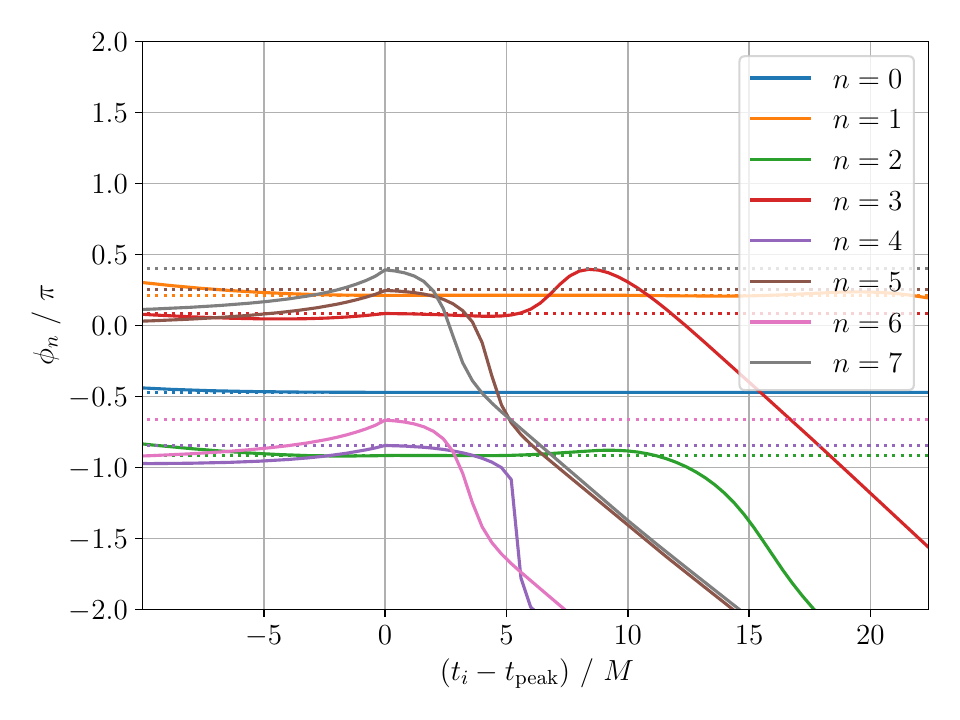}
    \includegraphics[width=\columnwidth]{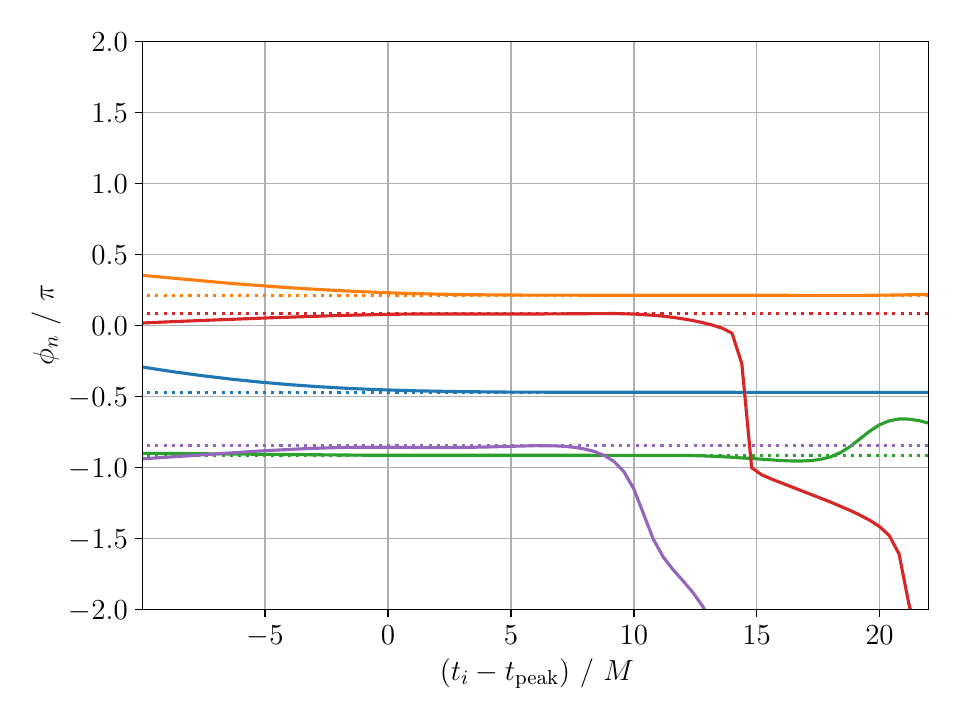}
    \caption{Comparison of the phase $\phi_n$ obtained by fitting the mock waveform $\Psi^{(c)}_{[0,7]}$ by the conventional method (top) with the fitting function $\psi^\fit_{[0,7]}$ and the iterative method (bottom) with the fitting function $\psi^\fit_{[K,N]}$ with $K=0,\cdots, 4$ and $N=K+3$. The dotted lines represent the fiducial values.}
    \label{fig:0305_phin_improve_mock_naive_N4}
\end{figure}

We present the relative error of $C_n$ and the rate of change $\gamma_n$ in the fourth and fifth panels of Fig.~\ref{fig:0305_M_A_Cre_gamma_mock_sub_0} for $N=4$ and Fig.~\ref{fig:0305_A_mock_sub_0} for the all cases $N=1,\cdots,7$, respectively.
As mentioned above, we determine the best-fit value by using the start time of the fit where $\gamma_n$ for the next-longest-lived mode takes the minimum. 
Comparing the fit of the mock waveform before and after the subtraction of the fundamental mode, the relative errors of the best-fit value are more or less the same for the longest-lived mode, the next-longest-lived mode, and so on.
In other words, focusing on the same $n$-th mode, the relative error reduces by virtue of the subtraction.
Also, we can see that the plateau is extended.
For $n=1$ mode, the region where $\gamma_1<10^{-2}$ is clearly extended from Fig.~\ref{fig:0305_M_A_Cre_gamma_mock_naive} to Fig.~\ref{fig:0305_M_A_Cre_gamma_mock_sub_0}.
For the relative error of $C_1$, the dip shows up in Fig.~\ref{fig:0305_M_A_Cre_gamma_mock_sub_0} at late time, but the relative errors of other modes do not take the minimum there.
We also observed such peculiar dip in the analysis of the mock waveform without noise in \S\ref{sec:mock}.
Even with the noise, our criteria still work and we can determine the optimal start time of the fit by the minimum of $\gamma_n$ for the next-longest-lived mode.

Having confirmed that our strategy works well, we iterate the subtraction of the longest-lived mode and obtained the best-fit value for the coefficient $C_n$ for $n=0,\cdots,4$.
We highlight the improvement of the fit of each mode by virtue of the subtraction in Fig.~\ref{fig:0305_n4_improve_mock_naive}, which should be contrasted with Fig.~\ref{fig:0305_n4_improve_mock_sub_const} for the case assuming the ideal subtraction and the absence of noise.
We can see that the plateau is extended in Fig.~\ref{fig:0305_n4_improve_mock_naive}, but the improvement is less significant than Fig.~\ref{fig:0305_n4_improve_mock_sub_const}.
Clearly, this is due to the existence of the constant, which places the upper bound for the plateau.

\begin{figure}[t]
    \centering
    \includegraphics[width=\columnwidth]{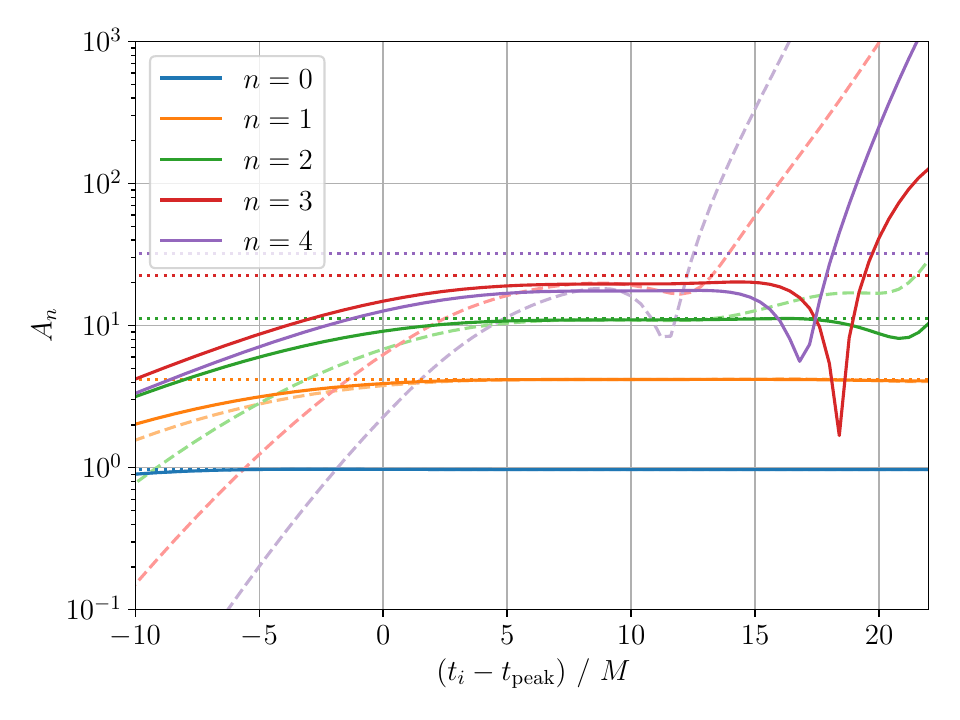}
    \includegraphics[width=\columnwidth]{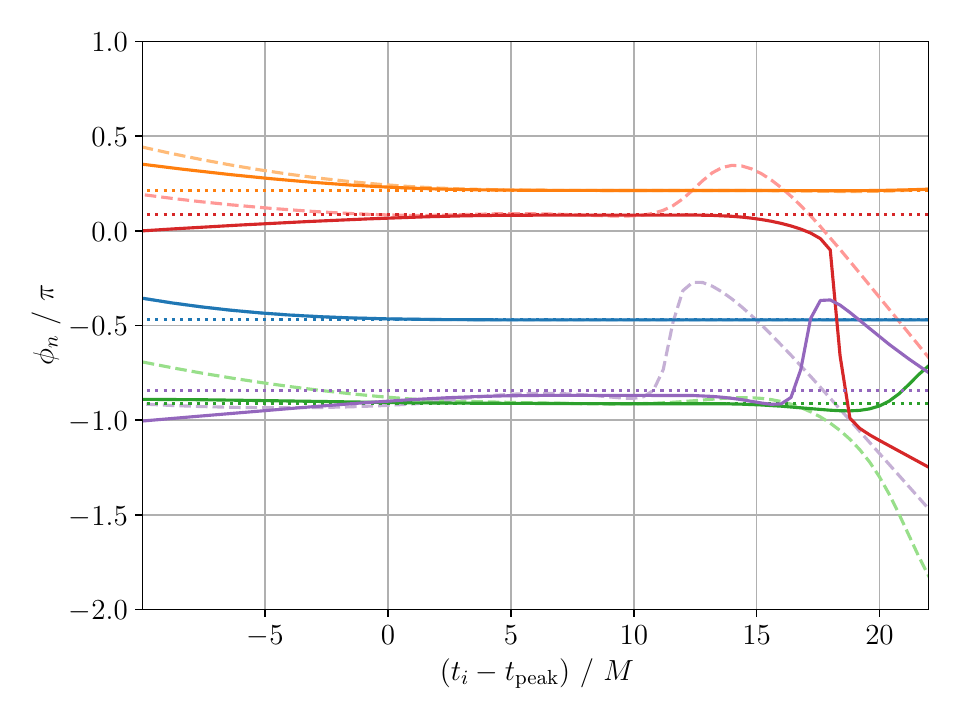}
    \caption{Comparison of the amplitude $A_n$ (top) and phase $\phi_n$ (bottom) obtained by fitting the mock waveform $\Psi^{(c)}_{[0,7]}$ by the conventional method (dashed) with the fitting function $\psi^\fit_{[0,4]}$ and the iterative method (solid) with the fitting function $\psi^\fit_{[K,4]}$ with $K=0,\cdots, 4$. The dotted lines represent the fiducial values.}
    \label{fig:0305_An_improve_mock_w_const_K0-4_N4}
\end{figure}

\begin{figure*}[t]
    \centering
    \includegraphics[width=\columnwidth]{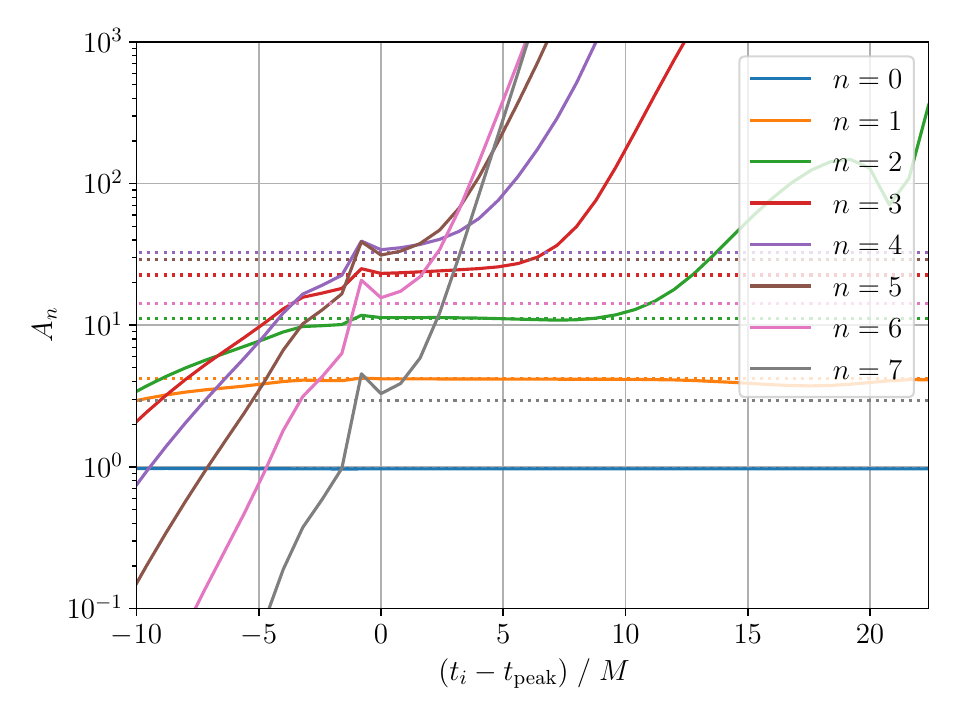}
    \includegraphics[width=\columnwidth]{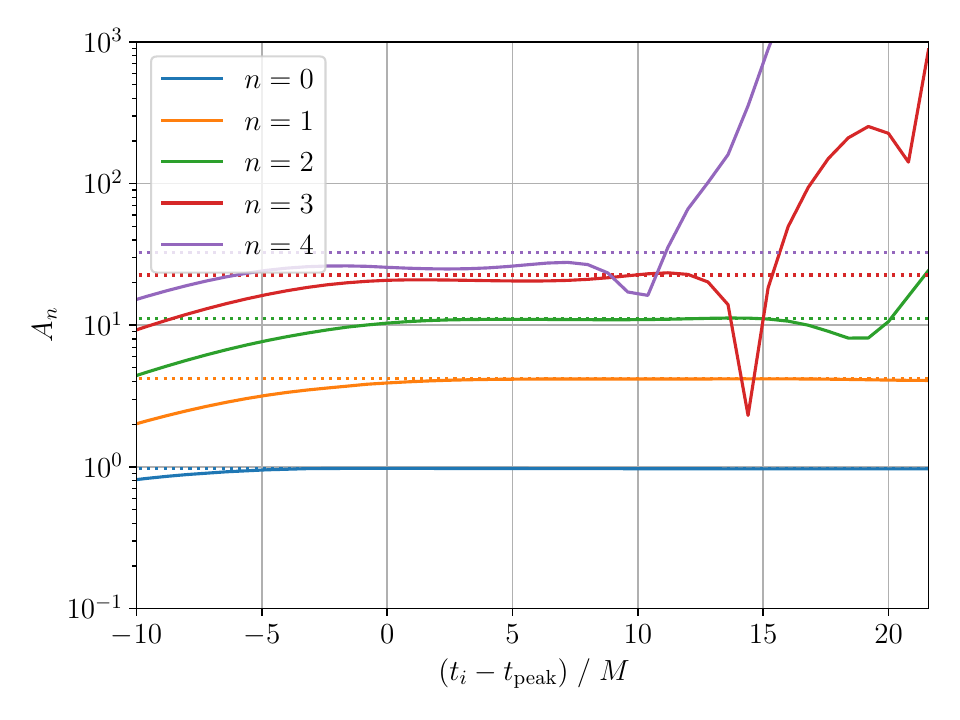}
    \includegraphics[width=\columnwidth]{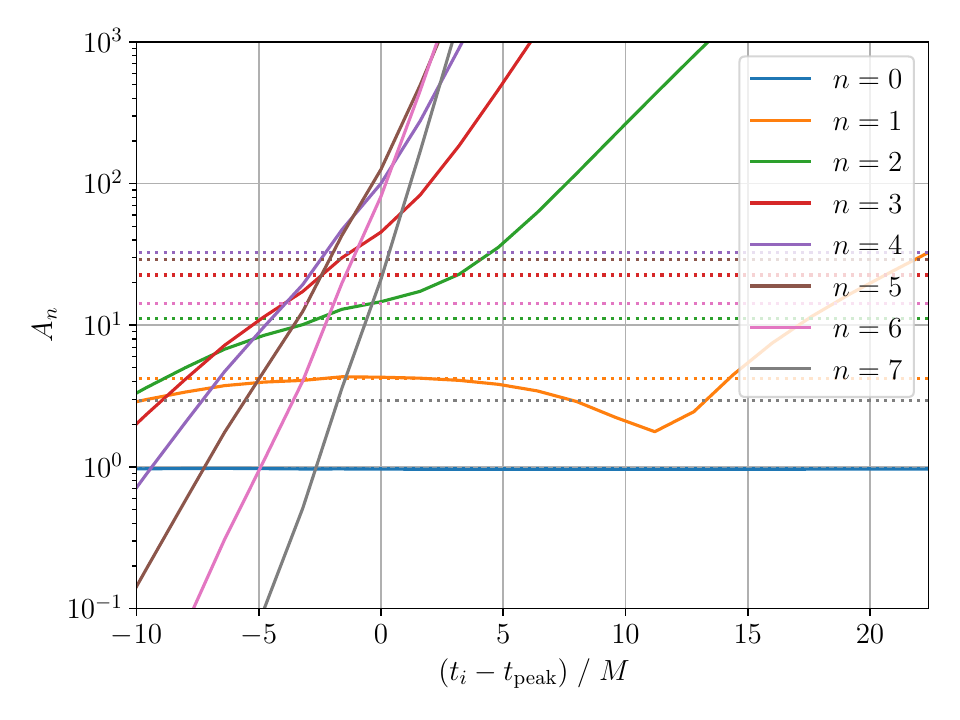}
    \includegraphics[width=\columnwidth]{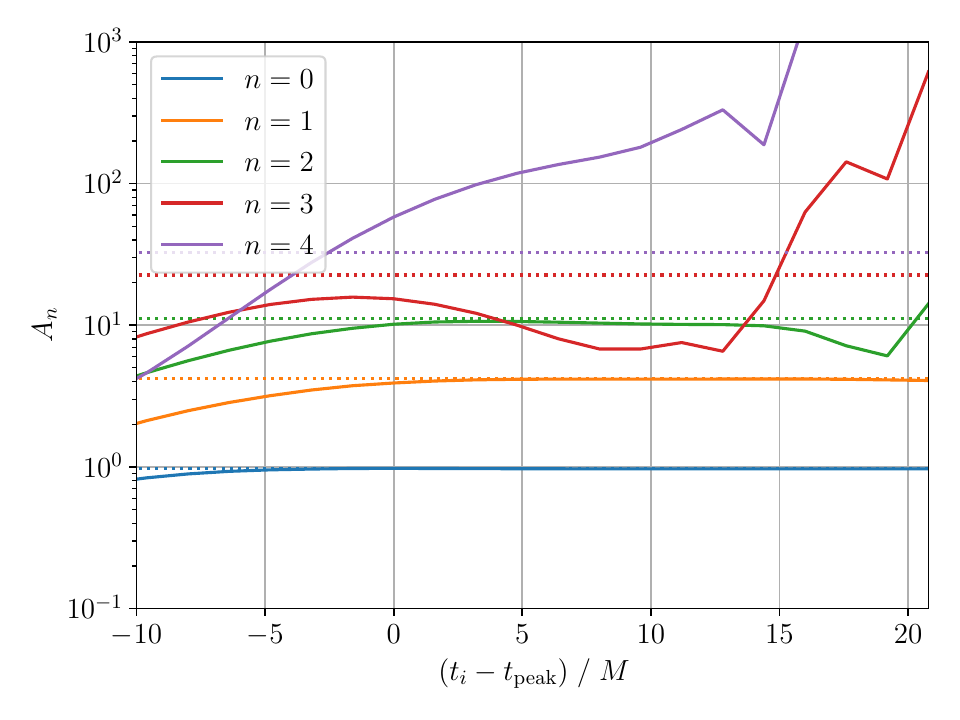}
    \includegraphics[width=\columnwidth]{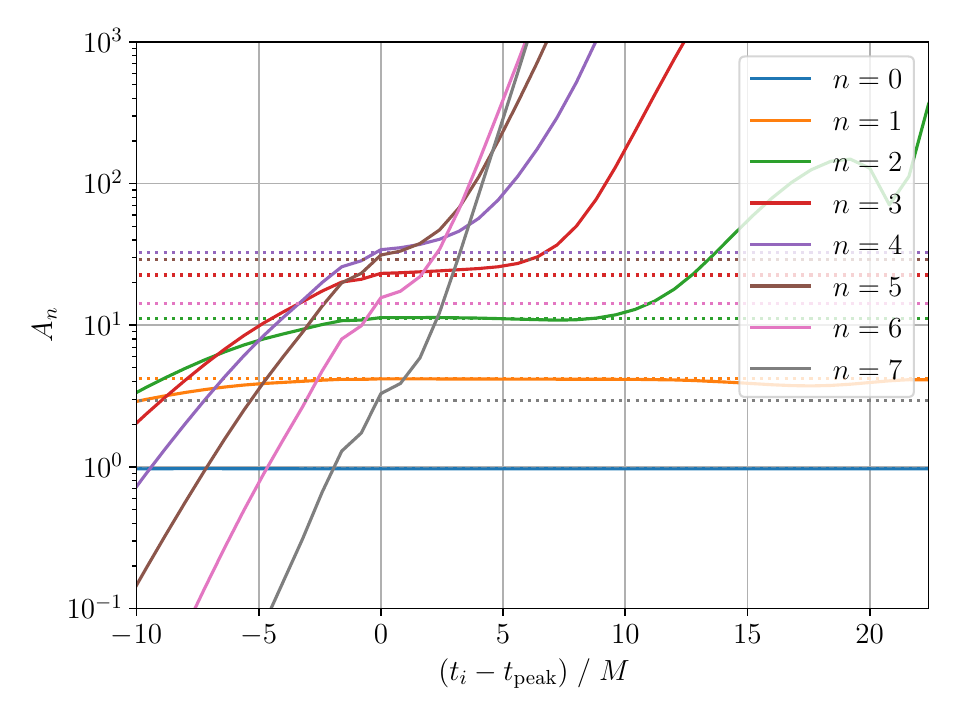}
    \includegraphics[width=\columnwidth]{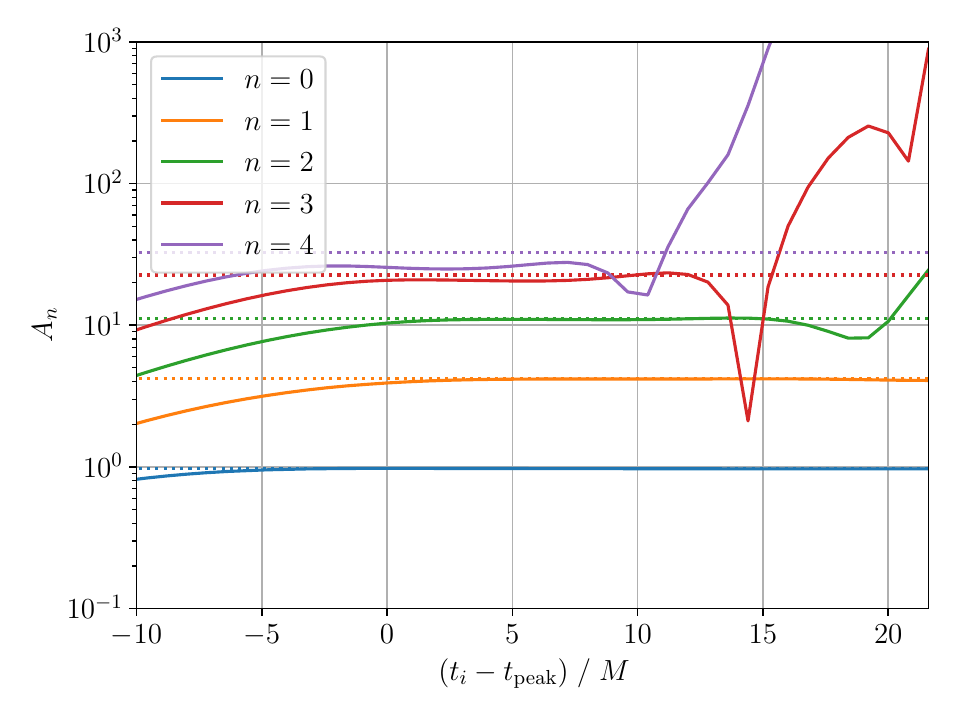}
    \caption{
    Same plots as in Fig.~\ref{fig:0305_An_improve_mock_naive_N4} for the mock waveform data $\Psi^{(c)}_{[0,7]}$ with reduced number of sampling points and its interpolation.  Conventional fit (left column) and iterative fit (right column) for data 1 (top row) with half sampling points, data 2 (middle row) with quarter sampling points, and data 3 (bottom row) generated by the interpolation of the data 2. 
    }
    \label{fig:0305_An_improve_mock_naive_N4_samplingrate}
\end{figure*}

\begin{figure*}[t]
    \centering
    \includegraphics[width=\columnwidth]{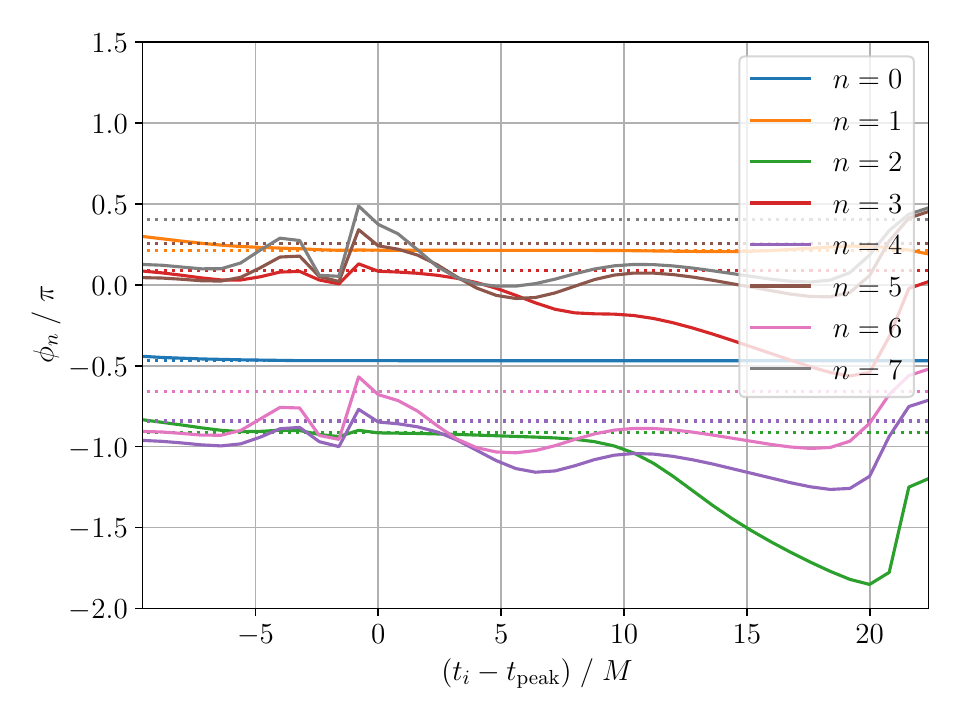}
    \includegraphics[width=\columnwidth]{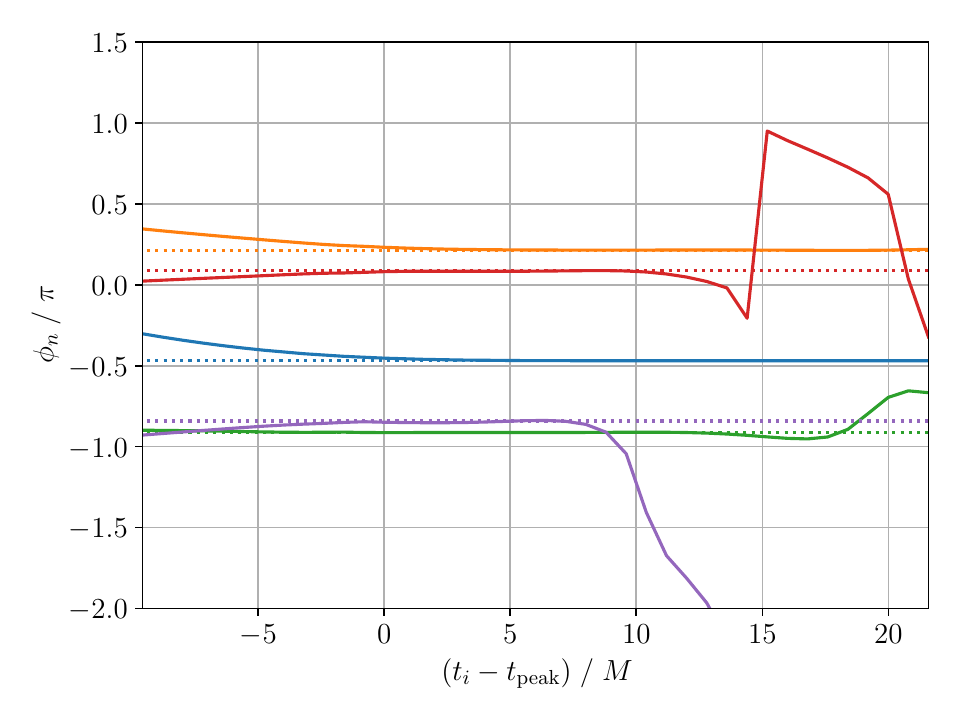}
    \includegraphics[width=\columnwidth]{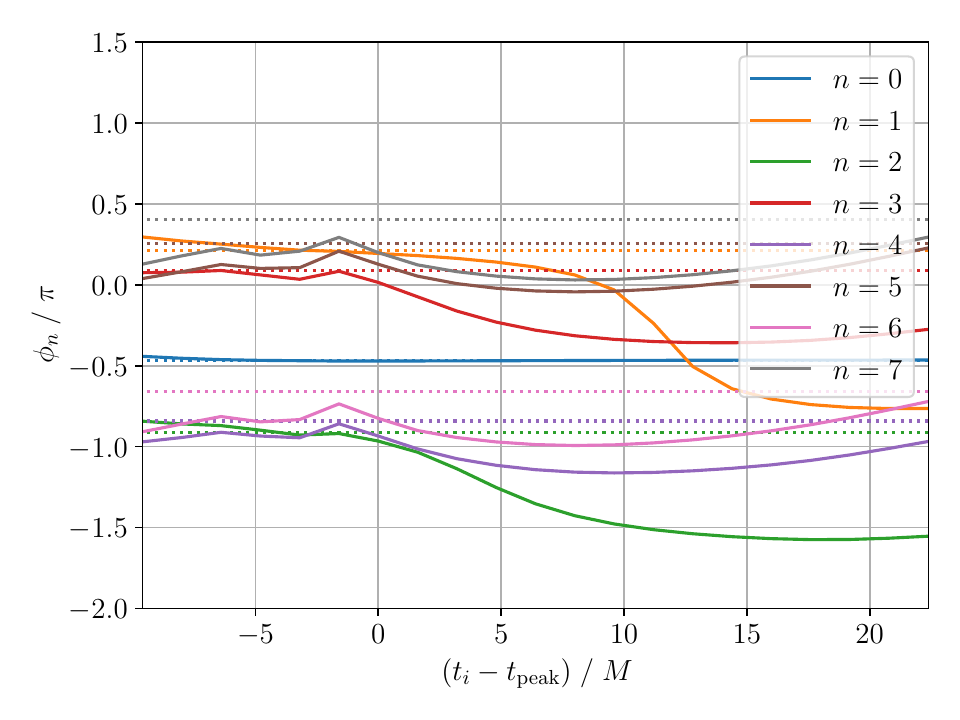}
    \includegraphics[width=\columnwidth]{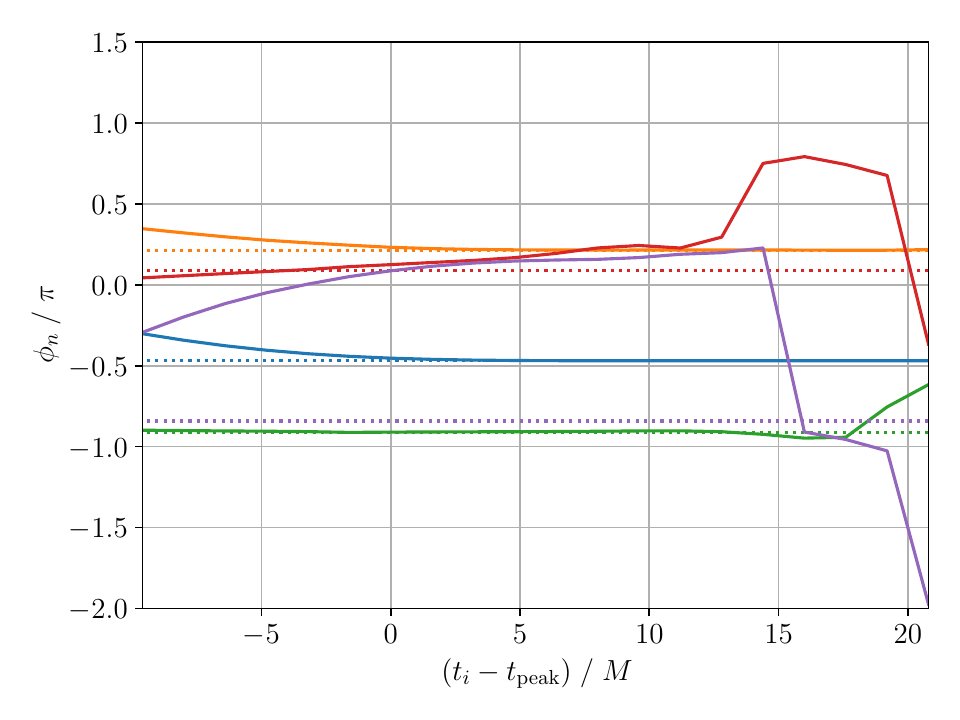}
    \includegraphics[width=\columnwidth]{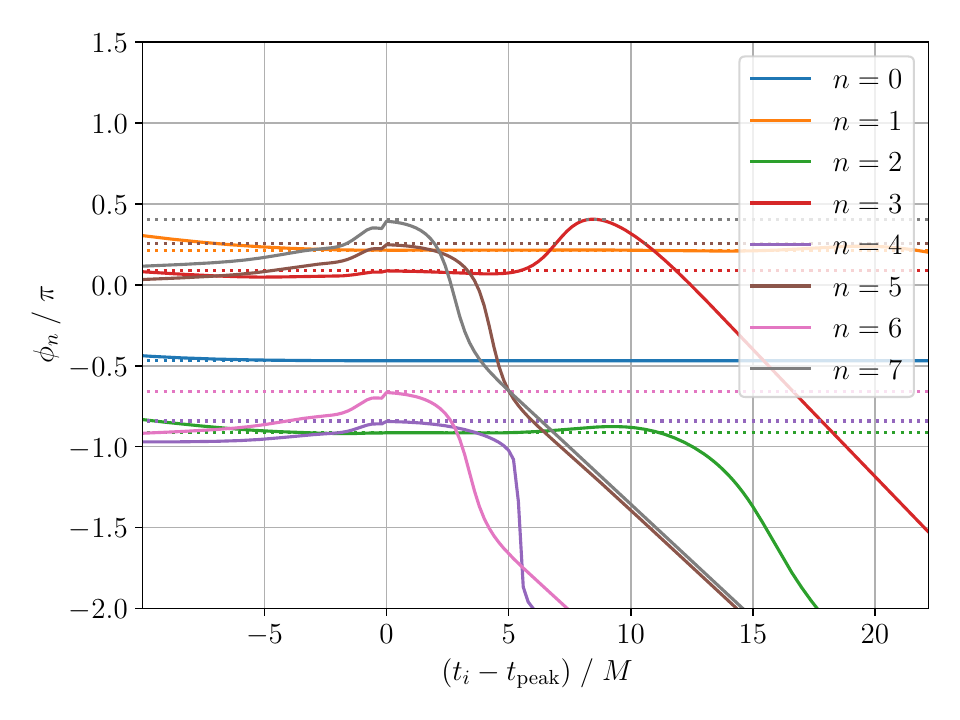}
    \includegraphics[width=\columnwidth]{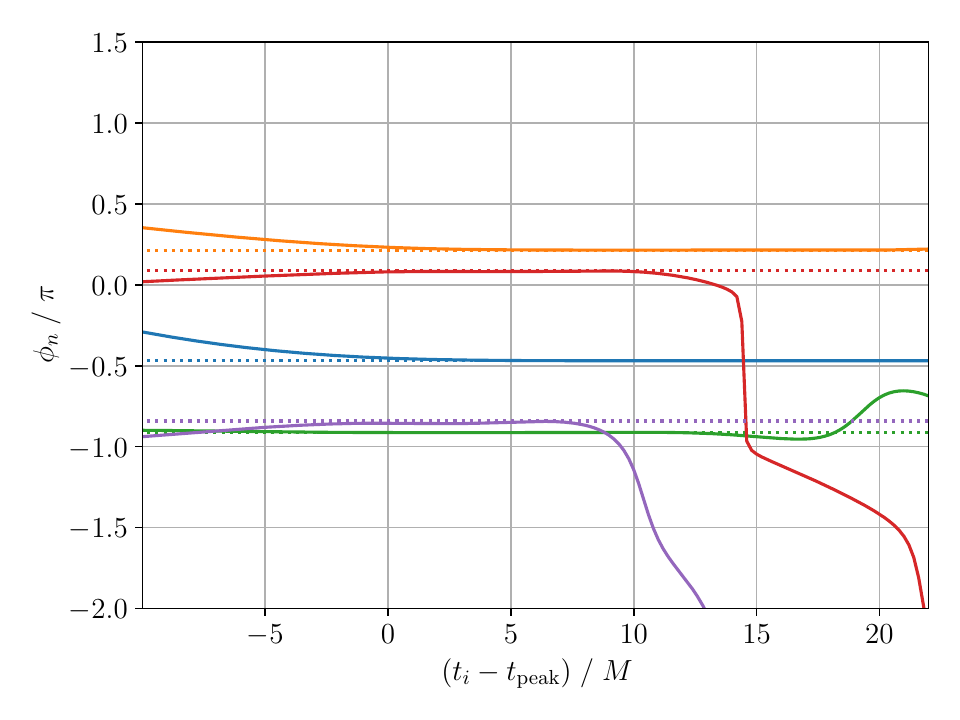}
    \caption{
    Same plots as in Fig.~\ref{fig:0305_phin_improve_mock_naive_N4} for the mock waveform data $\Psi^{(c)}_{[0,7]}$ with reduced number of sampling points and its interpolation.  Conventional fit (left column) and iterative fit (right column) for data 1 (top row) with half sampling points, data 2 (middle row) with quarter sampling points, and data 3 (bottom row) generated by the interpolation of the data 2. 
    }
    \label{fig:0305_phin_improve_mock_naive_N4_samplingrate}
\end{figure*}

Such behavior of the plateau can also be observed in Figs.~\ref{fig:0305_An_improve_mock_naive_N4}--\ref{fig:0305_An_improve_mock_w_const_K0-4_N4}, where we compare the conventional and iterative fit in the two ways described in \S\ref{sec:dosub}.
Although the iterative fit provided very long plateau for the case assuming the ideal subtraction and the absence of noise, in a more realistic case of the mock waveform with constant and performing numerical subtraction, the extension of the plateau by the iterative fit in Figs.~\ref{fig:0305_An_improve_mock_naive_N4}--\ref{fig:0305_An_improve_mock_w_const_K0-4_N4} is limited compared to Fig.~\ref{fig:0305_An_improve_mock_sub_const}--\ref{fig:0305_An_improve_mock_sub_const_K0-4_N4}.
Having said that, we can see that the iterative fitting method helps us to improve the stability of the fit of the overtones for $n\geq 2$ compared to the conventional fit.
Apparently, while the plateau is short, the conventional fit in Fig.~\ref{fig:0305_An_improve_mock_naive_N4} seems to succeed the extraction of the fiducial amplitude for the overtones.
This is because in this case we are fitting a superposition of the seven damped sinusoids by a superposition of the same number of damped sinusoids. 
Indeed, in Fig.~\ref{fig:0305_An_improve_mock_w_const_K0-4_N4} and also in Figs.~\ref{fig:0305_A_mock_naive} and \ref{fig:0305_A_mock_sub_0} in Appendix~\ref{sec:plots}, there is no plateau for overtones in the conventional method when the number of modes included in the fitting function is fewer.
As we discussed, in a realistic case, we need to fit a superposition of infinite number of damped sinusoids by a superposition of finite number of damped sinusoids. 
Also, the nonlinear effects become dominant at early time, and prevent the stable extraction of overtones. 
Hence, the overtones in the conventional fit are not likely to possess even short plateau. 
We shall see in \S\ref{sec:nr} that this is indeed the case for the fit of the numerical relativity waveform (see e.g., Figs.~\ref{fig:0305_An_improve_sub_const_N4} and \ref{fig:0305_An_improve_sub_const_K0-4_N4}). 

In Fig.~\ref{fig:0305_An_improve_mock_w_const_K0-4_N4} for the alternative way of comparison, the plateau is also extended by the iterative fit. 
However, from Fig.~\ref{fig:0305_An_improve_mock_w_const_K0-4_N4} we note that while the plateau in the phase $\phi_n$ is consistent with the fiducial value, the plateau in the amplitude $A_n$ is off from the fiducial value for higher overtones, with the difference in $A_4$ reaching approximately a factor of $2$.
On the other hand, in Fig.~\ref{fig:0305_An_improve_mock_naive_N4}, $A_4$ extracted by the iterative method is close to the fiducial value.
This is because we restrict ourselves to use the fitting function consisting of the modes up to fourth overtone in Fig.~\ref{fig:0305_An_improve_mock_w_const_K0-4_N4}, whereas we use the modes up to seventh overtone in Fig.~\ref{fig:0305_An_improve_mock_naive_N4}.

Finally, let us discuss the stability of the fit with respect to the sampling rate of the waveform data. 
So far, we use a discrete data $(t_a, \Psi^{(c)}_{[0,7]}(t_a) )$ of the mock waveform with the same sampling times as BBH:SXS:0305, which we denote waveform data 0.
The time data spacing is not exactly constant but remains $\Delta t_a /M \approx \mO(10^{-1})$ for the ringdown phase.
In general, even with the same waveform, sampling rate affects the fitting, i.e., different sampling rates can yield different fitting results~\cite{Wang:2023xsy}.
Based on the waveform data 0, $(t_a, \Psi^{(c)}_{[0,7]}(t_a))$ with $a=0,\cdots,N_{\rm samp}$, we generate the following three sets of data with reduced sampling rates:  
\begin{enumerate}
\item Waveform data 1: data with the half number of sampling points $\lfloor N_{\rm samp}/2 \rfloor$, using even-numbered samples of data 0.
\item Waveform data 2: data with the quarter number of sampling points $\lfloor N_{\rm samp}/4 \rfloor$, using even-numbered samples of data 1.
\item Waveform data 3: data with the number of sampling points $2\times \lfloor N_{\rm samp}/4 \rfloor - 1$ doubled by the cubic spline interpolation of data 2.
\end{enumerate}

In Figs.~\ref{fig:0305_An_improve_mock_naive_N4_samplingrate} and \ref{fig:0305_phin_improve_mock_naive_N4_samplingrate}, we show the amplitude $A_n$ and phase $\phi_n$ obtained by the conventional (left column) and iterative fitting (right column) between the above three different sampling rates of the mock waveform data. 
Compared to the results for the data 0 in Figs.~\ref{fig:0305_An_improve_mock_naive_N4} and \ref{fig:0305_phin_improve_mock_naive_N4}, the fitting becomes unstable as the sampling rate decreases.
Further, these figures highlight the efficiency of the iterative fitting.
For the data 1 (top rows), the plateau in the conventional fit disappears, whereas the iterative fit still keep almost same structure as the result for the data 0. 
Compared to the conventional fitting, the iterative fitting is able to extract the fiducial values in a stable manner even for data with low sampling rates.
Of course, as the number of sampling points is reduced, the iterative fitting eventually break down as well.
For the data 2 (middle rows), both fitting methods fail to extract the fiducial values.
One could encounter in such a situation for the analysis of the low resolution data, so we check the fitting of the data 3, generated by the interpolation of the data 2.
From the bottom rows of Fig.~\ref{fig:0305_An_improve_mock_naive_N4_samplingrate} and \ref{fig:0305_phin_improve_mock_naive_N4_samplingrate}, we see that the fitting of the interpolated data can recover the result of data 1.

The lessons from the mock waveform analysis in this section is that the constant, which mimics noise and/or tail in observed ringdown signal, is dominant for the late time and hence affects the fit if the start time of the fit is late.
The iterative fitting method helps us to extract higher overtones, although, due to the noise, the region where the meaningful data is available becomes shorter and shorter as we iterate the subtractions.
The noise places the upper bound for the plateau, and the improvement of the stability of the fit by the iterative subtractions is limited compared to the case without the noise.
Up to which overtone a sufficiently long plateau can be obtained depends on the magnitude of the noise and the damping time of the overtone.
While we dealt with the mock waveform with a constant in this section, these lessons are robust even for the random noises and/or power-law tail.
Further, we confirm that the iterative fitting method is robust to the data with low sampling rates, compared to the conventional fit.

\section{Numerical relativity waveform}
\label{sec:nr}

From the analyses of the mock waveform in \S\ref{sec:mock} and \ref{sec:dawnoise}, we figure out that we can improve the stability of the fit of the overtones by iteratively subtracting the longest-lived mode step by step from the ringdown waveform. 
It is because the longest-lived mode can be most stably fitted, and the next-longest-lived mode takes over the same feature after the subtraction.

In this section, we apply the iterative fitting method to numerical relativity waveform in the SXS catalog listed in Table~\ref{tab:SXS_BBH_spin} and explore the fit of the $(\ell,m)=(2,2)$ mode of these waveforms. 
The Kerr QNM frequencies corresponding to each value of the remnant dimensionless spin are listed in Table~\ref{tab:0305_QNMs}--\ref{tab:1108_QNMs} in Appendix~\ref{sec:qnms}.
We describe the estimation of the numerical errors in Appendix~\ref{sec:errors}.

First, we analyze SXS:BBH:0305, which is well-known for its similarity to GW150914~\cite{Abbott:2016blz}, the first observed gravitational waves from BBH.
Highlighting differences from the analysis in \S\ref{sec:mock} and \ref{sec:dawnoise}, the analysis of SXS:BBH:0305 in this section helps us to understand how the nonlinearity and/or spherical-spheroidal mixing affect the results of fitting.
Since previous work have mainly analyzed SXS:BBH:0305, the analysis of SXS:BBH:0305 is ideal for comparing and examining the validity of our iterative fitting method.

In addition to the SXS:BBH:0305, we also investigate high spin case SXS:BBH:0158 and low spin case SXS:BBH:0156 for comparison.
We choose them as the equal mass binaries having the highest/lowest remnant dimensionless spin among the SXS simulations with high resolutions.
Further, we also pick up another low spin simulation, SXS:BBH:1108, which is the simulation with the parameters similar to the observed event GW190814~\cite{LIGOScientific:2020zkf}.
While the numerical errors in these simulations are larger than in SXS:BBH:0305, it is interesting to investigate how the efficiency of the ringdown fitting depends on the black hole spin.

It is known that most of the Kerr QNM frequencies, except for the fifth overtone, have higher frequency and longer damping time for higher spin Kerr black holes, and they become eventually degenerate as they approach the accumulation point at the extremal limit~\cite{Detweiler:1980gk,Cardoso:2004hh,Hod:2008zz,Yang:2012pj,Yang:2013uba}.
On the one hand, slowly damping and fast oscillating ringdown signal would be more easily observed.  
On the other hand, damped sinusoids with almost degenerate frequencies and damping time would be hard to be resolved.
In contrast, the opposite applies to the QNMs for lower spin black holes.
They damp faster and oscillate more slowly, which would restrict the range of the time domain data available for the fitting.
However, the differences between the QNM frequencies are more distinct for lower spin black holes, and may be more easily distinguished.  
The fitting analysis of the waveforms with various values of black hole spin would clarify these points.

\begin{table}[t]
    \centering
    \caption{Remnant dimensionless spin $\chi_{\rm rem}$, large mass ratio $q$, and resolution level of the SXS numerical simulations~\cite{Mroue:2013xna,Boyle:2019kee} examined in \S\ref{sec:nr}.}
    \begin{tabular}{cccc}\hline\hline
       ID ~&~ $\chi_{\rm rem}$ ~&~ $q$ ~&~ Lev \\ \hline
       SXS:BBH:0305 ~&~ 0.6921 ~&~ 1.221 ~&~ 6 \\ 
       SXS:BBH:0158 ~&~ 0.9450 ~&~ 1.000 ~&~ 6 \\ 
       SXS:BBH:0156 ~&~ 0.3757 ~&~ 1.000 ~&~ 5 \\ 
       SXS:BBH:1108 ~&~ 0.2772 ~&~ 9.200 ~&~ 5 \\ 
       \hline\hline
    \end{tabular}
    \label{tab:SXS_BBH_spin}
\end{table}

We denote the best-fit waveform by using the iterative fitting method as
\be \psi^{\iter}_{[0,N]} = \sum_{n=0}^{N} C^{\iter}_n \psi_n . \ee
Here, $\psi_n = e^{-i\omega_n t}$ is the damped sinusoid with QNM frequencies $\omega_n$ of $(\ell,m)=(2,2)$ mode of gravitational waves emitted from a Kerr black hole with the remnant dimensionless spin of each SXS simulations, and $C^{\iter}_n$ is the best-fit value coefficient obtained by each step of the iterative procedure.
Namely, first, we fit the original SXS waveform $\Psi^\sxs = (h_+ - i h_\times)r/M$ by a fitting function $\psi^{\textrm{fit}}_{[0,7]}$ given in \eqref{eq:fit} with the fitting algorithm described in \S\ref{sec:fit}, and obtain the best-fit value $C^{\iter}_0$ for the fundamental mode by setting the start time of the fit as the time when the rate of change $\gamma_1$ significantly reduces.
We then subtract the fundamental mode and consider $\Psi^\sxs - \psi^{\iter}_{[0,0]}$, where the original first overtone is now the longest-lived mode.
Next, we fit the waveform $\Psi^\sxs - \psi^{\iter}_{[0,0]}$ by a fitting function the $\psi^{\textrm{fit}}_{[1,7]}$, and obtain the best-fit value $C^{\iter}_1$ for the first overtone.
With $C^{\iter}_1$, we can subtract the first overtone and consider $\Psi^\sxs - \psi^{\iter}_{[0,1]}$.
By iterating this procedure, we obtain the best-fit values $C^{\iter}_n$ and construct the fitted waveform $\psi^{\iter}_{[0,N]}$ step by step.

To further improve the stability of the fit, we also subtract a numerical constant $c$ which is dominant in the late time waveform.
Therefore, we insert a preliminary step for the identification and subtraction of $c$. 
After that we apply the subtraction procedure to the waveform $\Psi^\sxs-c$.

\subsection{SXS:BBH:0305 - GW150914-like simulation}

\begin{figure}[t]
    \centering
    \includegraphics[width=\columnwidth]{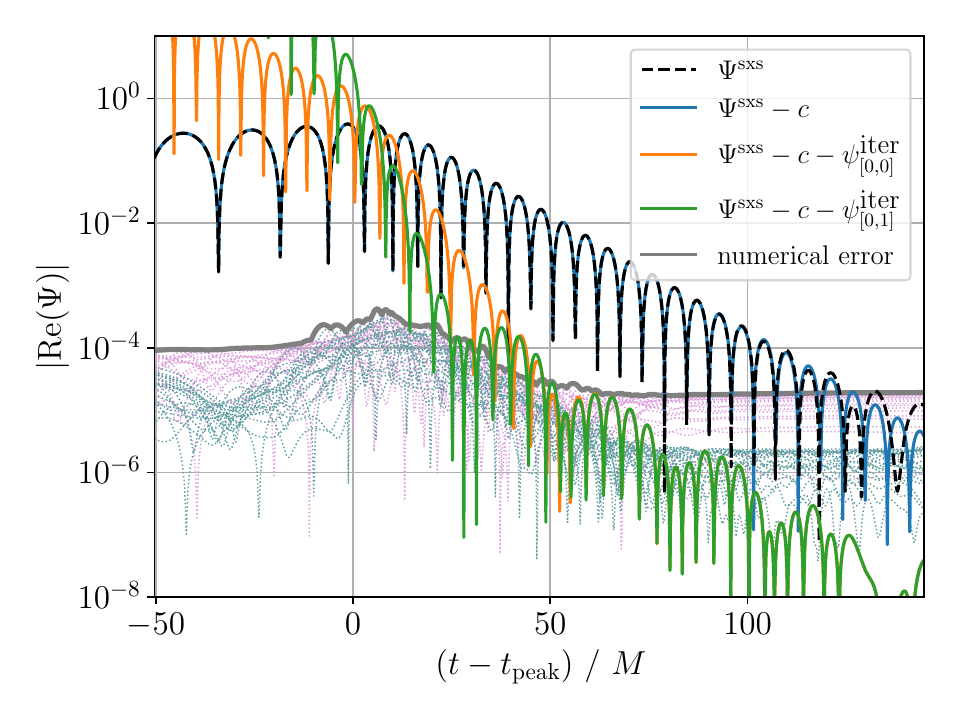}
    \caption{Numerical relativity waveform SXS:BBH:0305, waveform raw data (black dashed) and waveforms after the subtraction of the numerical constant (blue) with numerical errors (gray, see Appendix~\ref{sec:errors}).  We iteratively fit and subtract the longest-lived mode.  Waveform after the subtraction of the best-fit fundamental mode (orange) and then the best-fit first overtone (green) are shown.}
    \label{fig:0305_wave_naive}
\end{figure}

First, let us investigate the GW150914-like waveform SXS:BBH:0305.
Through the iterative fitting method, we consider several waveforms step by step.
Among them, we show first three waveforms in Fig.~\ref{fig:0305_wave_naive}.
The black curve depicts the raw data waveform $\Psi^\sxs$ of SXS:BBH:0305 from the SXS catalog. 
As we can see, this waveform includes the effect of a numerical constant at late time, after $(t-t_{\peak})/M\approx130$. 
We identify this constant value by averaging the  waveform for $(t-t_\peak)/M \geq 120$ after the subtraction of the fitted fundamental mode. 
We obtain $c=-(0.8+1.8i)\times10^{-5}$, which is consistent with the value used in the literature~\cite{Sago:2021gbq,Baibhav:2023clw}.
As we discussed in \S\ref{sec:dawnoise}, such a constant reduces the accuracy of the fit.
Therefore, we consider the waveform $\Psi^\sxs - c$, in which the numerical constant is subtracted from the original waveform.
The subtraction of the numerical constant was also performed in \cite{Giesler:2019uxc} and known to provide a result similar to the one with the mapping to the super rest frame~\cite{Moreschi:1988pc,Moreschi:1998mw,Dain:2000lij,Mitman:2021xkq,MaganaZertuche:2021syq}.

We confirm that the mismatch between $\Psi^\sxs$ and $\psi^\fit_{[0,N]}$ without the subtraction of the numerical constant is consistent with the results in \cite{Cook:2020otn}, where the lower bound of the mismatch is slightly lifted, and that the difference originates from the constant~\cite{MaganaZertuche:2021syq}.

The waveform after the subtraction of the constant is shown by blue curve in Fig.~\ref{fig:0305_wave_naive}. 
We can see an improvement that the damped sinusoid continues in the waveform after $(t-t_{\peak})/M=130$.
We then extract the coefficient for fundamental mode $C^{\iter}_{0}$ by fitting the waveform $\Psi^\sxs - c$ by the fitting function $\psi^\fit_{[0,3]}$ of a superposition of four QNM modes as discussed in \S\ref{sec:mock}.
We choose the start time of the fit as the time when $\gamma_1$ takes the minimum.
We denote the fitted fundamental mode as $\psi^{\iter}_{[0,0]} = C^{\iter}_{0} \psi_0$, and we shall subtract it from blue curve to obtain the orange curve in Fig.~\ref{fig:0305_wave_naive}, which is $\Psi^\sxs - c - \psi^{\iter}_{[0,0]}$.
Gray curve in Fig.~\ref{fig:0305_wave_naive} is the conservative estimation of the numerical errors (see Appendix~\ref{sec:errors} for details).

\begin{figure}[t]
    \centering
    \includegraphics[width=0.49\columnwidth]{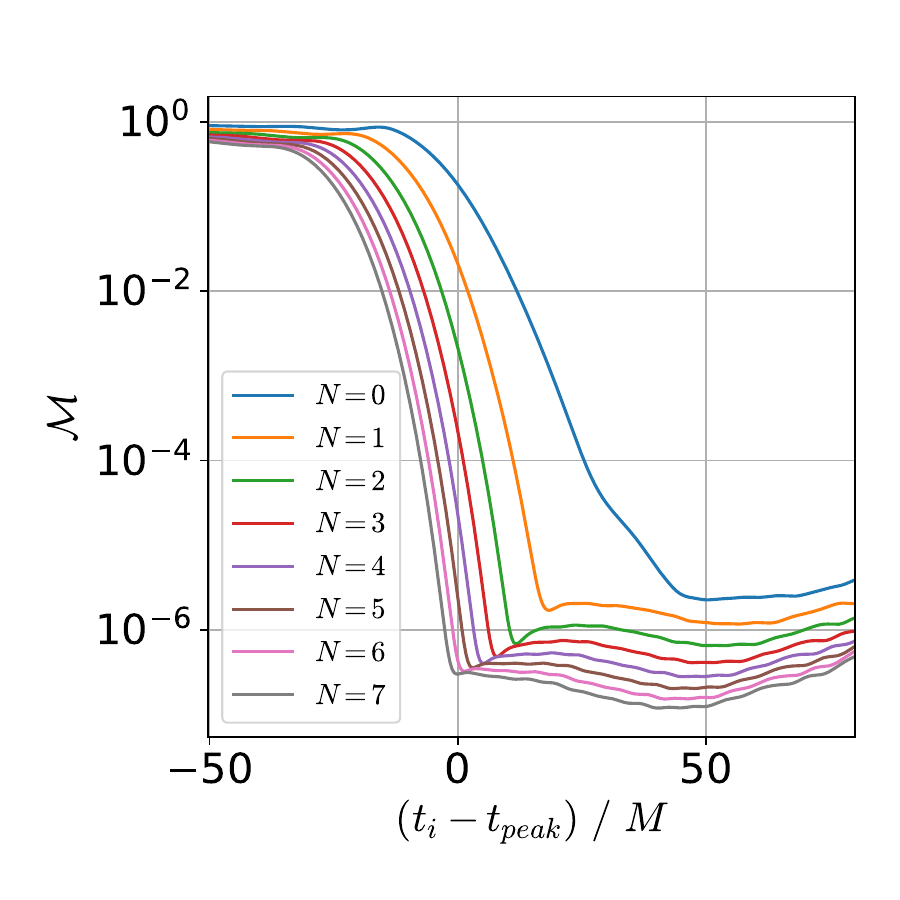}
    \includegraphics[width=0.49\columnwidth]{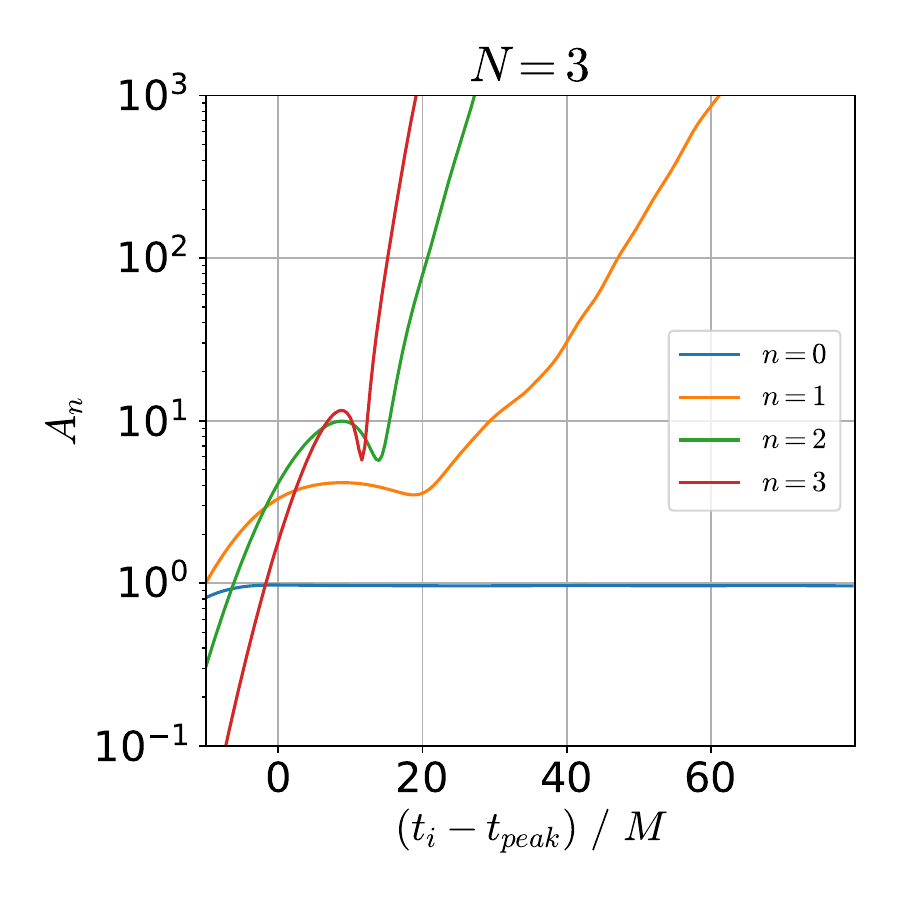}
    \includegraphics[width=0.49\columnwidth]{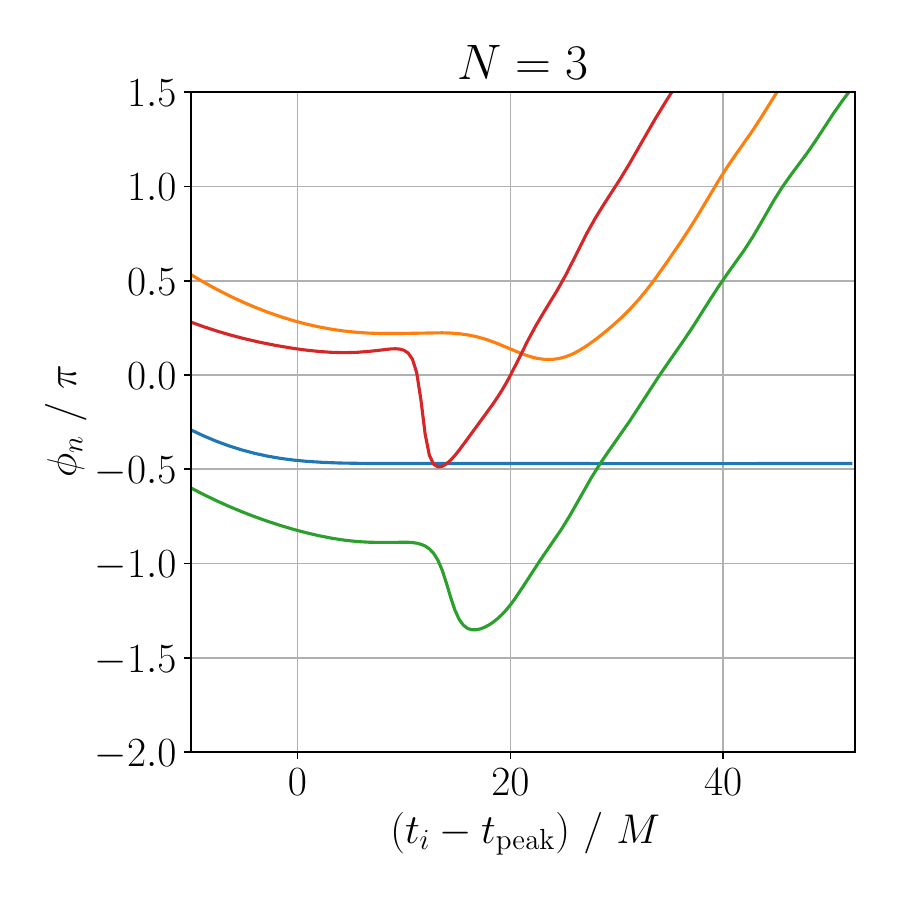}
    \includegraphics[width=0.49\columnwidth]{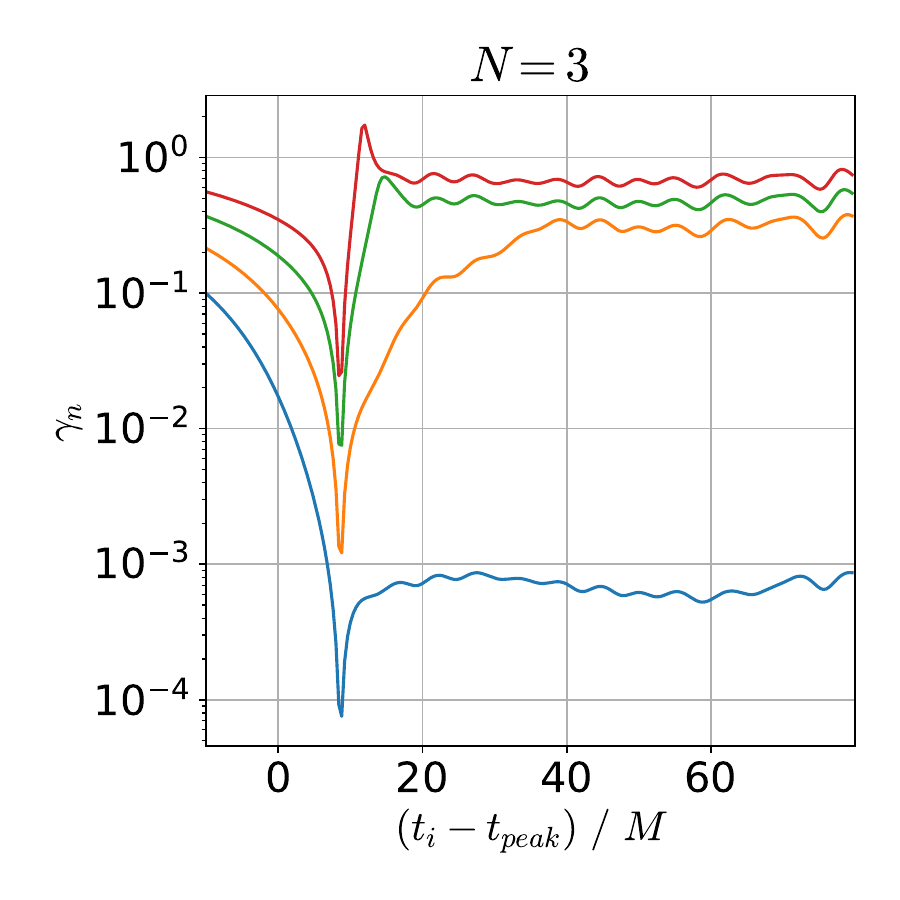}
    \caption{Mismatch $\M$ (left top), amplitude $A_n$ (right top), phase $\phi_n$ (left bottom), and rate of change $\gamma_n$ (right bottom) for the fit of SXS:BBH:0305 waveform $\Psi^\sxs - c$ by the fitting function $\psi^\fit_{[0,N]}$.}
    \label{fig:0305_M_A_gamma_sub_const}
\end{figure}

Before discussing the iterative subtraction of the longest-lived mode, let us take a closer look of the fit of $\Psi^\sxs - c$ by the fitting function $\psi^\fit_{[0,3]}$.
Figure~\ref{fig:0305_M_A_gamma_sub_const} shows the mismatch $\M$, amplitude $A_n$, and rate of change $\gamma_n$ as a function of the start time of the fit $t_i$.
It should be compared with Fig.~\ref{fig:0305_M_A_Cre_gamma_mock_naive}, but the panel for the relative error of $C_n$ is omitted here because there is no known fiducial values for the numerical relativity waveform.

The first panel of Fig.~\ref{fig:0305_M_A_gamma_sub_const} shows the mismatch $\M$ between the SXS:BBH:0305 waveform $\Psi^\sxs - c$ and the fitting function $\psi^\fit_{[0,N]}$, which is consistent with the results in \cite{Giesler:2019uxc}.
Each curve in Fig.~\ref{fig:0305_M_A_gamma_sub_const} represents what we used as the fitting function $\psi^\fit_{[0,N]}$.
The larger the number of $N$, the smaller the mismatch $\M$ with earlier $t_i$, which is the same trend as the mock waveform which we discussed in \S\ref{sec:mock} and \S\ref{sec:dawnoise}.
Compared to Fig.~\ref{fig:0305_M_mock_sub_const} for the mock waveform without constant, the mismatch $\M$ in Fig.~\ref{fig:0305_M_A_gamma_sub_const} has a lower bound in each curve. 

The second panel of Fig.~\ref{fig:0305_M_A_gamma_sub_const} shows that the amplitude $A_n$ as a function of the start time of the fit $t_i$, whose behavior is in agreement with \cite{Baibhav:2023clw}.
We also present the phase $\phi_n$ in the third panel of Fig.~\ref{fig:0305_M_A_gamma_sub_const}.
Unlike the first panel, each curve represents the $n$-th overtone fitted by the fitting function $\psi^\fit_{[0,3]}$.
As expected, the fit of the fundamental mode is robust and not sensitive to the choice of the start time of the fit, which is also consistent with what we learned from the mock waveform analysis. 
On the other hand, for the overtones $n\geq 1$, $A_n$ and $\phi_n$ diverge, similar to Fig.~\ref{fig:0305_M_A_Cre_gamma_mock_naive} for the mock waveform with constant.
Even though we subtracted the numerical constant observed in the late time of the SXS waveform, the fit is not as stable as the fit of the mock waveform analysis without a constant.
We can see that the start time of the fit $t_i$ of the onset of divergence is earlier than the case of Fig.~\ref{fig:0305_M_A_Cre_gamma_mock_naive}.
This may be due to the spherical-spheroidal mode mixing in the SXS waveform, which plays a similar role to random noise, reducing the accuracy of the fit.

The fourth panel of Fig.~\ref{fig:0305_M_A_gamma_sub_const} shows the rate of change $\gamma_n$.
The each curve represents the $n$-th overtone same as the second panel.
In parallel to $A_n$ and $\phi_n$, the behavior of $\gamma_n$ is also similar to Fig.~\ref{fig:0305_M_A_Cre_gamma_mock_naive} for the mock waveform with constant.
$\gamma_0$ remains small even for late $t_i$, which suggests that the plateau continues up to late start time of the fit and the fit is stable. 
On the other hand, for $n\geq 1$, $\gamma_n$ approaches to a constant, which roughly coincides with $|\Im(\omega_n)|$, signaling the overfit.

As we discussed in \S\ref{sec:mock} and \S\ref{sec:dawnoise}, we use the minimum of $\gamma_n$ as a criterion to extract the best-fit value of $C_n$.
In each step of the iterative procedure, to obtain the best-fit value of the longest-lived mode, we adopt the start time of the fit $t_i$, when $\gamma_n$ for the next-longest-lived mode has the minimum.
The reason why we focus on the minimum of $\gamma_n$ for the next-longest-lived mode rather than the of the longest-lived mode was that the latter can be isolated from the minima for other $\gamma_n$, as in the case of the mock waveform without constant. 
Of course, it might be the case that $\gamma_n$ for the next-longest-lived mode would have the minimum isolated form other minima.  
Ultimately one needs to check the behavior of $\gamma_n$ as a function of $t_i$ to decide the optimal start time of the fit.
In the case of the SXS:BBH:0305 waveform, we can see that the all $\gamma_n$ takes the minimum value at $(t_i-t_{\peak})/M\approx 11$, which we regard as the optimal start time of the fit.
We then adopt $C_0$ with the optimal start time of the fit as the best-fit value $C^{\iter}_0$.

After we obtain the best-fit value $C^{\iter}_0$, we subtract the fundamental mode from the blue curve to obtain the orange curve in Fig.~\ref{fig:0305_wave_naive}, which is $\Psi^\sxs - c - \psi^{\iter}_{[0,0]}$.
Up to $(t-t_{\peak})/M\approx50$, we see that the orange curve can be mainly described by a single damped sinusoid, which is the first overtone, but after that, we can see a peculiar behavior like the beat, which we discuss in detail in the next section.

\begin{figure}[t]
    \centering
    \includegraphics[width=0.49\columnwidth]{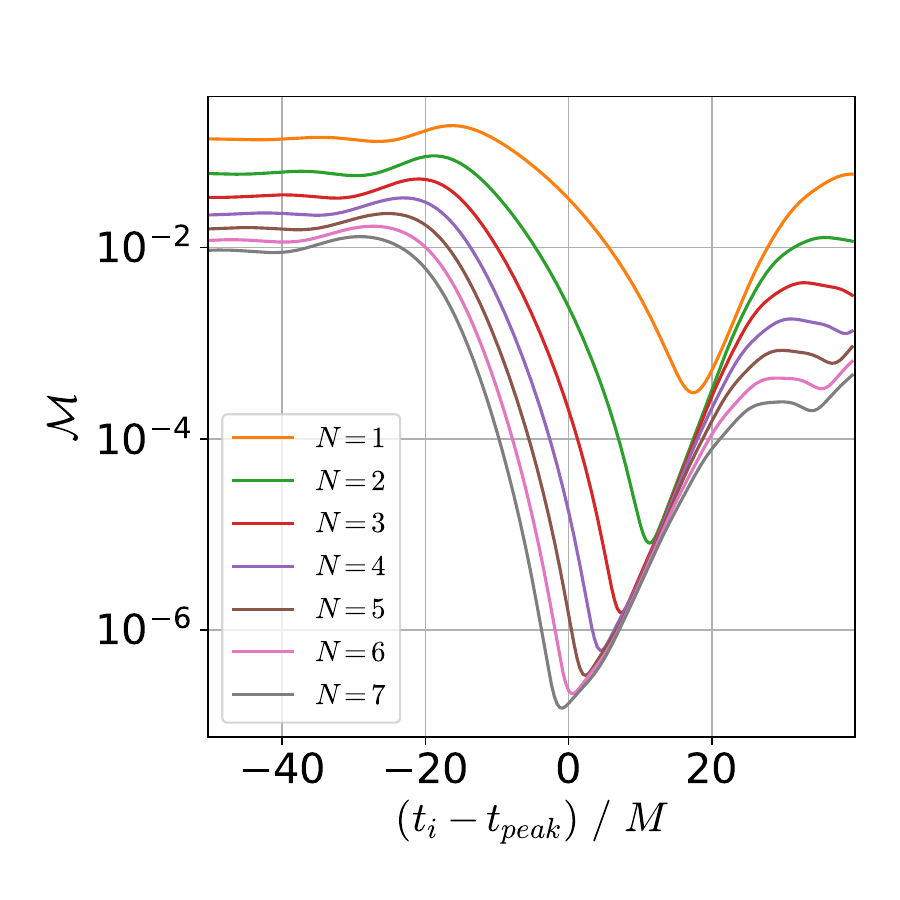}
    \includegraphics[width=0.49\columnwidth]{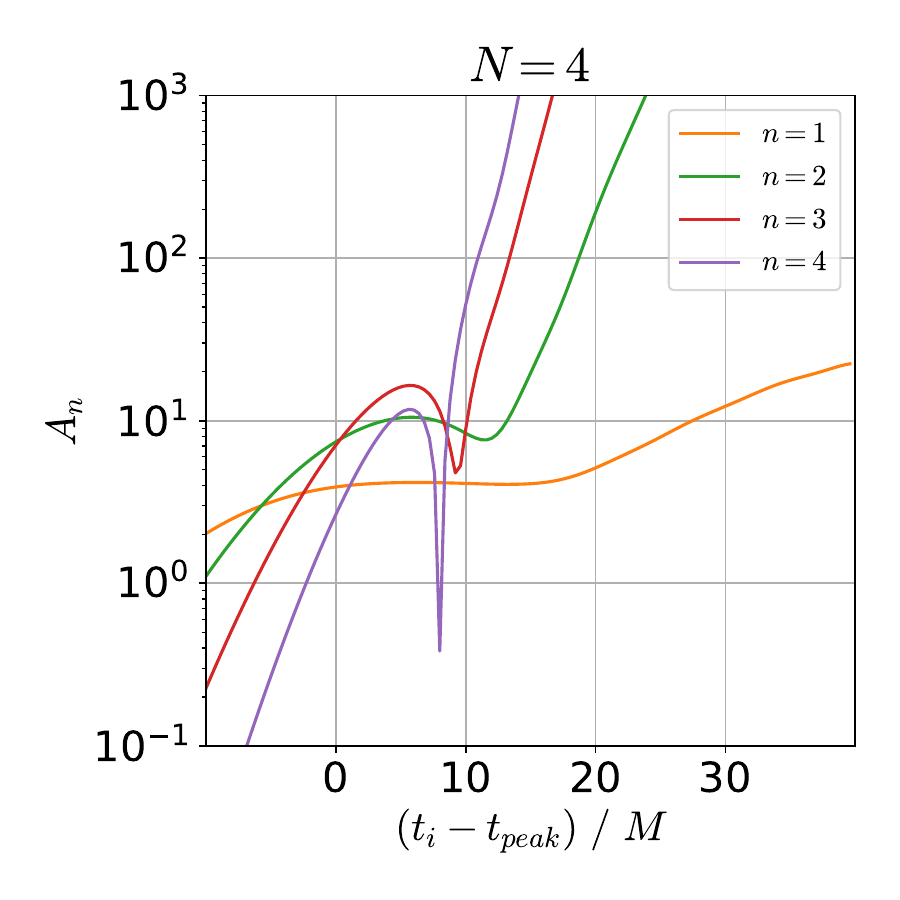}
    \includegraphics[width=0.49\columnwidth]{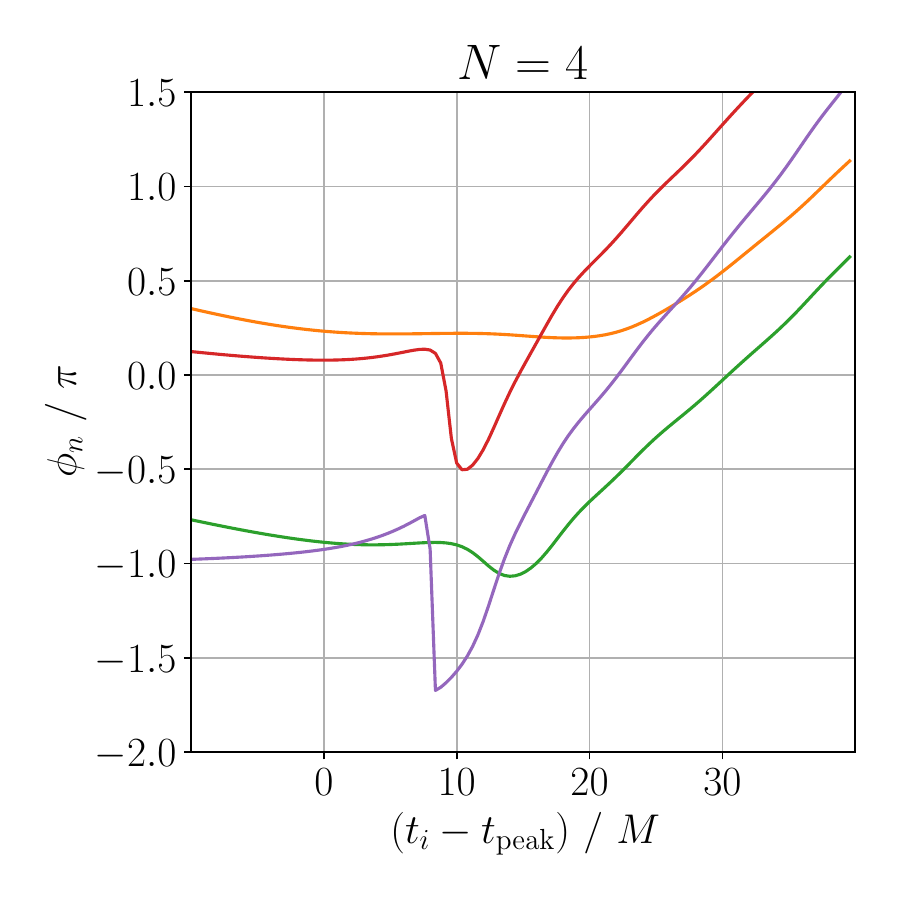}
    \includegraphics[width=0.49\columnwidth]{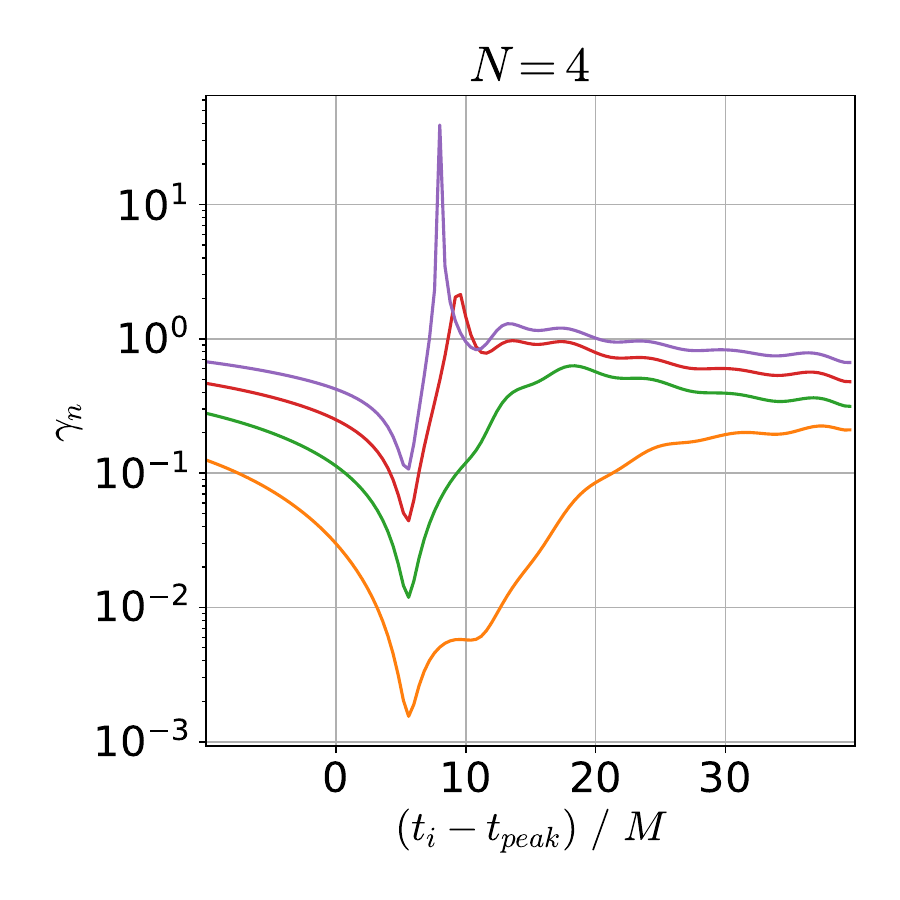}
    \caption{Mismatch $\M$ (left top), amplitude $A_n$ (right top), phase $\phi_n$ (left bottom), and rate of change $\gamma_n$ (right bottom) for the fit of SXS:BBH:0305 waveform $\Psi^\sxs - c - \psi^\iter_{[0,0]}$ after the subtraction of the fundamental mode by the fitting function $\psi^\fit_{[1,N]}$.
    }
    \label{fig:0305_M_A_gamma_sub_const0}
\end{figure}

In Fig.~\ref{fig:0305_M_A_gamma_sub_const0}, we present the result of the fit of $\Psi^{\rm{SXS}}-c-\psi^{\iter}_{[0,0]}$ waveform colored by orange in Fig~\ref{fig:0305_wave_naive} by the fitting function $\psi^\fit_{[1,N]}$.
Similar to Fig.~\ref{fig:0305_M_A_gamma_sub_const}, we show mismatch $\M$, the amplitude $A_n$, and the rate of change $\gamma_n$. 

By subtracting the fundamental mode, the mismatch shown in the first panel of Fig.~\ref{fig:0305_M_A_gamma_sub_const0} changed significantly from Fig.~\ref{fig:0305_M_A_gamma_sub_const}.
In Fig.~\ref{fig:0305_M_A_gamma_sub_const}, the mismatch is bounded from below but the lower bound remain almost constant with respect to $t_i$.
However, in Fig.~\ref{fig:0305_M_A_gamma_sub_const0}, the lower bound of the mismatch increases as $t_i$ increases.
This trend is close to the case of the mock waveform with noise rather than the mock waveform without noise.
It implies that even after subtracting the late time constant, the waveform still contains an effective noise such as the residual fundamental mode and spherical-spheroidal mixing.

\begin{figure*}[t]
    \centering
    \includegraphics[width=0.25\textwidth]{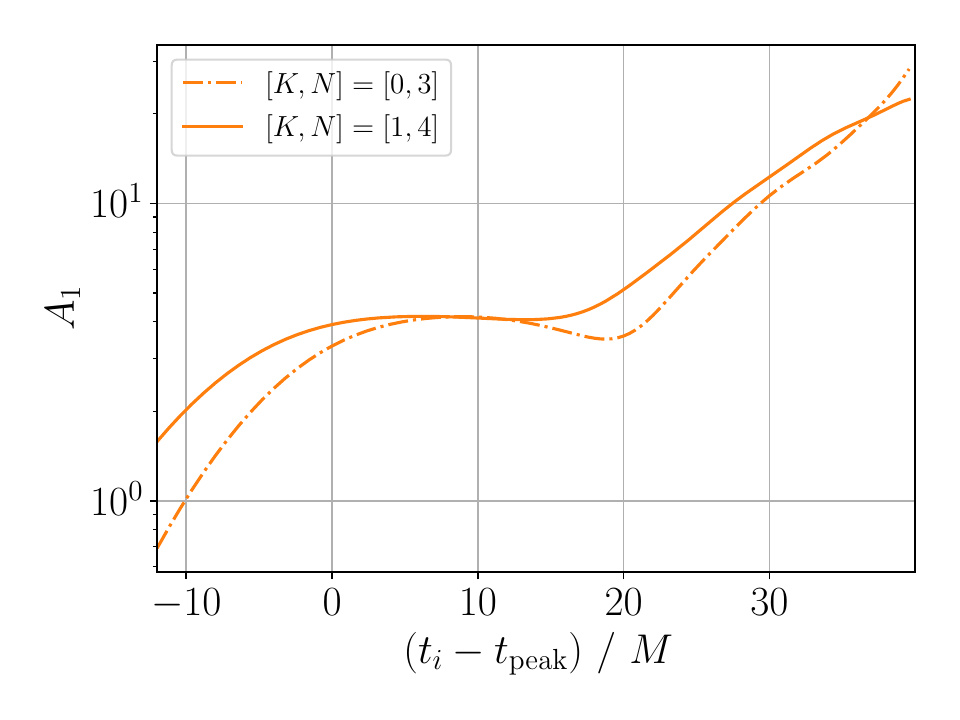}
    \includegraphics[width=0.25\textwidth]{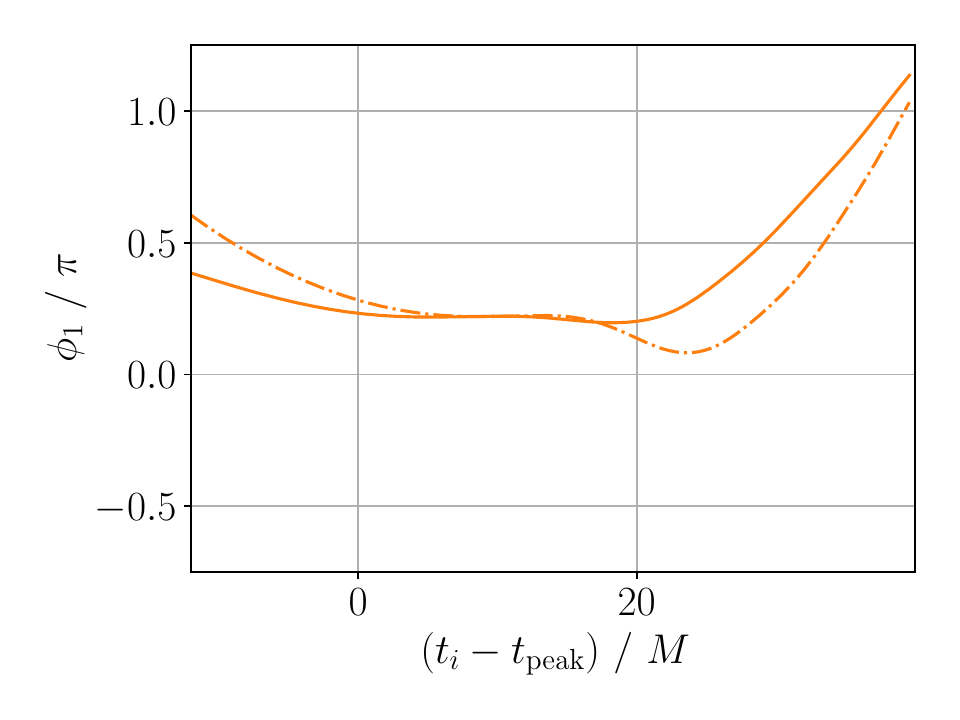}
    \includegraphics[width=0.25\textwidth]{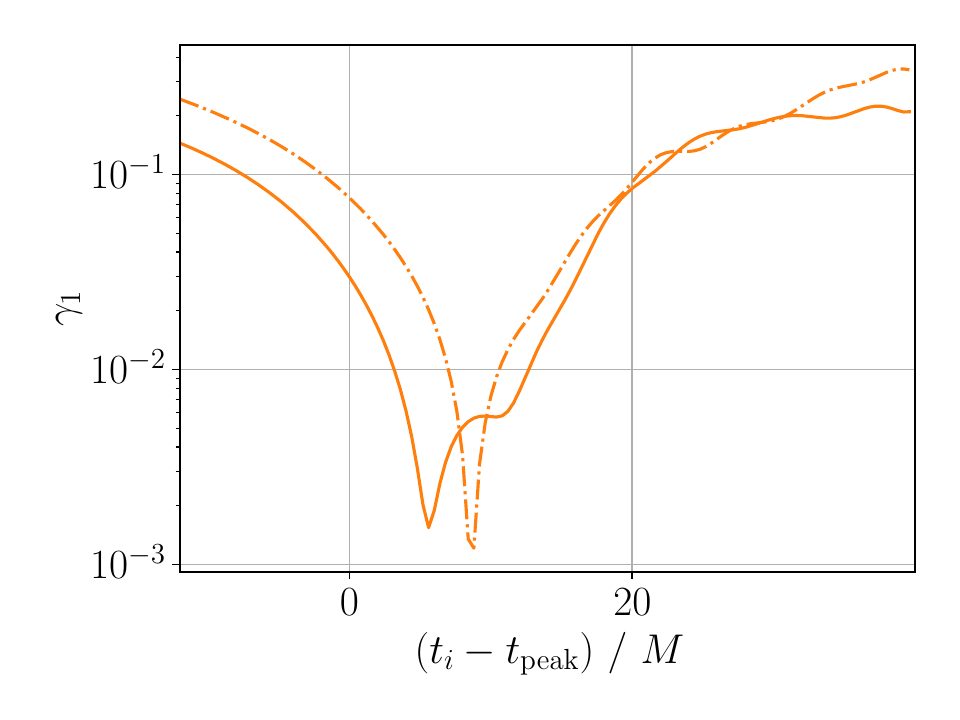}
    
    \includegraphics[width=0.25\textwidth]{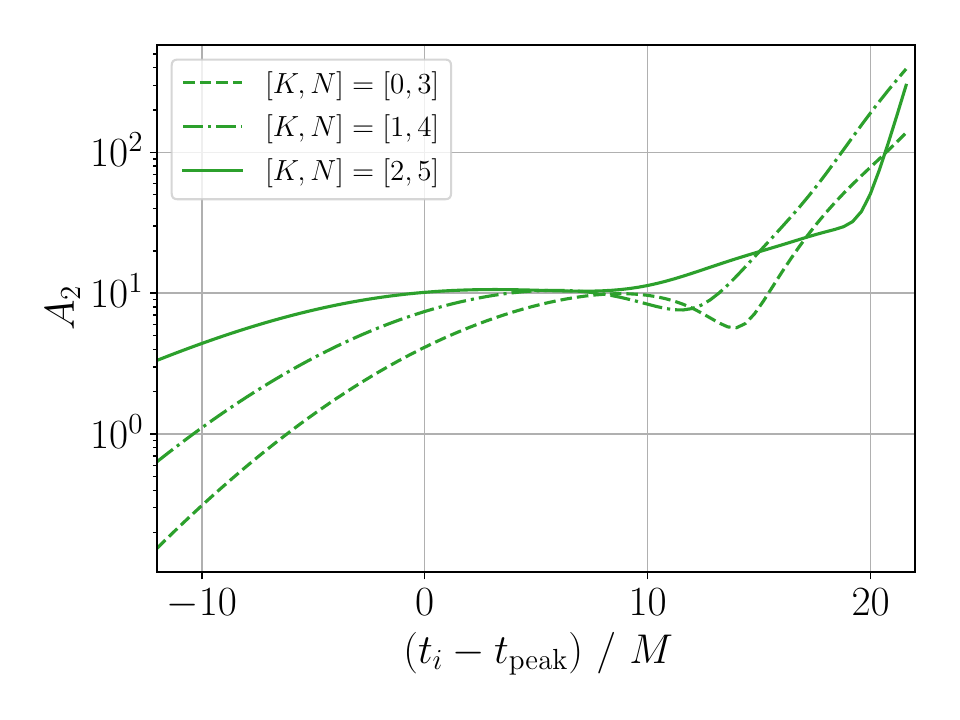}
    \includegraphics[width=0.25\textwidth]{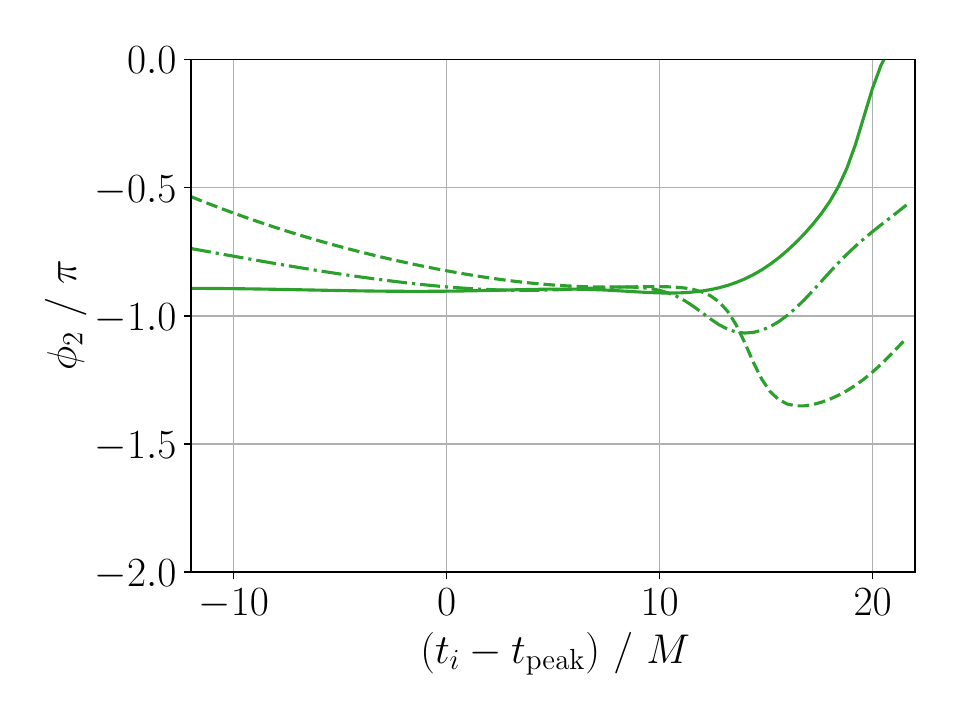}
    \includegraphics[width=0.25\textwidth]{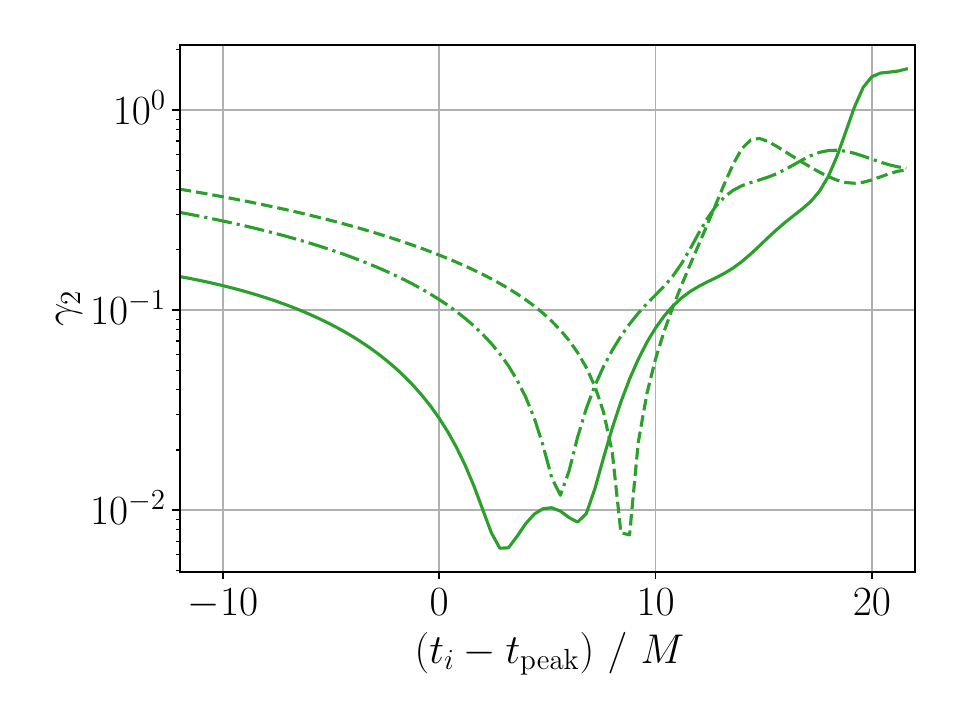}

    \includegraphics[width=0.25\textwidth]{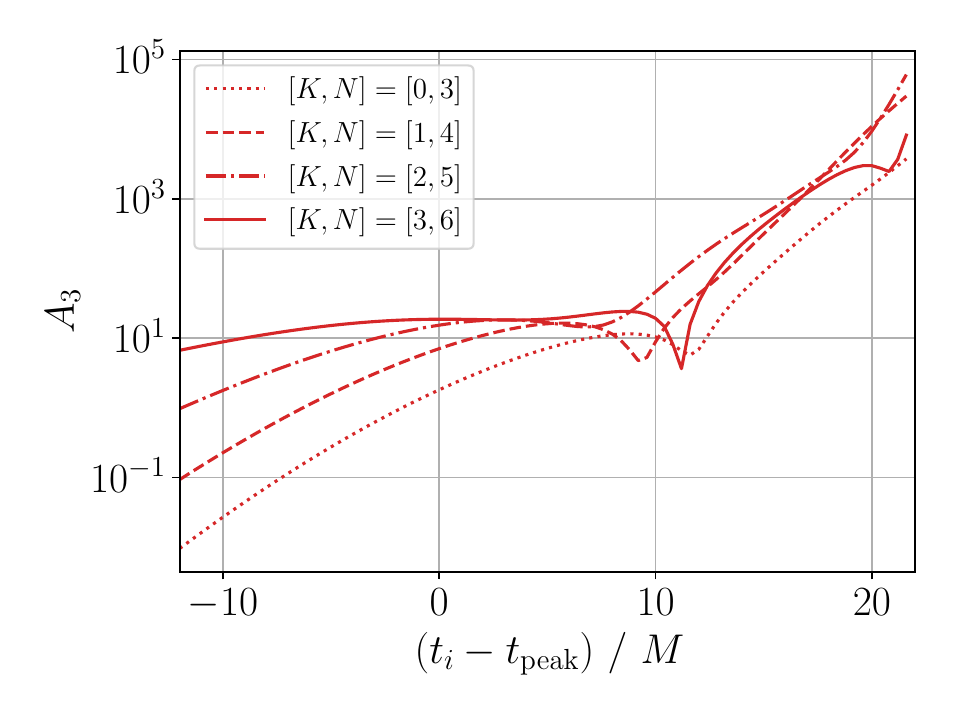}
    \includegraphics[width=0.25\textwidth]{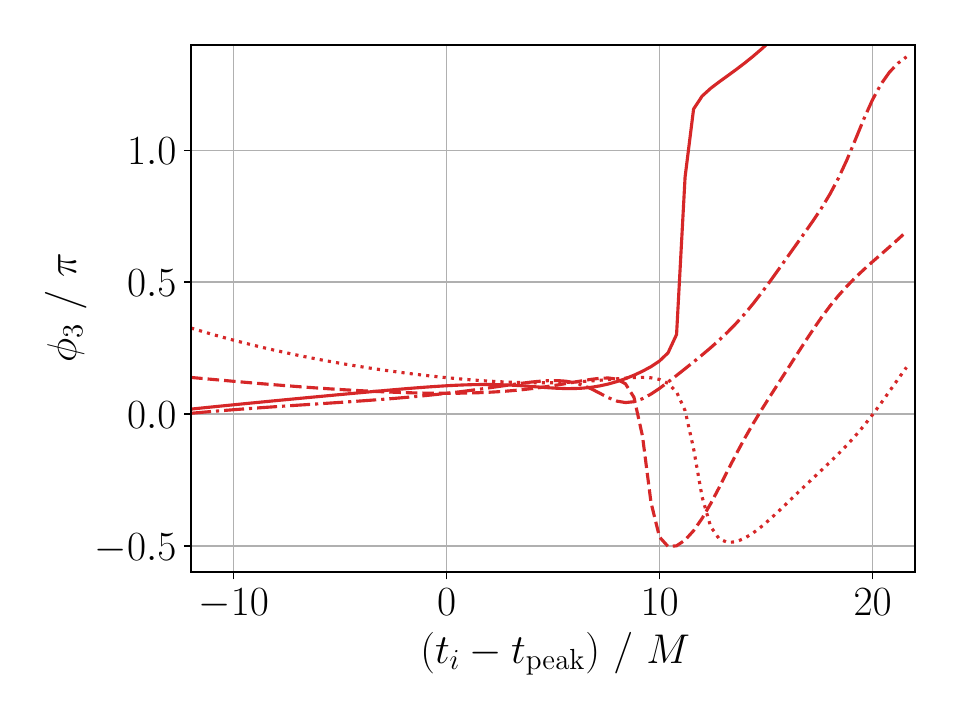}
    \includegraphics[width=0.25\textwidth]{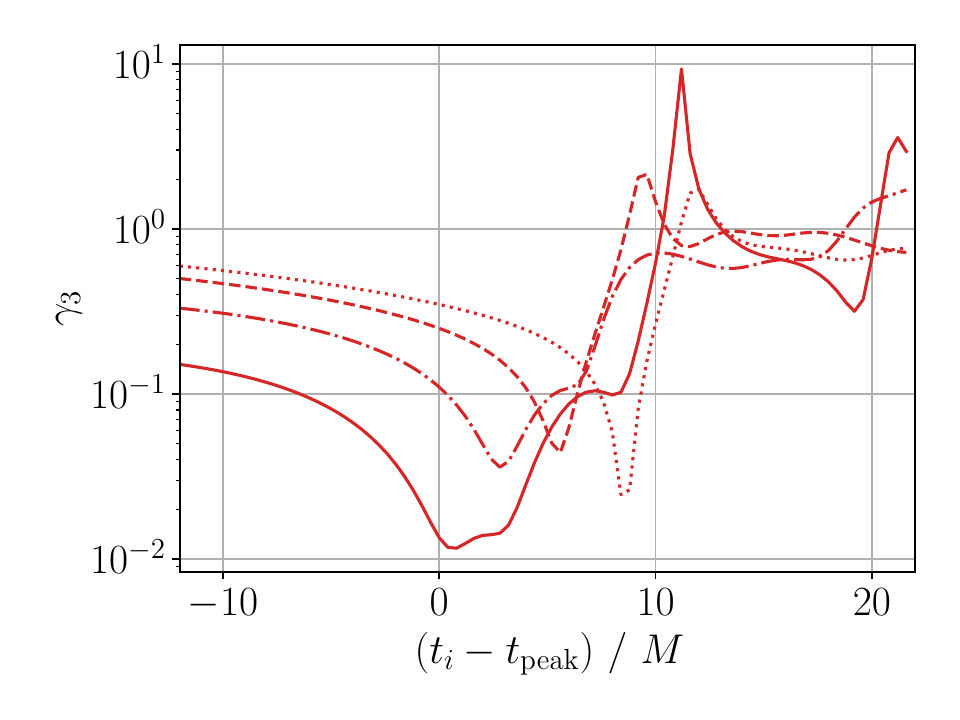}
    
    \includegraphics[width=0.25\textwidth]{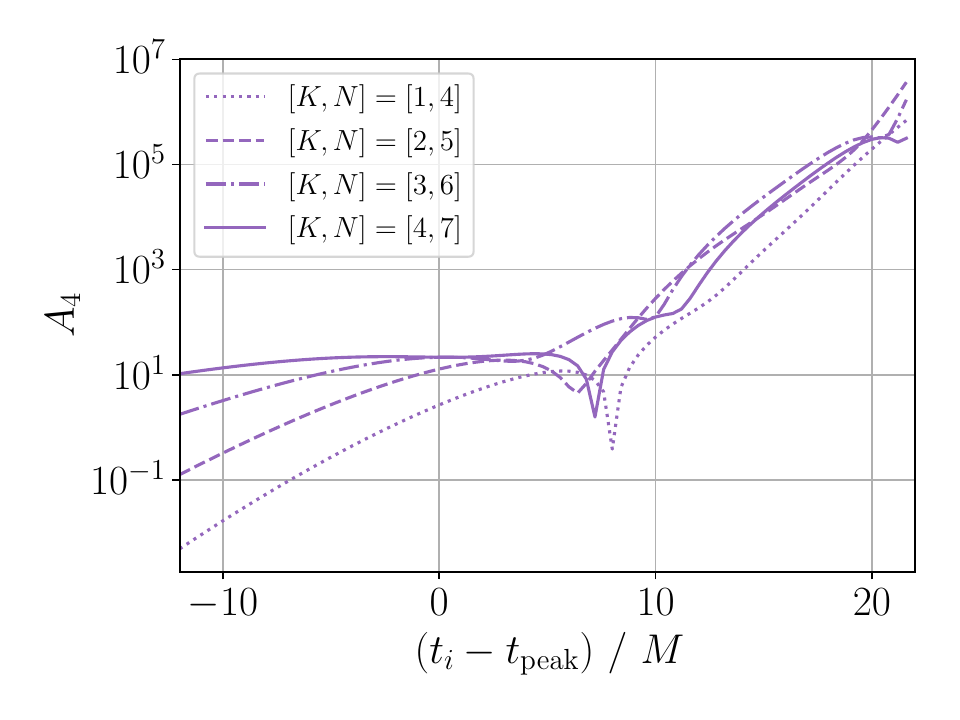}
    \includegraphics[width=0.25\textwidth]{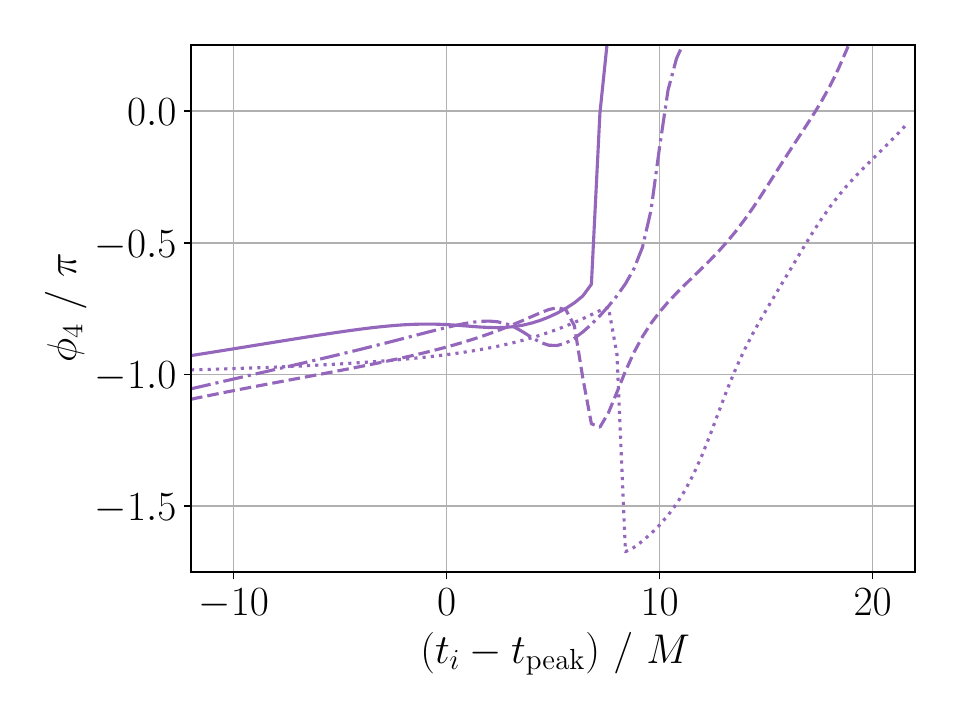}
    \includegraphics[width=0.25\textwidth]{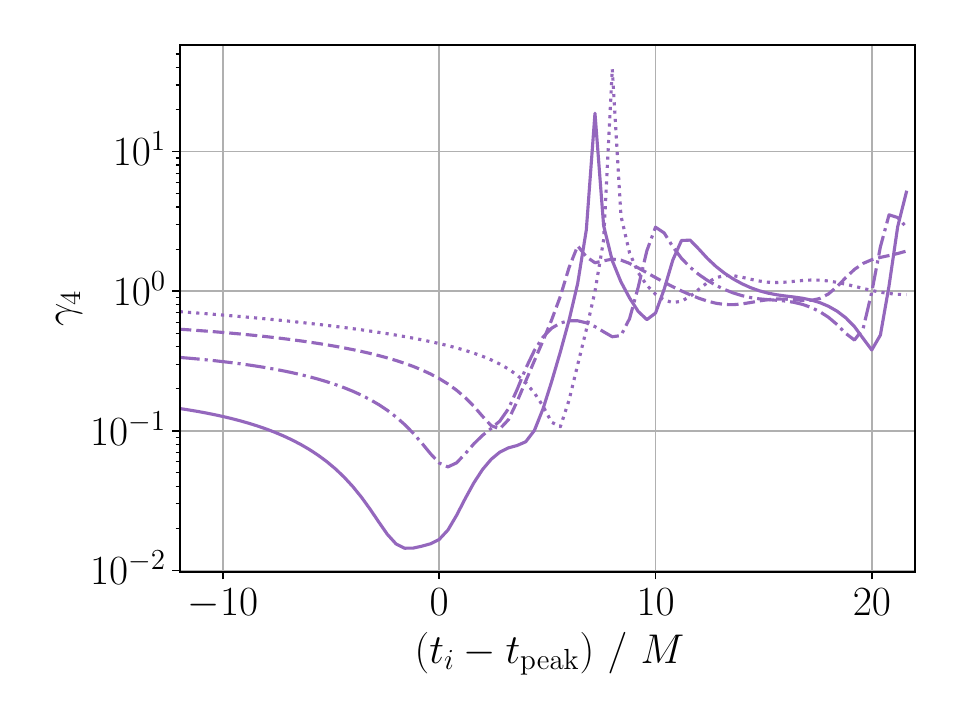}
    
    \includegraphics[width=0.25\textwidth]{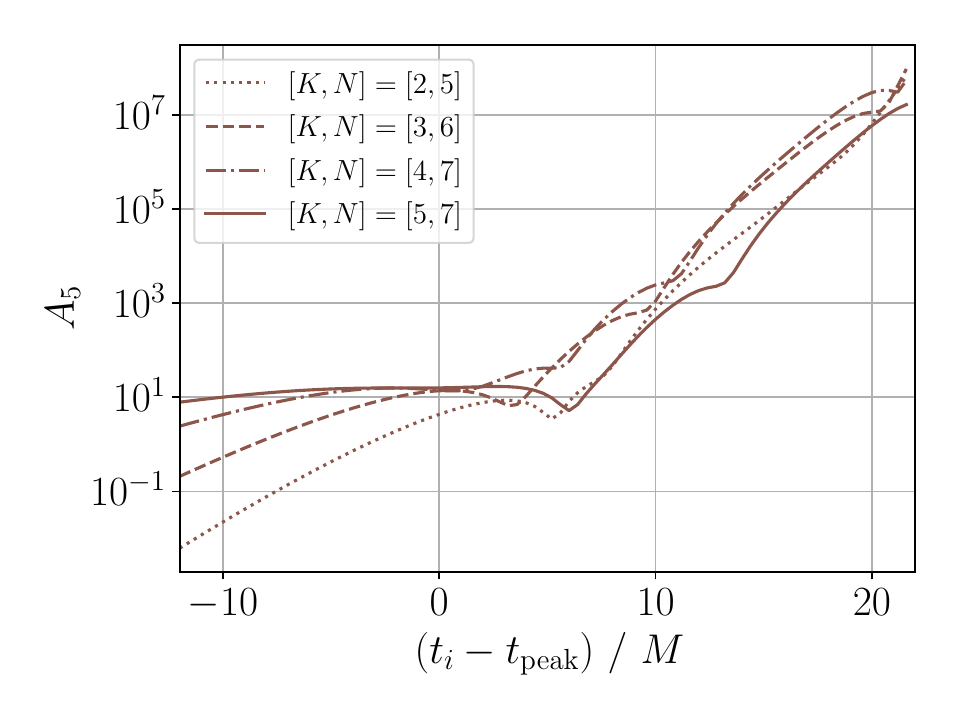}
    \includegraphics[width=0.25\textwidth]{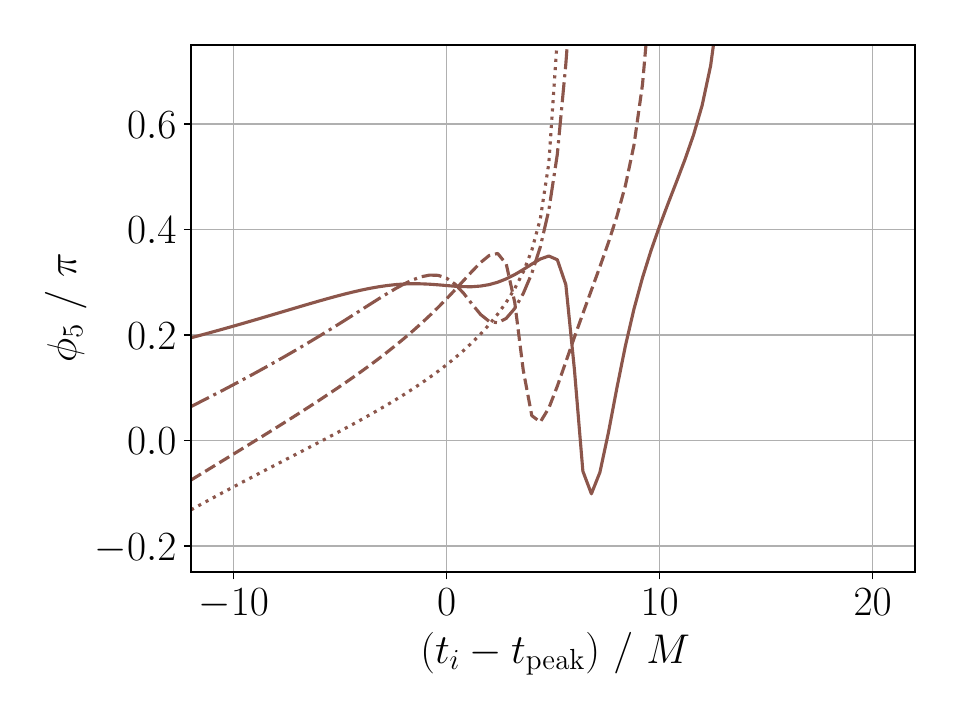}
    \includegraphics[width=0.25\textwidth]{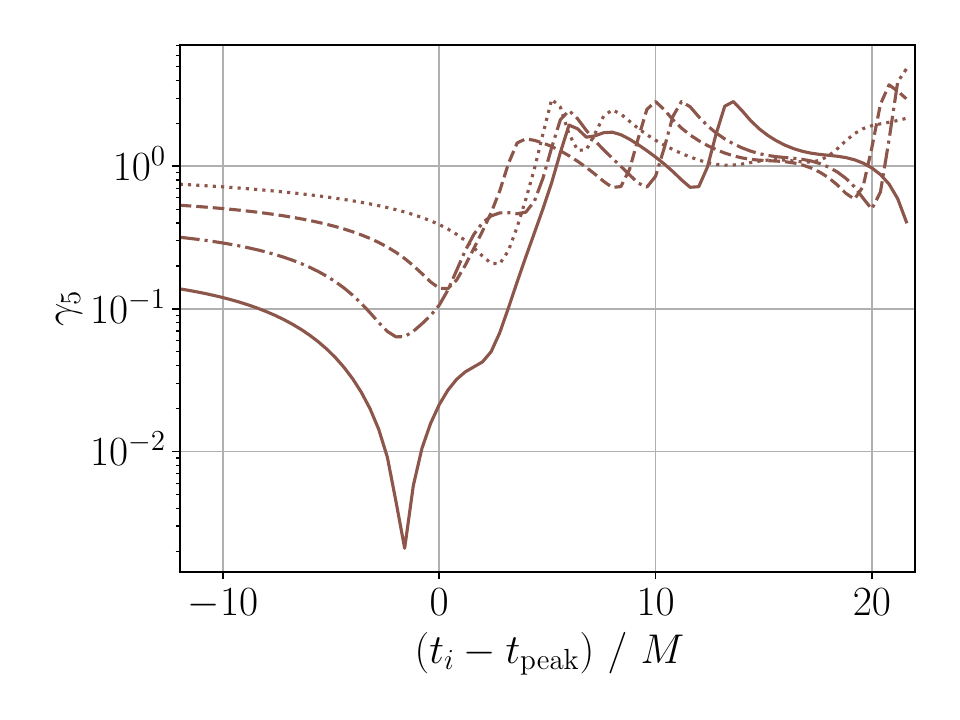}
    
    \includegraphics[width=0.25\textwidth]{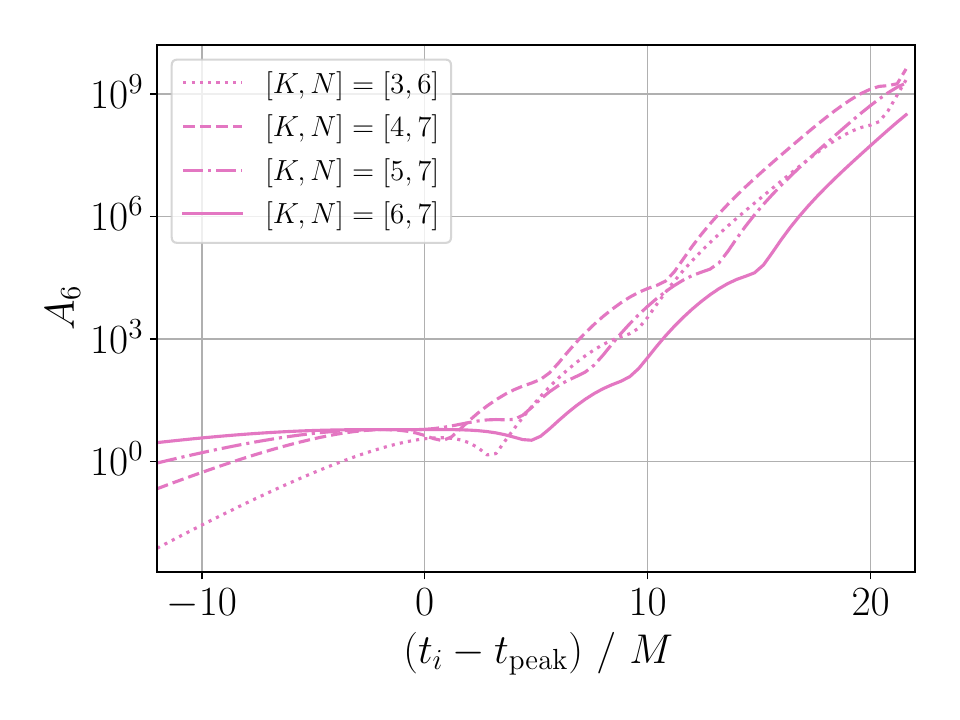}
    \includegraphics[width=0.25\textwidth]{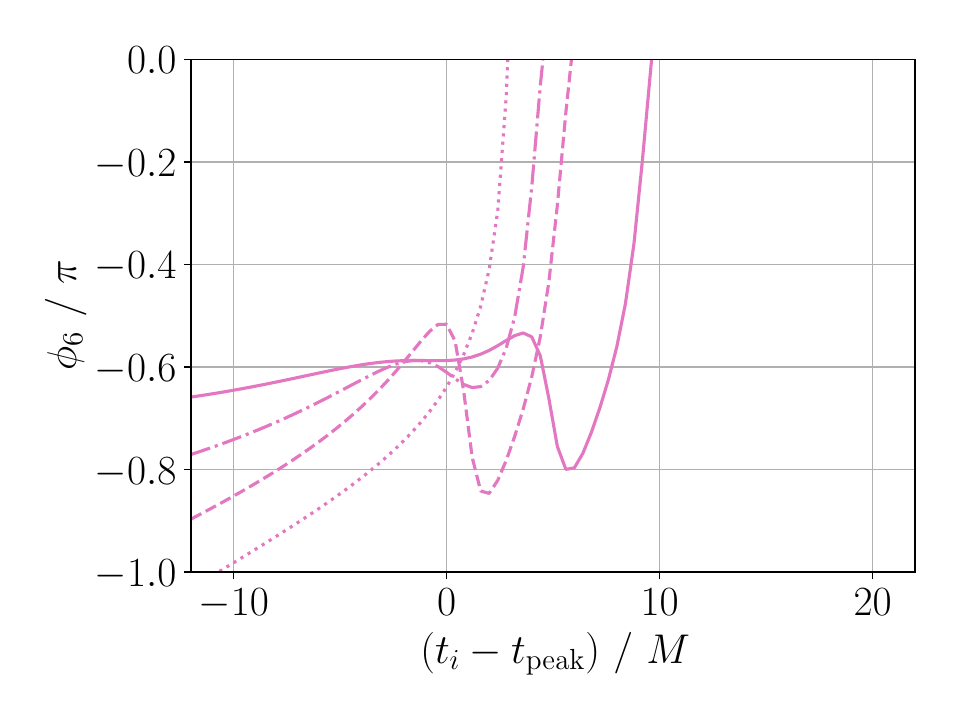}
    \includegraphics[width=0.25\textwidth]{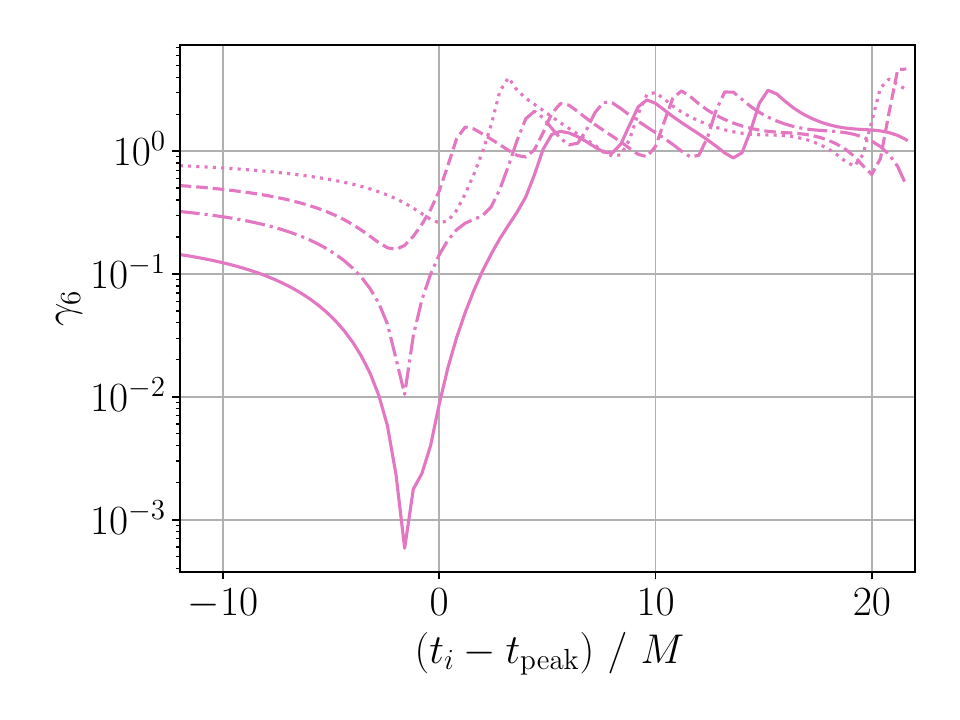}
    
    \includegraphics[width=0.25\textwidth]{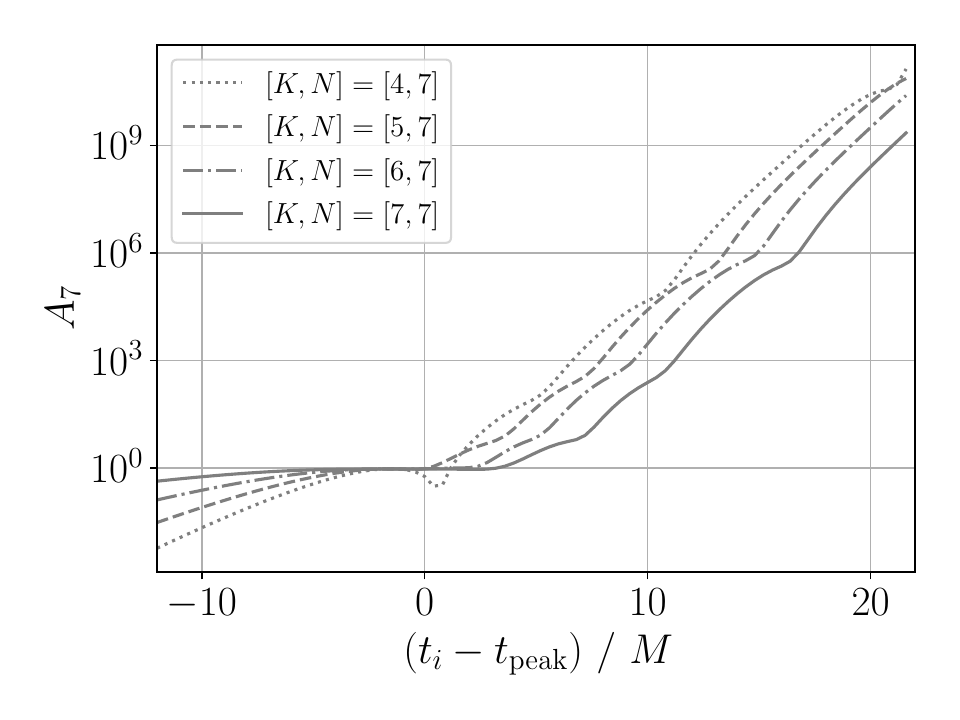}
    \includegraphics[width=0.25\textwidth]{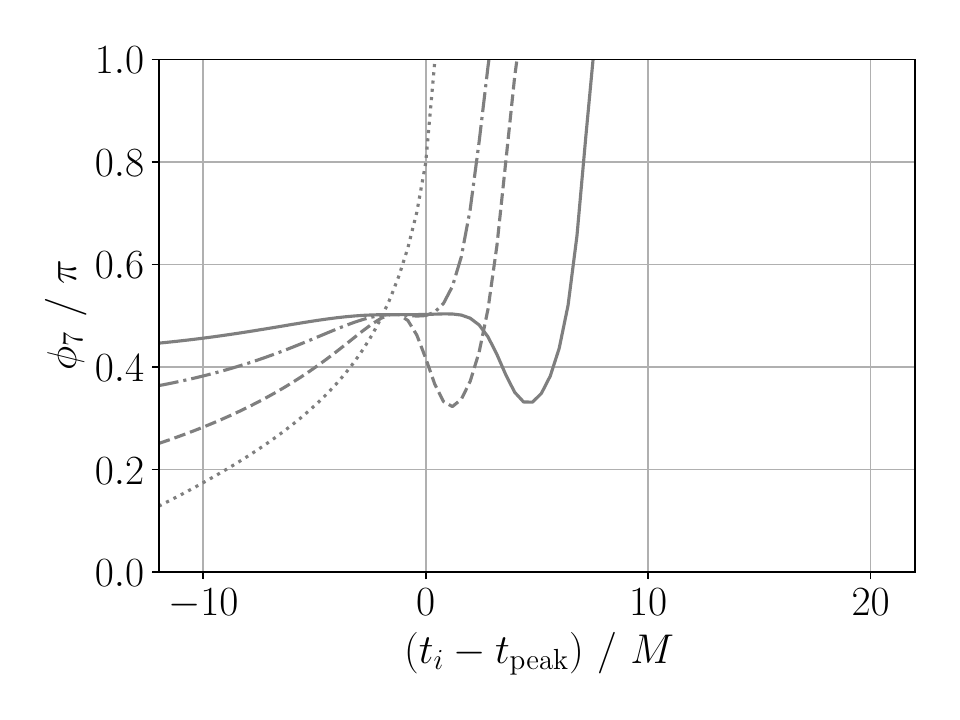}
    \includegraphics[width=0.25\textwidth]{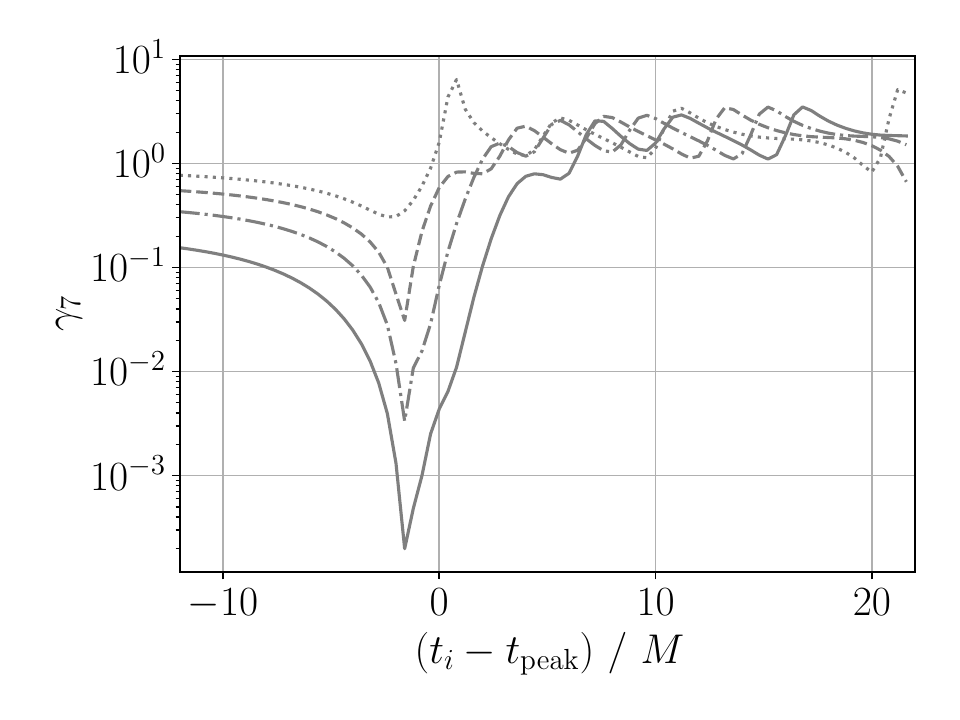}
    \caption{The improvement of the amplitude $A_n$ (left column), phase $\phi_n$ (right column) from $n=1$ mode (first row) to $n=7$ mode (seventh row) for the iterative fit of SXS:BBH:0305 by the fitting function $\psi^{\fit}_{[K,N]}$ with $K=0,\cdots, 7$ and $N={\rm min}(K+3,~ 7)$.}
    \label{fig:0305_n4_improve_sub_const}
\end{figure*}

The amplitude $A_n$ and the rate of change $\gamma_n$ for each overtone in the fitting function $\psi^\fit_{[1,4]}$ are shown in the second and fourth panels of Fig.~\ref{fig:0305_M_A_gamma_sub_const0}, respectively.
By virtue of the subtraction, the stability of the first overtone $n=1$ is slightly improved from Fig.~\ref{fig:0305_M_A_gamma_sub_const}.
For instance, while the minimum value of $\gamma_1$ is about $10^{-3}$ and is almost the same as Fig.~\ref{fig:0305_M_A_gamma_sub_const}, the region where $\gamma_n\lesssim 10^{-2}$ is longer due to the subtraction of the fundamental mode. 
The location of the plateau tends to move to earlier $t_i$ after subtraction.
As expected, the improvement is not as manifest as in the mock analysis without constant.

\begin{figure*}[t]
    \centering
    \includegraphics[width=\columnwidth]{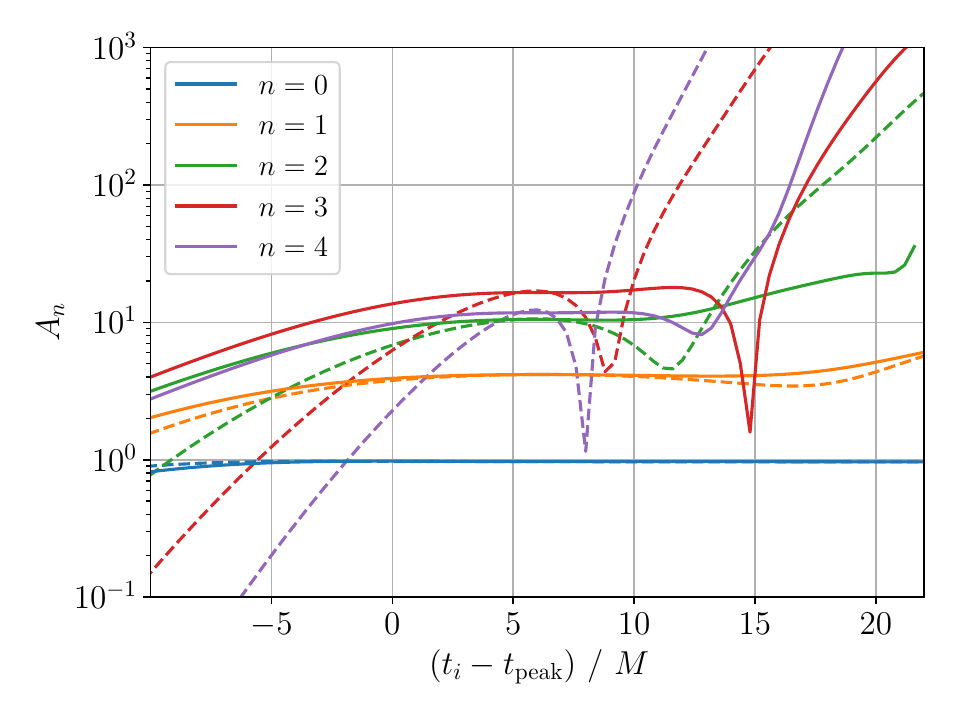}
    \includegraphics[width=\columnwidth]{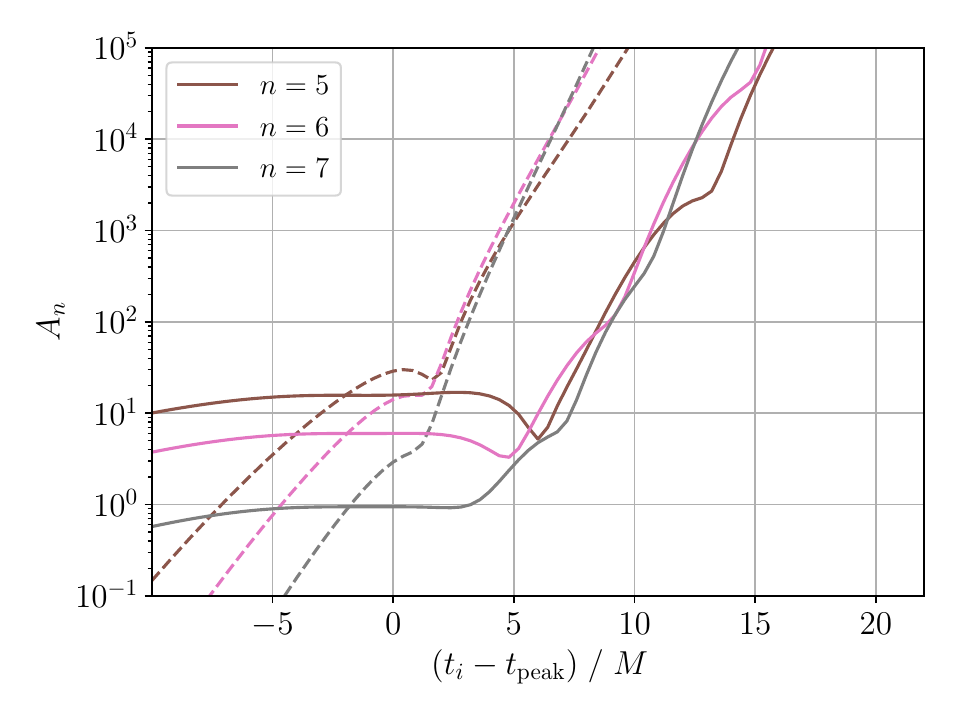}
    \includegraphics[width=\columnwidth]{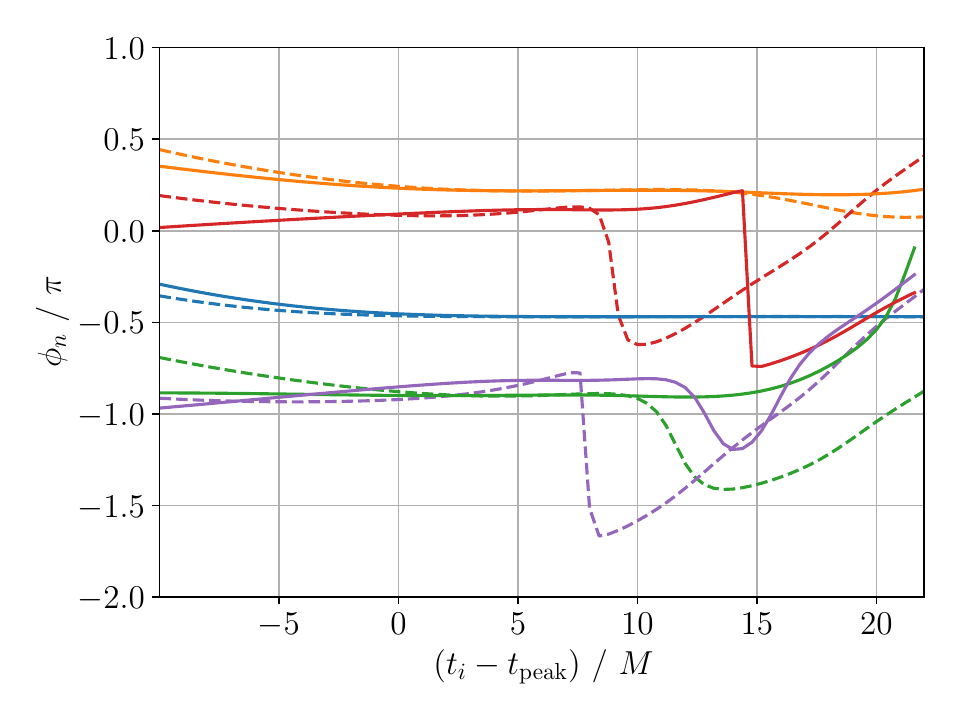}
    \includegraphics[width=\columnwidth]{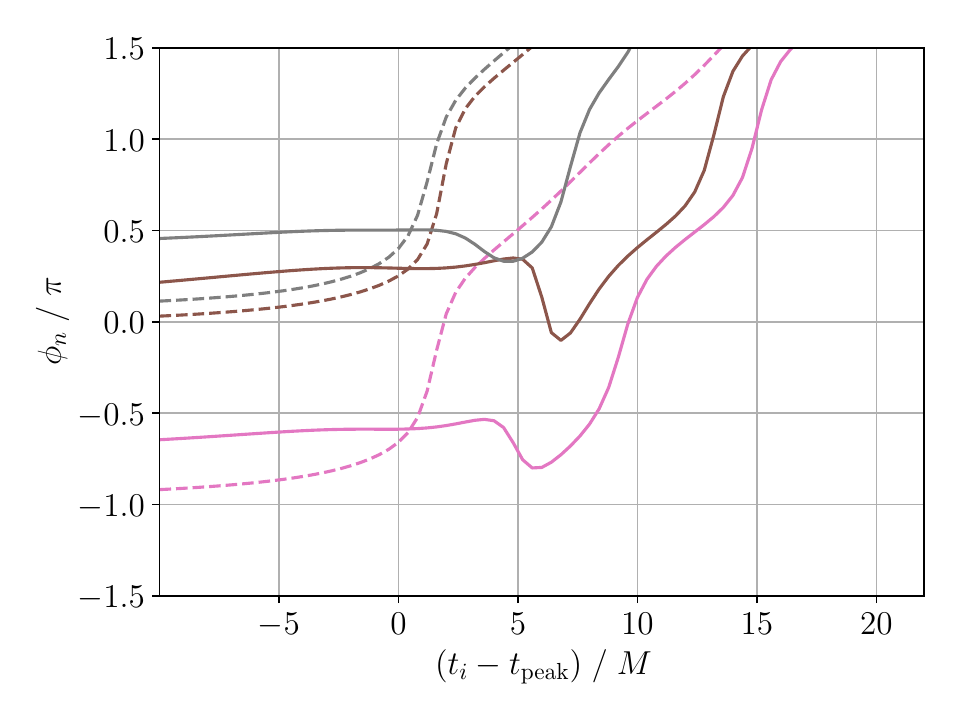}
    \caption{Comparison of the amplitude $A_n$ (top row) and phase $\phi_n$ (bottom row) obtained by fitting SXS:BBH:0305 by the conventional method (dashed) with the fitting function $\psi^\fit_{[0,7]}$ and the iterative method (solid) with the fitting function $\psi^\fit_{[K,N]}$ with $K=0,\cdots, 7$ and $N={\rm min}(K+3,~ 7)$.}
    \label{fig:0305_An_improve_sub_const_N4}
\end{figure*}

\begin{figure*}[t]
    \centering
    \includegraphics[width=0.8\columnwidth]{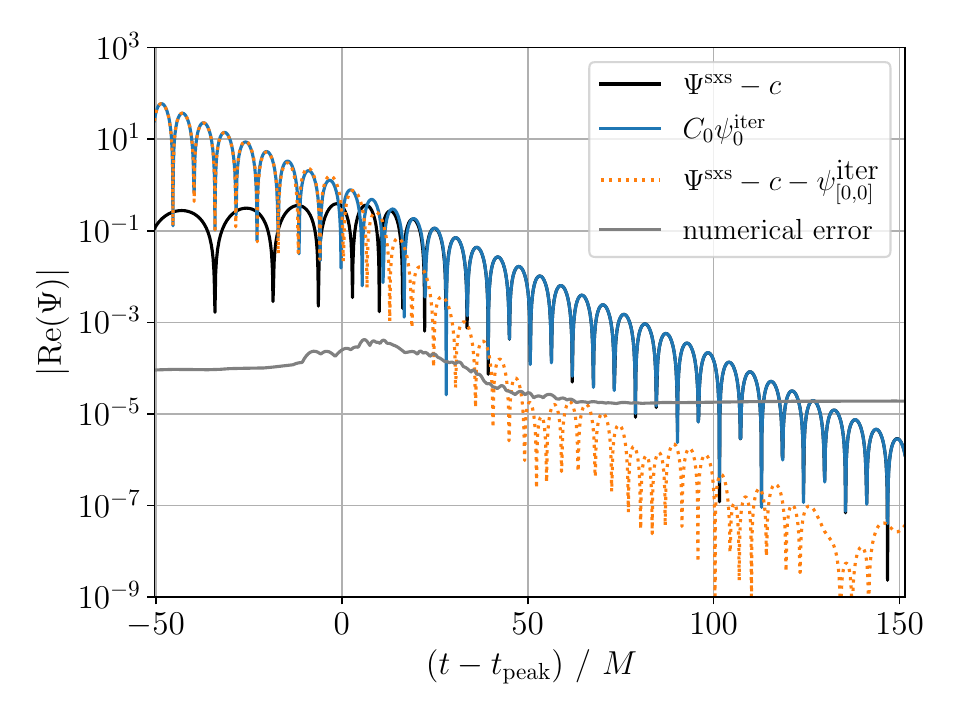}
    \includegraphics[width=0.8\columnwidth]{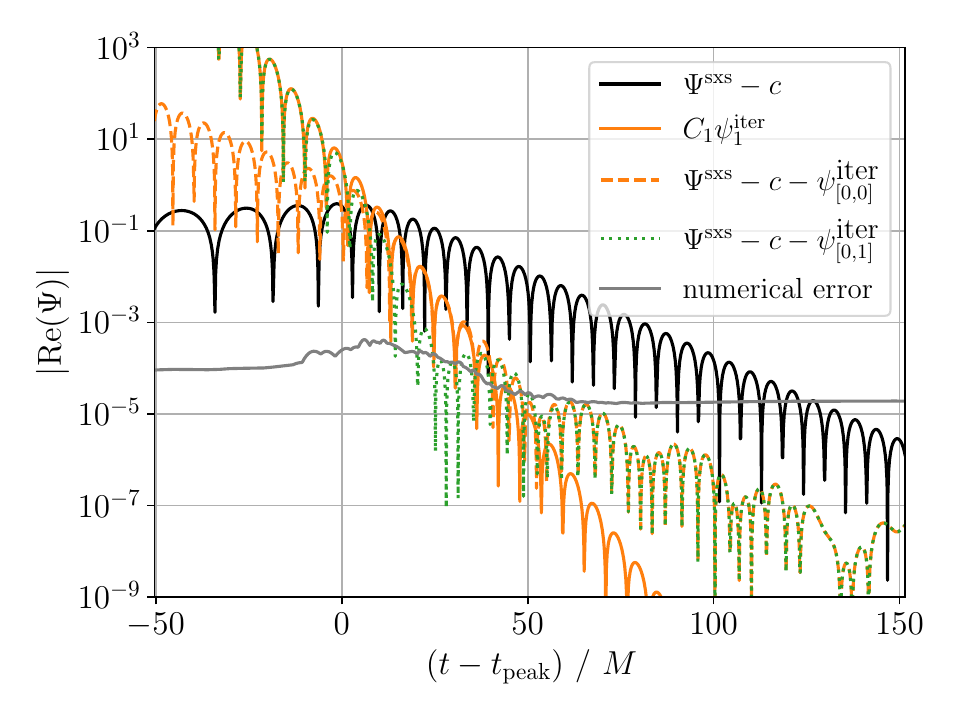}
    \includegraphics[width=0.8\columnwidth]{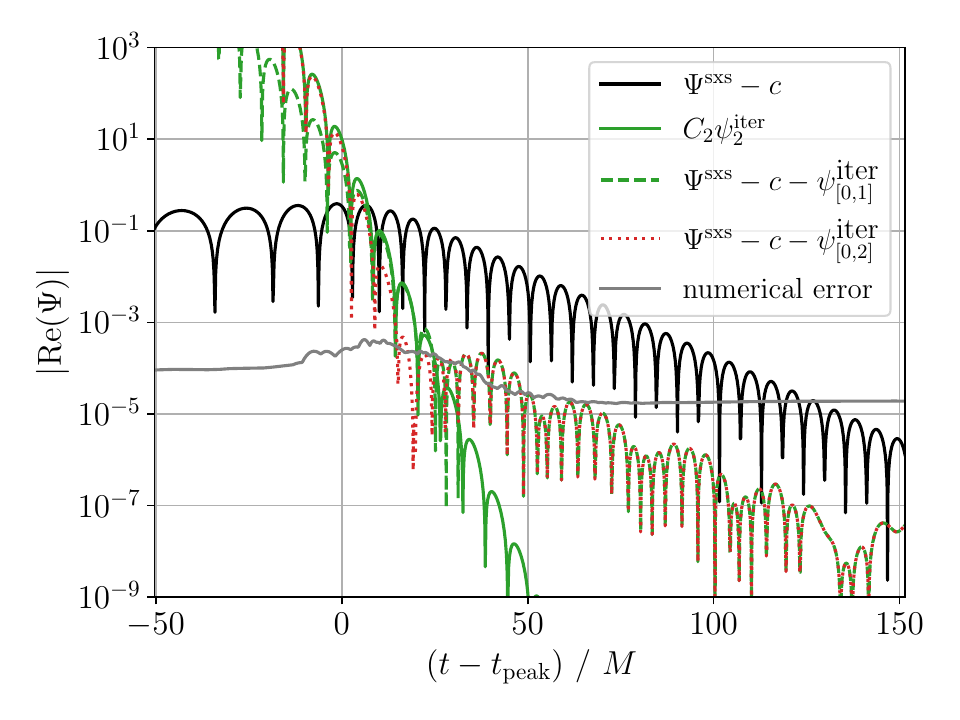}
    \includegraphics[width=0.8\columnwidth]{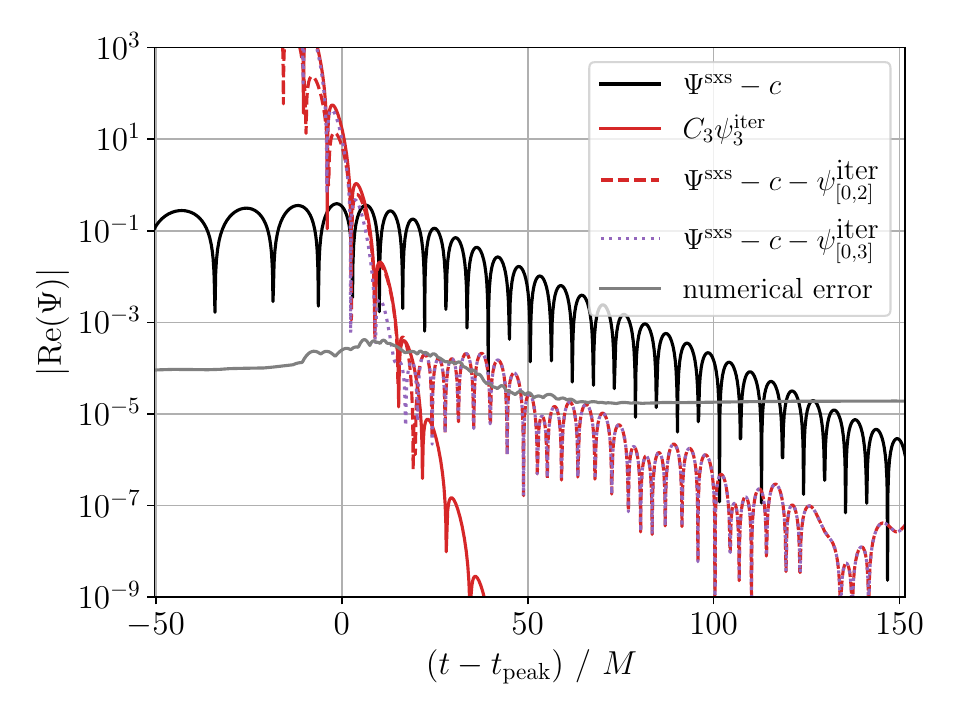}
    \includegraphics[width=0.8\columnwidth]{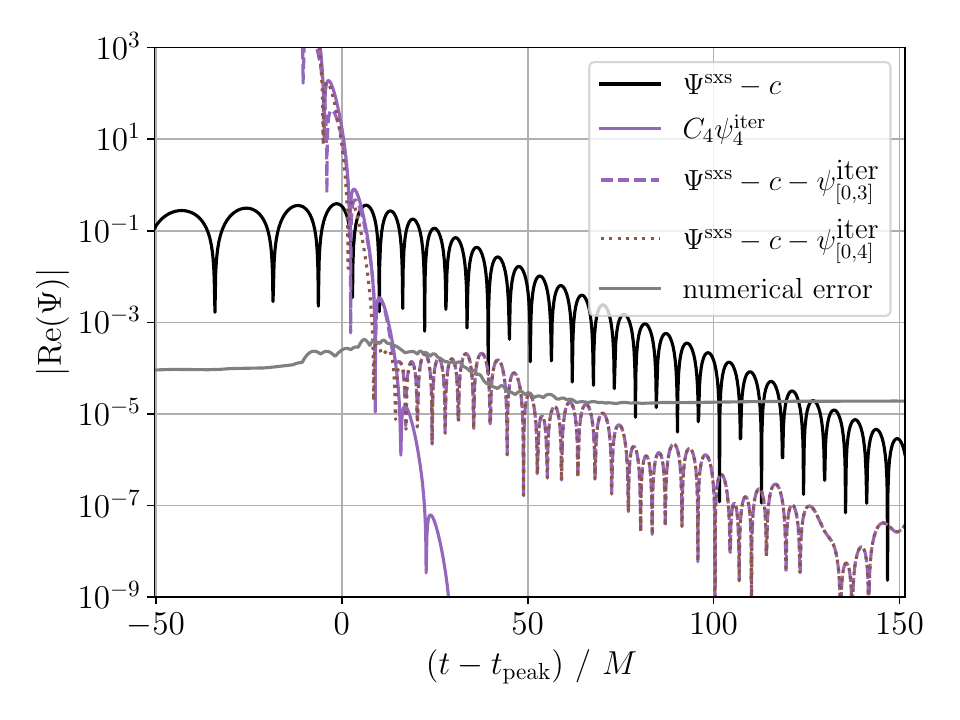}
    \includegraphics[width=0.8\columnwidth]{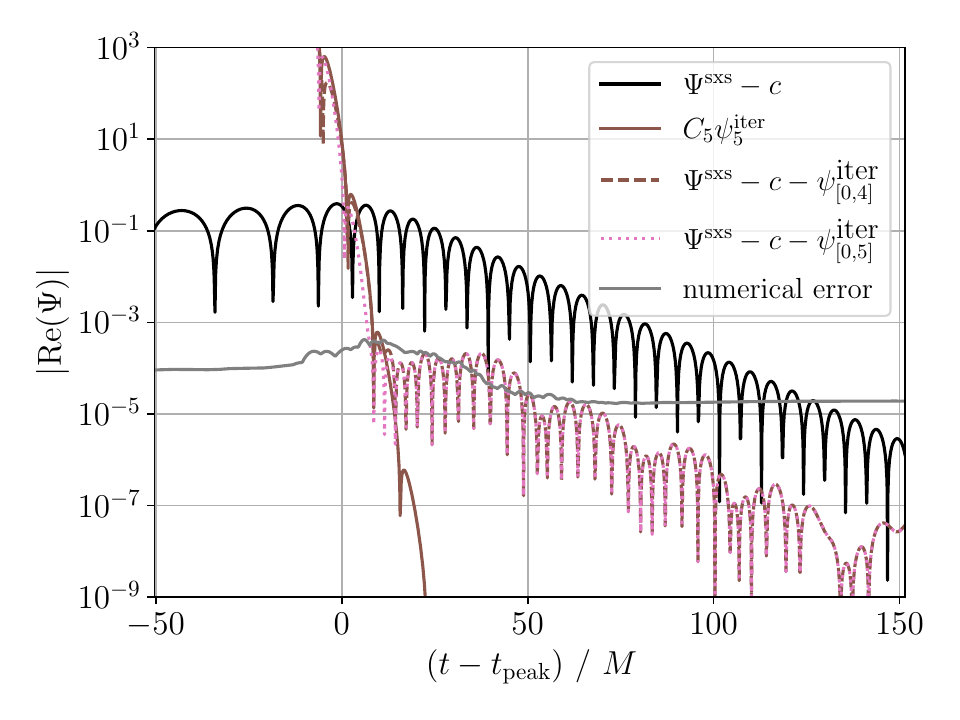}
    \includegraphics[width=0.8\columnwidth]{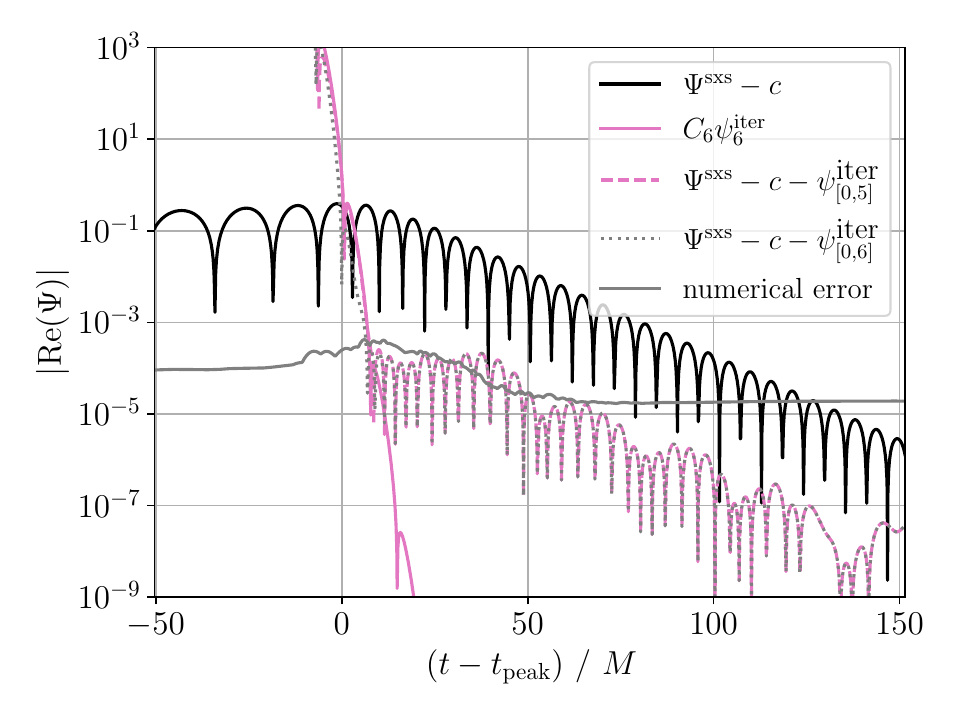}
    \includegraphics[width=0.8\columnwidth]{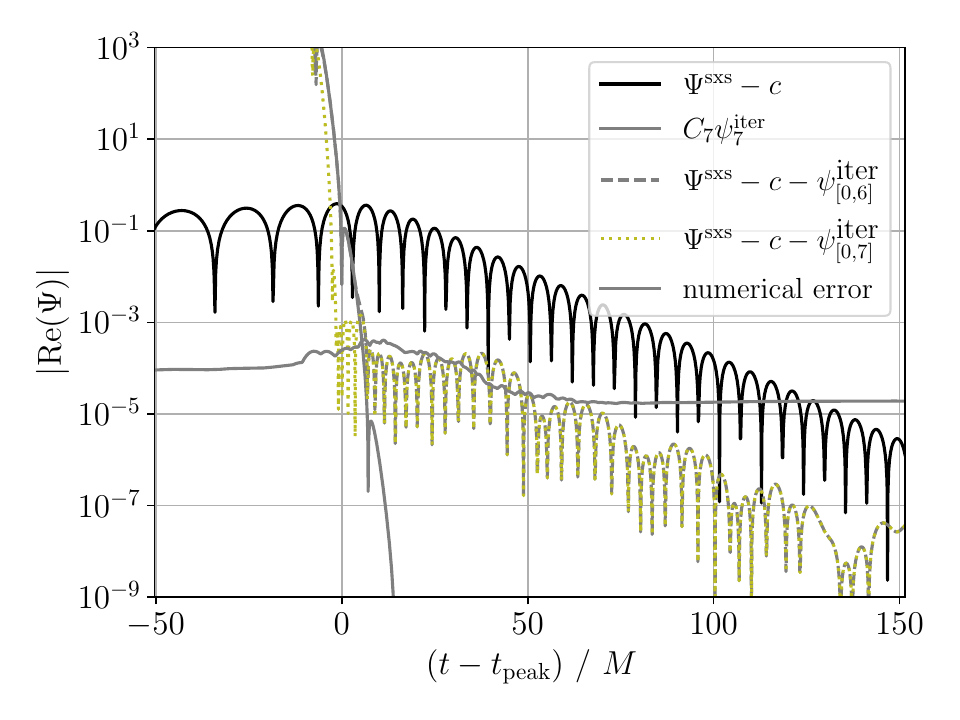}
    \caption{SXS:BBH:0305 waveform (black solid), the extracted longest-lived mode at each iterative step (colored solid) by fitting the waveform obtained by the previous step (dashed), and the waveform after the subtraction of the extracted mode (dotted).}
    \label{fig:fitting_error_0305}
\end{figure*}

\begin{figure}[t]
    \centering
    \includegraphics[width=\columnwidth]{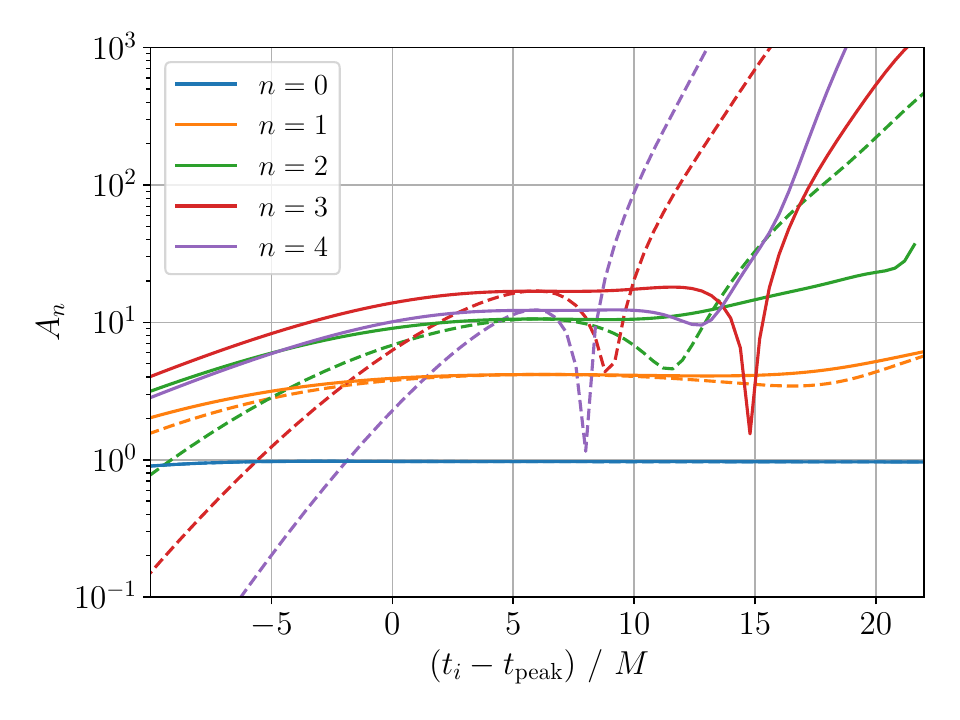}
    \includegraphics[width=\columnwidth]{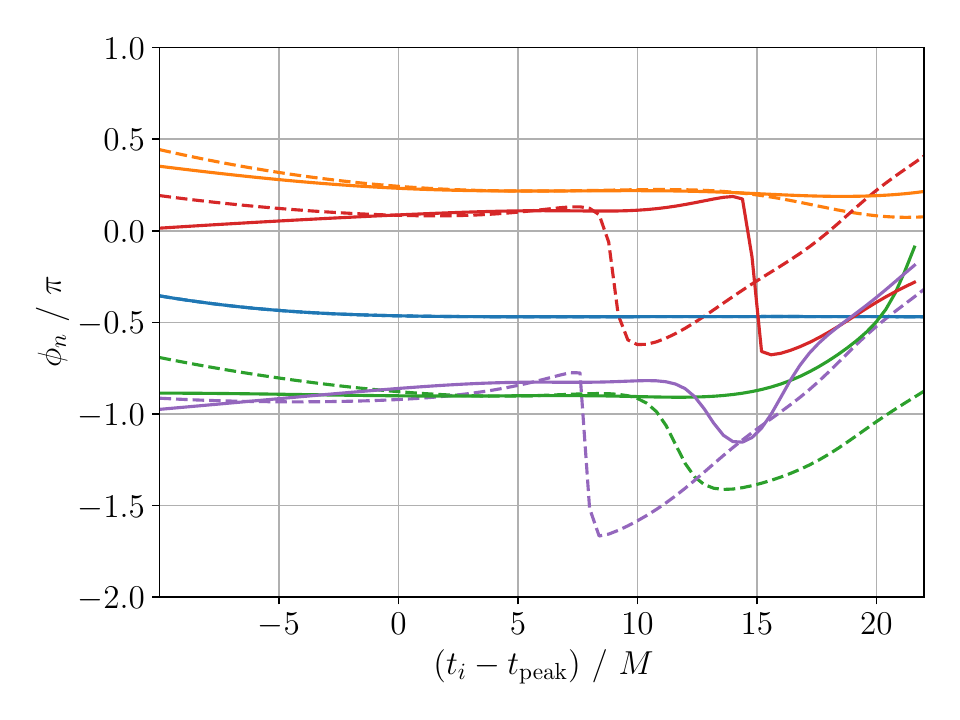}
    \caption{Comparison of the amplitude $A_n$ (top) and phase $\phi_n$ (bottom) obtained by fitting SXS:BBH:0305 by the conventional method (dashed) with the fitting function $\psi^\fit_{[0,4]}$ and the iterative method (solid) with the fitting function $\psi^\fit_{[K,4]}$ with $K=0,\cdots, 4$.}
    \label{fig:0305_An_improve_sub_const_K0-4_N4}
\end{figure}

From the fitting of $\Psi^{\rm{SXS}}-c-\psi^{\iter}_{[0,0]}$, we extract the best-fit value $C^{\iter}_{1}$.
We then subtract the best-fit first overtone from the waveform, after which we obtain $\Psi^{\rm{SXS}}-c-\psi^{\iter}_{[1,0]}$, which is shown by the green curve in Fig~\ref{fig:0305_wave_naive}.
By fitting this waveform, next we extract the best-fit value $C^{\iter}_{2}$ for the second overtone.
We can continue the iterative fitting so long as the available data interval is longer than the time scale of the effective longest-lived mode.

Thus, starting from the original SXS:BBH:0305 waveform, we iteratively fit and subtract the longest-lived mode step by step and obtain the best-fit values $C^{\iter}_n$, focusing on $(\ell,m)=(2,2)$ QNMs. 
This procedure accumulates the slight improvements of the stability of the fit of the overtones.
The results are summarized in Figs.~\ref{fig:0305_n4_improve_sub_const}--\ref{fig:0305_An_improve_sub_const_K0-4_N4}.
Figure~\ref{fig:0305_n4_improve_sub_const} shows the results of the fit by the fitting function $\psi^{\fit}_{[K,N]}$ with $K=0,\cdots, 7$ and $N={\rm min}(K+3,~ 7)$.
The amplitude $A_n$, phase $\phi_n$, and rate of change $\gamma_n$ are presented in each row for $n=1,\cdots,7$.
For $n=1,\cdots,4$ modes, we use a superposition of four QNMs as a fitting function, since the superposition of the four modes is efficient to extract the overtones as we examined above. 
Here, as a reference, we also provide the result for $n=5,6,7$ modes, which are fitted by fewer number of modes.
The type of each curve indicates the fit at each step of the iterative fitting procedure.
The solid curve represents the fit when the mode of interest is the longest-lived mode, from which we extract the best-fit value $C^\iter_n$.  
The dot-dashed, dashed, and dotted curves represents the fit at the intermediate steps when the mode of interest is the second, third, and fourth longest-lived mode, respectively.
Regarding the amplitude $A_n$, we see that the plateau is extended and the fit becomes more stable as the mode of interest approaches the longest-lived mode.
In particular, looking at the $n=2$ mode, we find that the plateau of $A_2$ is almost nonexistent in the first fit where it is the third longest-lived mode, whereas the plateau extends roughly from $t_i\simeq 0$ to $10$ for the final fit where the $n=2$ mode is the longest-lived mode. 
This is also the case for the phase $\phi_n$, which is also becoming more stable by iterative subtractions.
The phase $\phi_n$ sometimes fluctuates widely, but the iterative fitting method reduces the behavior.
The $\gamma_n$ in the third column of Fig.~\ref{fig:0305_n4_improve_sub_const} allows us to quantitatively confirm that the plateau is extended.
As a demonstration, let us focus on the range of $t_i/M$ when $\gamma_n<10^{-1}$ is satisfied.
For the $n=2$ mode, in the first fit, such a range is only about 5.
After iterating subtractions, the range is extended to 15 in the final fit, which is three times longer than without the subtraction procedure.
For the other modes, we can also see that the extension of the plateau can be evaluated qualitatively by $A_n, \phi_n$ and quantitatively by $\gamma_n$.
The iterative fitting method significantly improves the stable extraction of the overtones.
This also applies to $n=5,6,7$ modes which are fitted by a fewer number of modes.

In Fig.~\ref{fig:0305_An_improve_sub_const_N4}, we highlight the difference between the conventional fit and the iterative fit.
The dashed curves represent the conventional fit, which are obtained by a single fit of the waveform $\Psi^\sxs - c$ with the start time of the fit $t_i=t_\peak$ and the fitting function $\psi^\fit_{[0,7]}$ including up to seventh overtone, same as \cite{Giesler:2019uxc}.
The behavior of the dashed curves are consistent with Fig.~6 in \cite{Baibhav:2023clw}.
While the fundamental mode and the first overtone are fitted in a stable manner, the fit of the overtones $n\geq 2$ are sensitive to the choice of the start time of the fit, suggesting that the fit is not stable.
The stability of the overtones is improved in the iterative fit, as shown by the solid curves in Fig.~\ref{fig:0305_An_improve_sub_const_N4}.
From Fig.~\ref{fig:0305_An_improve_sub_const_N4} and the $\gamma_n$ in Fig.~\ref{fig:0305_n4_improve_sub_const}, we note that the plateau begins from earlier $t_i$ for higher overtones, which makes sense given that the higher overtones are responsible to early time ringdown.

It might appear from Fig.~\ref{fig:0305_An_improve_sub_const_N4} that the higher overtones up to $n=7$ may be extracted in a stable manner.
However, we should interpret the results carefully by checking the waveform at each iterative subtraction. 
Figure~\ref{fig:fitting_error_0305} represents SXS:BBH:0305 waveform (black solid), the extracted longest-lived mode at each iterative step (colored solid) by fitting the waveform obtained by the previous step (dashed), and the waveform after the subtraction of the extracted mode (dotted).
For lower overtones, we can observe several oscillations before the waveform is contaminated by numerical errors.
On the other hand, for higher overtones, since they decay more quickly, the waveform is contaminated by numerical errors before exhibiting several oscillations.
Therefore, the range of data of waveform over the numerical errors which we can use for the fitting becomes shorter.
Here, as a rough criterion, let us adopt the number of oscillations between the peak time and the time when numerical errors start to become dominant. 
We see that for $n\geq 5$, the waveform does not demonstrate a single period of damped sinusoid between the time range.
Of course, the peak time itself does not necessarily mean the starting time of the ringdown, so a more physical criterion needs to be established.
Nevertheless, we should bear in mind that even if the plateau is extended for $n\geq 5$ in Fig.~\ref{fig:0305_An_improve_sub_const_N4}, it is necessary to carefully determine up to which overtones can be extracted without overfit. 
Below, we focus on the modes up to $n=4$ for the iterative fitting method.

In Fig.~\ref{fig:0305_An_improve_sub_const_K0-4_N4}, we compare the results in the alternative way, where we use the fitting function consisting of the modes up to the fourth overtone for both conventional and iterative fit. 
Again, the plateau is extended by the iterative fit compared to the conventional fit.
The improvement is also consistent with what we expected from the improvement in Figs.~\ref{fig:0305_An_improve_mock_naive_N4}--\ref{fig:0305_An_improve_mock_w_const_K0-4_N4} for the mock waveform analysis.
As we can see from Figs.~\ref{fig:0305_An_improve_sub_const_N4} and \ref{fig:0305_An_improve_sub_const_K0-4_N4}, the difference between the extracted values in the two ways becomes larger for higher overtones, which is also in parallel to Figs.~\ref{fig:0305_An_improve_mock_naive_N4}--\ref{fig:0305_An_improve_mock_w_const_K0-4_N4} for the case of mock waveform with noise.
Based on the comparison of Figs.~\ref{fig:0305_An_improve_mock_naive_N4}--\ref{fig:0305_An_improve_mock_w_const_K0-4_N4} and the fiducial value, we expect that the plateau in Fig.~\ref{fig:0305_An_improve_sub_const_N4} gives a closer estimation of the underlying coefficient of each mode contained in the waveform.

Therefore, for the following, we mainly focus on the former way of comparison, i.e., Fig.~\ref{fig:0305_An_improve_sub_const_N4} with the modes up to $n=4$.
While there exists the above subtleties, we confirm that the lessons from the mock waveform analyses indeed work for the numerical relativity waveform and the subtractions helps us to extract the overtones in a more stable manner.

\begin{table*}[t]
    \centering
    \caption{Best-fit values for the amplitude $A_n$ and phase $\phi_n$ by fitting the SXS:BBH:0305 waveform subtracting the numerical constant with the conventional and iterative methods. 
    For the conventional fit, we set the fit start time $(t_i-t_\peak)/M=0$ and the fit end time $(t_e-t_\peak)/M=150$.}
    \begin{tabular}{c|cc|cccc}
    \hline\hline
        QNM & \multicolumn{2}{|c|}{Conventional} &  \multicolumn{4}{c}{Iterative} \\
        $n$ ~&~ $A_n$ ~&~ $\phi_{n}/\pi$ ~&~ $A_n$ ~&~ $\phi_{n}/\pi$ ~&~ $(t_i-t_\peak)/M$ ~&~ $(t_e-t_\peak)/M$ \\
        \hline
        $0$ ~&~ $0.9683$ ~&~ $-0.4697$ ~&~ $0.9692$ ~&~ $-0.4696$ ~&~ $8.8$ ~&~ $150$ \\
        $1$ ~&~ $4.180$ ~&~ $0.2134$ ~&~ $4.169$ ~&~ $0.2177$ ~&~ $5.6$ ~&~ $50$ \\
        $2$ ~&~ $11.21$ ~&~ $-0.9134$ ~&~ $10.63$ ~&~ $-0.8995$ ~&~ $2.8$ ~&~ $25$ \\
        $3$ ~&~ $22.59$ ~&~ $0.08648$ ~&~ $18.73$ ~&~ $0.1090$ ~&~ $0.40$ ~&~ $20$ \\
        $4$ ~&~ $32.36$ ~&~ $-0.8415$ ~&~ $21.99$ ~&~ $-0.8115$ ~&~ $-2.0$ ~&~ $15$ \\
        $5$ ~&~ $28.93$ ~&~ $0.2520$ ~&~ $(15.72)$ ~&~ $(0.2971)$ ~&~ $(-1.6)$ ~&~ $(10)$ \\
        $6$ ~&~ $14.14$ ~&~ $-0.6589$ ~&~ $(6.002)$ ~&~ $(-0.5872)$ ~&~ $(-1.6)$ ~&~ $(8.0)$ \\
        $7$ ~&~ $2.921$ ~&~ $0.4015$ ~&~ $(0.9460)$ ~&~ $(0.5023)$ ~&~ $(-1.6)$ ~&~ $(8.0)$ \\
        \hline\hline
    \end{tabular}
    \label{tab:0305_bestfit_An_phin}
\end{table*}

The best-fit values obtained by the conventional fit and iterative fit of the waveform $\Psi^\sxs - c$ in Fig.~\ref{fig:0305_An_improve_sub_const_N4} are listed in Table~\ref{tab:0305_bestfit_An_phin}.
For the conventional fit, we fix $t_i=t_\peak$ and $(t_e-t_\peak)/M=150$ and obtain the amplitude and phase by fitting $\Psi^\sxs - c$ by the fitting function $\psi^\fit_{[0,7]}$.
While there is a slight difference, e.g., the fitted waveform is $\Psi^\sxs$ in \cite{Giesler:2019uxc} but $\Psi^\sxs - c$ for our analysis, other setup of the fitting is the same and the resultant values of $A_n$ and $\phi_n$ are consistent with \cite{Giesler:2019uxc}.
For $t_i$ and $t_e$ we show only two digits as we adopt the closest value in a given grid of SXS waveform.

On the other hand, for the iterative fit, we iteratively adopt the fitting function $\psi^\fit_{[K,N]}$ with $K=0,\cdots, 7$ and $N={\rm min}(K+3,~ 7)$ as in Fig.~\ref{fig:0305_An_improve_sub_const_N4}. 
We optimize the range $[t_i,t_e]$ of the data for the fit as follows:  
When extracting $n$-th overtone, we determine $t_i$ as the fit start time when $\gamma_{m}$ with $m={\rm min}(n+1,~7)$ is minimum as we described above and $t_e$ as the time around which the corresponding waveform in Fig.~\ref{fig:fitting_error_0305} is comparable to the numerical errors, though we confirm that the result is not so sensitive to $t_e$ even if we extend it into the regime where the numerical errors are dominant.
The choice of $t_i$ is more important, and we expect that $t_i$ is earlier for higher overtones as higher overtones are dominant at earlier times.
Indeed, from Table~\ref{tab:0305_bestfit_An_phin} we confirm the trend that $t_i$ for the higher overtones are getting earlier.
This trend stops when $t_i < t_\peak$ for $n\geq 4$.

As we explained above, for $n\geq 5$, the iterative method is less efficient since the subtracted waveform is contaminated by numerical errors before exhibiting a single period of oscillation after the peak time.
Therefore, we consider that the extracted values $A_n$ and $\phi_n$ for $n\geq 5$ may be less meaningful, and present them in parentheses in Table~\ref{tab:0305_bestfit_An_phin} as a reference value.
Still, a criterion up to which overtones can be extracted without overfit is nontrivial.
In addition to the number of oscillations, another simple criterion would be whether $t_i$ for the best fit is earlier than some threshold value. 
If we adopt the peak time as a threshold, $n=4$ would not be extracted.
However, while $t_i$ for the best fit decreases and becomes even before the peak time, note that the plateau in Fig.~\ref{fig:0305_An_improve_sub_const_N4} extends after the peak time and hence the values evaluated with $t_i$ after the peak time are the same order.
For instance, $(A_4,\phi_4/\pi)=(21.46,-0.8107)$ for $t_i=t_\peak$ whereas $(A_4,\phi_4/\pi)=(21.99,-0.8115)$ for $(t_i-t_\peak)/M=-2$.
Note that this criterion is relevant to the debate on the start time of ringdown, which varies from before to $\sim 10M$ after the peak time, making it difficult to draw conclusions at this stage.

Let us focus on the amplitudes and phases up to $n=4$.
From Table~\ref{tab:0305_bestfit_An_phin}, we can see that the difference between the two fitting methods becomes larger for higher overtones.
Specifically, for the fundamental mode, the relative difference between conventional and iterative method in $C_n$ is about $10^{-3}$, but it reaches about 20\% for $n=3$, and 40\% for $n=4$.
From the analysis of the mock waveform shown in Fig.~\ref{fig:0305_An_improve_mock_naive_N4}, where the fiducial values are stably and correctly extracted by the iterative fit, we expect that the best-fit values obtained by the iterative fit capture the contribution of each mode more appropriately.

\subsection{Spherical-spheroidal mixing in SXS:BBH:0305}
\label{sec:0305mix}

As we saw above, in Fig.~\ref{fig:0305_M_A_gamma_sub_const0}, the subtraction of the lower modes reveals the existence of the beat. 
To determine whether the beat originates from the numerical errors, we evaluate the numerical errors by comparing the numerical relativity waveform obtained with different resolution and extrapolation order, which are shown as green and magenta thin dotted curves, respectively (see Appendix~\ref{sec:errors}). 
The envelope of the all error curves is shown as a gray curve, which we adopt as the conservative estimation of the numerical errors.
This is because for the ringdown analysis, higher order extrapolation tends to lead an overfit, and hence a mild extrapolation order is recommended~\cite{Boyle:2019kee}.
We note that the rough amplitude of the first beat in the waveform $\Psi^\sxs - c - \psi^{\iter}_{[0,0]}$ around $(t-t_{\peak})/M\approx 60$ is larger than the errors estimated by the next-highest resolution. 
Further, the beat shows up more clearly above the estimated numerical errors in the waveform $\Psi^\sxs - c - \psi^{\iter}_{[1,0]}$ after the subtraction of the first overtone, as shown by the green curve in Fig.~\ref{fig:0305_wave_naive}.
Therefore, it would be reasonable to attribute the beat to the effect originating from the spherical-spheroidal mixing rather than the numerical errors. 
Indeed, in the case of the mock waveform, after the subtraction of the fundamental mode we see the first overtone and such behavior does not show up. 
The appearance of the beat is peculiar to the numerical relativity waveform.

Actually, the QNM frequency of $(\ell,m,n)=(3,2,0)$ mode is close to the $(2,2,0)$ fundamental mode (see Table~\ref{tab:0305_QNMs}).
Considering the possibility that the fundamental mode is not completely subtracted, it is expected that the beat between the residual of $(2,2,0)$ mode and $(3,2,0)$ mode via the spherical-spheroidal mixing occurs.
Indeed, the beat frequency is given by $M(\omega_{220}-\omega_{320}) = -0.2268+0.003404 i$, which is consistent with the time scale of the beat appearing in Fig.~\ref{fig:0305_wave_naive}.

Let us discuss the impact of the spherical-spheroidal mixing in the fit of the $(\ell,m)=(2,2)$ mode of the waveform SXS:BBH:0305.
By taking into account the dominant mixing of $(\ell,m,n)=(3,2,0)$ mode, we consider the following fitting function
\be \label{eq:fitw320} \psi_{[K,N]+320}^\fit = \sum_{n=K}^{N} C_{22n} \psi_{22n} + C_{320} \psi_{320}. \ee
In this subsection, we recover the subscript $(\ell,m)$.
In the iterative fitting procedure, we subtract the best-fit longest-lived mode in each step.
Therefore, following the ordering $|\Im \omega_{220}| < |\Im \omega_{320}| < |\Im \omega_{221}| < \cdots$, we subtract $(2,2,0)$, $(3,2,0)$, $(2,2,1)$, and so on.
For the each step of the iteration, we denote the sum of the best-fit effective longest-lived modes as $\psi^{\iter}_{[0,0]}$, $\psi^{\iter}_{[0,0]+320}$, $\psi^{\iter}_{[0,1]+320}$ and so on.

\begin{figure}[t]
    \centering
    \includegraphics[width=\columnwidth]{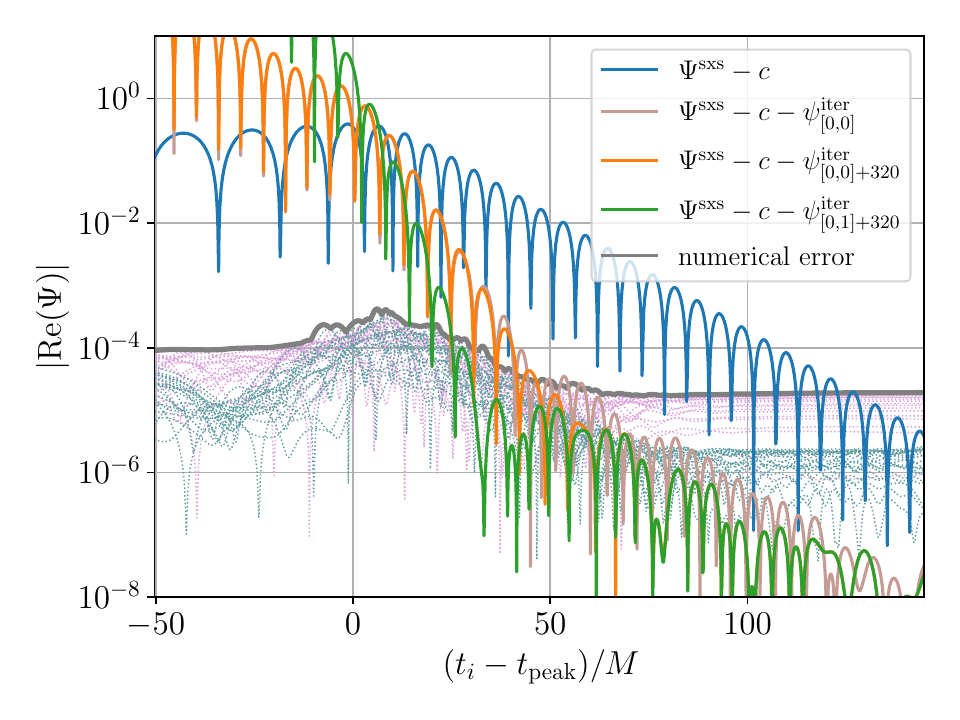}
    \caption{Numerical relativity waveform SXS:BBH:0305. Waveforms after the subtraction of the numerical constant (blue), the best-fit values of $(\ell,m,n)=(2,2,0)$ mode (light brown), $(3,2,0)$ mode (orange), and then $(2,2,1)$ mode (green).}
    \label{fig:0305_wave_naive320}
\end{figure}

We present the waveform of the $(\ell,m)=(2,2)$ mode of the SXS:BBH:0305 simulation after subtracting the numerical constant (blue) in Fig.~\ref{fig:0305_wave_naive320}, together with the waveform after the subtraction of $(\ell,m,n)=(2,2,0)$ mode (light brown), $(3,2,0)$ mode (orange), and then $(2,2,1)$ mode (green) as well. 
After the subtraction of the best-fit value of the $(3,2,0)$ mode, we can observe oscillations originating from $(2,2,1)$ mode a little longer up to $(t_i-t_\peak)/M\sim 60$, and the beat disappears.
Further, after the subtraction of the best-fit value of the $(2,2,1)$ mode, the beat in the green curve in Fig.~\ref{fig:0305_wave_naive} disappears in Fig~\ref{fig:0305_wave_naive320}.
Therefore, it is reasonable to attribute the origin of the beat to the spherical-spheroidal mixing.

\begin{figure}[t]
    \centering
    \includegraphics[width=\columnwidth]{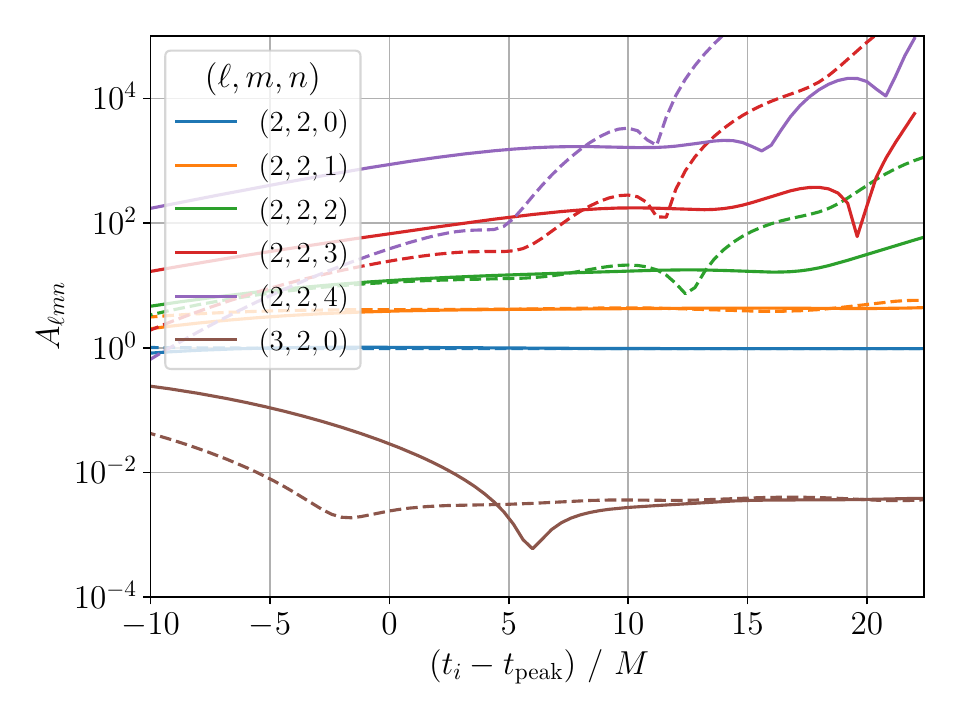}
    \includegraphics[width=\columnwidth]{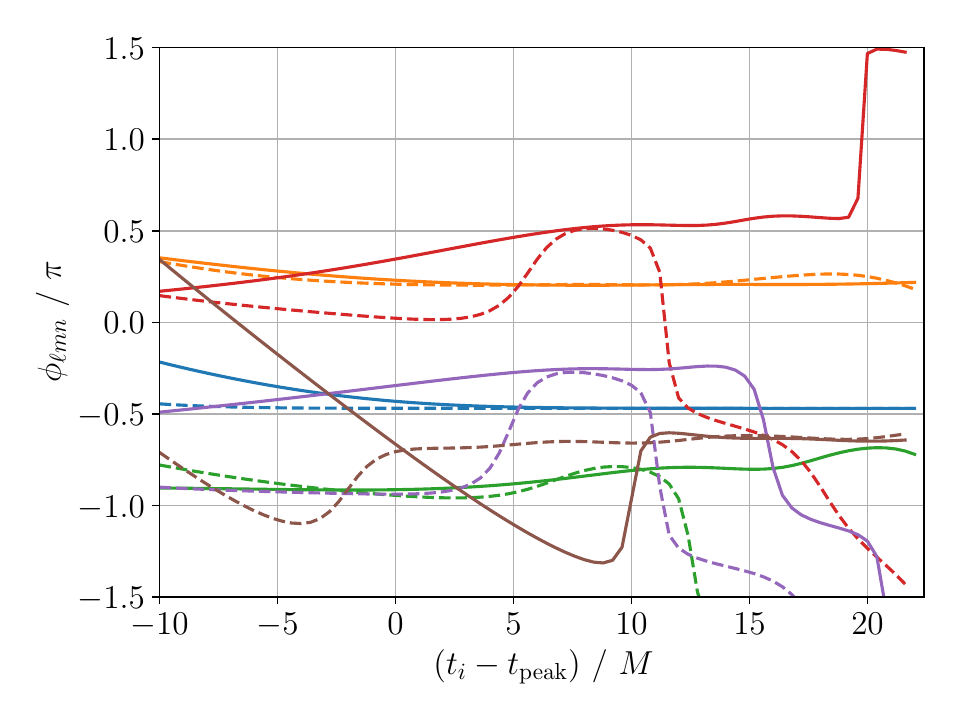}
    \caption{Comparison of the conventional fit (dashed) and iterative fit (solid) of SXS:BBH:0305 waveform including $(\ell,m,n)=(3,2,0)$ mode.}
    \label{fig:0305_An_improve_sub_320const_N4}
\end{figure}

We present the comparison of the results obtained by the conventional fit and iterative fit when taking into account the $(3,2,0)$ mode in Fig.~\ref{fig:0305_An_improve_sub_320const_N4}, and the best-fit values in Table~\ref{tab:0305_320_bestfit_An_phin}. 
Naively, taking the spherical-spheroidal mixing into account, the stability of the fit of overtones may be improved.
While the contribution of the $(3,2,0)$ mode is as small as $A_{320}=\mO(10^{-3})$, it affects the fit of the higher overtones of the $(2,2,n)$ modes.
Compared to Table~\ref{tab:0305_bestfit_An_phin} without the $(3,2,0)$ mode, while there is relatively small difference for $n\leq 2$, the $(2,2,n)$ modes with $n\geq 3$ differ significantly.
As another difference, $t_i$ for the best fit is shifted to later time.
Note that the validity of $n=3,4$ are marginal in the sense that the fitting interval is shorter than the lower modes and hence the fit of higher overtones is still challenging as we discussed above.

Our analysis shows that the iterative fitting method is efficient to isolate the spherical-spheroidal mixing.
The stability of the fit of the lower modes is improved when the spherical-spheroidal mixing compared to the conventional fit. 
However, after the inclusion of the mixing, the fit of the higher overtones becomes more sensitive and the best-fit values deviate significantly. 
Unfortunately, the analysis of the mixing is limited by the numerical errors.
While our analysis suggests the efficiency of the iterative fitting method to study the spherical-spheroidal mixing, a more detailed analysis with reduced errors would be necessary to obtain a rigorous understanding.

\begin{table*}[t]
    \centering
    \caption{Best-fit values for the amplitude $A_n$ and phase $\phi_n$ by fitting the $(\ell,m)=(2,2)$ mode of SXS:BBH:0305 waveform subtracting the numerical constant with the conventional and iterative methods, taking into account the spherical-spheroidal mixing with $(\ell,m,n)=(3,2,0)$ mode.
    For the conventional fit, we set $(t_i-t_\peak)/M=0$ and $(t_e-t_\peak)/M=150$.}
    \begin{tabular}{c|cc|cccc}
    \hline\hline
        QNM & \multicolumn{2}{|c|}{Conventional} &  \multicolumn{4}{c}{Iterative} \\
        $n$ ~&~ $A_n$ ~&~ $\phi_{n}/\pi$ ~&~ $A_n$ ~&~ $\phi_{n}/\pi$ ~&~ $(t_i-t_\peak)/M$ ~&~ $(t_e-t_\peak)/M$ \\
        \hline
        $(2,2,0)$ ~&~ $0.9672$ ~&~ $-0.4691$ ~&~ $0.9667$ ~&~ $-0.4695$ ~&~ $19.6$ ~&~ $150$ \\
        $(2,3,0)$ ~&~ $0.002412$ ~&~ $-0.7058$ ~&~ $0.003610$ ~&~ $-0.6520$ ~&~ $16.8$ ~&~ $150$ \\
        $(2,2,1)$ ~&~ $4.095$ ~&~ $0.2084$ ~&~ $4.331$ ~&~ $0.2070$ ~&~ $14.4$ ~&~ $50$ \\
        $(2,2,2)$ ~&~ $11.21$ ~&~ $-0.9446$ ~&~ $17.75$ ~&~ $-0.7910$ ~&~ $12.4$ ~&~ $35$ \\
        $(2,2,3)$ ~&~ $24.52$ ~&~ $0.02240$ ~&~ $175.3$ ~&~ $0.5325$ ~&~ $10.0$ ~&~ $25$ \\
        $(2,2,4)$ ~&~ $38.68$ ~&~ $-0.9379$ ~&~ $1625$ ~&~ $-0.2576$ ~&~ $10.4$ ~&~ $25$ \\
        $(2,2,5)$ ~&~ $37.22$ ~&~ $0.1243$ ~&~ $-$ ~&~ $-$ ~&~ $-$ ~&~ $-$ \\
        $(2,2,6)$ ~&~ $19.17$ ~&~ $-0.8183$ ~&~ $-$ ~&~ $-$ ~&~ $-$ ~&~ $-$ \\
        $(2,2,7)$ ~&~ $4.125$ ~&~ $0.2134$ ~&~ $-$ ~&~ $-$ ~&~ $-$ ~&~ $-$ \\
        \hline\hline
    \end{tabular}
    \label{tab:0305_320_bestfit_An_phin}
\end{table*}

\subsection{SXS:BBH:0158 - high spin simulation}

Next, we explore the fit of the waveform SXS:BBH:0158, whose remnant dimensionless spin $\chi_{\rm rem}=0.9450$, as a representative example of the high spin simulation.
As we mentioned above, as we increase the black hole spin, the QNMs damp more slowly and oscillate more rapidly, and most QNM frequencies become eventually degenerate.
By analyzing the high spin simulation, we shall see how these factors affect the extraction of overtones.

\begin{figure}[t]
    \centering
    \includegraphics[width=\columnwidth]{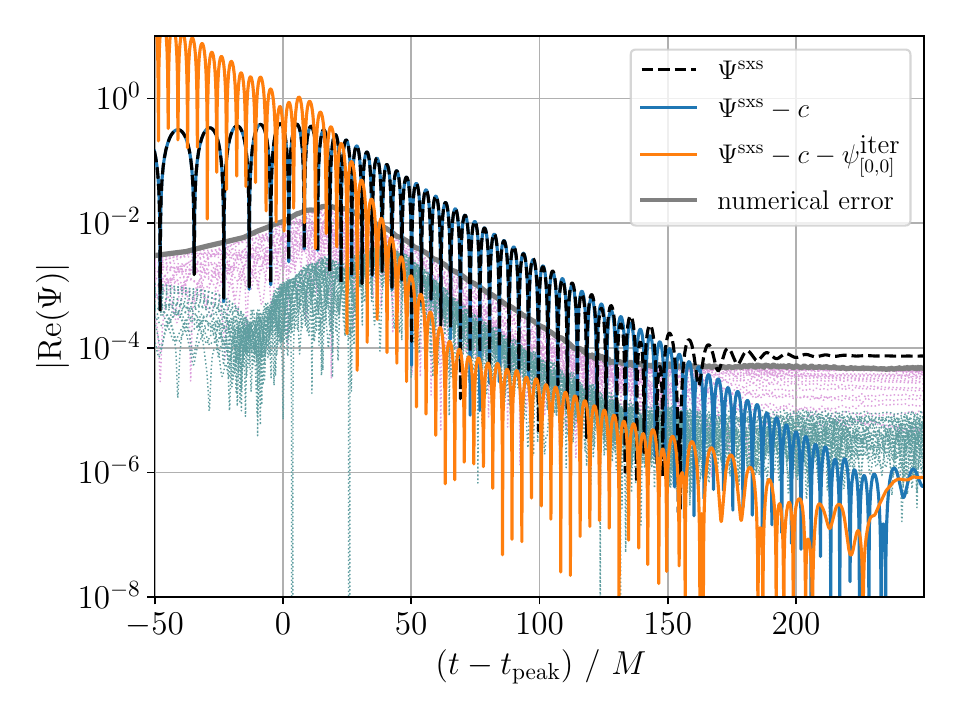}
    \caption{Numerical relativity waveform SXS:BBH:0158. Raw data (black dashed) and waveforms after the subtraction of the numerical constant (blue) and the best-fit fundamental mode (orange).}
    \label{fig:0158_wave_naive}
\end{figure}

\begin{figure*}[t]
    \centering
    \includegraphics[width=0.32\textwidth]{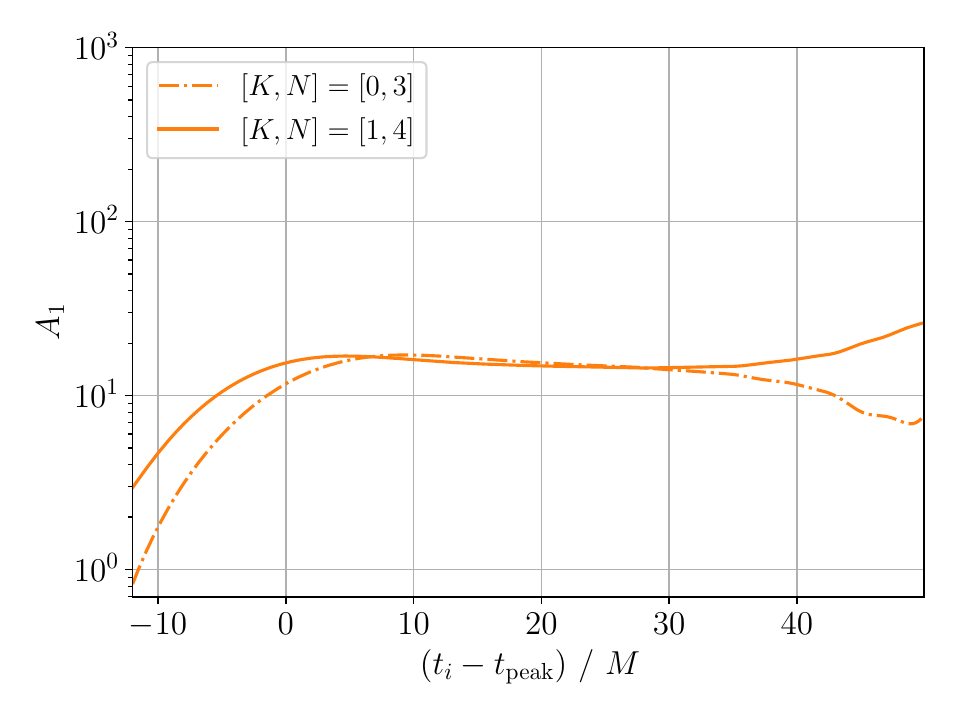}
    \includegraphics[width=0.32\textwidth]{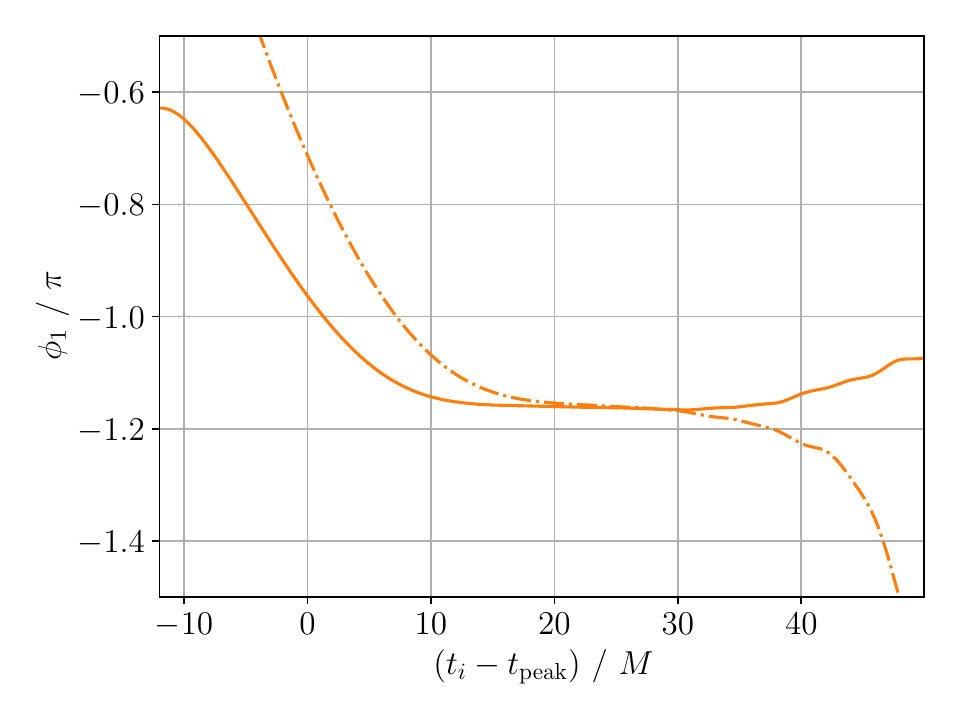}
    \includegraphics[width=0.32\textwidth]{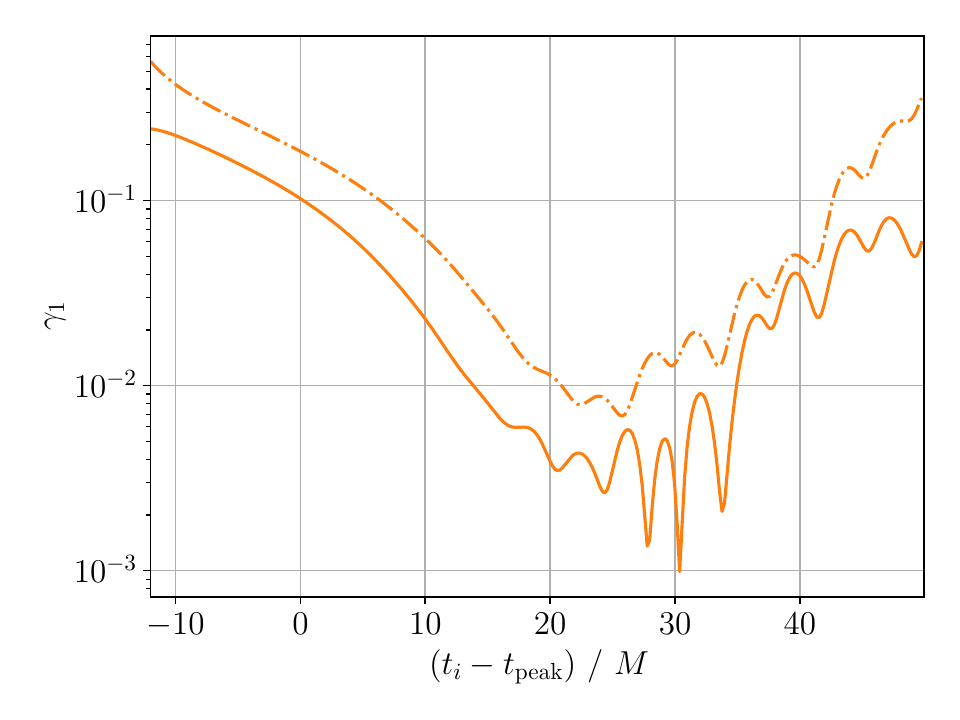}
    
    \includegraphics[width=0.32\textwidth]{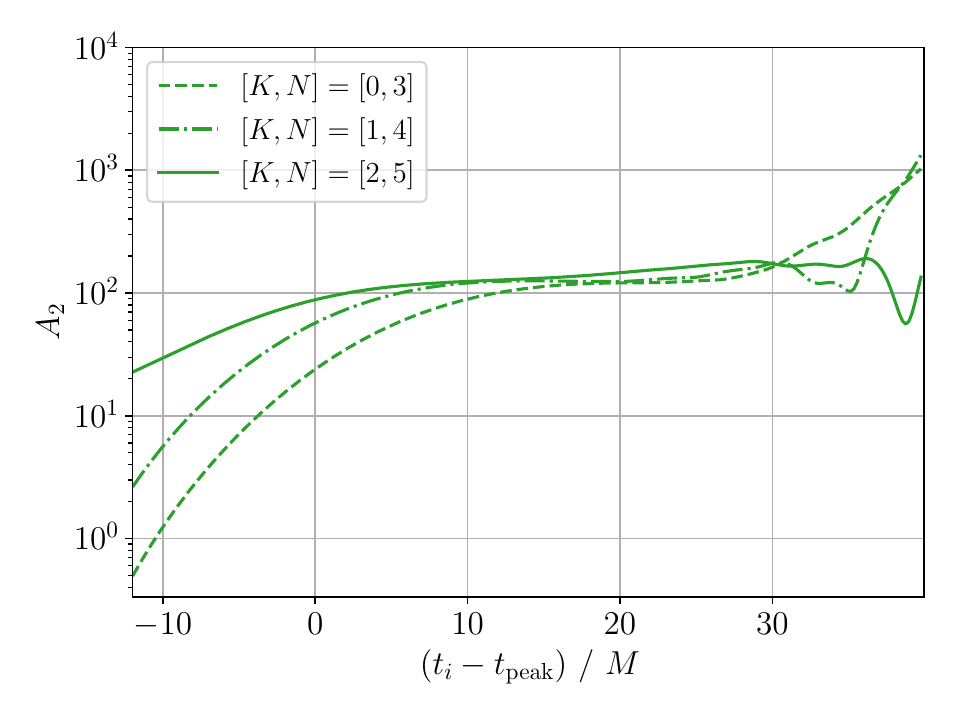}
    \includegraphics[width=0.32\textwidth]{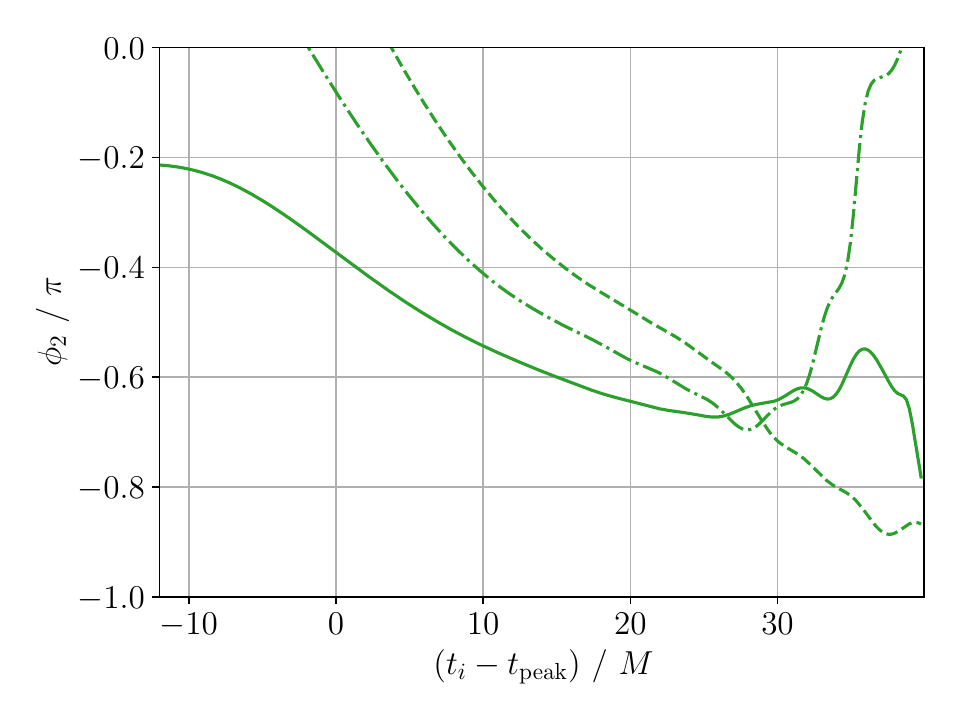}
    \includegraphics[width=0.32\textwidth]{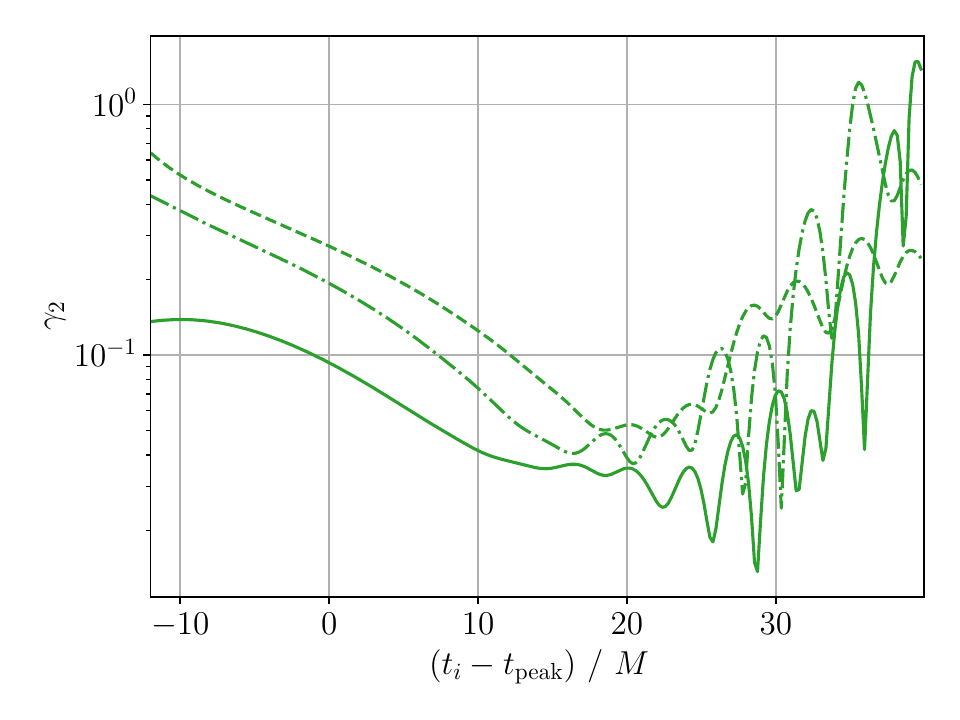}

    \includegraphics[width=0.32\textwidth]{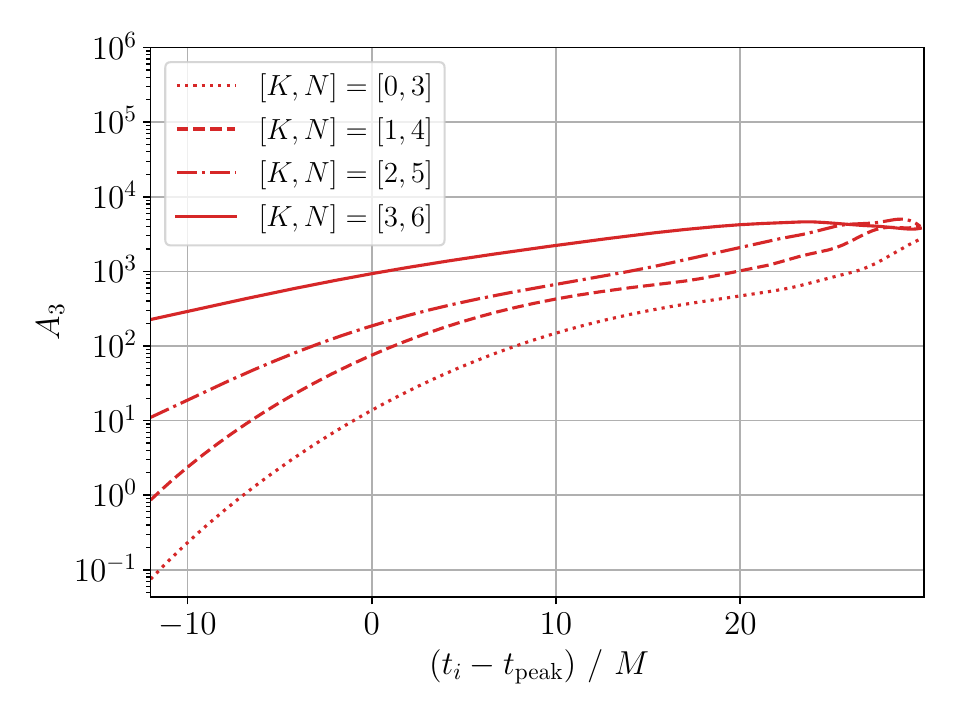}
    \includegraphics[width=0.32\textwidth]{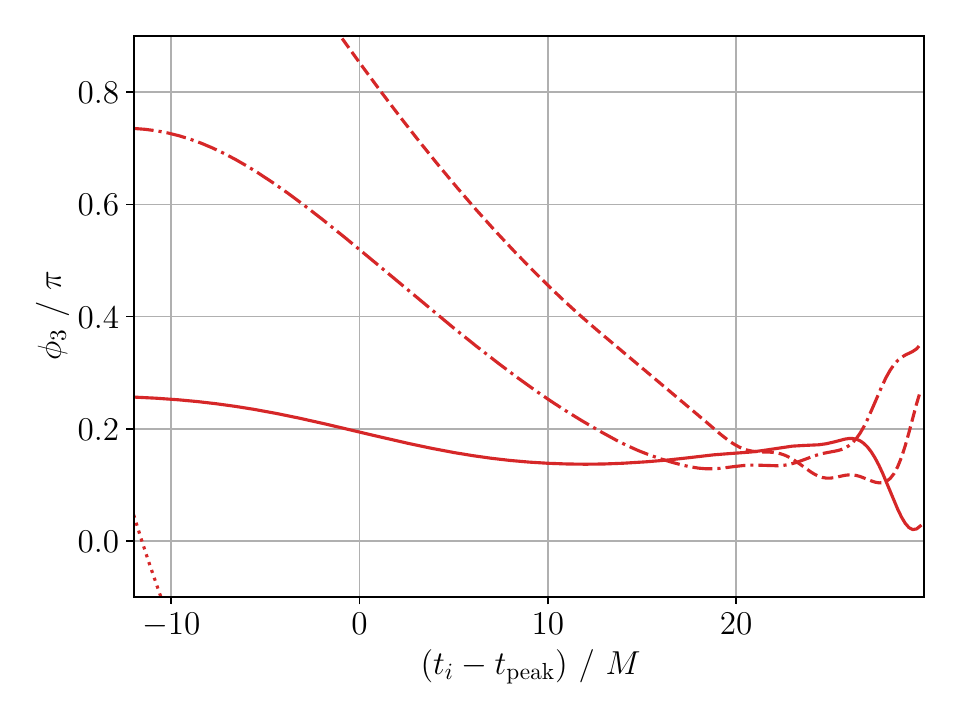}
    \includegraphics[width=0.32\textwidth]{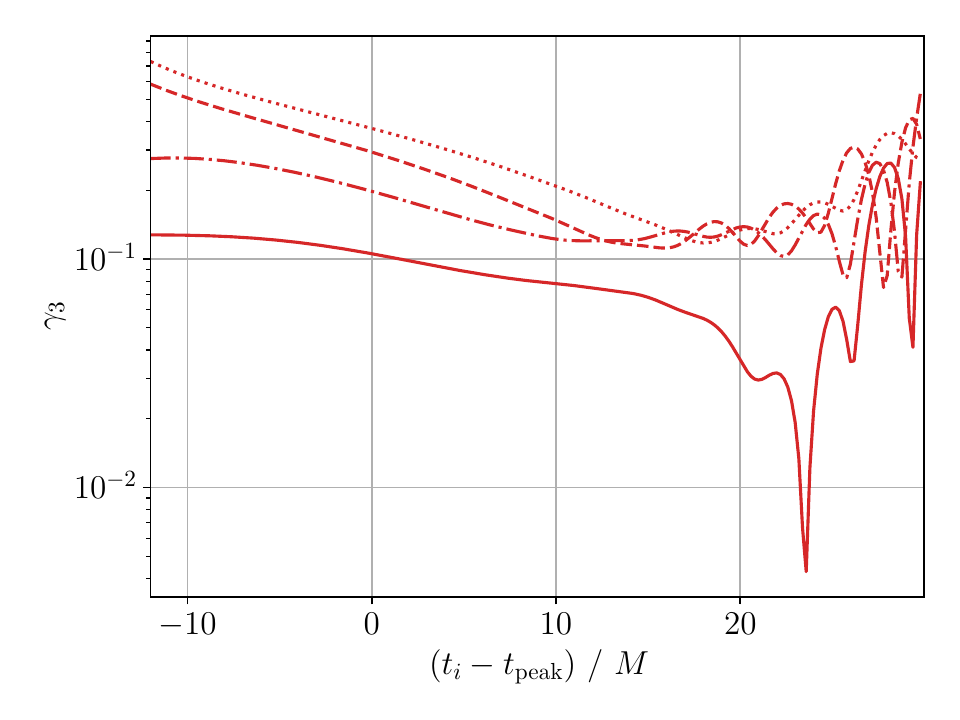}
    
    \includegraphics[width=0.32\textwidth]{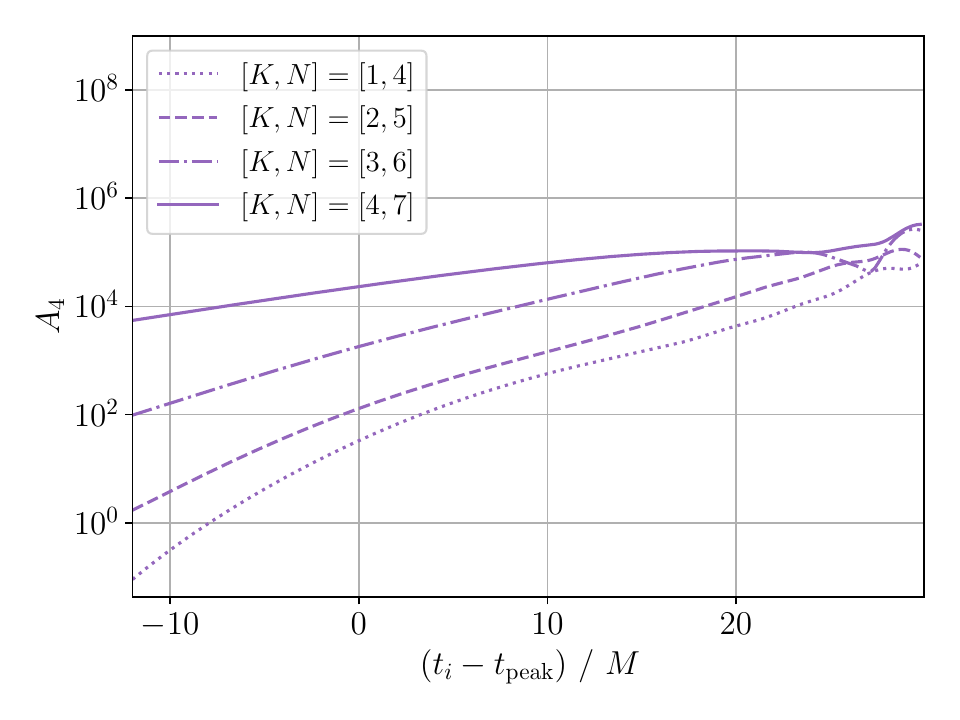}
    \includegraphics[width=0.32\textwidth]{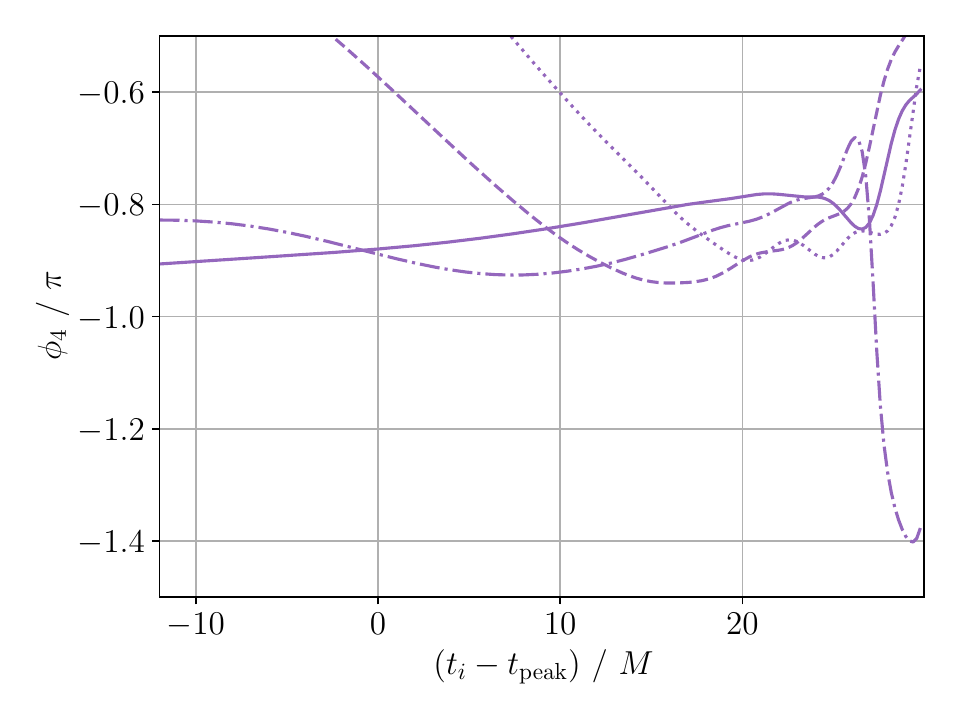}
    \includegraphics[width=0.32\textwidth]{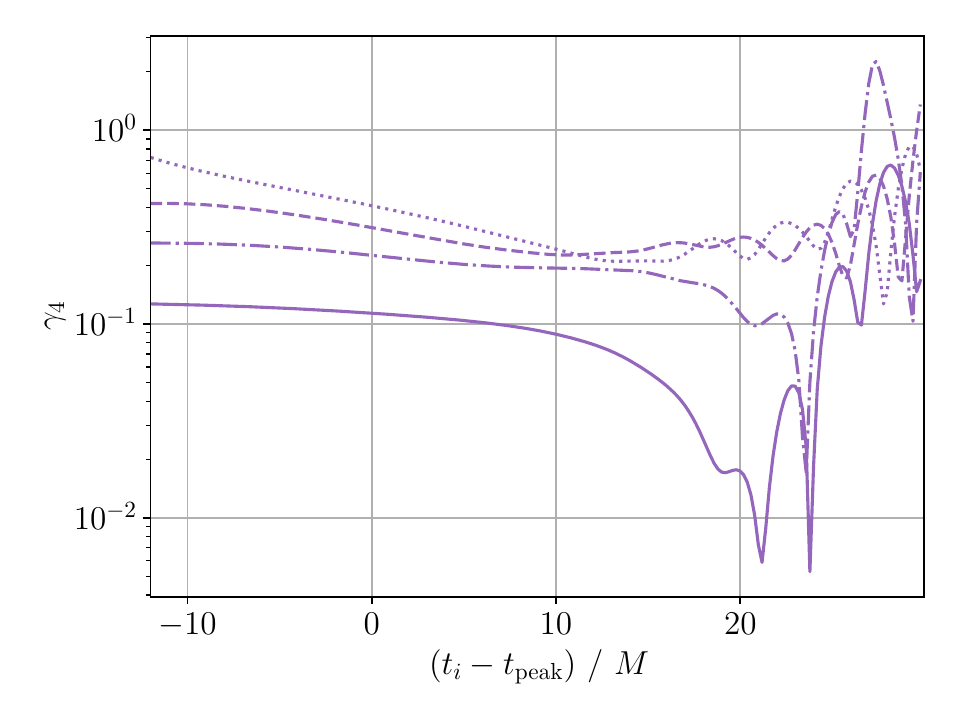}
    \caption{The improvement of the amplitude $A_n$ (left column), phase $\phi_n$ (right column) from $n=1$ mode (first row) to $n=4$ mode (fourth row) for the iterative fit of SXS:BBH:0158 by the fitting function $\psi^{\fit}_{[K,N]}$ with $N=K+3$.
    }
    \label{fig:0158_n4_improve_sub_const}
\end{figure*}

We present the waveform of SXS:BBH:0158 in Fig.~\ref{fig:0158_wave_naive}.
It is clear that the raw data waveform shown by the black curve is contaminated by a numerical constant at late time, whose value is about $c=(7.4-4.8i)\times 10^{-5}$.
While the waveform data of SXS:BBH:0158 are available up to $(t-t_{\rm peak})/M=400$, we did not show the full waveform since the late time waveform simply remains the constant and does not have physical meaning. 
For the following we use the waveform $\Psi^\sxs-c$ shown by the blue curve.

As expected, we can observe more oscillations compared to Fig.~\ref{fig:0305_wave_naive}, which would make the fitting easier than the case of SXS:BBH:0305.
We present the result of the fitting in Fig.~\ref{fig:0158_M_A_gamma_sub_const} in Appendix~\ref{sec:plots}, which should be contrasted with Figs.~\ref{fig:0305_M_A_gamma_sub_const} and \ref{fig:0305_M_A_gamma_sub_const0}.
As we did in the SXS:BBH:0305 analysis, we fit the waveform $\Psi^\sxs-c$ and obtain the best-fit value $C^{\iter}_0$ by using the fit starting from the time where $\gamma_1$ takes the minimum.
We then subtract the fundamental mode from the blue curve to obtain the orange curve in Fig.~\ref{fig:0158_wave_naive}, the latter of which is $\Psi^{\sxs} - c - \psi^{\iter}_{[0,0]}$.
Up to $(t-t_{\peak})/M\approx70$, we see that the orange curve can be mainly described by a single damped sinusoid, which is the first overtone.
The damping time seems changing for $(t-t_{\peak})/M\gtrsim 70$ but we expect that the waveform is subject to the numerical errors there.

We iterate the fit and subtraction and collect the best-fit value for the longest-lived mode at each step as $C^{\iter}_n$. 
The results are shown in Fig.~\ref{fig:0158_n4_improve_sub_const}.
We can see that the plateau of $A_n$ and $\phi_n$ is longer than the case of SXS:BBH:0305.
Even before the subtraction of the longest-lived mode, we can observe a mild plateau up to $n=2$, while there is no plateau on the second overtone in SXS:BBH:0305 before the subtractions.
The stability of the fit vary greatly depending on the ringdown waveform.

We can evaluate the stability or the length of the plateau quantitatively by the plot of $\gamma_n$ in Fig.~\ref{fig:0158_n4_improve_sub_const}.
The region where $\gamma_n$ remains small is longer than the case of SXS:BBH:0305.
Specifically, for the fit of $\Psi^\sxs-c$, the range of $t_i$ where $\gamma_1<10^{-1}$ is about $-5<(t_i-t_\peak)/M<20$ for SXS:BBH:0305, but it is about $10<(t_i-t_\peak)/M<45$ for SXS:BBH:0158, i.e., it is extended about $\Delta t /M \simeq 10$.
Furthermore, the region of $t_i$ where $\gamma_2<10^{-1}$ is about $5<(t_i-t_\peak)/M<10$ for SXS:BBH:0305, while it is about $10<(t_i-t_\peak)/M<30$ for SXS:BBH:0158, which is about four times larger than the range for SXS:BBH:0305.
Therefore, as expected, we find that the fit tends to be more stable for higher spin black holes, as the QNMs have slower damping time and higher real frequency.

The iterative fitting method further improves the stability of the fit, and the plateau of $A_n$ and $\phi_n$ is extended.  
From the middle column in Fig.~\ref{fig:0158_n4_improve_sub_const}, we see that the stability of $\phi_n$ improves significantly in all panels.
Especially, the improvement of $\phi_3$ (red) and $\phi_4$ (purple) are manifest.
While $A_3$ and $A_4$ are also flattened, they are still tilted. 
We expect that additional subtractions would make them sufficiently flat, for which we need the waveform with smaller numerical errors.

\begin{figure}[t]
    \centering
    \includegraphics[width=\columnwidth]{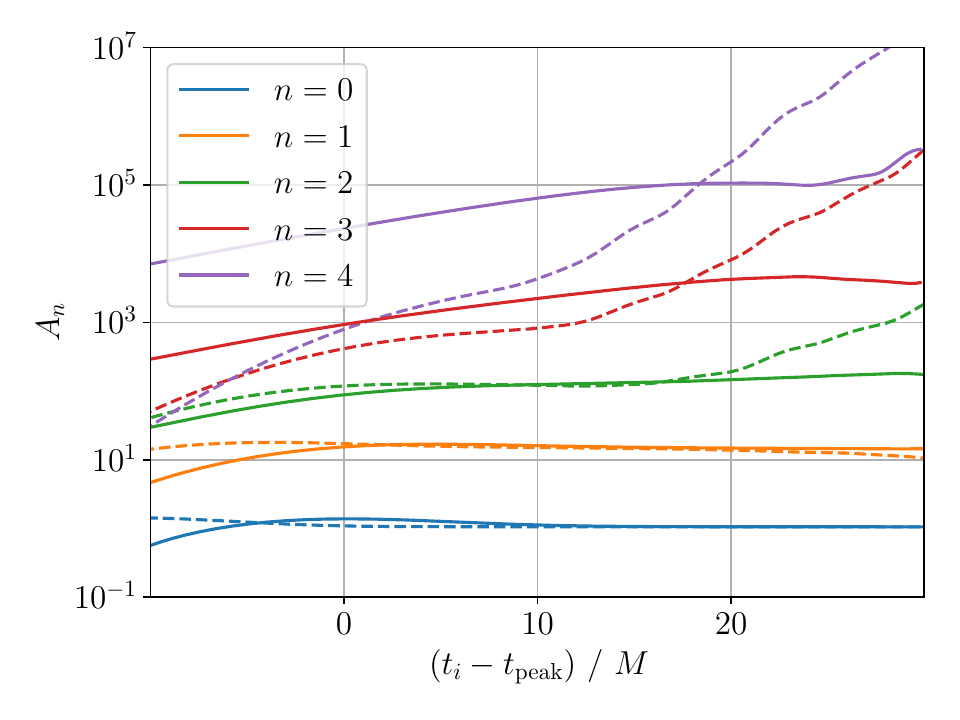}
    \includegraphics[width=\columnwidth]{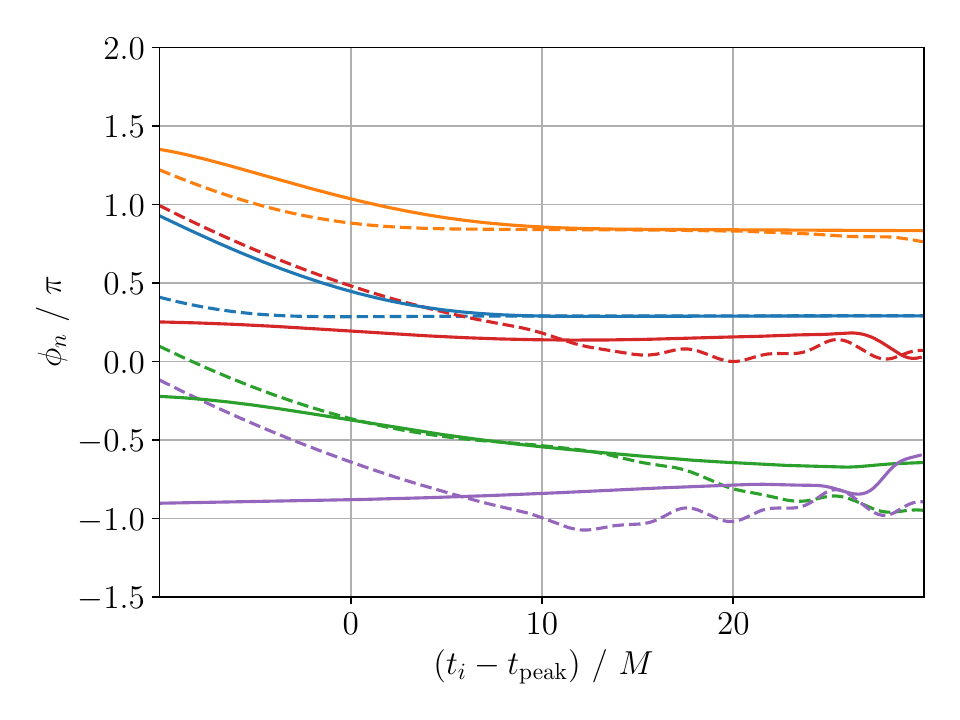}
    \caption{Comparison of the conventional fit (dashed) and iterative fit (solid) of SXS:BBH:0158 waveform.}
    \label{fig:0158_An_improve_sub_const_N4}
\end{figure}

We represent a comparison of the $A_n$ and $\phi_n$ obtained by the conventional fit and the iterative fit in Fig.~\ref{fig:0158_An_improve_sub_const_N4}.
It shows that the iterative fit provides more stable fit than the conventional fit for all overtones up to the fourth overtone.
On the other hand, we can see that some values of $A_n$ obtained by each fit are significantly different between the conventional fit and our method, especially for the higher overtone.
It can also be confirmed from Table~\ref{tab:0158_bestfit_An_phin}, where we compare the best-fit values.\footnote{For the SXS:BBH:0158, we set $(t_i-t_\peak)/M=10$ for the conventional fit, while we set $t_i=t_\peak$ for other simulations. 
This is because, as we can see from the top left panel in Fig.~\ref{fig:0158_M_A_gamma_sub_const} in Appendix~\ref{sec:plots}, the decrease of the mismatch still continues at $t_i=t_\peak$, but almost saturate around $(t_i-t_\peak)/M=10$. 
We also confirm that the decrease of $\gamma_n$ also almost saturate around $(t_i-t_\peak)/M=10$.
Hence, we adopt $(t_i-t_\peak)/M=10$ as the start time of the fit of the conventional fit, and present the best-fit values extracted with $(t_i-t_\peak)/M=10$ in Table~\ref{tab:0158_bestfit_An_phin}.}
It should be contrasted to the case of SXS:BBH:0305, where the conventional fit and the iterative fit provide the best-fit values of the same order.
The two fitting methods do not necessarily provide a similar best-fit values.  
The consistency is not a priori guaranteed, but depends on each specific waveform.
From the point of the stability of the fit, it would be plausible to expect that the best-fit values obtained from the iterative fit provides a more reasonable extraction.
Note also that the best-fit values of overtones up to $n=4$ are extracted at relatively large $t_i$.
Therefore, if the numerical errors are smaller, one may be able to extract overtones $n\geq 5$.

\begin{table*}[t]
    \centering
    \caption{Best-fit values for the amplitude $A_n$ and phase $\phi_n$ by fitting the SXS:BBH:0158 waveform subtracting the numerical constant with the conventional and iterative methods. 
    For the conventional fit, we set $(t_i-t_\peak)/M=10$ and $(t_e-t_\peak)/M=220$.}
    \begin{tabular}{c|cc|cccc}
    \hline\hline
        QNM & \multicolumn{2}{|c|}{Conventional} &  \multicolumn{4}{c}{Iterative} \\
        $n$ ~&~ $A_n$ ~&~ $\phi_{n}/\pi$ ~&~ $A_n$ ~&~ $\phi_{n}/\pi$ ~&~ $(t_i-t_\peak)/M$ ~&~ $(t_e-t_\peak)/M$ \\
        \hline
        $0$ ~&~ $1.057$ ~&~ $0.2905$ ~&~ $1.057$ ~&~ $0.2909$ ~&~ $25.77$ ~&~ $220$ \\
        $1$ ~&~ $15.05$ ~&~ $0.8419$ ~&~ $14.46$ ~&~ $0.8334$ ~&~ $30.37$ ~&~ $70$ \\
        $2$ ~&~ $121.6$ ~&~ $-0.5377$ ~&~ $169.3$ ~&~ $-0.6724$ ~&~ $25.77$ ~&~ $50$ \\
        $3$ ~&~ $832.7$ ~&~ $0.1747$ ~&~ $4616$ ~&~ $0.1703$ ~&~ $23.58$ ~&~ $40$ \\
        $4$ ~&~ $4501$ ~&~ $0.9967$ ~&~ $9.868\times 10^4$ ~&~ $-0.7867$ ~&~ $23.78$ ~&~ $40$ \\
        $5$ ~&~ $6685$ ~&~ $-0.3755$ ~&~ $-$ ~&~ $-$ ~&~ $-$ ~&~ $-$ \\
        $6$ ~&~ $1.554\times 10^4$ ~&~ $-0.1409$ ~&~ $-$ ~&~ $-$ ~&~ $-$ ~&~ $-$ \\
        $7$ ~&~ $2.130\times 10^4$ ~&~ $0.6683$ ~&~ $-$ ~&~ $-$ ~&~ $-$ ~&~ $-$ \\
    \hline\hline
    \end{tabular}
    \label{tab:0158_bestfit_An_phin}
\end{table*}

\subsection{SXS:BBH:0156 - low spin simulation}

\begin{figure}[t]
    \centering
    \includegraphics[width=\columnwidth]{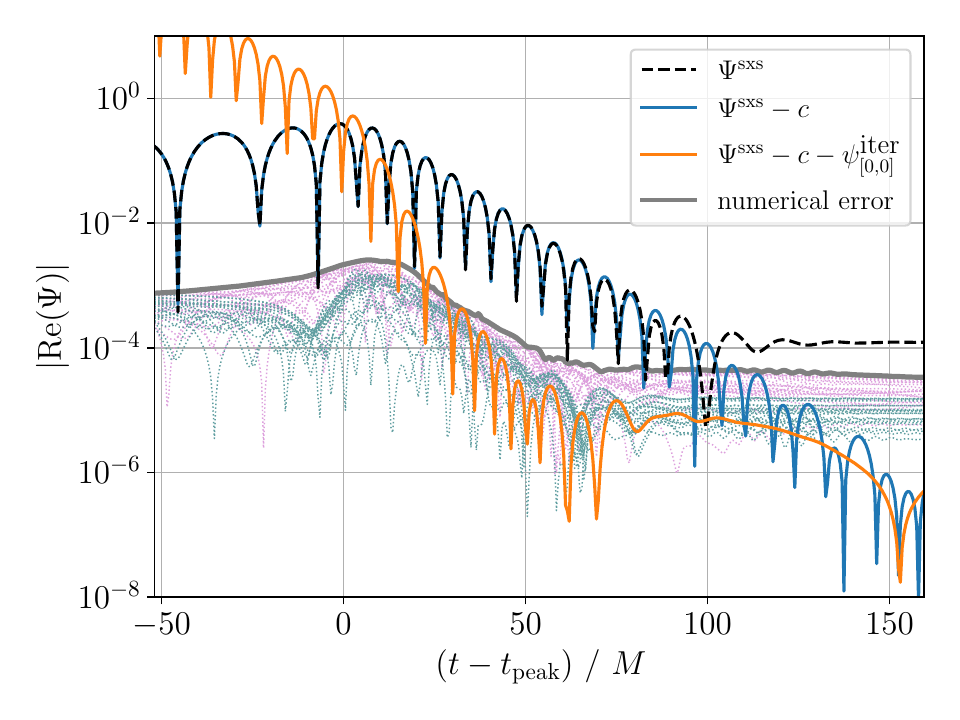}
    \caption{Numerical relativity waveform SXS:BBH:0156. Raw data (black dashed) and waveforms after the subtraction of the numerical constant (blue) and the best-fit fundamental mode (orange).}
    \label{fig:0156_wave_naive}
\end{figure}

\begin{figure*}[t]
    \centering
    \includegraphics[width=0.32\textwidth]{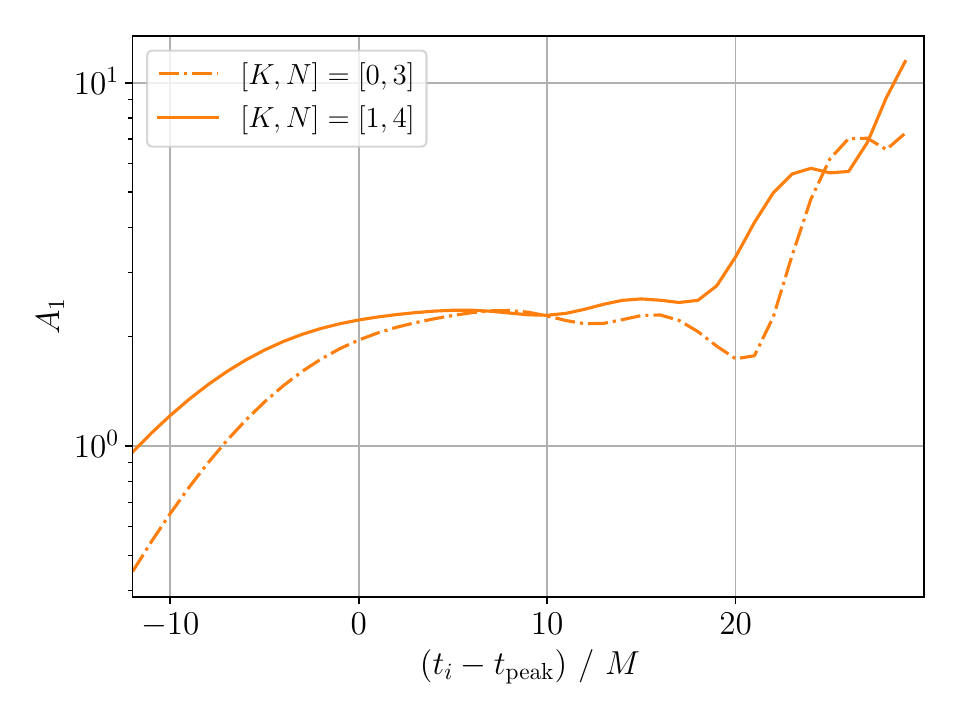}
    \includegraphics[width=0.32\textwidth]{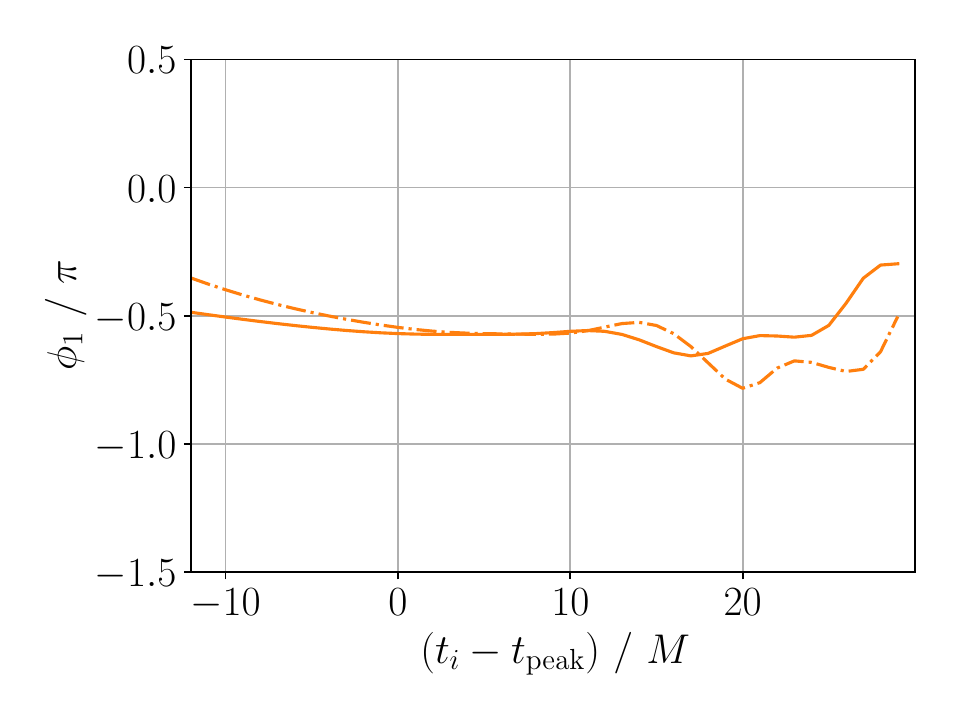}
    \includegraphics[width=0.32\textwidth]{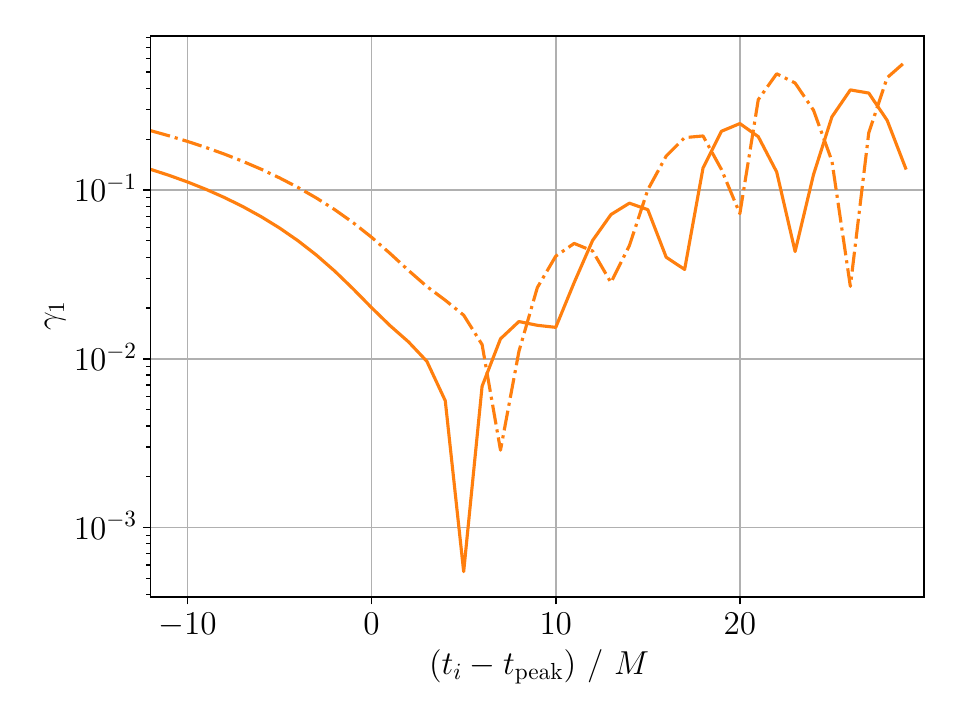}
    
    \includegraphics[width=0.32\textwidth]{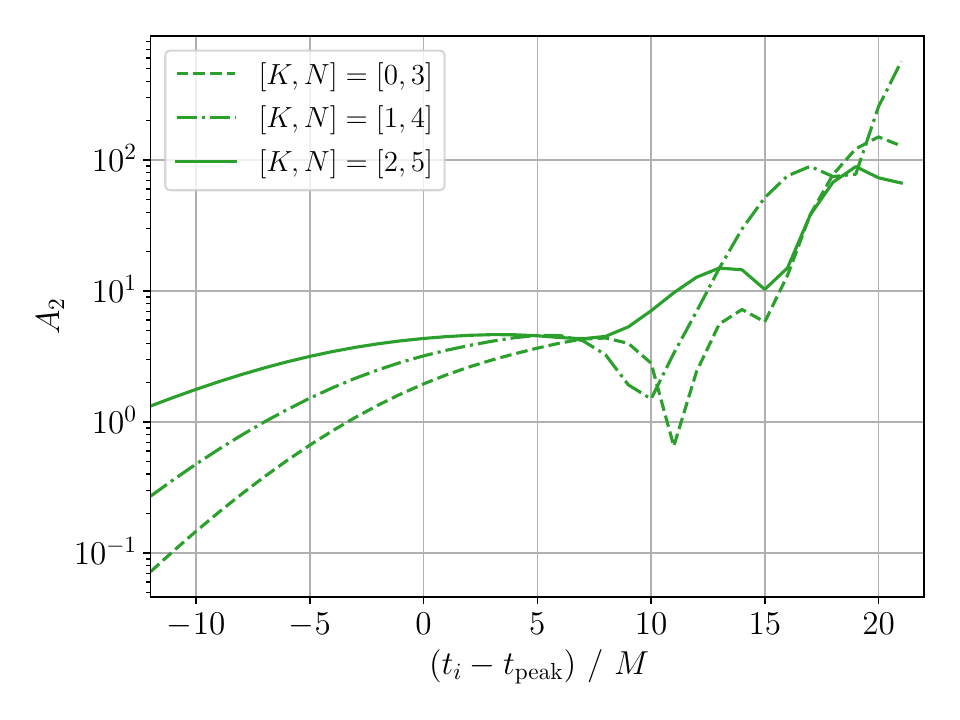}
    \includegraphics[width=0.32\textwidth]{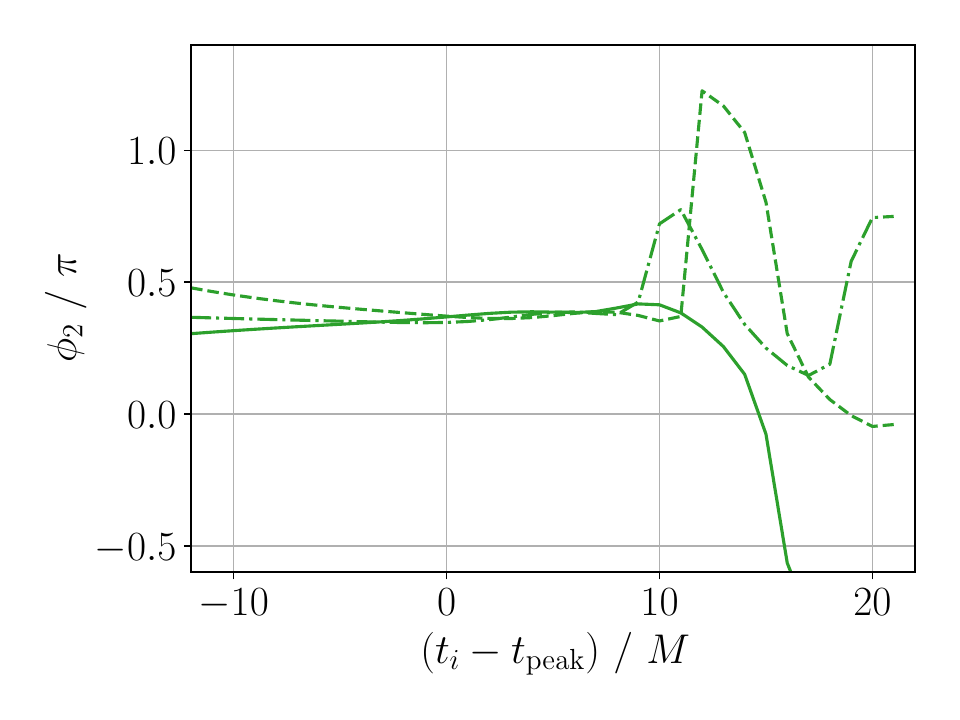}
    \includegraphics[width=0.32\textwidth]{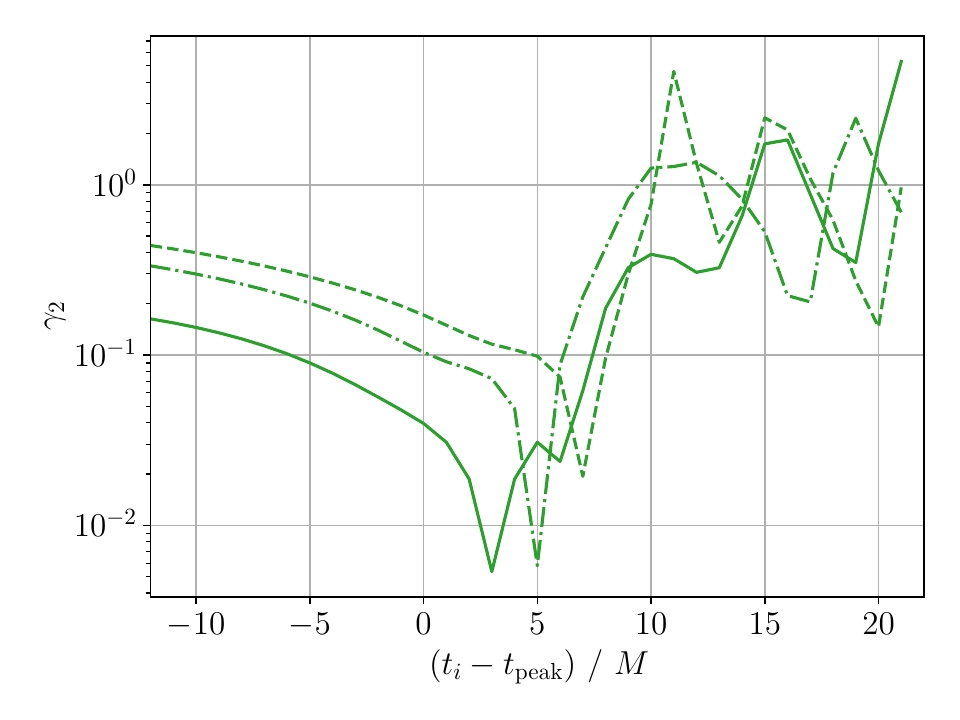}

    \includegraphics[width=0.32\textwidth]{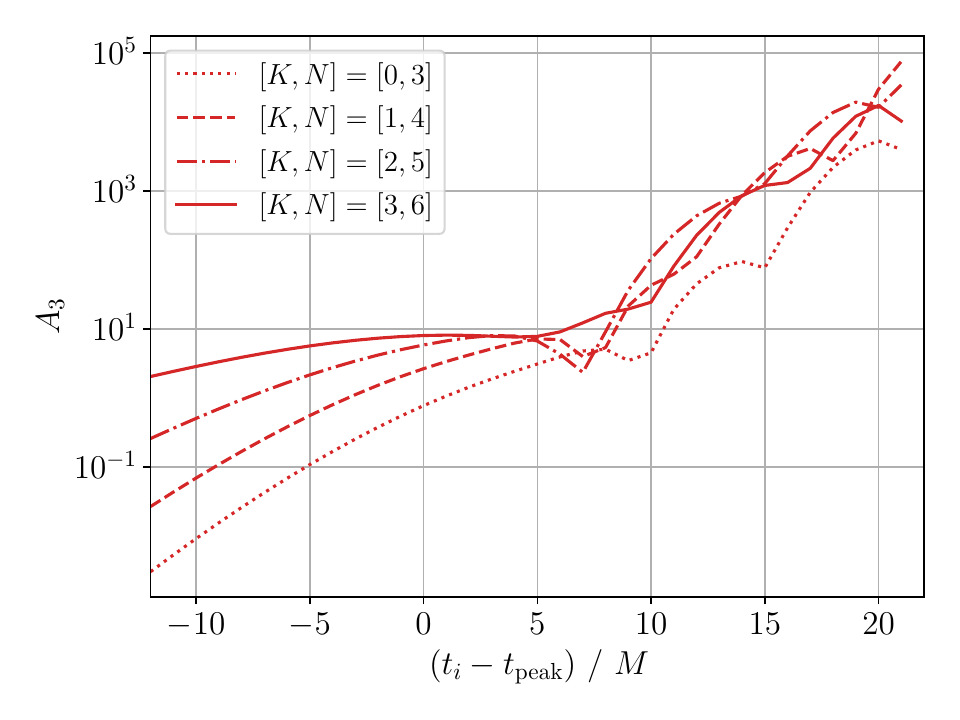}
    \includegraphics[width=0.32\textwidth]{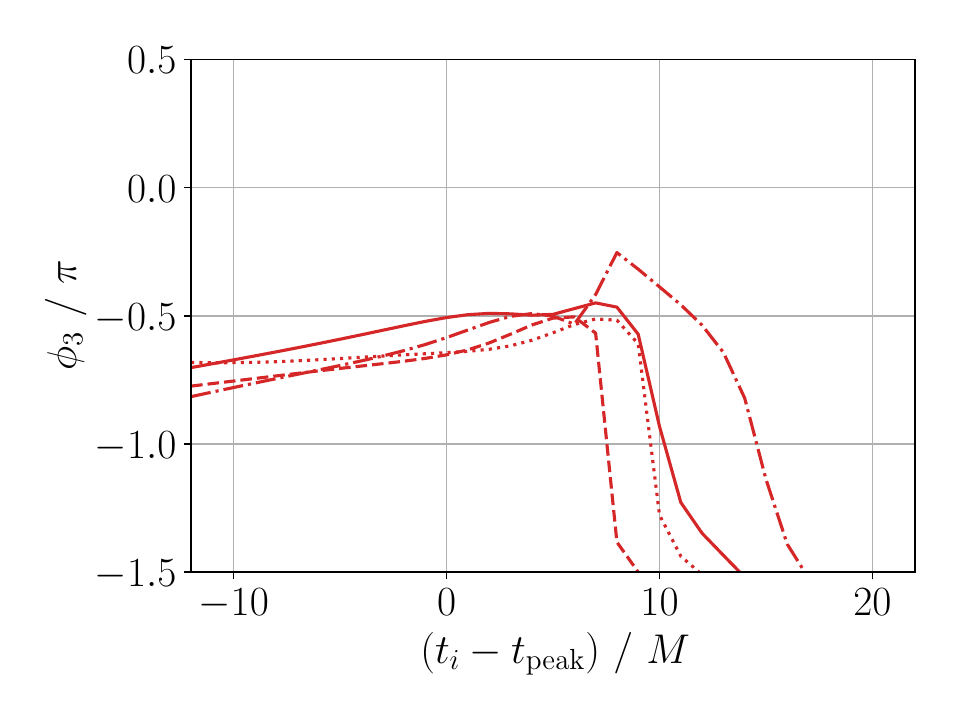}
    \includegraphics[width=0.32\textwidth]{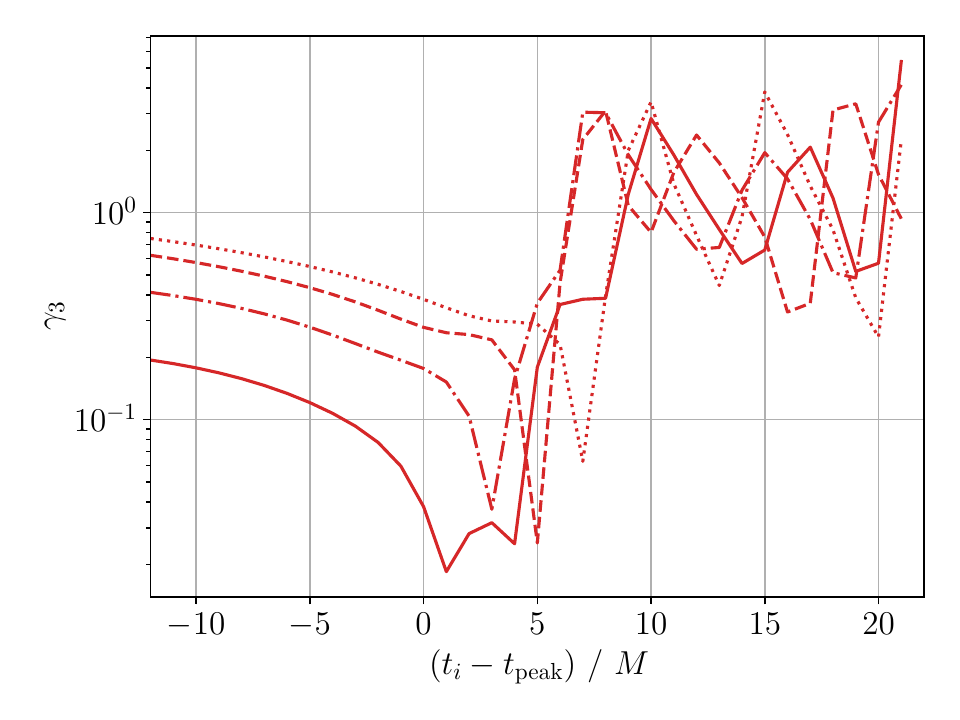}
    
    \includegraphics[width=0.32\textwidth]{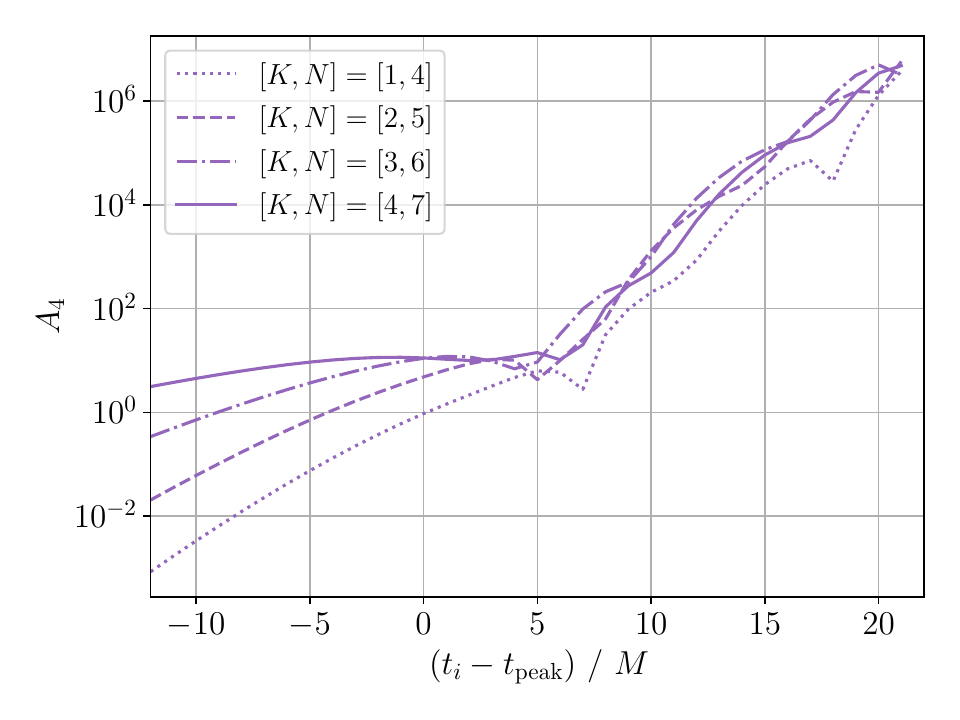}
    \includegraphics[width=0.32\textwidth]{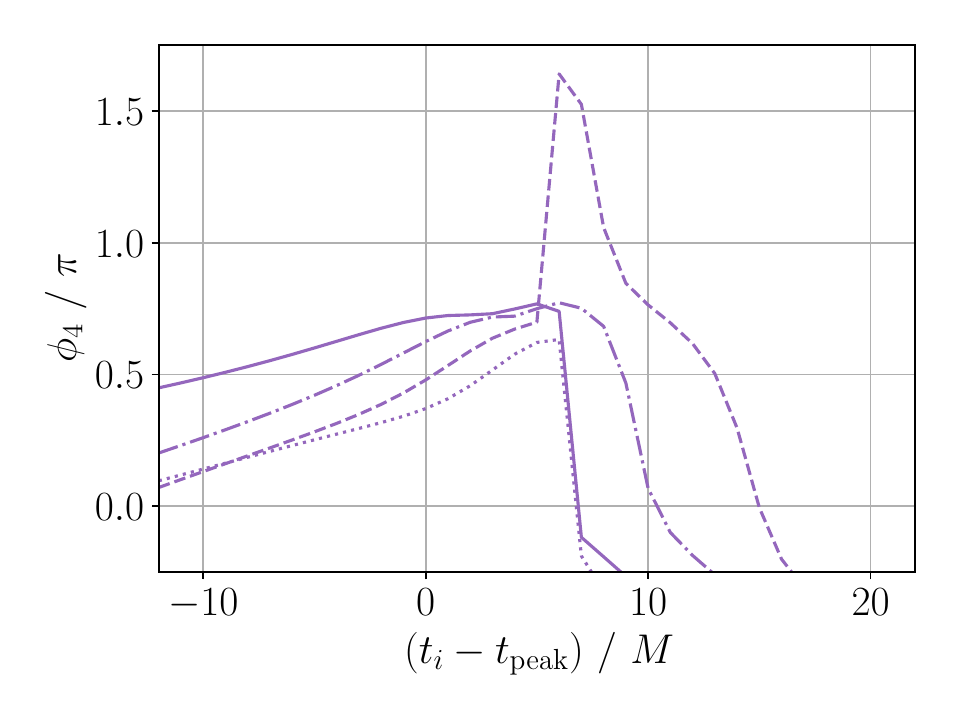}
    \includegraphics[width=0.32\textwidth]{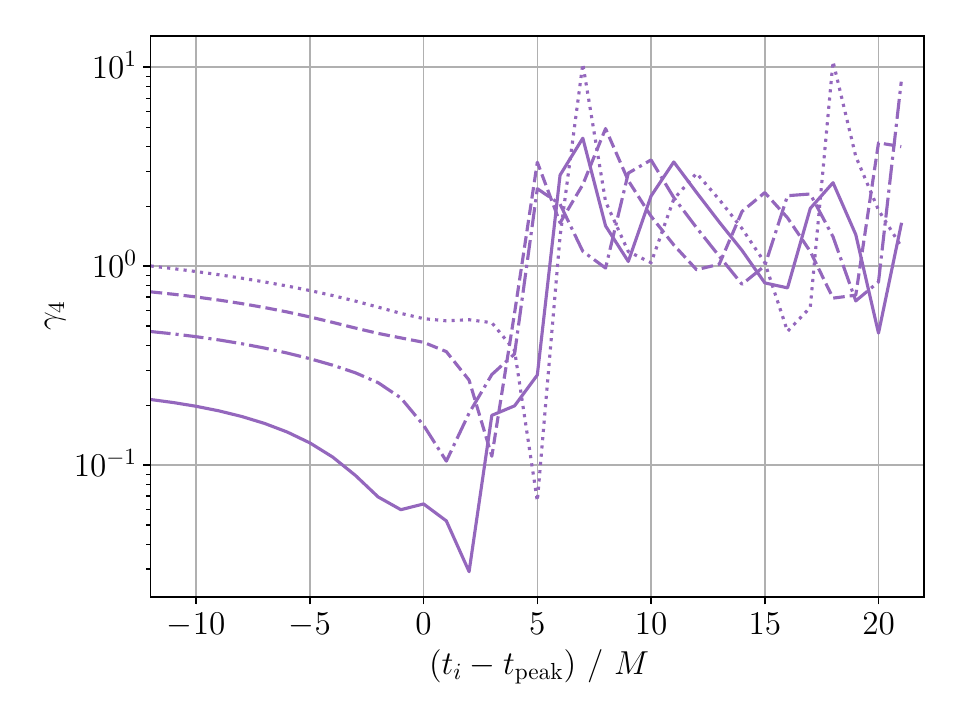}
    \caption{The improvement of the amplitude $A_n$ (left column), phase $\phi_n$ (right column) from $n=1$ mode (first row) to $n=4$ mode (fourth row) for the iterative fit of SXS:BBH:0156 by the fitting function $\psi^{\fit}_{[K,N]}$ with $N=K+3$.
    }
    \label{fig:0156_n4_improve_sub_const}
\end{figure*}

Let us investigate the fit of the waveform of the SXS:BBH:0156 with $\chi_{\rm rem}=0.3757$ shown in Fig.~\ref{fig:0156_wave_naive} as an example of the low spin simulation. 
Again, while the waveform data of SXS:BBH:0156 are available up to $(t-t_{\rm peak})/M=400$, we did not show the full waveform since at late time it remains roughly constant $c=(1-0.3i)\times 10^{-4}$ and fluctuate around it.
As expected, compared to SXS:BBH:0305 and SXS:BBH:0156, we can observe a fewer number of oscillations before the contamination by the constant, which originates from the short damping time of the QNM frequencies of low spin Kerr black holes.
Since we do not have sufficient number of sampling points, we use the cubic spline interpolation to increase the number of sampling points by two times for the regime $(t-t_{\rm peak})/M\geq -60$. 
For this regime, the raw data spacing is almost constant $\Delta t/M \approx 0.5$, so after the interpolation, we have $\Delta t/M \approx 0.25$.

\begin{figure}[t]
    \centering
    \includegraphics[width=\columnwidth]{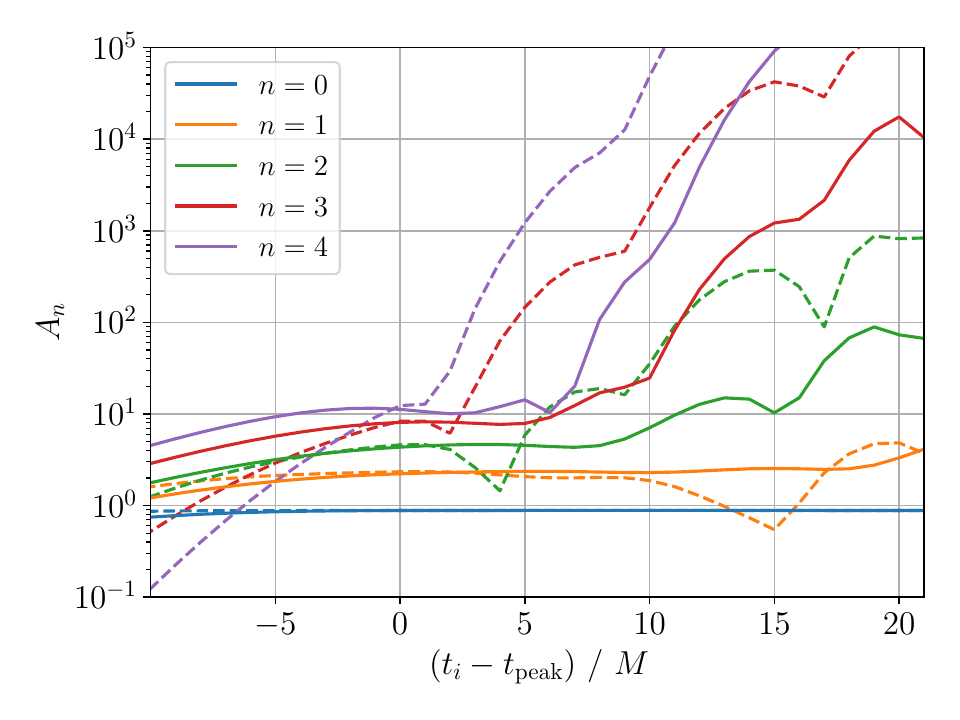}
    \includegraphics[width=\columnwidth]{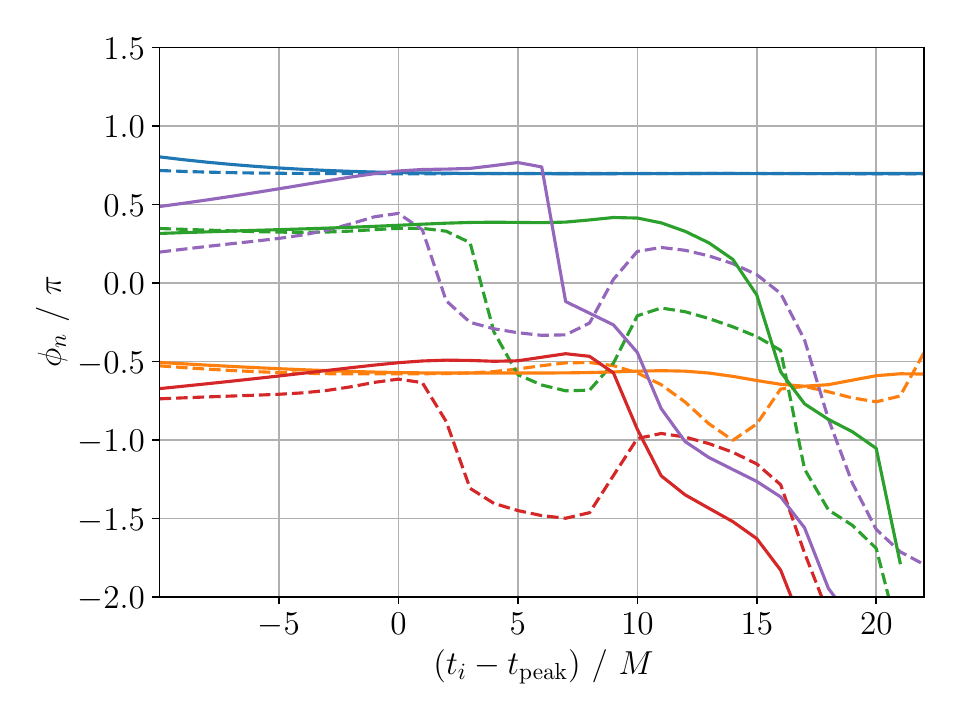}
    \caption{Comparison of the conventional fit (dashed) and iterative fit (solid) of SXS:BBH:0156 waveform.}
    \label{fig:0156_An_improve_sub_const_N4}
\end{figure}

Since the interval of available data is shorter, we expect that the fit of the low spin simulation is more challenging.
Nevertheless, from Figs.~\ref{fig:0156_n4_improve_sub_const} and \ref{fig:0156_An_improve_sub_const_N4}, we see that relatively clear plateaus show up.
Actually, while the range of the plateau is short, $A_n$ remains almost constant, in contrast to the fact that for the high spin SXS:BBH:0158 waveform, $A_n$ shows a continuous mild growth as a function of $t_i$.

After the subtraction of the third overtone, the waveform exhibits about half period of oscillation between the peak time and the time when the waveform becomes comparable to numerical errors.
A half period would be more or less a threshold of the necessary range of data to extract damped sinusoid.
Further, the time of the strain peak may not exactly coincide with the time of the onset of the QNM oscillations.
They may be triggered even before the peak time.
Taking into account the waveform before the peak time, we can include more than half period of oscillation to the fitting analysis.
Indeed, we see that the plateau in Fig.~\ref{fig:0156_An_improve_sub_const_N4} is extended before the strain peak time.

In Table~\ref{tab:0156_bestfit_An_phin}, We compare the best-fit values obtained by the conventional fit and iterative fit.
We see that in this case the two best-fit values are the same order.
Note that extracted amplitudes and phases slightly change when the number of sampling points changes.

\begin{table*}[t]
    \centering
    \caption{Best-fit values for the amplitude $A_n$ and phase $\phi_n$ by fitting the SXS:BBH:0156 waveform subtracting the numerical constant with the conventional and iterative methods. 
    For the conventional fit, we set $(t_i-t_\peak)/M=0$ and $(t_e-t_\peak)/M=150$.}
    \begin{tabular}{c|cc|cccc}
    \hline\hline
        QNM & \multicolumn{2}{|c|}{Conventional} &  \multicolumn{4}{c}{Iterative} \\
        $n$ ~&~ $A_n$ ~&~ $\phi_{n}/\pi$ ~&~ $A_n$ ~&~ $\phi_{n}/\pi$ ~&~ $(t_i-t_\peak)/M$ ~&~ $(t_e-t_\peak)/M$ \\
        \hline
        $0$ ~&~ $0.8840$ ~&~ $0.6966$ ~&~ $0.8850$ ~&~ $0.6965$ ~&~ $7.0$ ~&~ $150$ \\
        $1$ ~&~ $2.339$ ~&~ $-0.5778$ ~&~ $2.365$ ~&~ $-0.5728$ ~&~ $5.0$ ~&~ $40$ \\
        $2$ ~&~ $4.600$ ~&~ $0.3485$ ~&~ $4.647$ ~&~ $0.3861$ ~&~ $3.0$ ~&~ $25$ \\
        $3$ ~&~ $8.320$ ~&~ $-0.6116$ ~&~ $8.213$ ~&~ $-0.4959$ ~&~ $1.0$ ~&~ $25$ \\
        $4$ ~&~ $12.35$ ~&~ $0.4438$ ~&~ $10.05$ ~&~ $0.7258$ ~&~ $2.0$ ~&~ $25$ \\
        $5$ ~&~ $12.33$ ~&~ $-0.5226$ ~&~ $-$ ~&~ $-$ ~&~ $-$ ~&~ $-$ \\
        $6$ ~&~ $6.926$ ~&~ $0.4875$ ~&~ $-$ ~&~ $-$ ~&~ $-$ ~&~ $-$ \\
        $7$ ~&~ $1.640$ ~&~ $-0.5222$ ~&~ $-$ ~&~ $-$ ~&~ $-$ ~&~ $-$ \\
    \hline\hline
    \end{tabular}
    \label{tab:0156_bestfit_An_phin}
\end{table*}

\subsection{SXS:BBH:1108 - GW190814-like simulation}

Finally, we consider the SXS:BBH:1108 waveform, which is related to the GW190814.
GW190814~\cite{LIGOScientific:2020zkf} is gravitational wave signal from a coalescence of $23M_\odot$ black hole with a $2.6M_\odot$ compact object, the latter of which is either a light black hole or a heavy neutron star.
The source has the most unequal mass ratio among the gravitational waves observed so far, and the dimensionless spin of the primary black hole is constrained to $\leq 0.07$.
Assuming the secondary component is a black hole, SXS:BBH:1108 serves as a waveform consistent with GW190814.

\begin{figure}[t]
    \centering
    \includegraphics[width=\columnwidth]{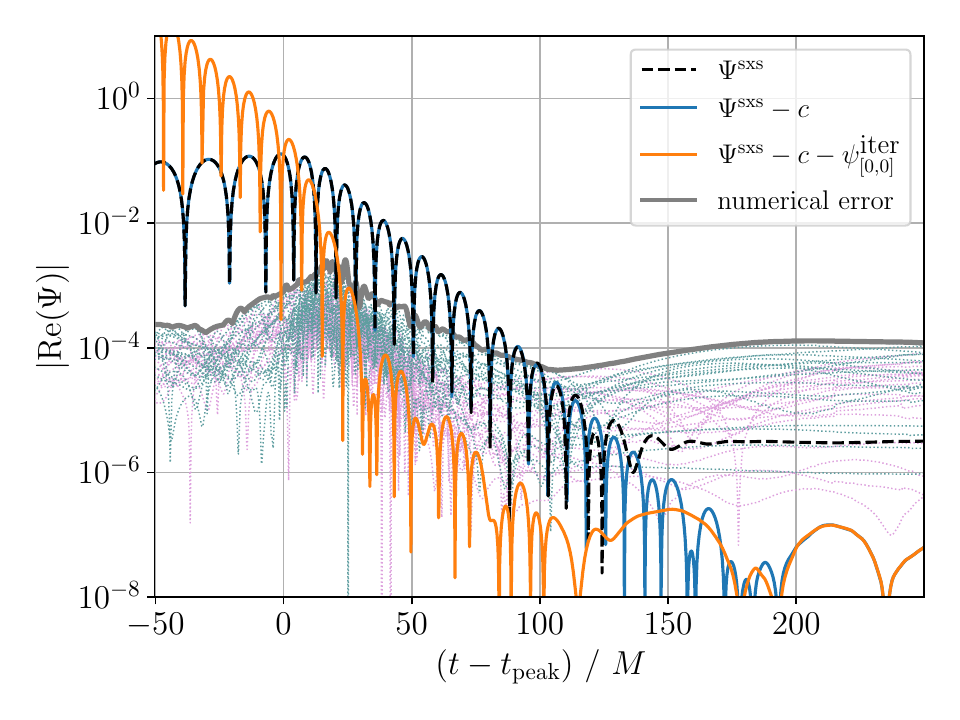}
    \caption{Numerical relativity waveform SXS:BBH:1108. Raw data (black dashed) and waveforms after the subtraction of the numerical constant (blue) and the best-fit fundamental mode (orange).}
    \label{fig:1108_wave_naive}
\end{figure}

Compared to the three numerical relativity waveforms considered so far, the SXS:BBH:1108 waveform shown in Fig.~\ref{fig:1108_wave_naive} is contaminated by large numerical errors.
The waveform roughly approaches a constant $c=(-3.1-8.8i)\times10^{-6}$, but it is below the estimated numerical errors.
Also, the waveform damps quickly since the SXS:BBH:1108 is the low spin simulation with $\chi_{\rm rem}=0.2772$.
However, in the case of SXS:BBH:1108 the sampling points are not insufficient, so we do not use the spline interpolation.

\begin{figure*}[t]
    \centering
    \includegraphics[width=0.32\textwidth]{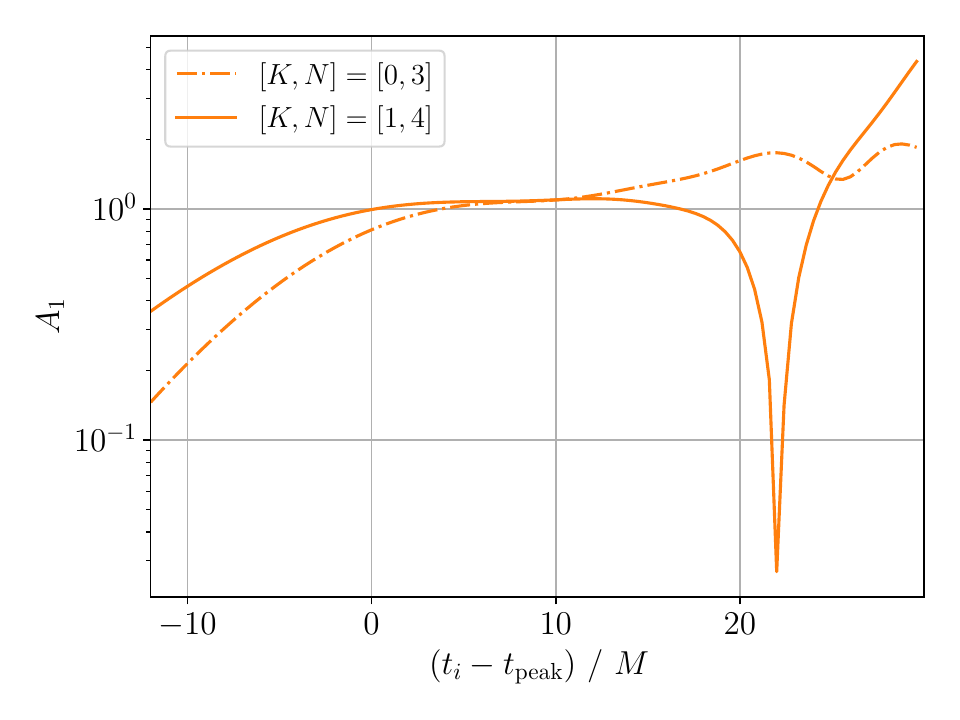}
    \includegraphics[width=0.32\textwidth]{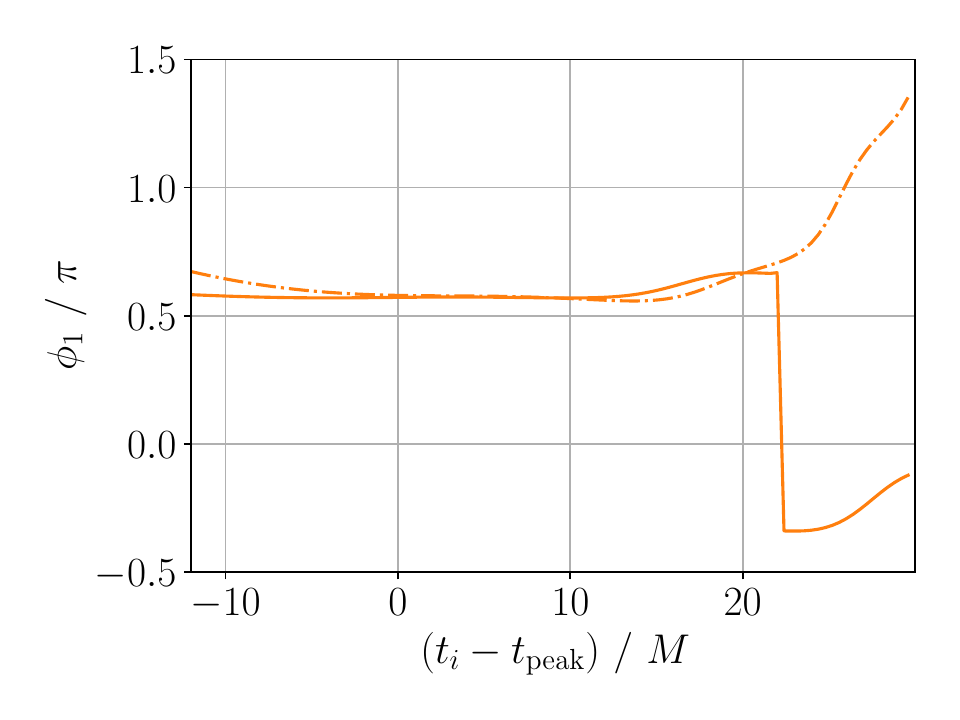}
    \includegraphics[width=0.32\textwidth]{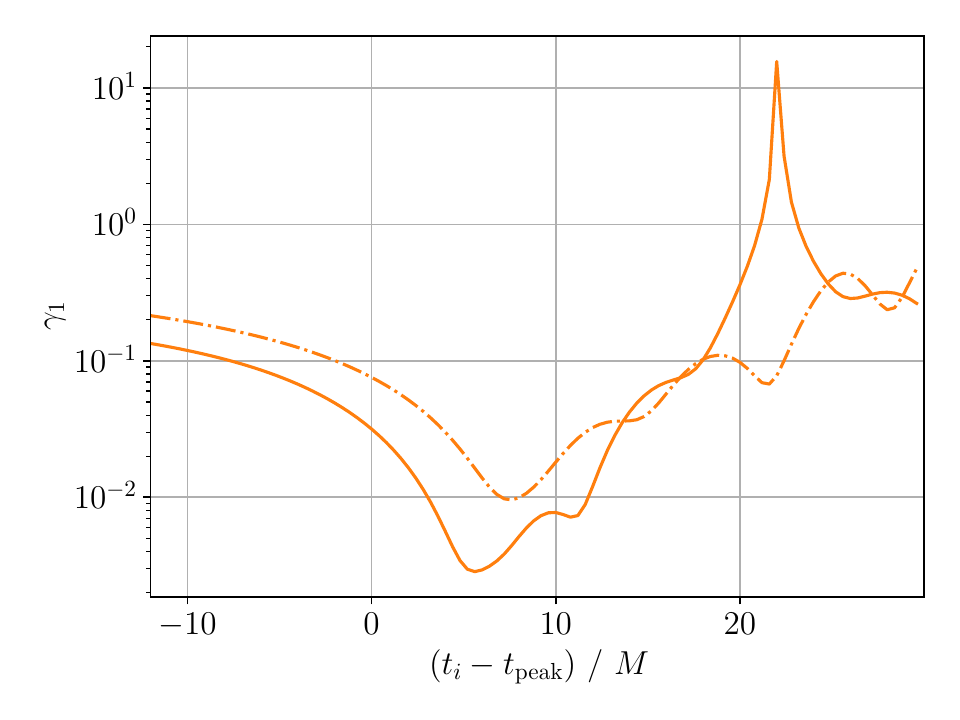}
    
    \includegraphics[width=0.32\textwidth]{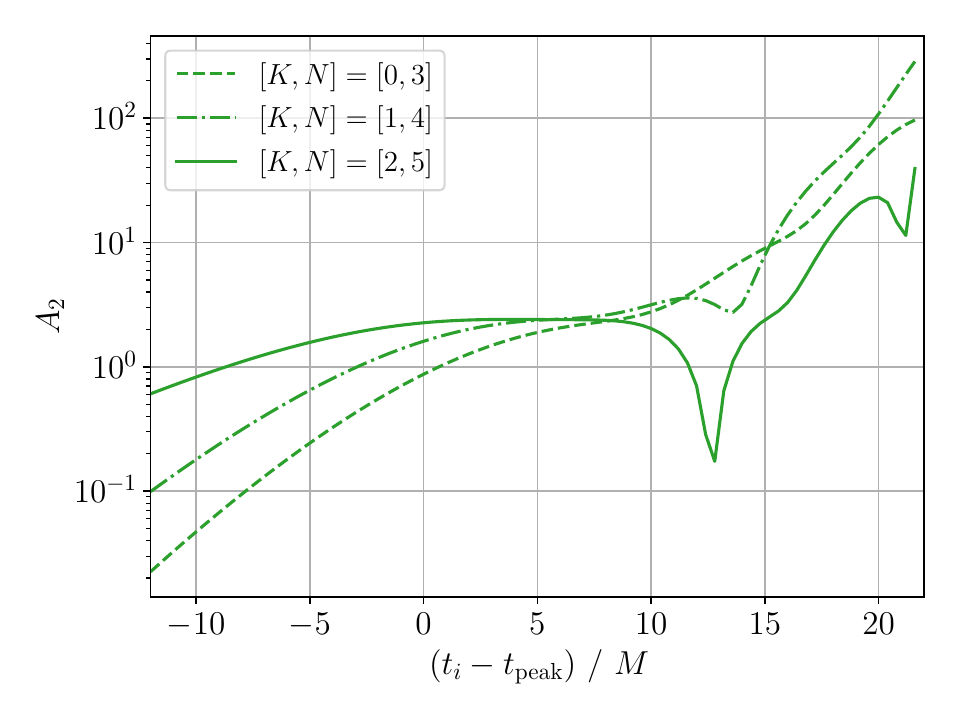}
    \includegraphics[width=0.32\textwidth]{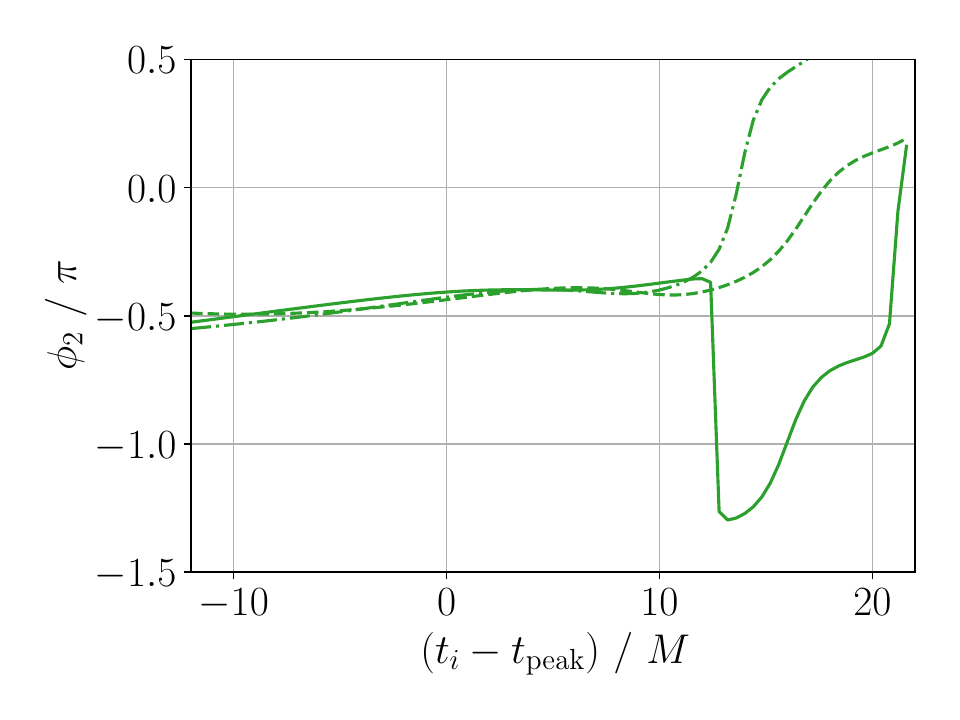}
    \includegraphics[width=0.32\textwidth]{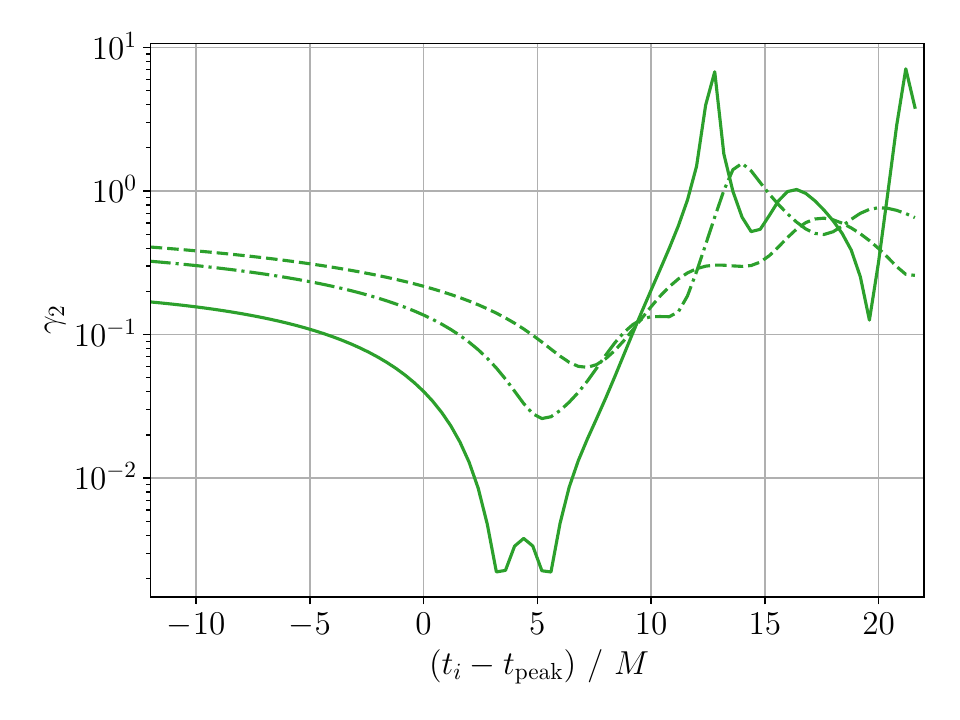}

    \includegraphics[width=0.32\textwidth]{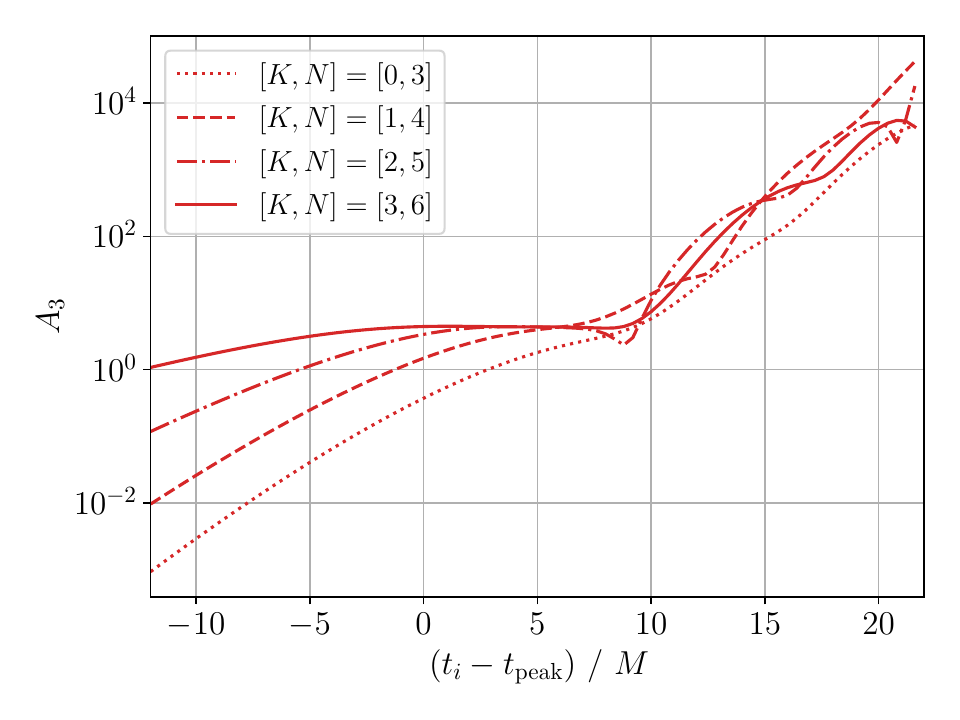}
    \includegraphics[width=0.32\textwidth]{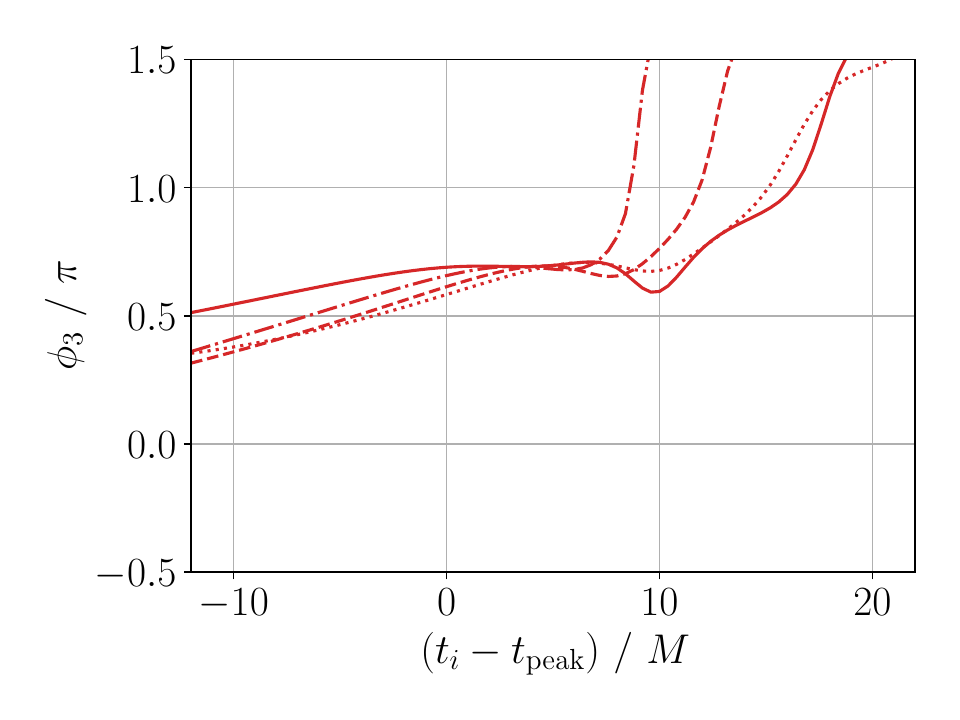}
    \includegraphics[width=0.32\textwidth]{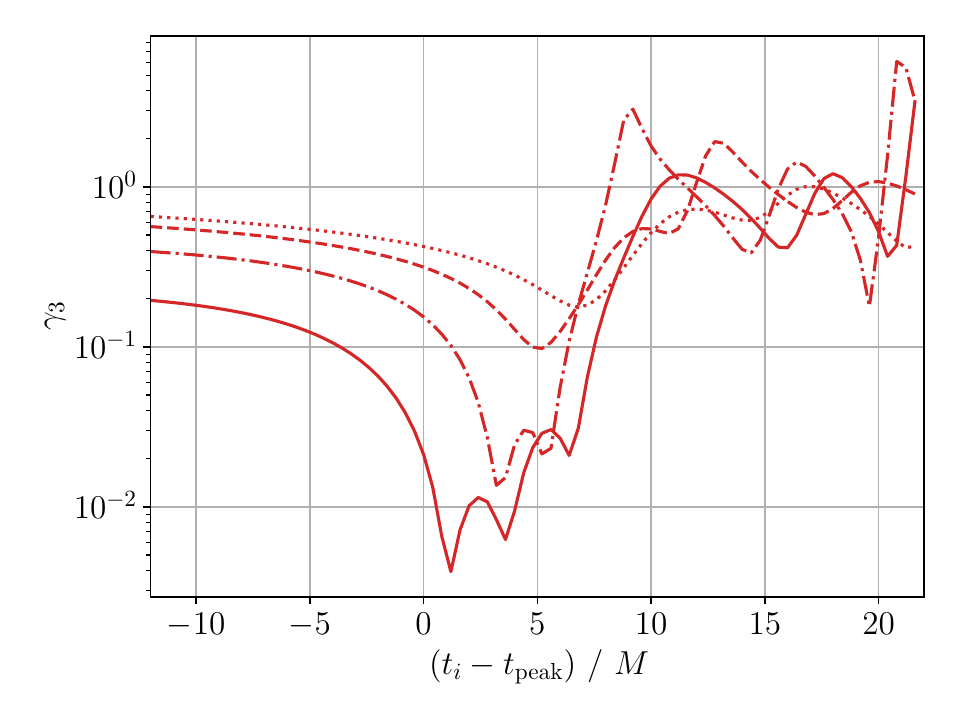}
    
    \includegraphics[width=0.32\textwidth]{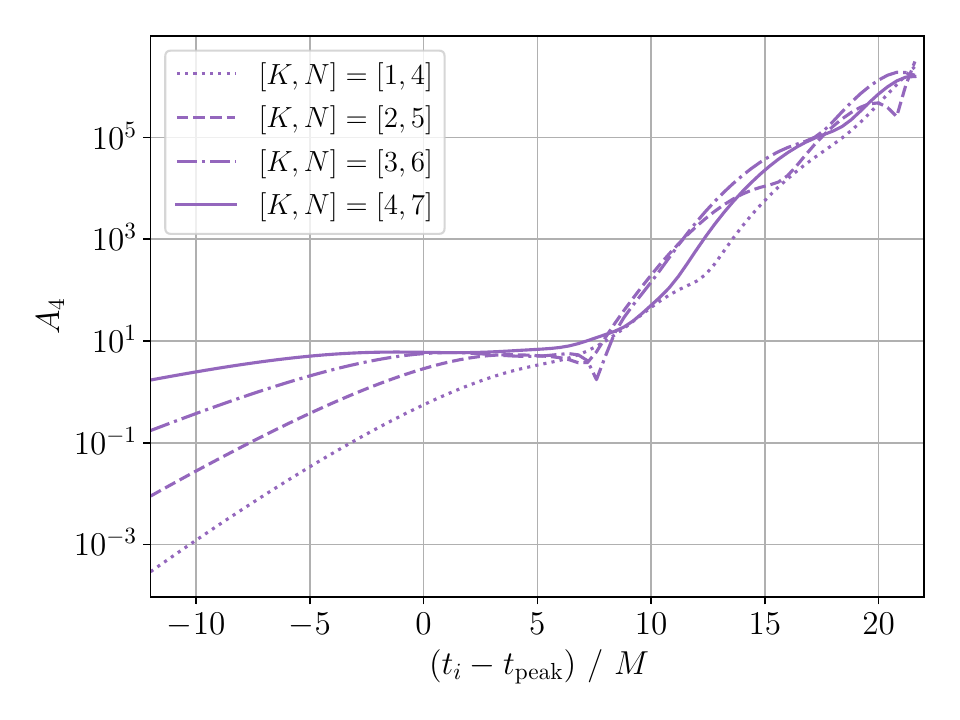}
    \includegraphics[width=0.32\textwidth]{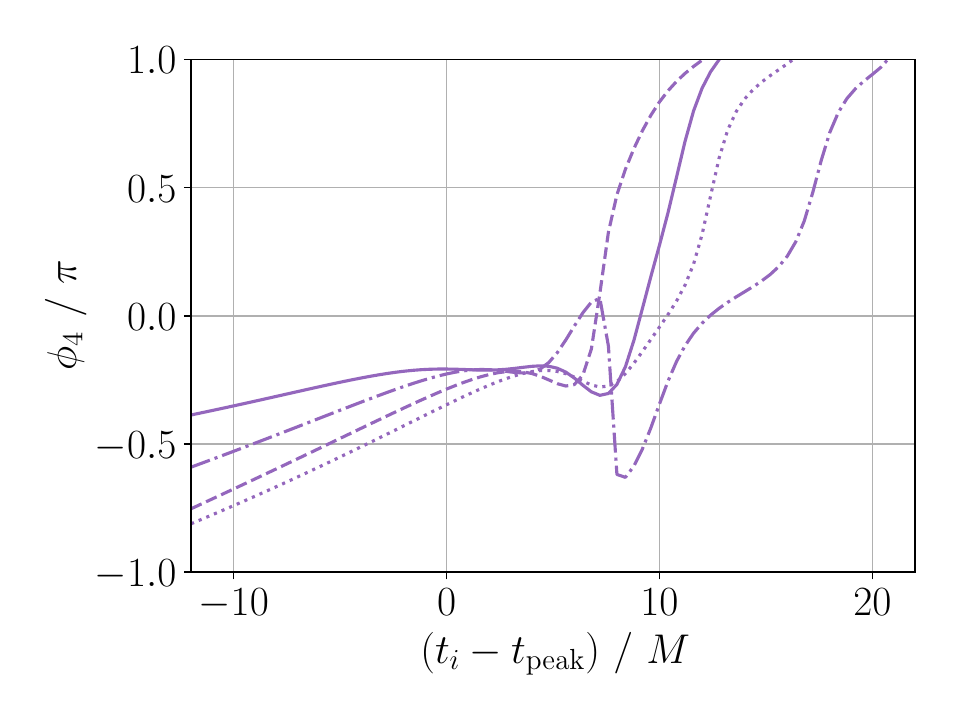}
    \includegraphics[width=0.32\textwidth]{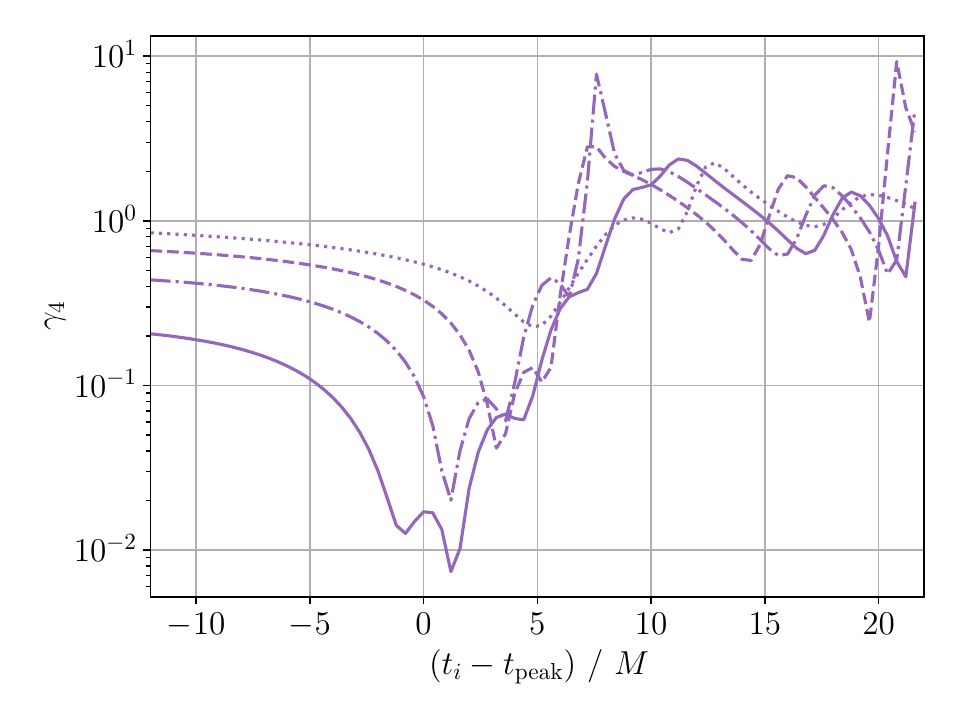}
    \caption{The improvement of the amplitude $A_n$ (left column), phase $\phi_n$ (right column) from $n=1$ mode (first row) to $n=4$ mode (fourth row) for the iterative fit of SXS:BBH:1108 by the fitting function $\psi^{\fit}_{[K,N]}$ with $N=K+3$.
    }
    \label{fig:1108_n4_improve_sub_const}
\end{figure*}

\begin{figure}[t]
    \centering
    \includegraphics[width=\columnwidth]{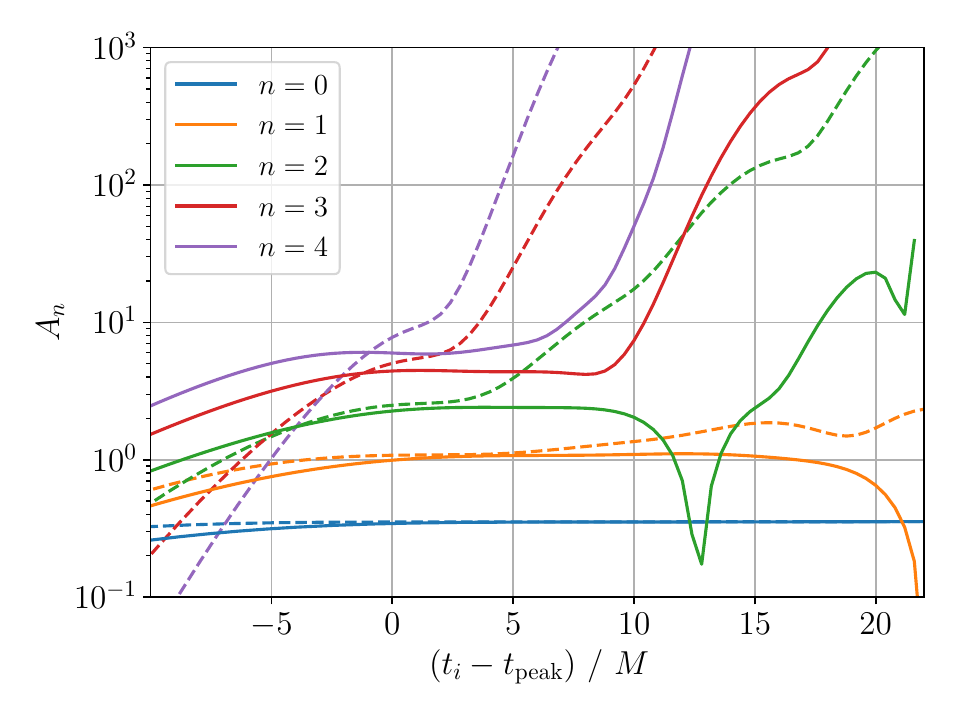}
    \includegraphics[width=\columnwidth]{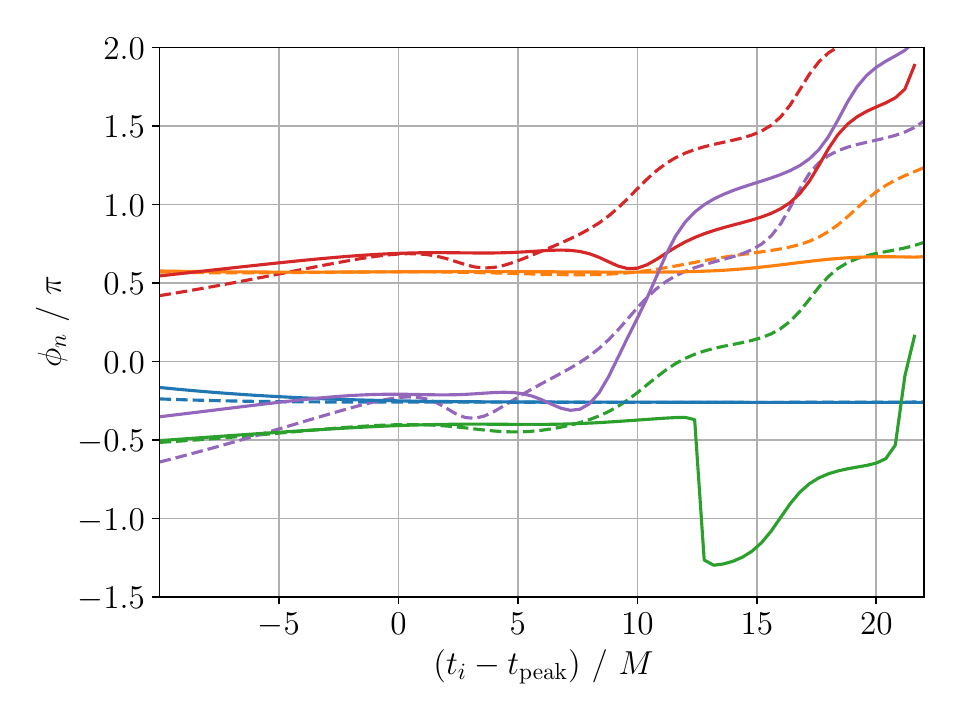}
    \caption{Comparison of the conventional fit (dashed) and iterative fit (solid) of SXS:BBH:1108 waveform.}
    \label{fig:1108_An_improve_sub_const_N4}
\end{figure}

The results of the fitting analysis are shown in Figs.~\ref{fig:1108_n4_improve_sub_const} and \ref{fig:1108_An_improve_sub_const_N4} and Table~\ref{tab:1108_bestfit_An_phin}.
Their qualitative behavior is the same as the results for the SXS:BBH:0156.
Our results imply that the extraction of the QNM depends mainly on the spin of the remnant black hole, rather than the mass ratio of the binary before merger.

\begin{table*}[t]
    \centering
    \caption{Best-fit values for the amplitude $A_n$ and phase $\phi_n$ by fitting the SXS:BBH:1108 waveform subtracting the numerical constant with the conventional and iterative methods. 
    For the conventional fit, we set $(t_i-t_\peak)/M=0$ and $(t_e-t_\peak)/M=150$.}
    \begin{tabular}{c|cc|cccc}
    \hline\hline
        QNM & \multicolumn{2}{|c|}{Conventional} &  \multicolumn{4}{c}{Iterative} \\
        $n$ ~&~ $A_n$ ~&~ $\phi_{n}/\pi$ ~&~ $A_n$ ~&~ $\phi_{n}/\pi$ ~&~ $(t_i-t_\peak)/M$ ~&~ $(t_e-t_\peak)/M$ \\
        \hline
        $0$ ~&~ $0.3523$ ~&~ $-0.2580$ ~&~ $0.3523$ ~&~ $-0.2579$ ~&~ $7.6$ ~&~ $150$ \\
        $1$ ~&~ $1.079$ ~&~ $0.5718$ ~&~ $1.073$ ~&~ $0.5728$ ~&~ $5.2$ ~&~ $40$ \\
        $2$ ~&~ $2.495$ ~&~ $-0.4017$ ~&~ $2.406$ ~&~ $-0.3983$ ~&~ $3.2$ ~&~ $25$ \\
        $3$ ~&~ $5.036$ ~&~ $0.6849$ ~&~ $4.468$ ~&~ $0.6935$ ~&~ $1.2$ ~&~ $25$ \\
        $4$ ~&~ $7.784$ ~&~ $-0.2290$ ~&~ $6.043$ ~&~ $-0.2086$ ~&~ $-0.7$ ~&~ $25$ \\
        $5$ ~&~ $7.954$ ~&~ $0.8437$ ~&~ $-$ ~&~ $-$ ~&~ $-$ ~&~ $-$ \\
        $6$ ~&~ $4.601$ ~&~ $-0.09626$ ~&~ $-$ ~&~ $-$ ~&~ $-$ ~&~ $-$ \\
        $7$ ~&~ $1.120$ ~&~ $0.9546$ ~&~ $-$ ~&~ $-$ ~&~ $-$ ~&~ $-$ \\
    \hline\hline
    \end{tabular}
    \label{tab:1108_bestfit_An_phin}
\end{table*}

\section{Conclusions and discussion}
\label{sec:conc}

The extraction of QNMs from black hole ringdown is a key to understand gravity at dynamical and strong field regime.
For the extraction of the first overtone from the observed ringdown gravitational waves, the conflicting results have been reported mainly due to the low signal-to-noise ratio of the observational data.
Even for the simulated waveform by numerical relativity calculations, the extraction of overtones is not straightforward, since the higher overtones more quickly damp and become subdominant compared to lower modes, and hence it tends to end up with the overfit.
In the present paper, we proposed an improved way of extracting the overtone QNMs.
In contrast to the conventional fitting method, where one extracts the coefficients of all the QNMs of interest from a single fitting, we considered an iterative procedure of fitting and subtraction to extract the longest-lived QNM at each step.
The iterative fitting method is summarized as follows:
\begin{enumerate}
\item Fit a given waveform by a fitting function composed of a superposition of the four longest-lived damped sinusoids with coefficients $C_n$ as fitting parameters. 
Set the end time of the fit $t_e$ around the time when the signal becomes comparable to noises.
\item Check if there exists a certain range of the start time of the fit $t_i$ when the rate of change $\gamma_n=|\f{1}{C_n}\f{d C_n}{dt_i}|$ remains small compared to $|\Im(\omega_n)|$ to guarantee the stable fit rather than the overfit.
\item Read off the best-fit value of the coefficient $C_n$ for the longest-lived mode by a fit starting from a time when the decrease of the rate of change $\gamma_n$ for the next-longest-lived mode saturates (not necessarily exactly the minimum since it tends to oscillate).
\item Subtract the best-fit longest-lived damped sinusoid from the waveform.
\item Iterate the above steps so long as the available data interval is longer than the time scale of the effective longest-lived mode.
\end{enumerate}

This method enables us to reveal the subdominant contribution of the higher overtones by iteratively subtracting the dominant contribution of the lower modes.
Through the analyses of the mock waveform and numerical relativity waveform, we clarified that by collecting the best-fit values of the longest-lived mode at each step we can improve the stability of the fit of overtones.
We confirmed that the coefficients for the higher overtones are sensitive to the start time of the fit, and they begin to diverge exponentially for the fit starting from late time.  
The divergence of the coefficient simply compensates the exponential damping, which can be clearly seen from the fact that the rate of change $\gamma_n$ roughly coincides with the damping rate $|\Im(\omega_n)|$, signaling the overfit.
Therefore, we should adopt the start time of the fit where $\gamma_n$ remains smaller than $|\Im(\omega_n)|$, for which it is important to include higher overtones for the fitting function to improve the stability of the fitting of the lower overtones.
Indeed, we found that as we increase the number of overtones, the plateau of the fit of the lower overtone coefficients are extended.
This is because the higher overtones are responsible for describing the early time waveform, and hence with the higher overtones included, the lower overtones are appropriately fitted at early time.
However, inclusion of too many modes in the fitting function tends to destabilize the fitting of the lower overtones.
This is why we adopted a fitting function composed of four longest-lived damped sinusoids.

A caveat for the above step 1 is that it is still unclear whether it is always optimal to adopt the four longest-lived modes.
We adopted the four modes based on the mock waveform related to SXS:BBH:0305, whose dimensionless remnant spin $\sim 0.7$ is more or less typical value for the remnant black hole.
In this sense, the four modes should be regarded as a representative number, and it would be safe to identify the optimal number for each waveform by checking the stability of the extracted coefficients.

Also, in the step 3, one may think that it would be more natural to use $\gamma_n$ for the longest-lived mode as an indicator of the stability of the fit.
The reason why we focus on the $\gamma_n$ for the next-longest-lived mode is based on our analysis of the mock waveform of the pure damped sinusoids, where the minimum of $\gamma_n$ for the longest-lived mode differs from the minimum of $\gamma_n$ for the other modes.
However, for waveforms with noises, the mimima of $\gamma_n$ tend to match each other.
In such cases, there is no reason to focus on $\gamma_n$ for the next-longest-lived mode, and we are simply adopting the start time of the fit $t_i$ where $\gamma_n$ are minimized.

We examined how the fit of the SXS numerical relativity waveform can be improved by the iterative extractions.
Iterating the subtractions helps to reveal the subdominant contribution including the spherical-spheroidal mixing and improve the stability of the fit of the overtones.
We clarified that for the GW150914-like numerical relativity waveform SXS:BBH:0305, the overtones with $n\geq 2$ can be extracted in a more stable manner, compared to the conventional fit where only $n=0,1$ can be stably extracted.
In addition to SXS:BBH:0305, we also investigated high spin simulation SXS:BBH:0158 and low spin simulations SXS:BBH:0156 and SXS:BBH:1108.
As expected, for the high spin case, we have more oscillating signals.
Even before the subtraction of the longest-lived mode, we can observe a mild plateau for the fit of overtones up to $n=2$, while there is no plateau on the second overtone in SXS:BBH:0305 before the subtraction.
We found that the best-fit values for SXS:BBH:0158 obtained by the iterative fit differ significantly from the conventional fit, especially for the higher overtones.
On the other hand, for the low spin case, we confirmed that the subtraction still improves the fit of overtones, while the range of the plateau is shorter than the high spin case.

We studied how the fit is affected by noises, which eventually hides the ringdown signal at late time.
While the stability of the fit is reduced by noises at late time, the iterative fit is efficient to extract overtones at early time.
Up to which overtone a sufficiently stably extracted depends on the magnitude of the noise and the damping time of the overtone.
A rough criterion would be given by the number of oscillations between the peak time and the time when the noise becomes dominant. 
For overtones whose number of oscillations is more than one, we expect its stable extraction. 

We also confirmed that the iterative fitting method is robust to the data with low sampling rates, compared to the conventional fit.
These natures are advantageous for the extraction of the QNMs from observed data. 
In particular, a recent analysis of the observed ringdown gravitational waves shows that the conventional analysis is sensitive to the sampling rate~\cite{Wang:2023xsy}.
Iterative fitting method would be more advantageous to apply to the observed data as it provides robust fitting results even for data with low sampling rates, compared to conventional fitting method.

We expect that the iterative fitting still works to the case of the QNM spectral instability under small perturbation of the effective potential.
Such a perturbation could drastically alter the QNM frequencies, but it modifies the ringdown signal only at the late time, so it is not fatal for the black hole spectroscopy program~\cite{Berti:2022xfj,Kyutoku:2022gbr}.
Since the iterative fitting method allows us a more direct way to extract the QNMs from the ringdown signal in the time domain, it is plausible that the method is robust against the spectral instability.

The fitting of the ringdown by QNMs is a nontrivial task and various approaches would be possible.
An approach closely related to ours is to consider a filter in the frequency domain, which removes the QNM poles of interest on the complex frequency plane~\cite{Ma:2022wpv,Ma:2023cwe,Ma:2023vvr}.
The filter also affects to the contribution from other modes, and hence the amplitudes and phases obtained by the fit of the filtered waveform are different from the original ones.
While one can reconvert them and estimate the original values, it might be the case that the modification associated with the filtering affects the fit of the subdominant modes.
For instance, if the amplitude of a subdominant mode becomes significantly small due to the filtering, it may be the case that the fit of the subdominant mode would become difficult or be destabilized. 
Since the amplitudes and phases are related to the excitation coefficient of QNMs~\cite{Leaver:1986gd,Andersson:1995zk,Andersson:1996cm,Glampedakis:2003dn,Berti:2006wq,Zhang:2013ksa}, it would be ideal if we could extract them without additional modifications by the mode cleaning process.
The iterative fitting method in the time domain which we consider in the present paper does not affect other modes, and allows us to extract subdominant modes in a stable manner. 

Another approach of a similar spirit would be the ``bootstrap procedure'' suggested in \cite{Baibhav:2023clw}, where one first identifies longer-lived modes by a theory-agnostic fitting, and then fixes their frequencies when searching for shorter-lived modes.
Very recently, this fitting strategy is further explored in \cite{Cheung:2023vki}, where after checking the stability of the amplitudes and phases of QNMs, the unstable modes are iteratively removed from the fitting function so that the all the modes are stably fitted.
Our strategy also adopts the iterative procedure but is complementary in the sense that we iteratively extract and subtract the most stable longest-lived mode one-by-one from the waveform and then consider the fitting of subdominant modes.

While we mainly focus on the extraction of overtones in the present paper, the extraction of subdominant modes with the iterative fitting method deserves further study as we clarified that the higher overtone coefficients are changed depending on whether we include the spherical-spheroidal mixing in the fitting function.  
Nonlinear effects predicted by second-order perturbation theory~\cite{Gleiser:1995gx,Ioka:2007ak,Nakano:2007cj,Okuzumi:2008ej,Brizuela:2009qd,Pazos:2010xf,Loutrel:2020wbw,Ripley:2020xby,Lagos:2022otp} would also be an interesting target~\cite{London:2014cma,MaganaZertuche:2021syq,Sberna:2021eui,Ma:2022wpv,Cheung:2022rbm,Mitman:2022qdl,Redondo-Yuste:2023seq,Bucciotti:2023ets}.
Further, while we focused on the frequency-fixed fitting of numerical relativity waveform in the present paper, we expect that our fitting method may also be efficient for a theory-agnostic fitting and/or observed ringdown signal.
For instance, we expect to improve the theory-agnostic fitting by performing, e.g., the Prony method first to identify the QNM frequency and coefficient of the longest-lived mode, subtracting it from the ringdown data, and iterating the procedure to extract higher overtones.
It would also be interesting to apply our fitting method to analyze the overtones in the ringdown gravitational waves emitted from a particle plunging into a black hole~\cite{Zhang:2013ksa,Hughes:2019zmt,Lim:2019xrb,Oshita:2021iyn,Lim:2022veo}.
We leave these topics for future work.

\acknowledgments
H.M.\ was supported by Japan Society for the Promotion of Science (JSPS) Grants-in-Aid for Scientific Research (KAKENHI) Grant No.~JP22K03639.

\appendix

\section{Kerr QNM frequencies}
\label{sec:qnms}

In this appendix, we provide the QNM frequencies corresponding to the SXS simulations used in the main text.
For SXS:BBH:0305, we provide the QNM frequencies for $(\ell,m)=(2,2)$ and $(3,2)$ modes in Table~\ref{tab:0305_QNMs}.
For other simulations, the QNM frequencies of $(\ell,m)=(2,2)$ mode are listed in Tables~\ref{tab:0158_QNMs}--\ref{tab:1108_QNMs}, respectively.
All QNM frequencies are generated by using the {\tt qnm} package~\cite{Stein:2019mop} with machine precision, and here we present values rounded to four digits as a reference.

Note that, since we use the normalization by the total binary mass $M$, to obtain the QNM frequencies describing the ringdown signal after merger, we multiply a factor $M/M_{\rm rem}$ to the Kerr QNM frequencies associated with the dimensionless remnant spin, where $M_{\rm rem}$ is the remnant mass after merger.
As a specific example, let us consider SXS:BBH:0305, whose dimensionless remnant spin is $a/M_{\rm rem} = 0.6921$.
The fundamental mode of the $(\ell,m)=(2,2)$ mode of the Kerr QNM frequencies for $a/M_{\rm rem} = 0.6921$ is given by $M_{\rm rem} \omega = 0.5291 - 0.08109i$.
Using the normalization by the total binary mass $M$ before merger, the QNM frequency is given by $M \omega = (0.5291 - 0.08109i)M/M_{\rm rem}$.
Since $M_{\rm rem}/M = 0.9520$ for SXS:BBH:0305, we obtain $M \omega = 0.5558 - 0.08517i$.
The QNM frequencies in Tables~\ref{tab:0305_QNMs}--\ref{tab:1108_QNMs} are obtained in this way.

\begin{table}[t]
    \centering
    \caption{Kerr QNM frequencies corresponding to SXS:BBH:0305 for $(\ell,m)=(2,2)$ and $(3,2)$.}
    \begin{tabular}{ccc}
    \hline\hline
        $n$ ~&~ $M\omega_{22n}$ ~&~ $M\omega_{32n}$ \\
        \hline
    	$0$ ~&~ $0.5558-0.08517i$ ~&~ $0.7941-0.08875i$ \\
    	$1$ ~&~ $0.5435-0.2575i$ ~&~ $0.7854-0.2675i$ \\
    	$2$ ~&~ $0.5208-0.4348i$ ~&~ $0.7691-0.4500i$ \\
    	$3$ ~&~ $0.4905-0.6165i$ ~&~ $0.7472-0.6379i$ \\
    	$4$ ~&~ $0.4580-0.7958i$ ~&~ $0.7220-0.8318i$ \\
    	$5$ ~&~ $0.4399-0.9702i$ ~&~ $0.6957-1.031i$ \\
    	$6$ ~&~ $0.4396-1.158i$ ~&~ $0.6699-1.236i$ \\
    	$7$ ~&~ $0.4410-1.361i$ ~&~ $0.6455-1.444i$ \\
        \hline\hline
    \end{tabular}
    \label{tab:0305_QNMs}
\end{table}

\begin{table}[t]
    \centering
    \caption{Kerr QNM frequencies corresponding to SXS:BBH:0158 for $(\ell,m)=(2,2)$.}
    \begin{tabular}{cc}
    \hline\hline
        $n$ ~&~ $M\omega_{22n}$ \\
        \hline
        $0$ ~&~ $0.8271-0.06149i$ \\
        $1$ ~&~ $0.8249-0.1848i$ \\
        $2$ ~&~ $0.8207-0.3091i$ \\
        $3$ ~&~ $0.8145-0.4352i$ \\
        $4$ ~&~ $0.8065-0.5643i$ \\
        $5$ ~&~ $0.5838-0.8198i$ \\
        $6$ ~&~ $0.7968-0.6985i$ \\
        $7$ ~&~ $0.7898-0.8404i$ \\
        \hline\hline
    \end{tabular}
    \label{tab:0158_QNMs}
\end{table}

\begin{table}[t]
    \centering
    \caption{Kerr QNM frequencies corresponding to SXS:BBH:0156 for $(\ell,m)=(2,2)$. }
    \begin{tabular}{cc}
    \hline\hline
        $n$ ~&~ $M\omega_{22n}$ \\
        \hline
    	$0$ ~&~ $0.4489-0.08999i$ \\
    	$1$ ~&~ $0.4287-0.2744i$ \\
    	$2$ ~&~ $0.3937-0.4703i$ \\
    	$3$ ~&~ $0.3527-0.6794i$ \\
    	$4$ ~&~ $0.3135-0.8967i$ \\
    	$5$ ~&~ $0.2816-1.116i$ \\
    	$6$ ~&~ $0.2615-1.337i$ \\
    	$7$ ~&~ $0.2518-1.564i$ \\
        \hline\hline
    \end{tabular}
    \label{tab:0156_QNMs}
\end{table}

\begin{table}[t]
    \centering
    \caption{Kerr QNM frequencies corresponding to SXS:BBH:1108 for $(\ell,m)=(2,2)$.}
    \begin{tabular}{cc}
    \hline\hline
        $n$ ~&~ $M\omega_{22n}$ \\
        \hline
    	$0$ ~&~ $0.4191-0.08868i$ \\
    	$1$ ~&~ $0.3973-0.2711i$ \\
    	$2$ ~&~ $0.3598-0.4671i$ \\
    	$3$ ~&~ $0.3167-0.6786i$ \\
    	$4$ ~&~ $0.2759-0.9002i$ \\
    	$5$ ~&~ $0.2413-1.125i$ \\
    	$6$ ~&~ $0.2158-1.348i$ \\
    	$7$ ~&~ $0.2016-1.573i$ \\
        \hline\hline
    \end{tabular}
    \label{tab:1108_QNMs}
\end{table}

\section{Supplemental plots}
\label{sec:plots}

\begin{figure*}[t]
    \centering
    \includegraphics[height=0.25\textheight]{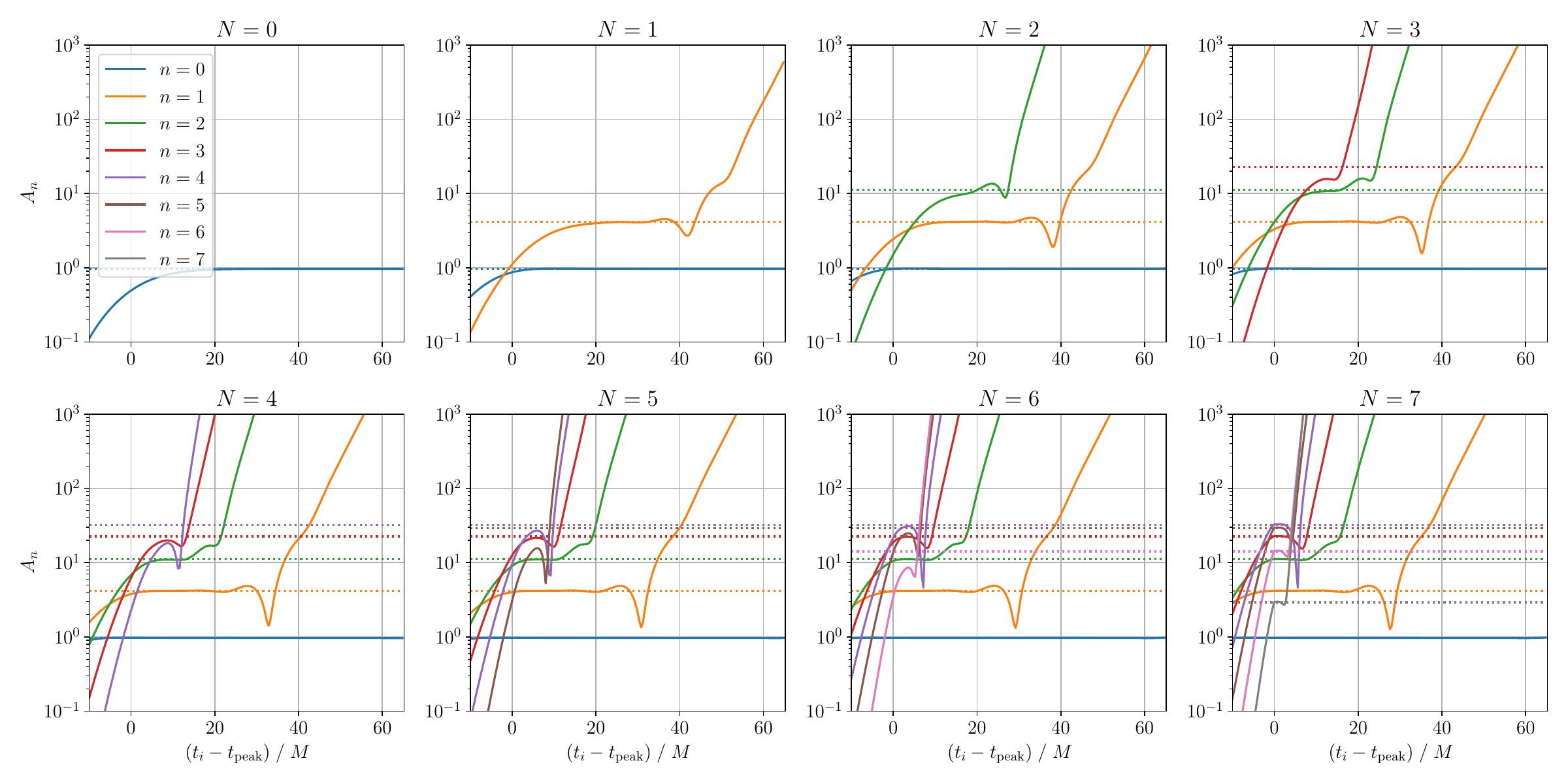}
    \includegraphics[height=0.25\textheight]{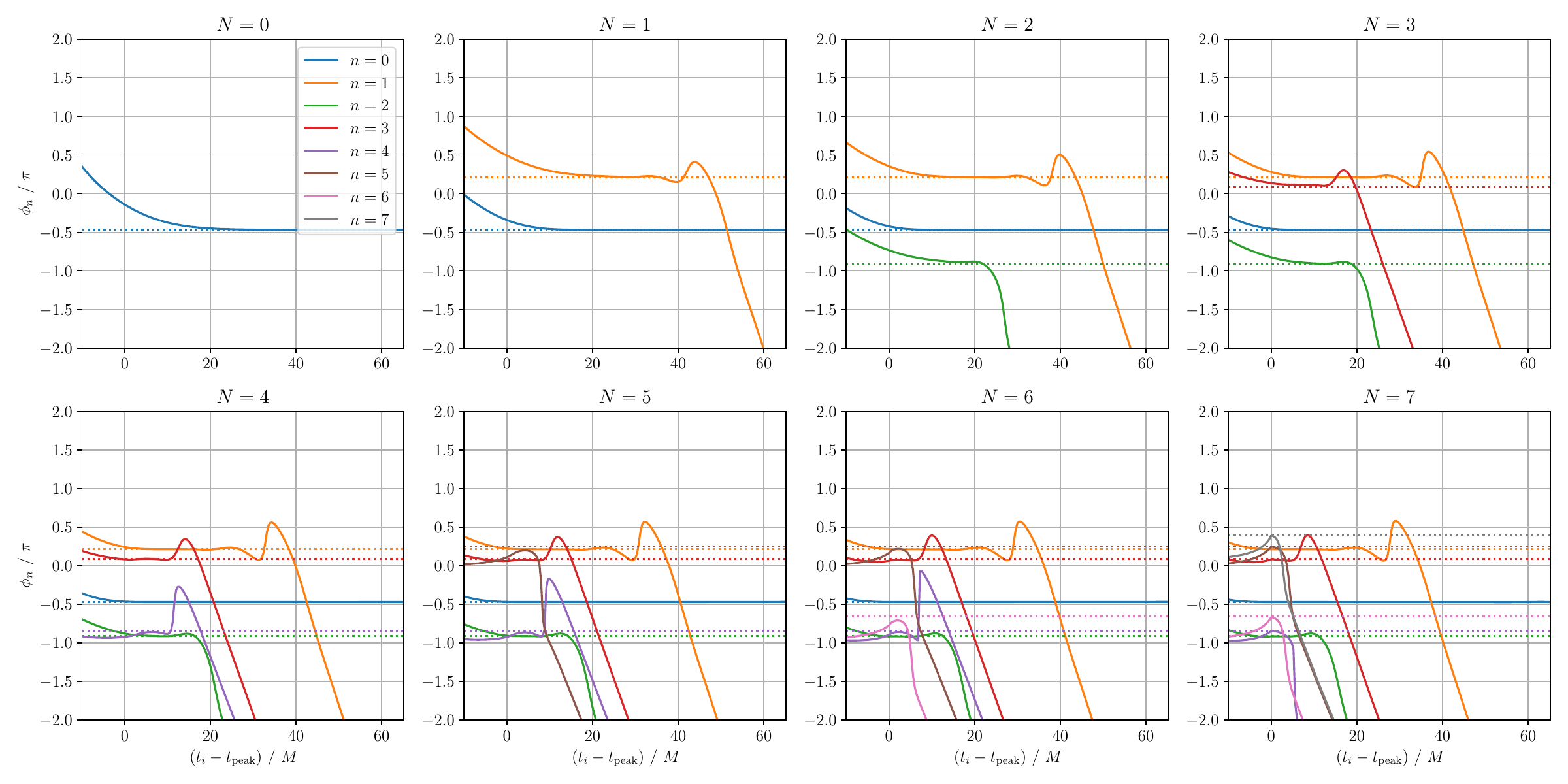}
    \includegraphics[height=0.25\textheight]{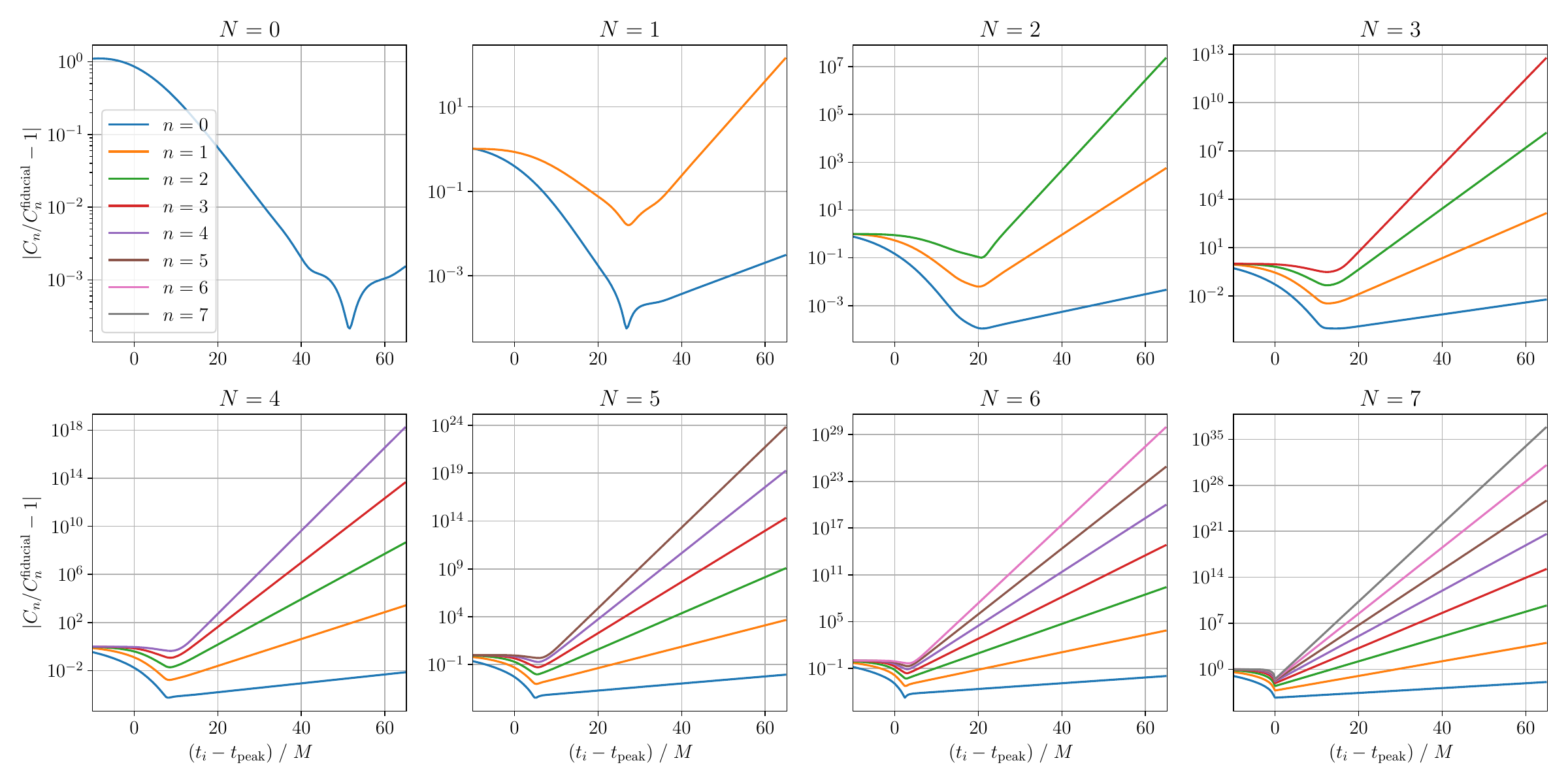}
    \includegraphics[height=0.25\textheight]{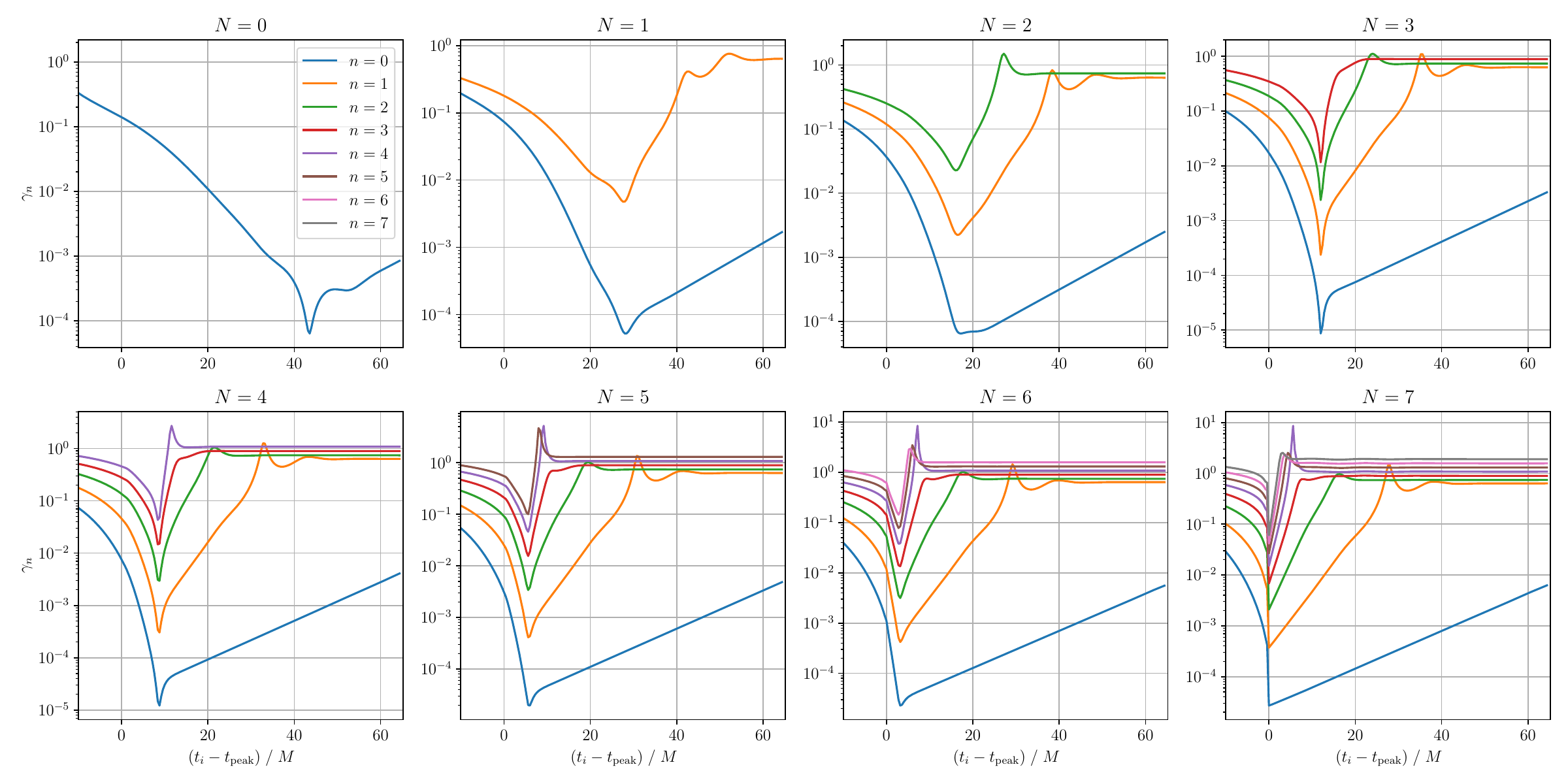}
    \caption{Amplitude $A_n$, phase $\phi_n$, relative error of $C_n$, and rate of change $\gamma_n$ for the fit of the mock waveform $\Psi_{[0,7]}^{(c)}$ by the fitting function $\psi_{[0,N]}^{\fit}$.}
    \label{fig:0305_A_mock_naive}
\end{figure*}

\begin{figure*}[t]
    \centering
    \includegraphics[height=0.25\textheight]{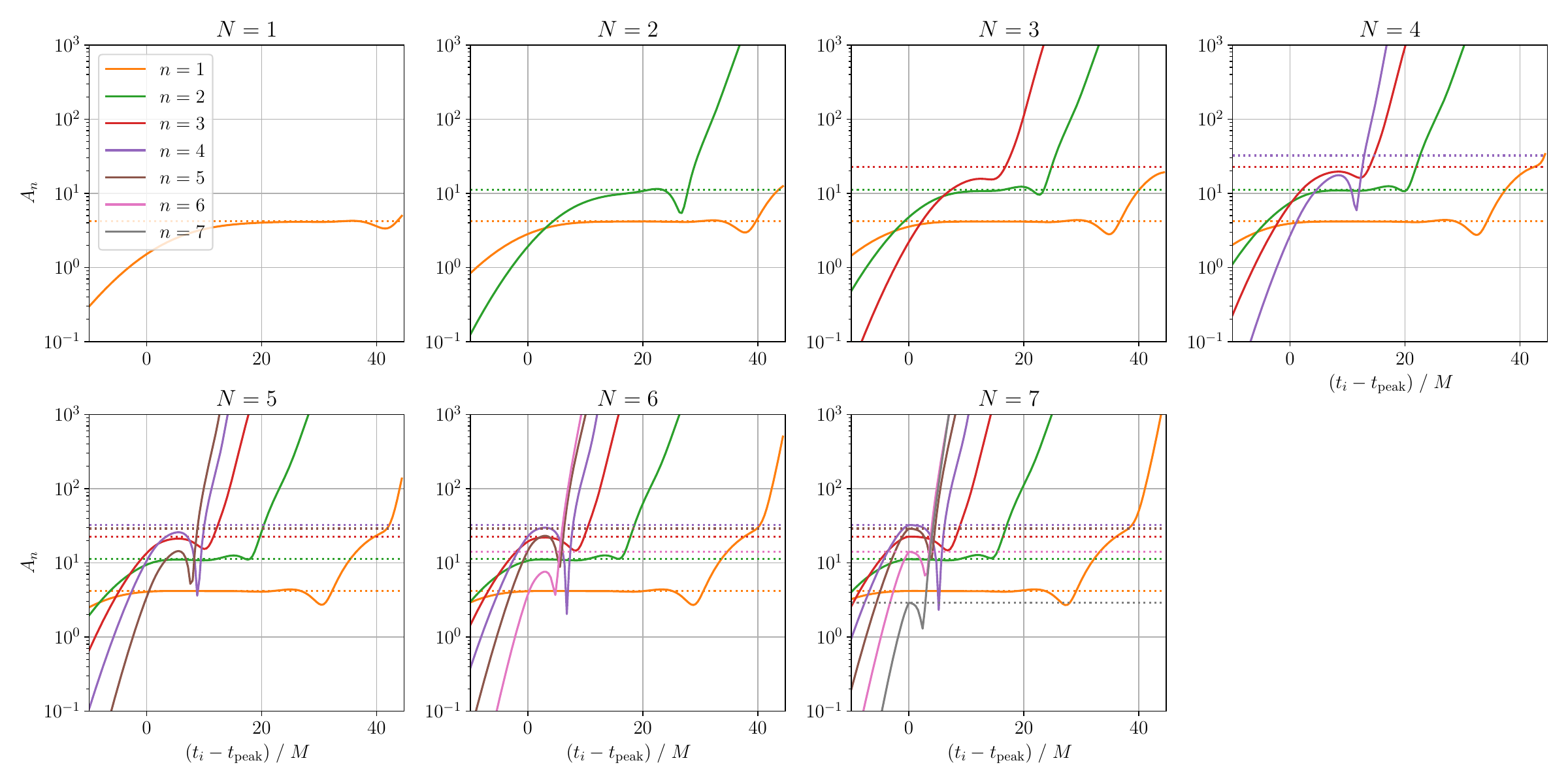}
    \includegraphics[height=0.25\textheight]{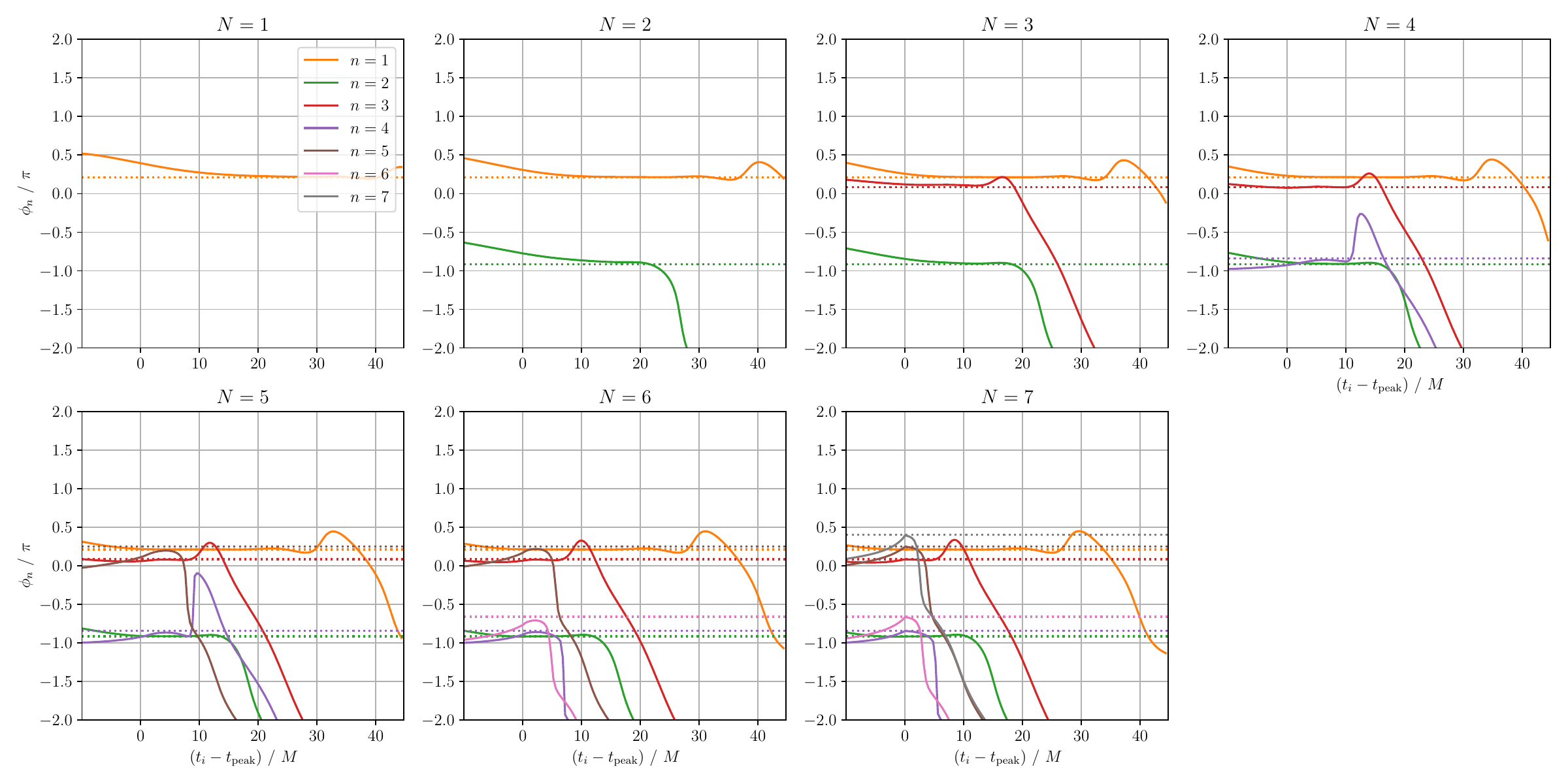}
    \includegraphics[height=0.25\textheight]{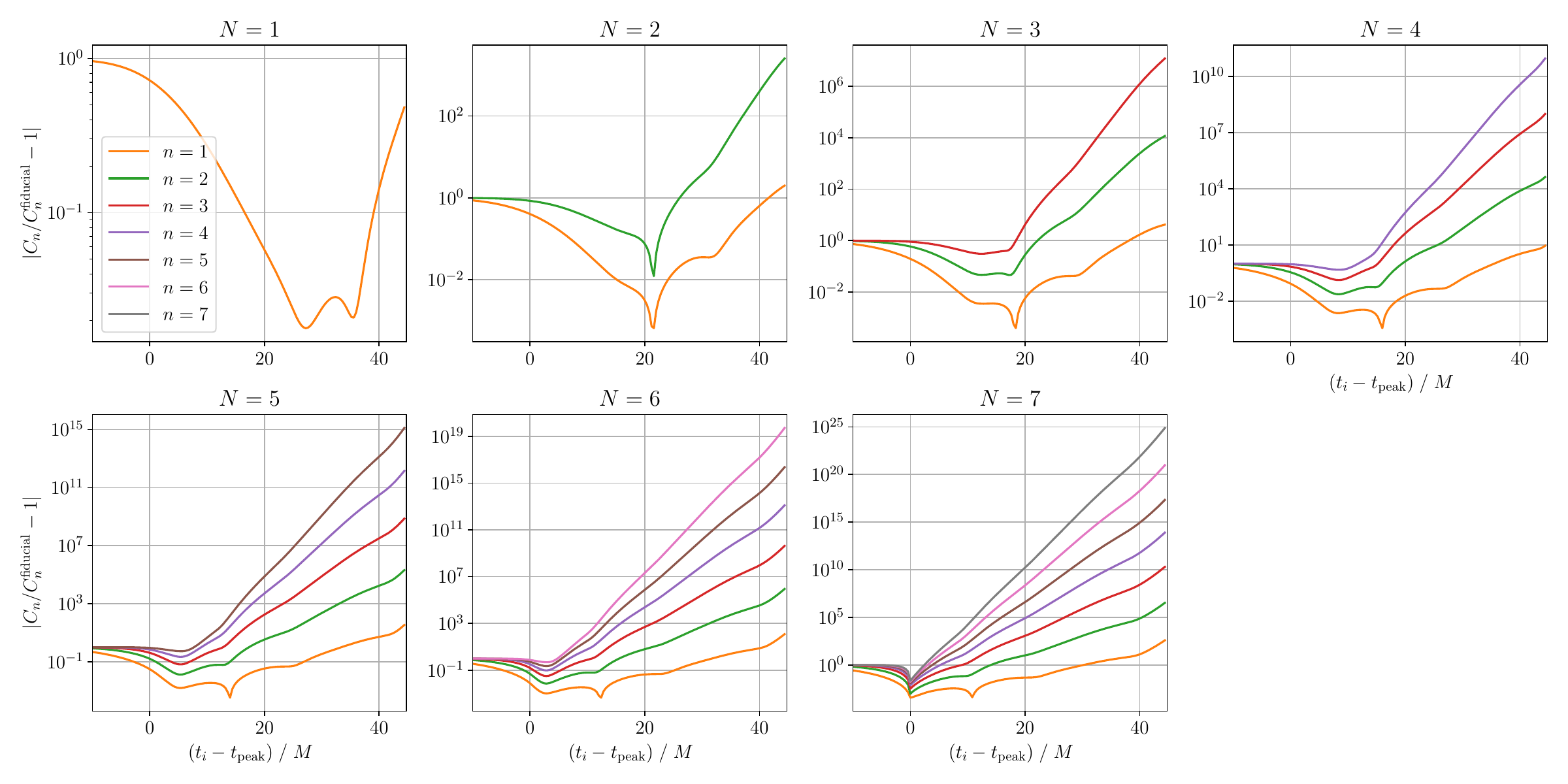}
    \includegraphics[height=0.25\textheight]{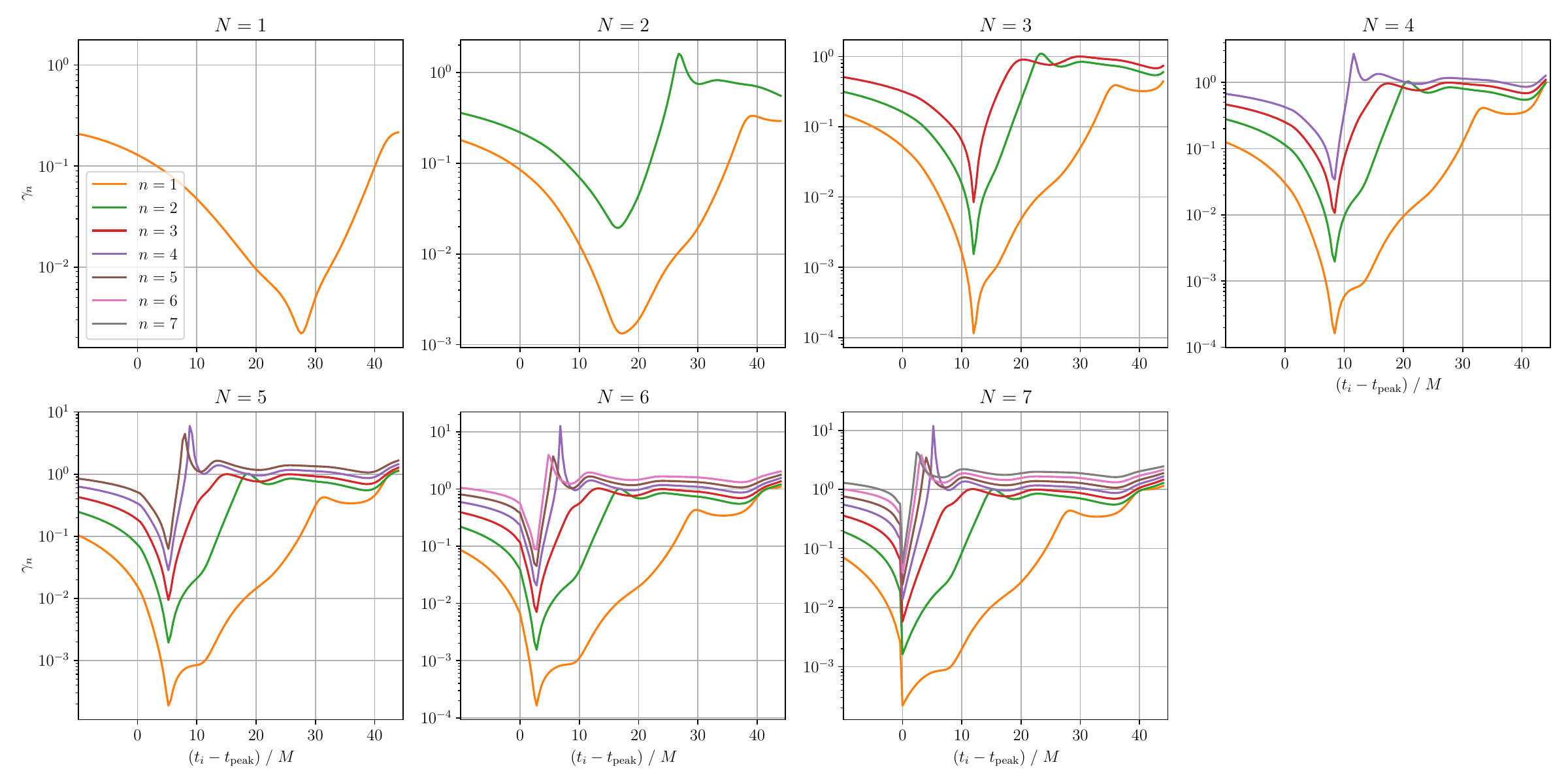}
    \caption{Amplitude $A_n$, phase $\phi_n$, relative error of $C_n$, and rate of change $\gamma_n$ for the fit of the mock waveform $\Psi_{[0,7]}^{(c)}-C^{\iter}_{0}\psi_{0}$ by the fitting function $\psi_{[1,N]}^{\fit}$.}
    \label{fig:0305_A_mock_sub_0}
\end{figure*}

In this appendix, we supplement additional figures which are not included in the main text.
Figure~\ref{fig:0305_A_mock_naive} shows the results of the fit of the mock waveform with constant $\Psi_{[0,7]}^{(c)}$ considered in \S\ref{sec:dawnoise}, and 
Fig.~\ref{fig:0305_A_mock_sub_0} shows those after the subtraction of the fundamental mode.
Comparing them to Figs.~\ref{fig:0305_A_mock_sub_const}--\ref{fig:0305_gamma_mock_sub_const} and Figs.~\ref{fig:0305_A_mock_sub_const0}--\ref{fig:0305_gamma_mock_sub_const0} for the case of the mock waveform without constant, we see that the constant reduces the stability of the fit.

\begin{figure}[t]
    \centering
    \includegraphics[width=0.49\columnwidth]{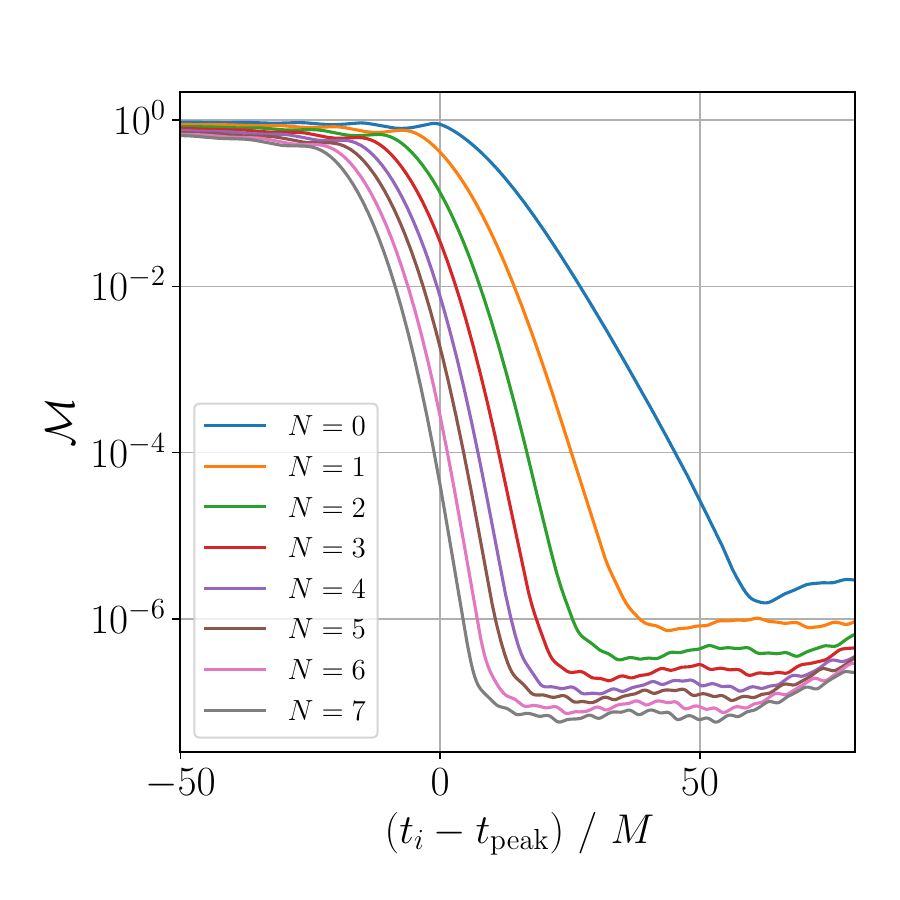}
    \includegraphics[width=0.49\columnwidth]{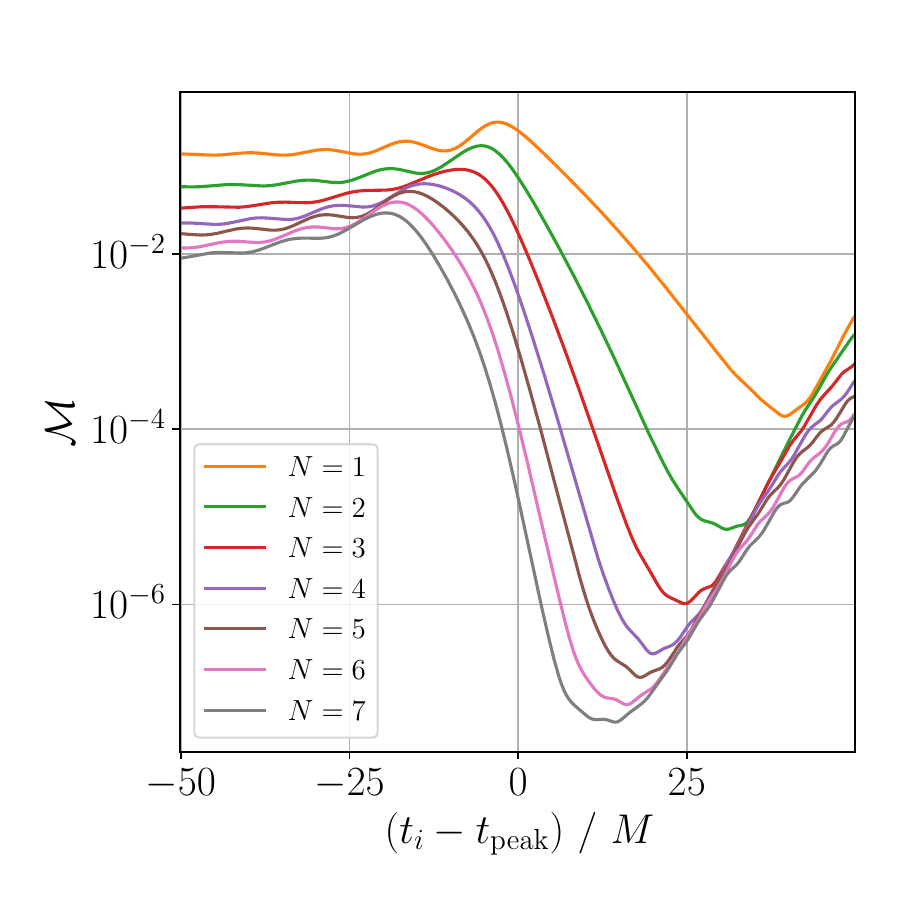}
    \includegraphics[width=0.49\columnwidth]{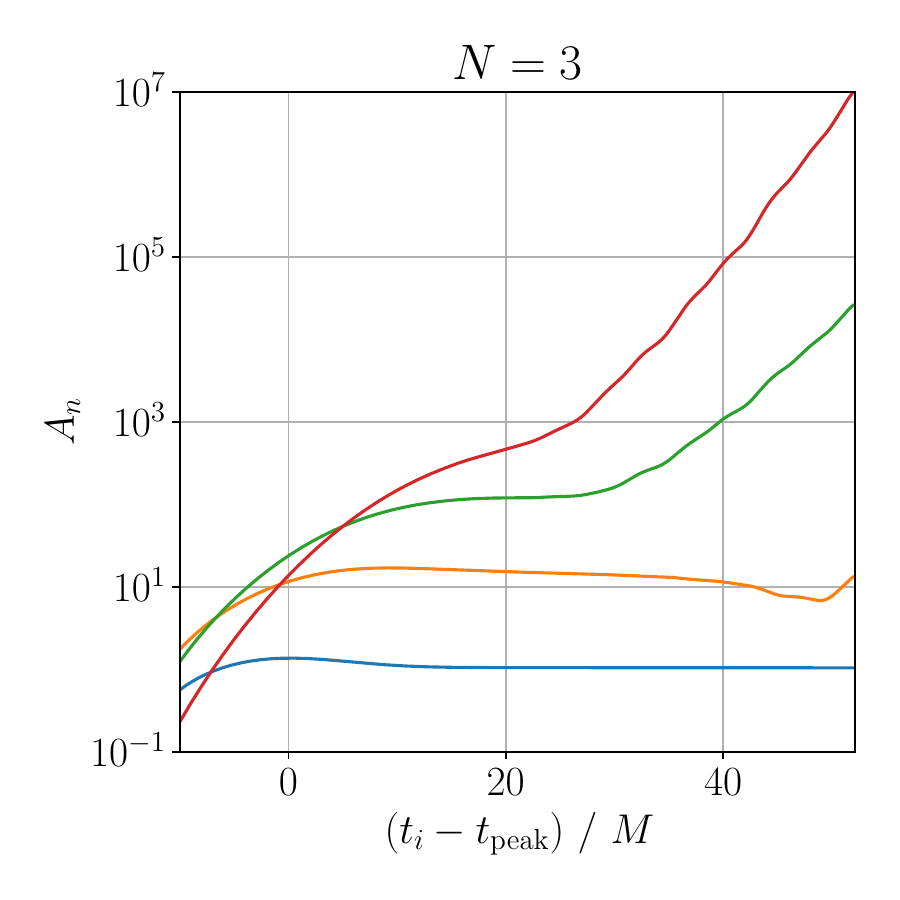}
    \includegraphics[width=0.49\columnwidth]{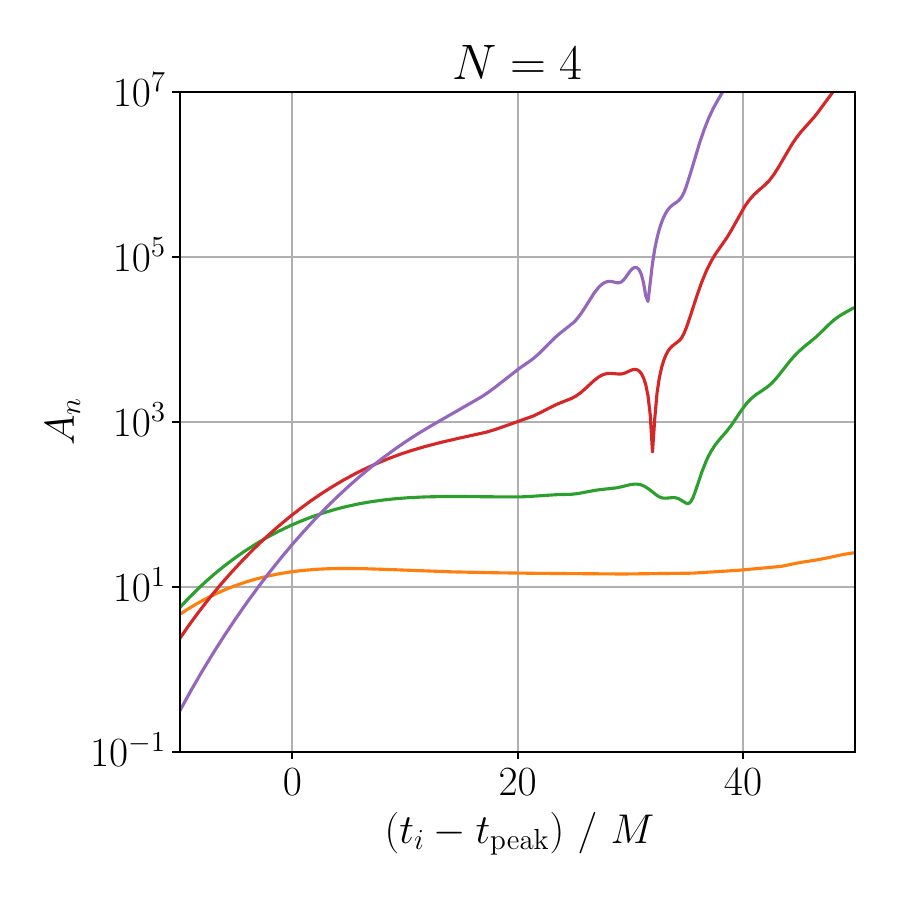}
    \includegraphics[width=0.49\columnwidth]{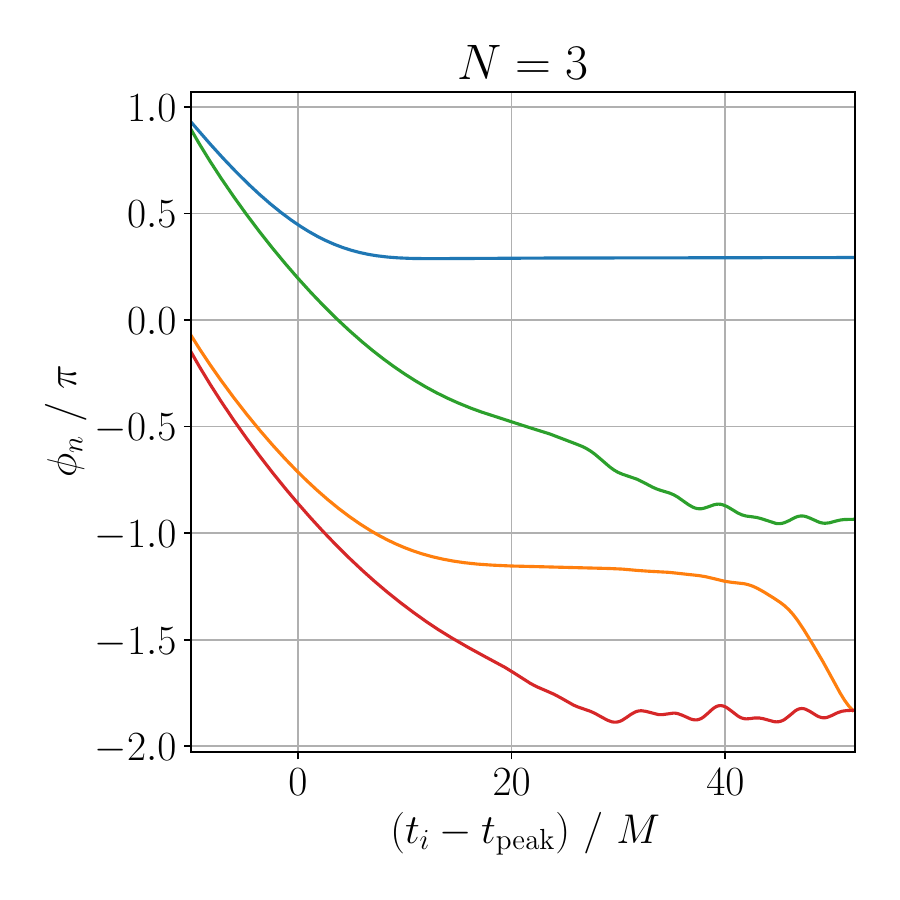}
    \includegraphics[width=0.49\columnwidth]{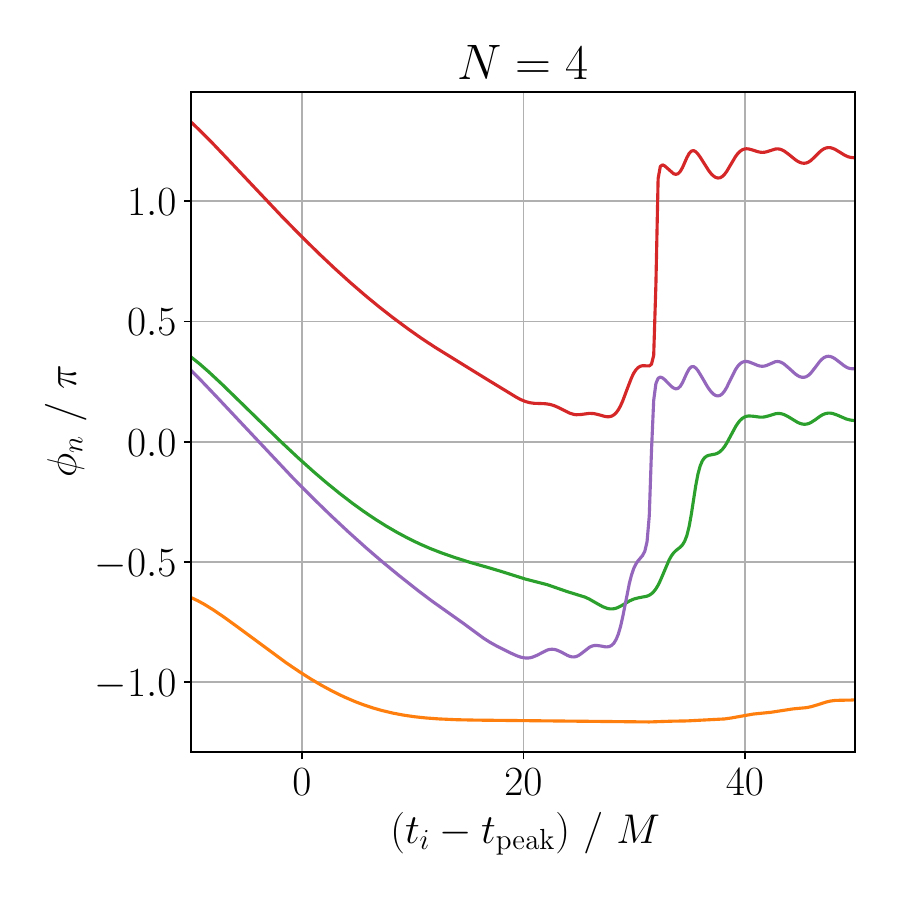}
    \includegraphics[width=0.49\columnwidth]{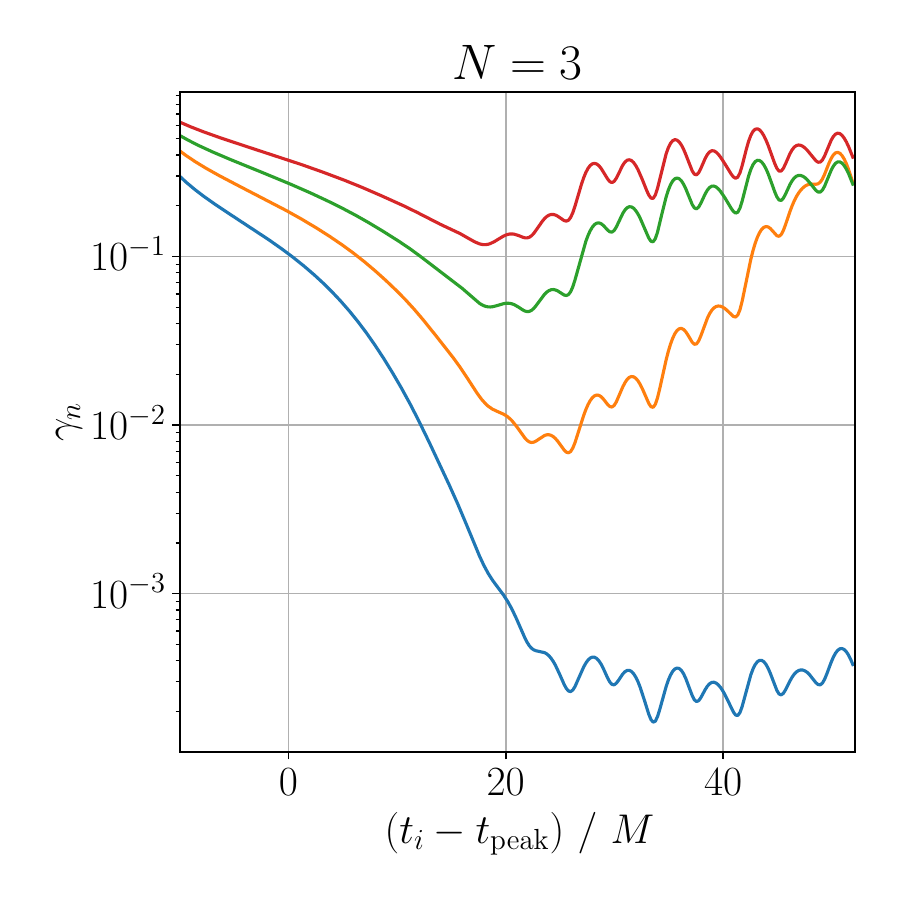}
    \includegraphics[width=0.49\columnwidth]{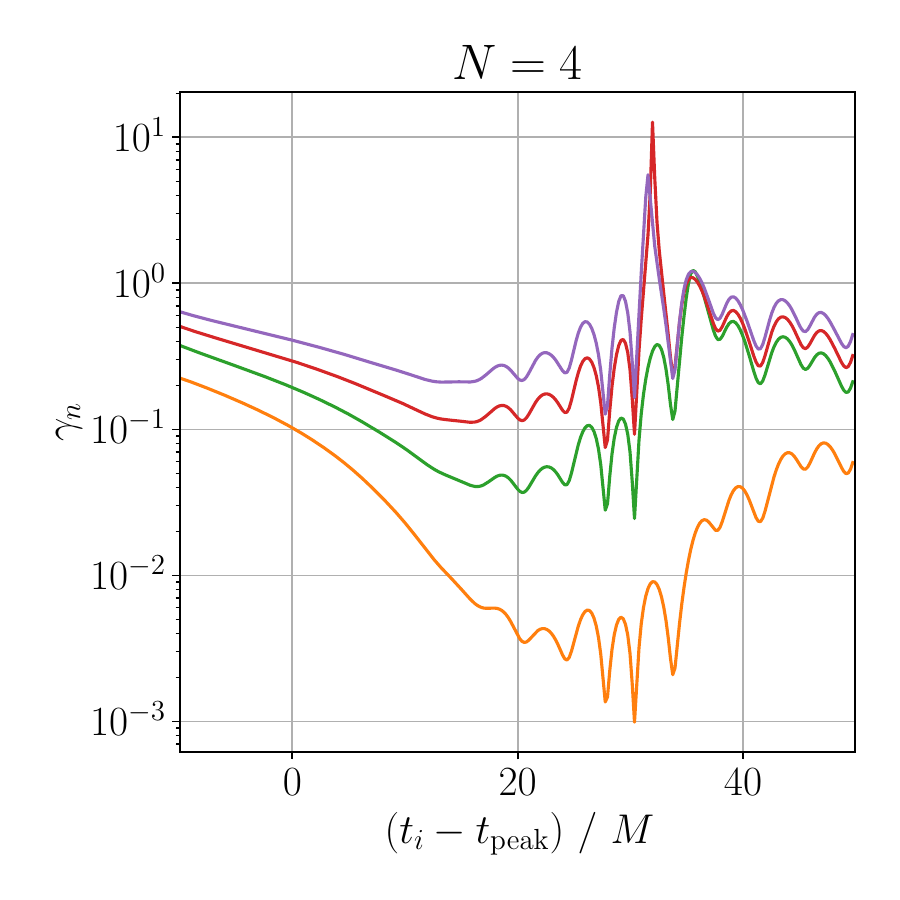}
    \caption{Mismatch $\M$ (first row), amplitude $A_n$ (second row), phase $\phi_n$ (third row), and rate of change $\gamma_n$ (fourth row) for the fit of SXS:BBH:0158 waveform $\Psi^\sxs - c$ by the fitting function $\psi^\fit_{[0,N]}$ (left column), and $\Psi^\sxs - c - \psi^\iter_{[0,0]}$ after the subtraction of the fundamental mode by the fitting function $\psi^\fit_{[1,N]}$ (right column).
    }
    \label{fig:0158_M_A_gamma_sub_const}
\end{figure}

\begin{figure}[t]
    \centering
    \includegraphics[width=0.49\columnwidth]{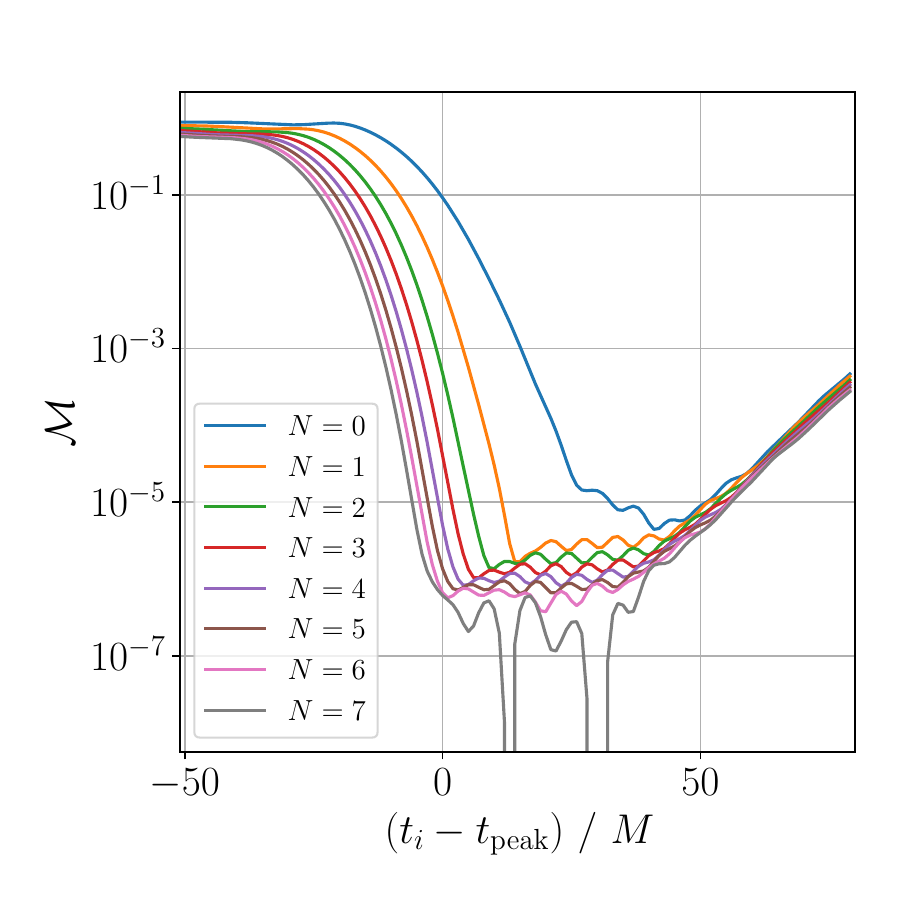}
    \includegraphics[width=0.49\columnwidth]{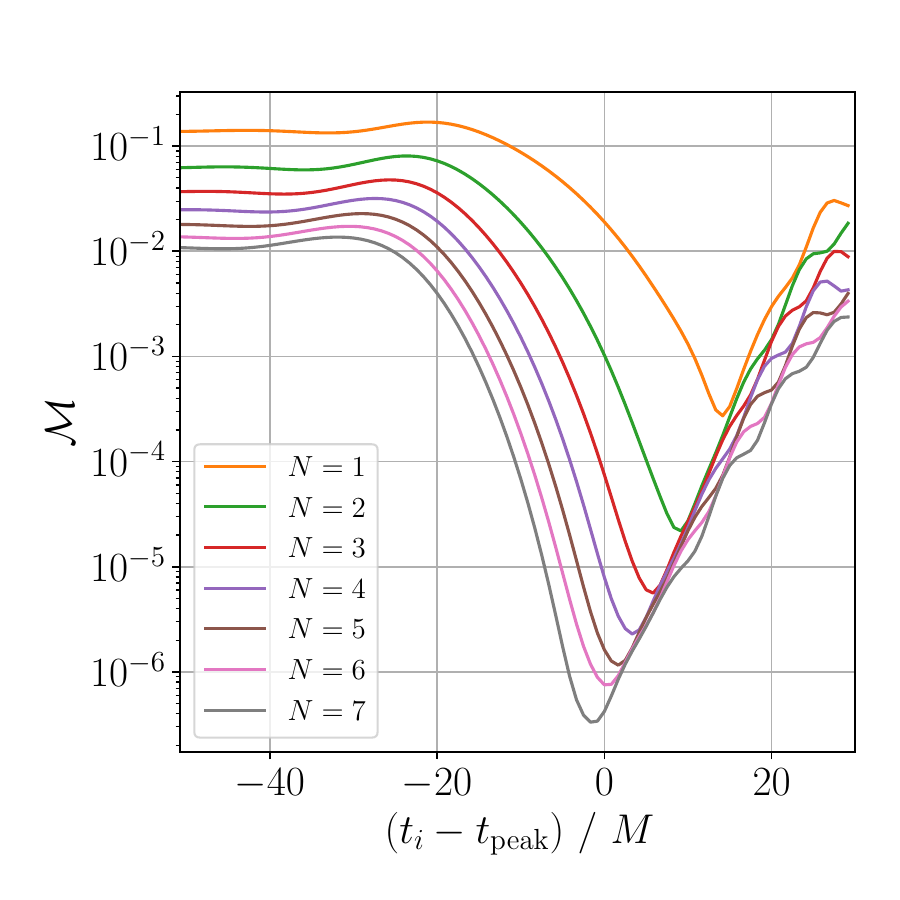}
    \includegraphics[width=0.49\columnwidth]{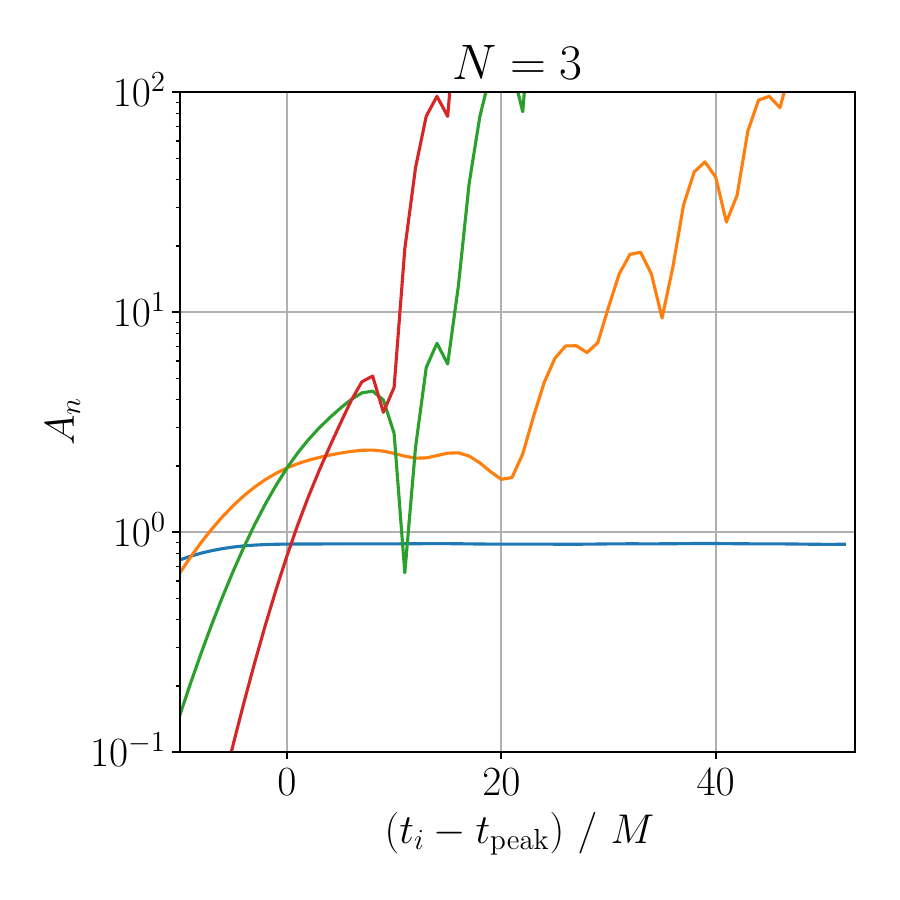}
    \includegraphics[width=0.49\columnwidth]{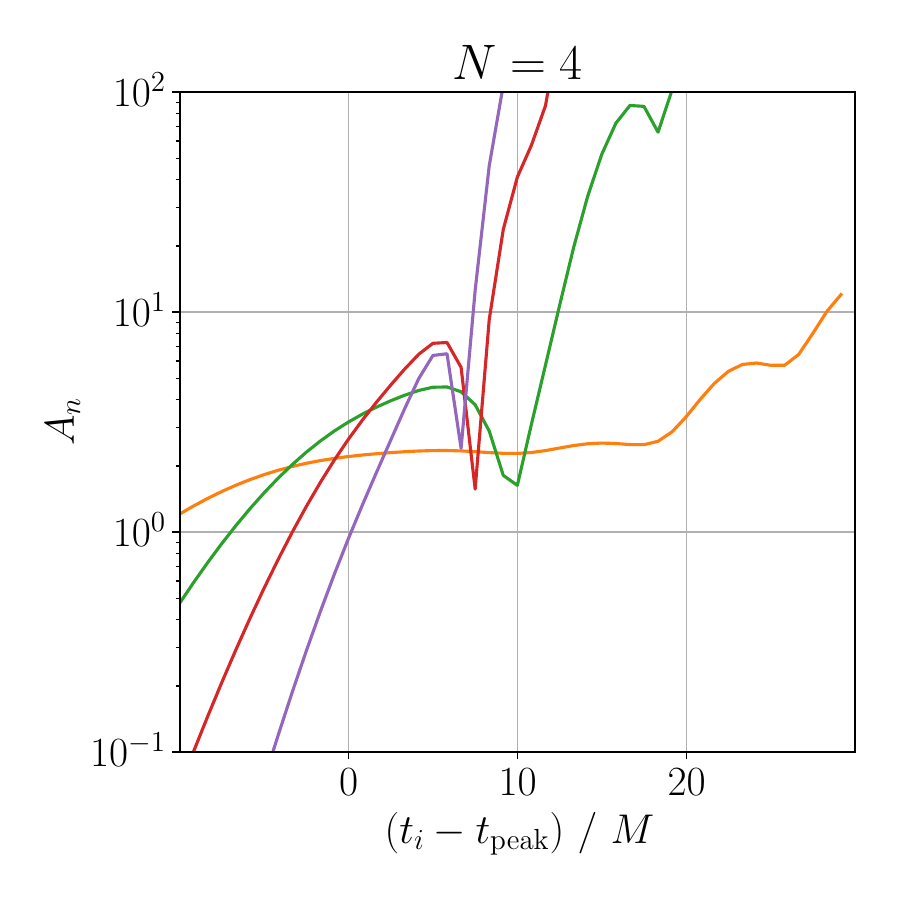}
    \includegraphics[width=0.49\columnwidth]{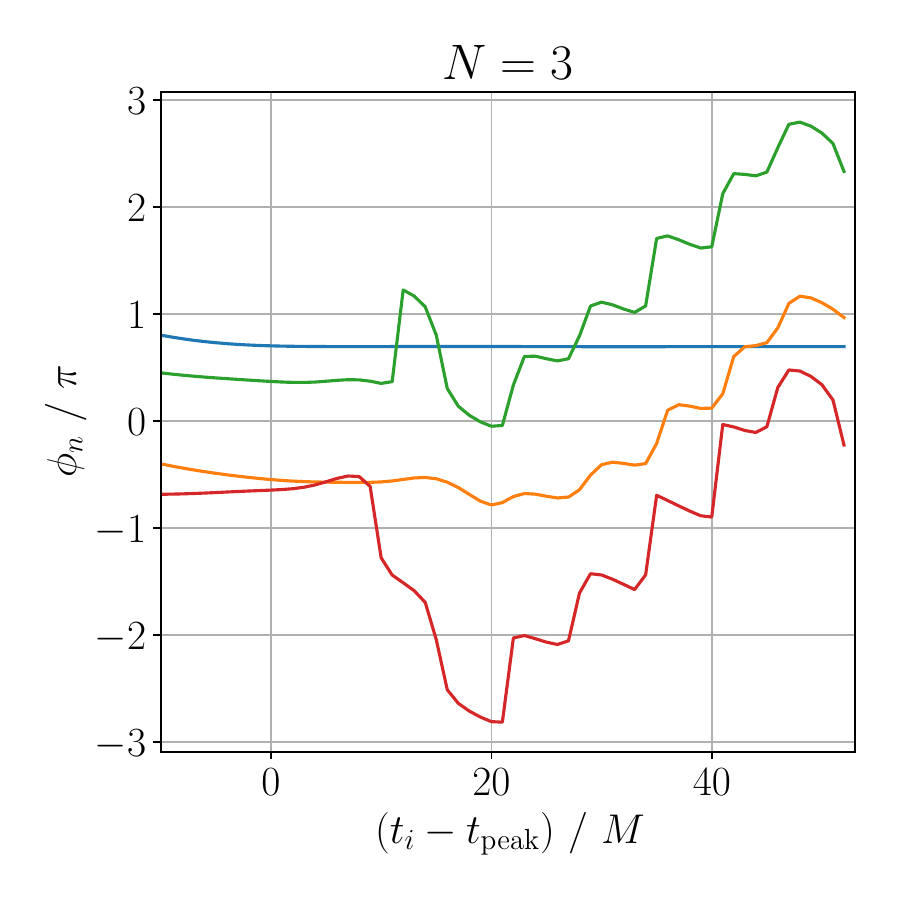}
    \includegraphics[width=0.49\columnwidth]{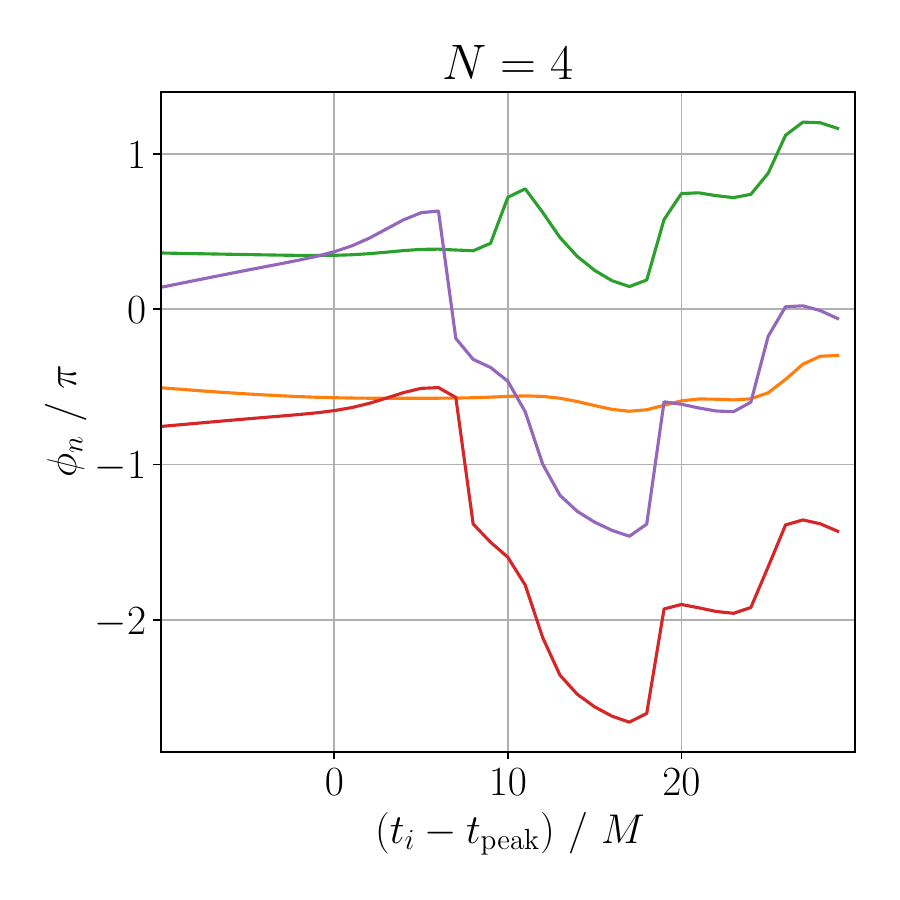}
    \includegraphics[width=0.49\columnwidth]{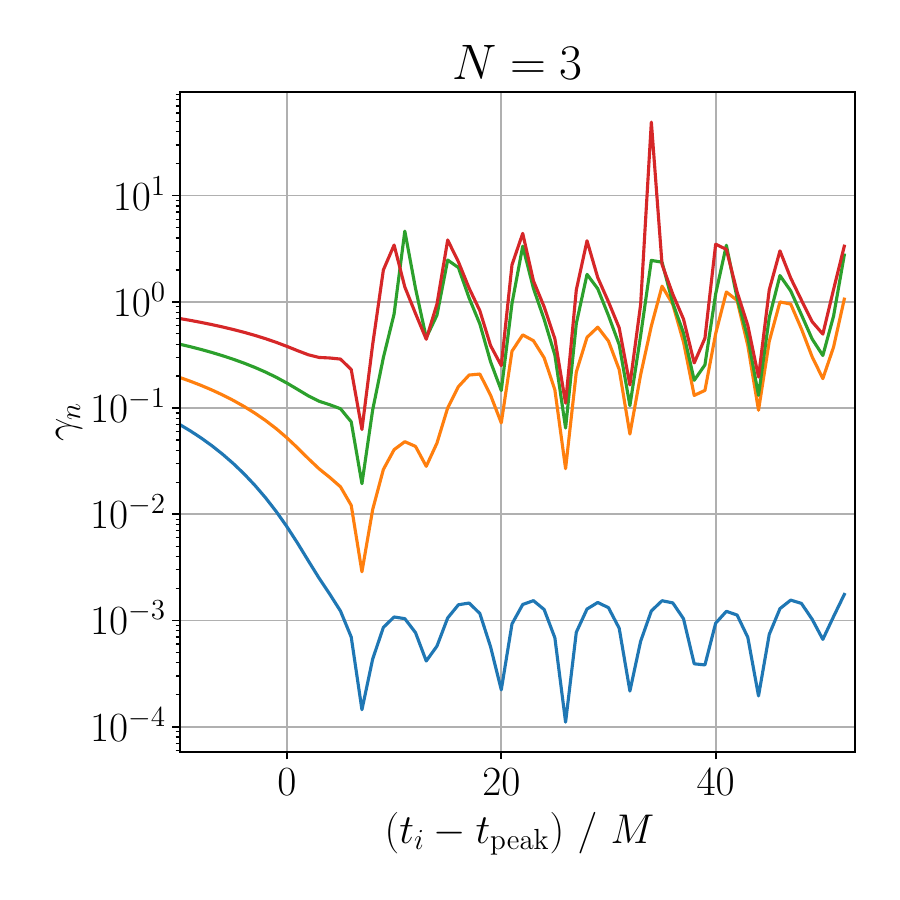}
    \includegraphics[width=0.49\columnwidth]{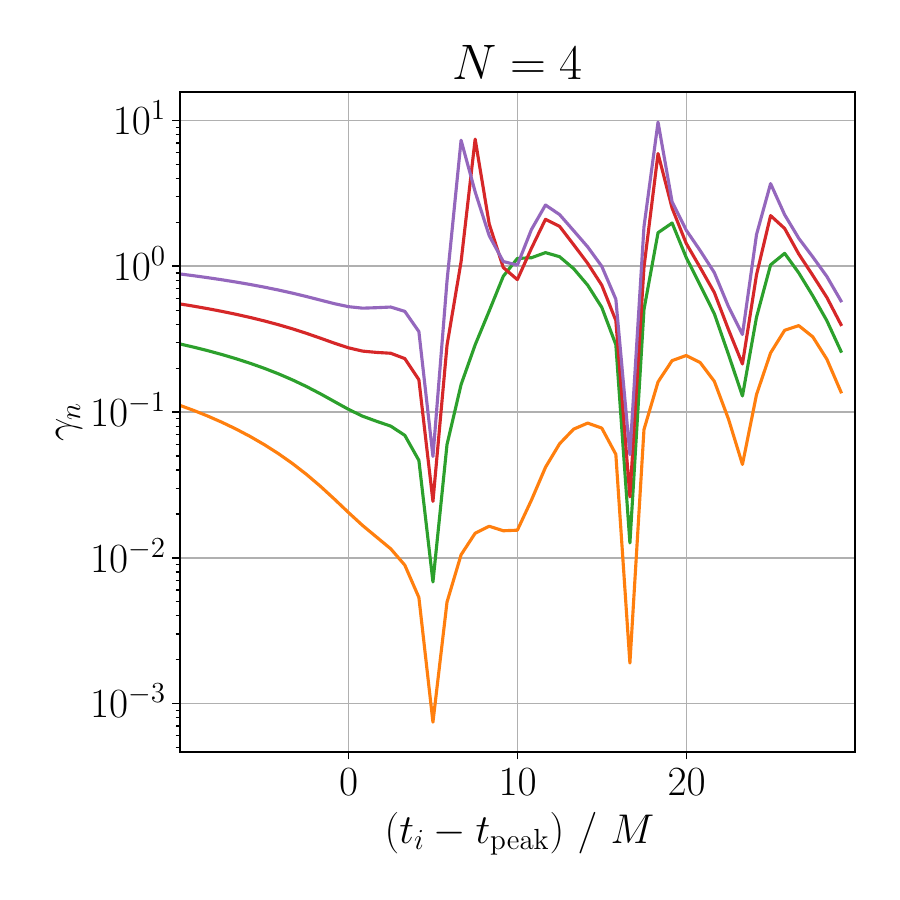}
    \caption{Mismatch $\M$ (first row), amplitude $A_n$ (second row), phase $\phi_n$ (third row), and rate of change $\gamma_n$ (fourth row) for the fit of SXS:BBH:0156 waveform $\Psi^\sxs - c$ by the fitting function $\psi^\fit_{[0,N]}$ (left column), and $\Psi^\sxs - c - \psi^\iter_{[0,0]}$ after the subtraction of the fundamental mode by the fitting function $\psi^\fit_{[1,N]}$ (right column).
    }
    \label{fig:0156_M_A_gamma_sub_const}
\end{figure}

\begin{figure}[t]
    \centering
    \includegraphics[width=0.49\columnwidth]{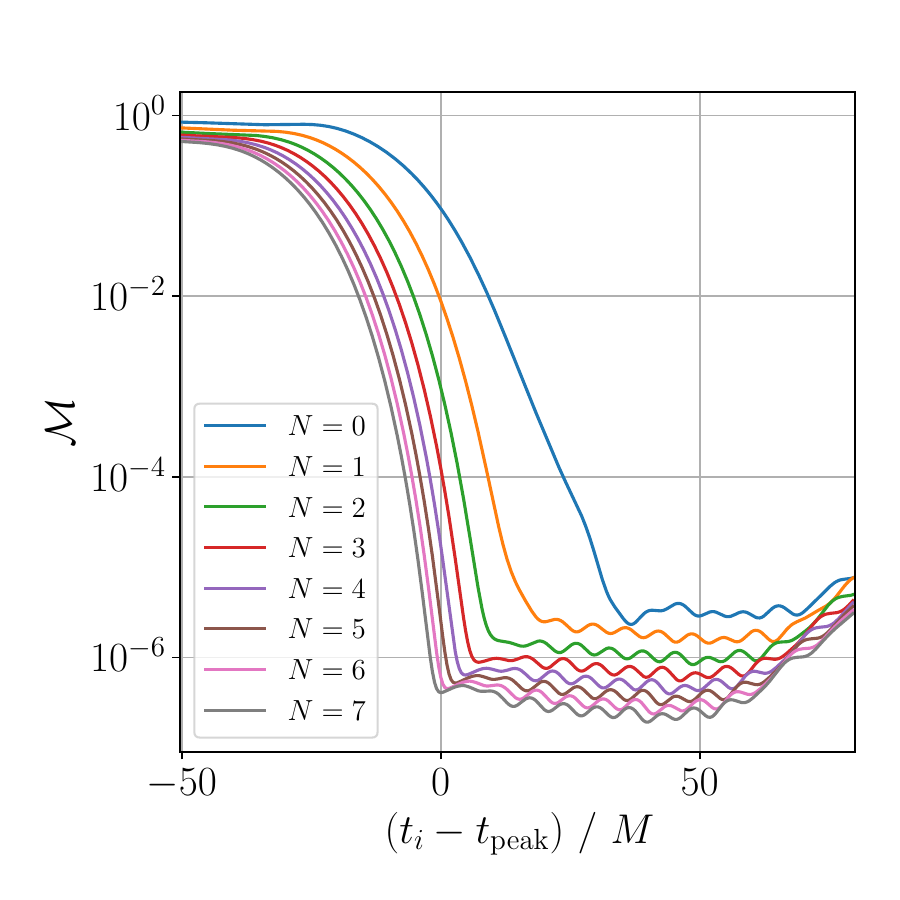}
    \includegraphics[width=0.49\columnwidth]{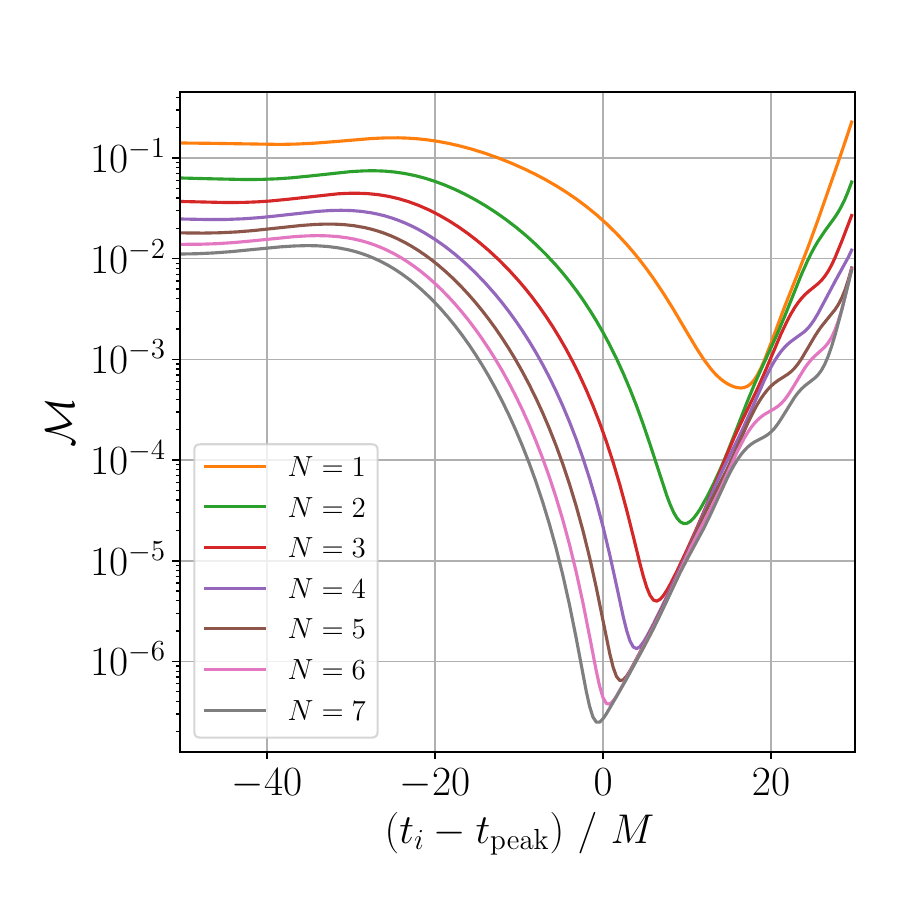}
    \includegraphics[width=0.49\columnwidth]{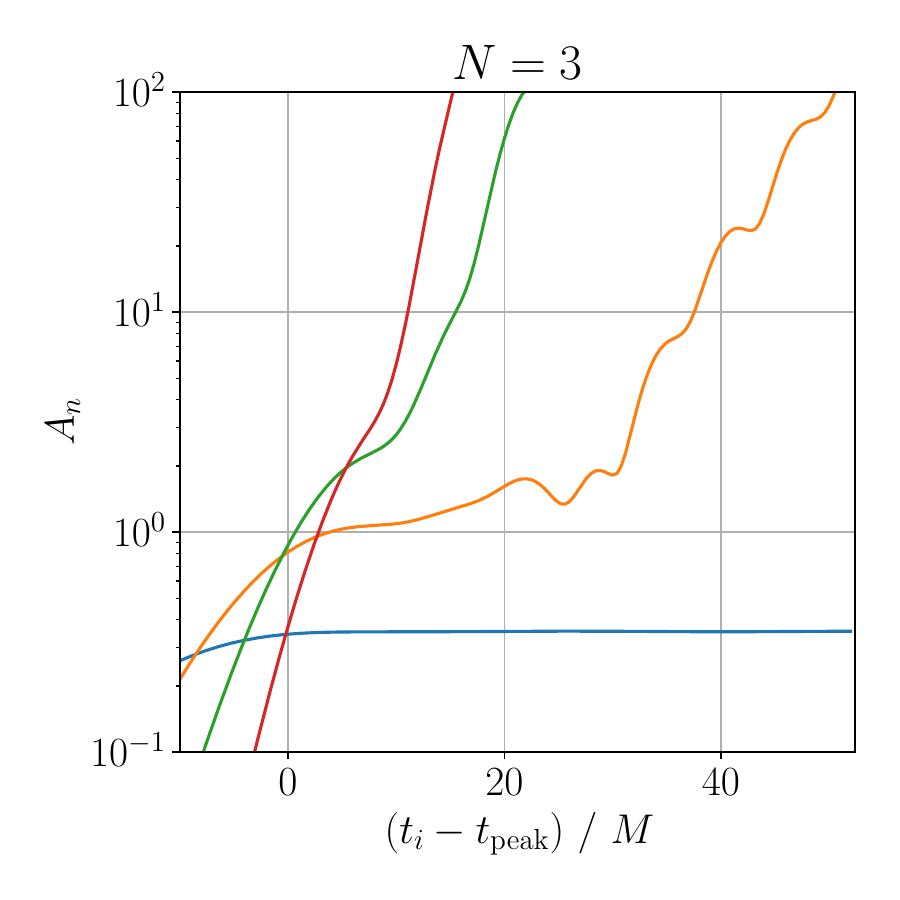}
    \includegraphics[width=0.49\columnwidth]{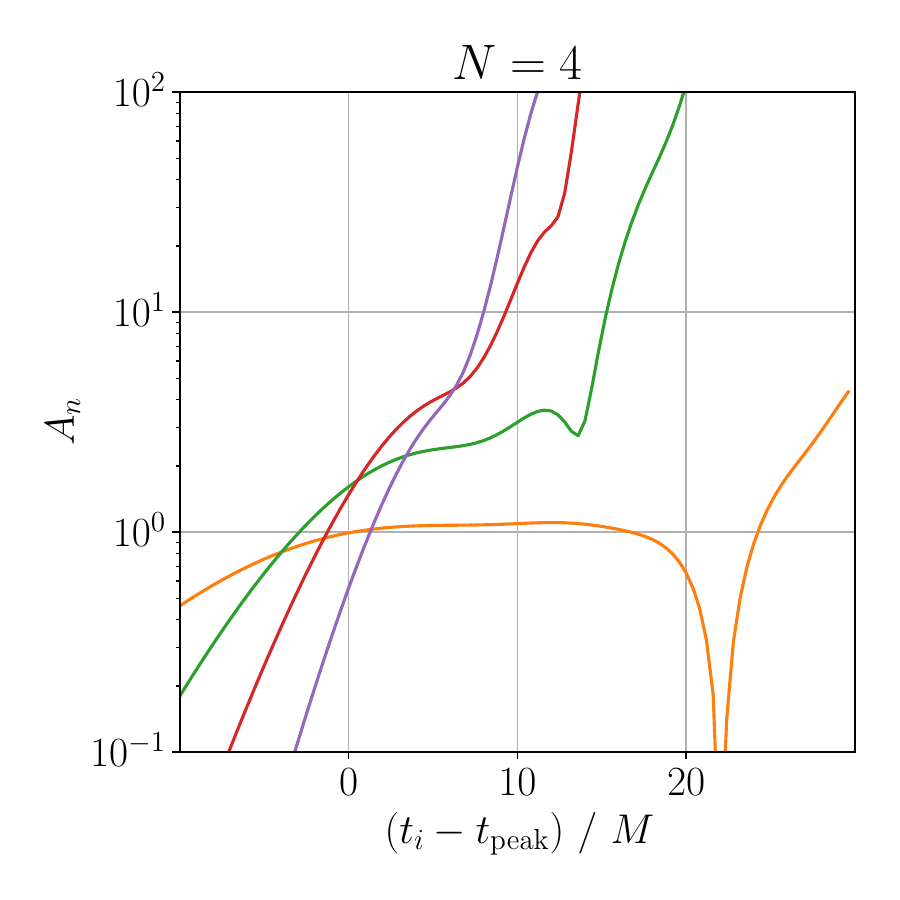}
    \includegraphics[width=0.49\columnwidth]{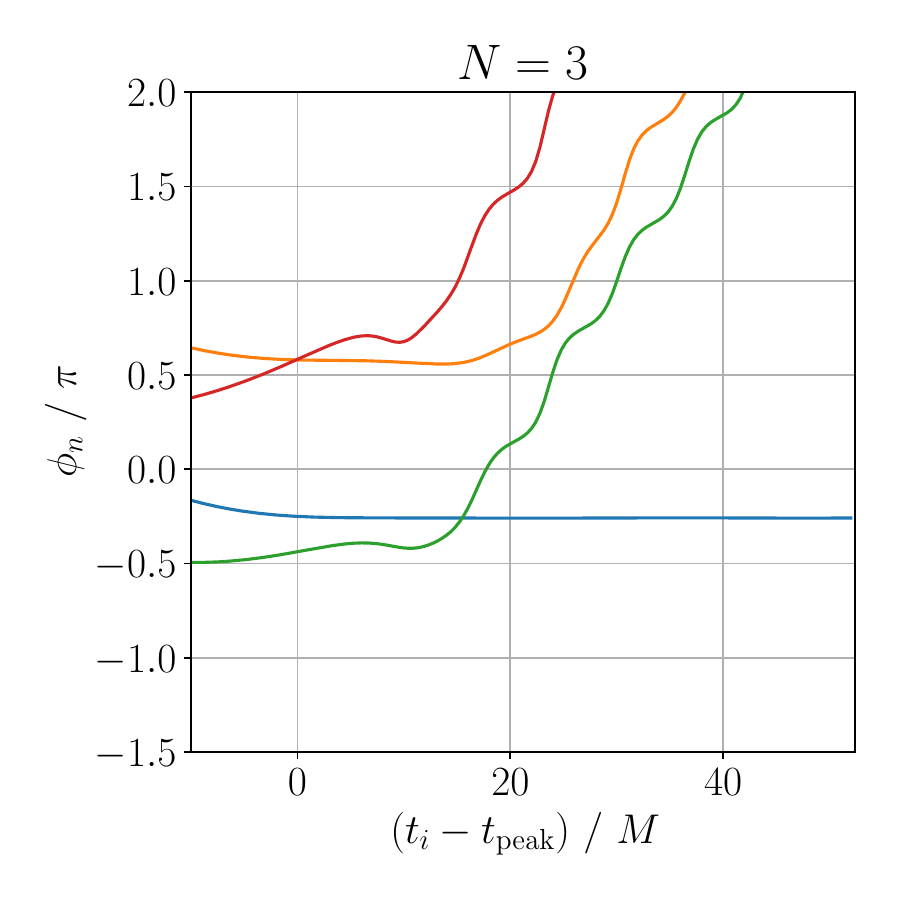}
    \includegraphics[width=0.49\columnwidth]{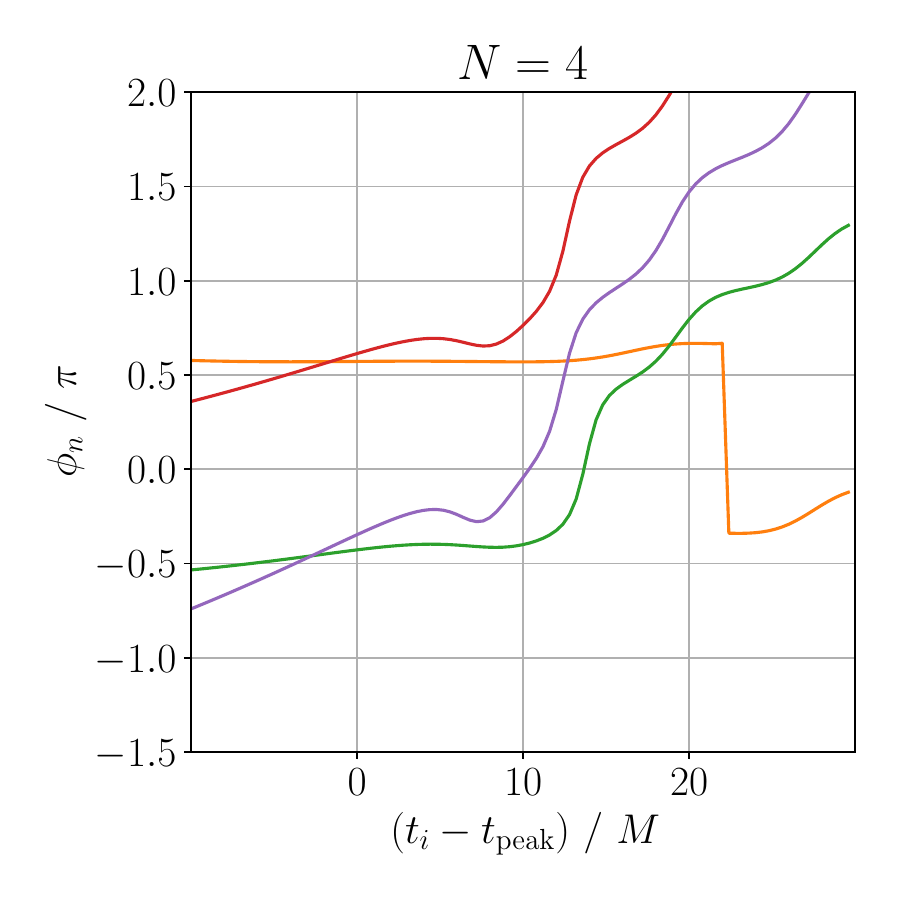}
    \includegraphics[width=0.49\columnwidth]{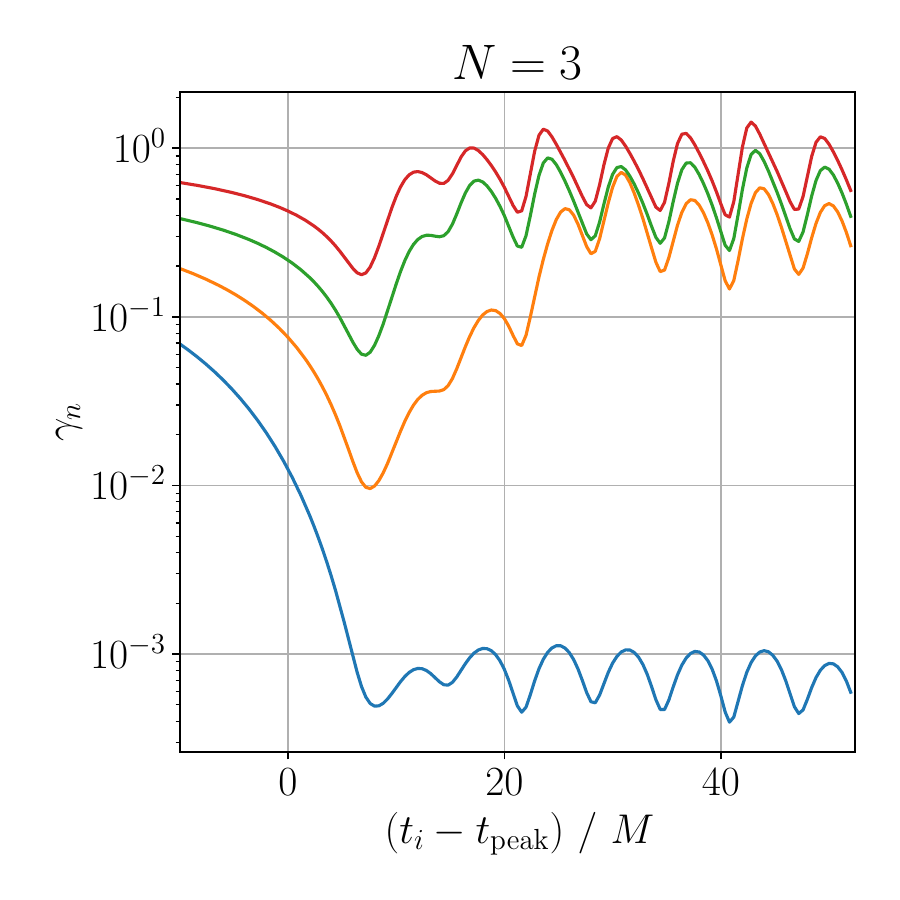}
    \includegraphics[width=0.49\columnwidth]{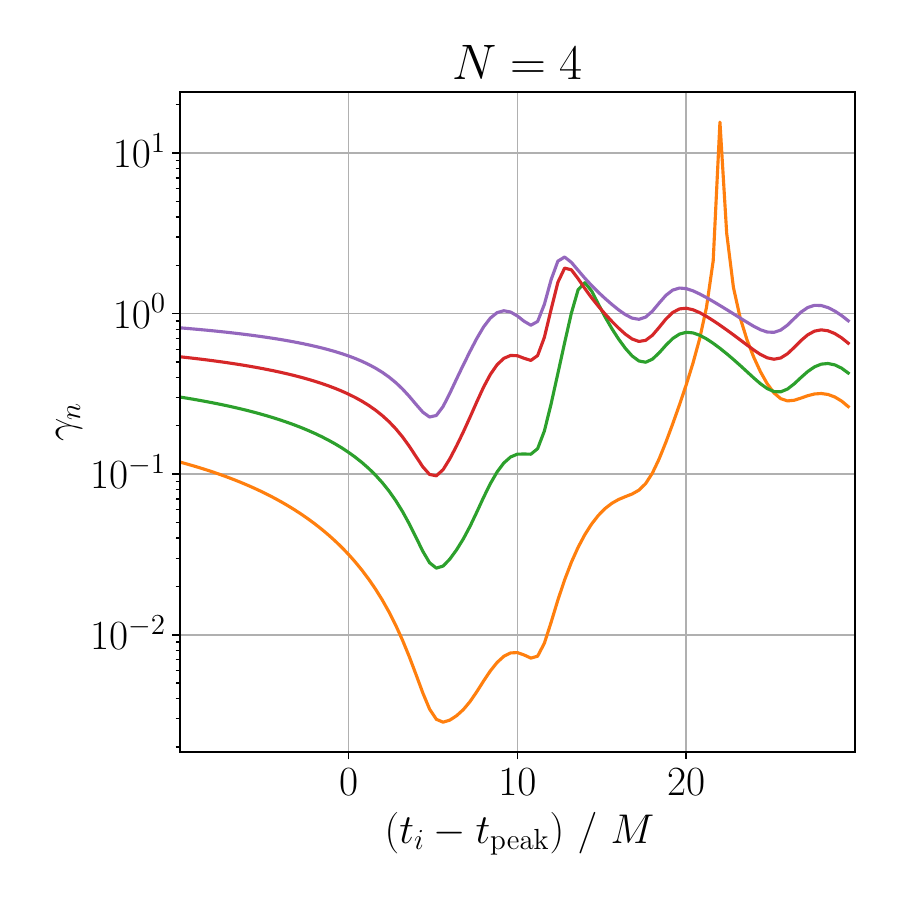}
    \caption{Mismatch $\M$ (first row), amplitude $A_n$ (second row), phase $\phi_n$ (third row), and rate of change $\gamma_n$ (fourth row) for the fit of SXS:BBH:1108 waveform $\Psi^\sxs - c$ by the fitting function $\psi^\fit_{[0,N]}$ (left column), and $\Psi^\sxs - c - \psi^\iter_{[0,0]}$ after the subtraction of the fundamental mode by the fitting function $\psi^\fit_{[1,N]}$ (right column).
    }
    \label{fig:1108_M_A_gamma_sub_const}
\end{figure}

Figures~\ref{fig:0158_M_A_gamma_sub_const}--\ref{fig:1108_M_A_gamma_sub_const} show the results of the fit of SXS:BBH:0158, SXS:BBH:0156, and SXS:BBH:1108, respectively, with $\Psi^\sxs - c$ (left column), and $\Psi^\sxs - c - \psi^\iter_{[0,0]}$ (right column) after the subtraction of the fundamental mode.
By comparing with the results for SXS:BBH:0158 and the ones for SXS:BBH:0305 in Figs.~\ref{fig:0305_M_A_gamma_sub_const} and \ref{fig:0305_M_A_gamma_sub_const0}, we can examine how the stability of the fit is changed for the case of high spin.
Specifically, it can be seen that the start time of the fit $t_i$ at which the mismatch value reaches the lower bound is generally later.
In particular, comparing the fit of $\Psi^\sxs - c$ with $N=7$, for SXS:BBH:0305 the mismatch is quite small and takes the minimum around $(t_i-t_\peak)/M=0$, whereas for SXS:BBH:0158, the mismatch has not yet fallen completely around $(t_i-t_\peak)/M=0$ but around $(t_i-t_\peak)/M=10$.
To make the mismatch for SXS:BBH:0158 take the minimum at the fit starting from the peak time, it would be necessary to superpose more overtones in the fitting function.
However, while the mismatch would be reduced, such higher overtones cannot be extracted in a stable manner but would end up with overfit.
As argued in the main text, the plateau of $A_n$ and $\phi_n$ is extended than that in SXS:BBH:0305.
Regarding the rate of change, for the fundamental mode (blue), $\gamma_0$ remain the order of $10^{-4}$ for both cases, whereas for the overtones, the region where $\gamma_1$ remains small is extended.
The left column of Fig.~\ref{fig:0158_M_A_gamma_sub_const} show the result of the fit after the subtraction of the fundamental mode, where we can observe a similar change from Fig.~\ref{fig:0305_M_A_gamma_sub_const} to \ref{fig:0305_M_A_gamma_sub_const0}.

On the other hand, from the results for the low spin cases depicted in Figs.~\ref{fig:0156_M_A_gamma_sub_const} and \ref{fig:1108_M_A_gamma_sub_const}, we see that that the trend is opposite to the case of the high spin of SXS:BBH:0158. 
For instance, the mismatch is already reduced and reaches the lower bound at $t_\peak$ with a fewer number of overtones. 
The plateau is short, and the fit of the low spin event is more challenging.

\section{Estimation of numerical errors}
\label{sec:errors}

In this appendix we explain how we estimate the numerical errors used in the main text. 
When using a SXS simulation, we adopt the highest resolution level and the extrapolation order $N_{\ext}=2$ as the base waveform $\Psi^{(0)}$.
To estimate the numerical errors, we compare the base waveform with the following two kinds of waveforms: $\Psi^{(1)}$ with the next-highest resolution level and $N_{\ext}=2$, and $\Psi^{(2)}$ with the highest resolution level and $N_{\ext}=3$.
However, we should bear in mind that the higher extrapolation order is not necessarily good especially for the analysis of the ringdown~\cite{Boyle:2019kee}, so the comparison with the waveform $\Psi^{(2)}$ is treated as a conservative estimation of the numerical errors.

To estimate the numerical errors between $\Psi^{(0)}$ and $\Psi^{(j)}$ with $j=1,2$, we closely follow the procedure described in Appendix~B in \cite{Boyle:2019kee}.
First, we truncate the first $500M$ of the waveform data to avoid the initial transients.
Then, we define the time $t_{\rm begin}$ and $t_{\rm end}$ as the earliest and latest time that is covered by the both waveforms of interest.  
We would like to compare the two waveforms in this time range, but we should note that in general a mismatch between two waveforms could originate merely from an overall coordinate rotation and/or an overall time shift.
Therefore, we choose an coordinate system, where the mismatch between the two waveforms is minimum.
First, we choose a $z$-axis so that the orbital angular momentum ${\bf L} = \mu {\bf r}\times {\bf v}$, where $\mu$ is the reduced mass and ${\bf r}$ and ${\bf v}$ are relative distance and velocity, lies along the $+z$ direction at a reference time $t=t_{\rm begin}+1000M$. 
We consider the waveforms observed from a particular direction and apply the Planck-taper window function given in Eq.~(7) in \cite{McKechan:2010kp} with $t_1=t_{\rm begin}$, $t_4=t_{\rm end}$, $t_2$ being the 10th zero-crossing time of $\Re(\Psi_{22})$ after $t = t_1$, and $t_3$ being $50M$ after the peak of the waveform.
We then calculate a mismatch between $\Psi^{(0)}(t,\varphi)$ and $\Psi^{(j)}(t+\delta t^{(j)},\varphi+\delta\varphi^{(j)})$, and optimize $\delta t^{(j)}$ and $\delta \varphi^{(j)}$ so that the mismatch is minimized.
Since the sampling points are different for the two waveforms in general, to evaluate the mismatch, we first interpolate the data, and then perform numerical integration.
After optimization, we adopt the difference $|\Psi^{(0)}-\Psi^{(j)}|$ as estimation of the numerical errors.

We perform the estimation of numerical errors for various directions of observation.
Instead of the latitude-longitude lattice, we adopt a spherical Fibonacci lattice, which is known to provide an almost isotropic distributions~\cite{Gonzalez2010}.  
In the spherical coordinates, the $k$-th direction $(k=-N,\cdots,N)$ is given by 
\be \varphi = \f{2\pi k}{\tau}, \quad \theta = \f{\pi}{2} - \arcsin\mk{\f{2k}{2N+1}} \ee
where $\tau=(1+\sqrt{5})/2$ is the golden ratio.  
We adopt $N=10$ throughout the present paper, i.e, we estimate the numerical errors for the observations from almost evenly distributed 21 directions.

After generating the spherical Fibonacci lattice, we implement the above process of the choice of the $z$-axis and the rotation of $\delta\varphi^{(j)}$ by multiplying the rotation matrices as follows.
Let ${\bf r}_k$ be the locations of spherical Fibonacci lattice on the unit sphere, and 
$R_a (\alpha)$ be the rotation matrix which rotates a vector by an angle $\alpha$ around the $a$-axis.
First, we define
\be {\bf r}^{(0)}_k = R_z(\varphi^{(0)}_L)R_y(\theta^{(0)}_L) {\bf r}_k, \ee
where $\theta^{(0)}_L$ and $\varphi^{(0)}_L$ be the polar and azimuthal angles of the orbital angular momentum ${\bf L}^{(0)}$, respectively.
${\bf r}^{(0)}_k$ yields the spherical Fibonacci lattice with the $z$-axis in the direction of the orbital angular momentum.
We use the polar and azimuthal angles of ${\bf r}^{(0)}_k$ to calculate the strain $\Psi^{(0)}$.
Second, we define
\be {\bf r}^{(j)}_k = R_z(\varphi^{(j)}_L)R_y(\theta^{(j)}_L)R_z(\delta \varphi^{(j)}) {\bf r}_k, \ee
where $\theta^{(j)}_L$ and $\varphi^{(j)}_L$ be the polar and azimuthal angles of the orbital angular momentum ${\bf L}^{(j)}$, respectively.
Compared to ${\bf r}^{(0)}_k$, ${\bf r}^{(j)}_k$ yields the lattice rotated by the angle $\delta \varphi^{(j)}$ around the direction of the orbital angular momentum.
We use the polar and azimuthal angles of ${\bf r}^{(j)}_k$ and the time shifted by $\delta t^{(j)}$ to calculate the strain $\Psi^{(j)}$.
We then optimize $\delta \varphi^{(j)}$ and $\delta t^{(j)}$ to minimize the mismatch between $\Psi^{(0)}$ and $\Psi^{(j)}$, after which we evaluate the difference $|\Psi^{(0)}-\Psi^{(j)}|$ as the estimation of the numerical errors.

Having said that, since for the four simulations we considered in the main text, the orbital angular momentum is along the $+z$ direction, we did not multiply the rotation matrices $R_z(\varphi^{(0)}_L)R_y(\theta^{(0)}_L)$ and $R_z(\varphi^{(j)}_L)R_y(\theta^{(j)}_L)$ as they are close to identity matrix.

In Figs.~\ref{fig:0305_wave_naive}, \ref{fig:0158_wave_naive}, \ref{fig:0156_wave_naive}, \ref{fig:1108_wave_naive}, green and magenta thin dotted curves represent the errors estimated by $|\Psi^{(0)}-\Psi^{(1)}|$ and $|\Psi^{(0)}-\Psi^{(2)}|$, respectively.
As expected, the errors estimated by using the higher-order extrapolation is larger than the ones estimated by using the next-highest resolution.
Note that the error may be small for observations from particular directions.
The envelope of the all error curves is shown as a gray curve, which we adopt as the conservative estimation of the numerical errors.

\bibliographystyle{JHEPmod}
\bibliography{ref}
\end{document}